%% file: main_note.tex
\documentclass{cernatsnote}
\usepackage[colorinlistoftodos]{todonotes}
\usepackage{placeins}
\usepackage{graphicx}
\usepackage{smartdiagram}
\usepackage{hyperref}
\usesmartdiagramlibrary{additions}
\usepackage[capitalise]{cleveref}
\usepackage{booktabs}
\usepackage{xspace,upgreek}
\usepackage{cernunits}
\usepackage{fcc_calo_common}
\usepackage{units_definitions}
\usepackage{tikz}
\usepackage{authblk}

\def\bibfiles{\pwd/volume.bib}	

\usepackage{listings}
\usetikzlibrary{angles,quotes}

\title{Calorimeters for the FCC-hh}
\input{\pwd/tex/authors.tex}
\documentlabel{CERN-FCC-PHYS-2019-0003}
\date{\today}

\begin{document}

\newcommand*{\sqrtsfcc}{\ensuremath{\sqrt{s}=\text{100 TeV}}}
\newcommand*{\sqrtslhc}{\ensuremath{\sqrt{s}=\text{14 TeV}}}
\newcommand*{\instlumi}{\ensuremath{\mathcal{L}=3~10^{35} \mathrm{\;cm^{-2} s^{-1}}}}
\newcommand*{\intlumifcc}{\ensuremath{\mathcal{L}=30\text{ ab}^{-1}}}
\newcommand*{\invab}{\ensuremath{\mathrm{ab}^{-1}}}

\maketitle

\begin{abstract}

The future proton-proton collider (FCC-hh) will deliver collisions at a center of mass energy up to \sqrtsfcc\ at an unprecedented instantaneous luminosity of \instlumi, resulting in extremely challenging radiation and luminosity conditions. By delivering an integrated luminosity of few tens of \invab, the FCC-hh will provide an unrivalled discovery potential for new physics. Requiring high sensitivity for resonant searches at masses up to tens of TeV imposes strong constraints on the design of the calorimeters. Resonant searches in final states containing jets, taus and electrons require both excellent energy resolution at multi-TeV energies as well as outstanding ability to resolve highly collimated decay products resulting from extreme boosts. In addition, the FCC-hh provides the unique opportunity to precisely measure the Higgs self-coupling in the di-photon and b-jets channel. Excellent photon and jet energy resolution at low energies as well as excellent angular resolution for pion background rejection are required in this challenging environment.
This report describes the calorimeter studies for a multi-purpose detector at the FCC-hh. The calorimeter active components consist of Liquid Argon (LAr), scintillating plastic tiles (Tile) and Monolithic Active Pixel Sensors (MAPS) technologies. The technological choices, design considerations and achieved performances in full \textsc{Geant4} simulations are discussed and presented. The simulation studies are focused on the evaluation of the concepts. Standalone studies under laboratory conditions as well as first tests in realistic FCC-hh environment, including radiation hardness and pileup rejection capabilities, by making use of fast signals and high granularity have been performed. 
This report also includes the technical description of calorimeter components and possible R\&D directions to be undertaken.
These studies have been performed within the context of the preparation of the FCC conceptual design reports (CDRs).

\end{abstract}

\clearpage

\begingroup
\color{black}
\tableofcontents
\endgroup

\newpage

\section{Introduction}
\graphicspath{{calorimetry/img/intro/}{img/intro/}}
\input{\pwd/tex/introduction/intro.tex}

\input{\pwd/tex/introduction/requirements_general.tex}

\input{\pwd/tex/introduction/requirements_ecal.tex}

\input{\pwd/tex/introduction/requirements_hcal.tex}

\clearpage
\section{Layout of the Calorimeter System}
\label{sec:layout}
\graphicspath{{calorimetry/img/layout/}{img/layout/}}
\input{\pwd/tex/layout/overview.tex}

\input{\pwd/tex/layout/electromagnetic.tex}

\clearpage
\input{\pwd/tex/layout/hadronic.tex}

\clearpage
\section{Software Implementation}
\label{sec:implementation}
\graphicspath{{calorimetry/img/software/}{img/software/}}
\input{\pwd/tex/software/general.tex}
\input{\pwd/tex/software/reconstruction.tex}
\input{\pwd/tex/software/noise.tex}

\input{\pwd/tex/software/dnn.tex}

\clearpage
\section{Performance}
\label{sec:performance}
\graphicspath{{calorimetry/img/performance/}{img/performance/}}
\input{\pwd/tex/performance/egamma.tex}
\clearpage
\input{\pwd/tex/performance/hadronic.tex}

\clearpage
\input{\pwd/tex/performance/id.tex}
\clearpage
\input{\pwd/tex/performance/jets.tex}
\clearpage
\section{Alternative Technology for the EM Barrel Calorimeter}
\label{sec:alternative}
\graphicspath{{calorimetry/img/alternative/}{img/alternative/}}
\input{\pwd/tex/alternative/layout.tex}

\input{\pwd/tex/alternative/software.tex}
\input{\pwd/tex/alternative/performance.tex}

\clearpage
\section{Summary}

\input{\pwd/tex/summary/summary.tex}

%% \clearpage
%% \section{Appendix}
%% \label{sec:appendix}
%% \graphicspath{{calorimetry/img/appendix/}{img/apppendix/}}
%% \input{\pwd/tex/appendix/appendix.tex}

% -- Add volume bibliography and part specific bibliographies
\clearpage
\bibliographystyle{elsarticle-num}
\addcontentsline{toc}{section}{Bibliography}
\bibliography{\bibfiles}

\end{document}

%% file: tex/authors.tex
\author[${}^{1}$]{M.~Aleksa}
\author[${}^{2}$]{P.~Allport}
\author[${}^{2}$]{R.~Bosley}
\author[${}^{1,3}$]{J.~Faltova}
\author[${}^{4}$]{J.~Gentil} %% double checked with Ricardo
\author[${}^{4,5}$]{R.~Goncalo} %% double checked with Ricardo
\author[${}^{1}$]{C.~Helsens}
\author[${}^{1}$]{A.~Henriques}
\author[${}^{6}$]{ A.~Karyukhin} %% double checked with Oleg
\author[${}^{1}$]{J.~Kieseler}
\author[${}^{1}$]{C.~Neub\"user}
\author[${}^{1,4}$]{H.~F.~Pais Da Silva} %% contacted, and updated affiliation
\author[${}^{2}$]{T.~Price}
\author[${}^{7}$]{J.~Schliwinski}
\author[${}^{1}$]{M.~Selvaggi}
\author[${}^{1,8}$]{O.~Solovyanov} %% double checked with Oleg
\author[${}^{9}$]{R.~Stein}
\author[${}^{6}$]{N.~Topiline} %% double checked with Oleg
\author[${}^{2}$]{N.~K.~Watson}
\author[${}^{2}$]{A.~Winter}
\author[${}^{1}$]{A.~Zaborowska}

\affil[$$]{\footnotesize}
\affil[${}^{1}$]{\footnotesize CERN, Switzerland}
\affil[${}^{2}$]{University of Birmingham, United Kingdom}
\affil[${}^{3}$]{Charles University, Prague, Czech Republic}
\affil[${}^{4}$]{LIP Laboratorio de Instrumentacao e Fisica Experimental de Particulas, Portugal}
\affil[${}^{5}$]{University of Coimbra, Portugal}
\affil[${}^{6}$]{JINR Dubna, Russia}
\affil[${}^{7}$]{Humboldt University of Berlin, Germany}
\affil[${}^{8}$]{High Energy Physics of the National Research Centre Kurchatov Institute, Protvino, Russia}
\affil[${}^{9}$]{University of York, England}

%% file: tex/introduction/intro.tex
\subsection{The FCC-hh Detector}
The Future Circular Collider (FCC) is the ambitious project of an accelerator complex in the CERN area for the after LHC era. An electron-positron collider (FCC-ee) is considered as a possible first step to measure precisely the Higgs properties. The main drive on the complex tunnel and infrastructure is set by a 100\,TeV hadron circular collider (FCC-hh). Such center of mass energy can be achieved by means of a 100\,km tunnel and 16\,T bending dipole magnets. The FCC-hh will deliver a peak luminosity of \instlumi\ in its ultimate phase. This will result in O(20) \invab\ of integrated luminosity per experiment. Such high luminosity defines stringent requirements on the radiation hardness of the detector, in particular in the forward region at small angular distances from the beampipe.

The FCC-hh machine allows for a direct exploration of massive particles up to 40\,TeV~\cite{Golling:2016gvc}, improving by approximately one order of magnitude the LHC sensitivity for discovering heavy resonant states. In addition, during its lifetime the FCC-hh is expected to produce trillions of top quarks and tens of billions of Higgs bosons allowing for a rich standard model precision program~\cite{Contino:2016spe}. Most importantly, a 100\,TeV machine will be the only machine allowing for a percent level measurement of the Higgs self-coupling~\cite{Contino:2016spe}. It is therefore essential to design detectors that provide excellent energy resolution in a wide range of energy.

An experimental apparatus that operates within the FCC-hh must therefore operate optimally on two main fronts. Physics at the EW scale, in particular the Higgs, will produce objects in the detector with momenta in the range $\pt=20-100$\,GeV. The LHC detectors were built to produce an optimal response in such an energy range. In addition, a new regime, at the energy frontier, will be characterised by the energy scale of decay products originating from high mass resonances (potentially as high as $m_X=50$\,TeV). An FCC-hh detector must therefore be capable to reconstruct leptons, jets, and potentially $t$ and $H$, $W/Z$ bosons with momenta as large as $\pt=20$\,TeV. Thus the detector must provide accurate measurements not only in the high energy limit but also in the low energy regime.

\begin{figure}[h]
  \centering
	\includegraphics[width=\textwidth]{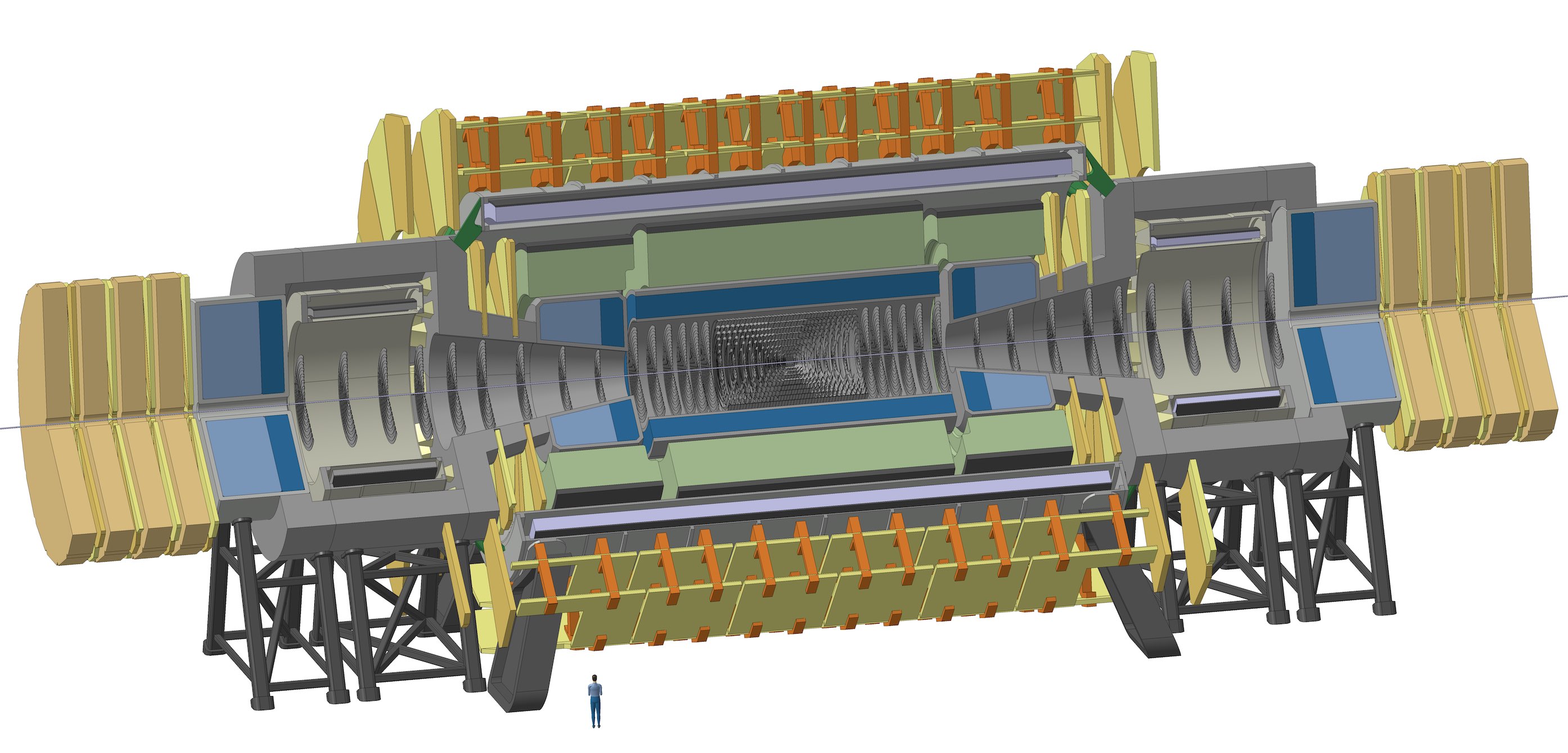}
	\caption{The layout of the FCC-hh reference detector~\cite{Abada2019}.}
    \label{fig:intro:refDet}
\end{figure}

Figure~\ref{fig:intro:refDet} shows the layout of the FCC-hh reference detector. This detector concept does not represent the final design, but rather respresents a concrete example that suits the performance and physics requirements and allows to identify areas where dedicated further R\&D efforts are needed. The detector has a diameter of 20\,m and a length of 50\,m, comparable to the dimensions of the ATLAS detector but much more heavy. The central detector with coverage of $|\eta| < 2.5$ houses the tracking, electromagnetic calorimetry and hadron calorimetry inside a 4\,T solenoid with a free bore diameter of 10\,m. In order to reach the required performance for $2.5 < |\eta| < 6$, the forward parts of the detector are displaced by 10\,m from the interaction point along the beam axis. Two forward magnet coils with an inner bore of 5\,m provide the required bending power. These forward magnets are also solenoids with a 4\,T field, providing a total solenoid volume of 32\,m length for high precision momentum spectroscopy up to rapidity values of $|\eta| \approx 4$ and tracking up to $|\eta| \approx 6$. The reference detector does not assume any shielding of the magnetic field. The tracker cavity has a radius of 1.7\,m with the outermost layer at around 1.6\,m from the beam in the central and the forward regions, providing the full spectrometer arm up to $|\eta| = 3$. The Electromagnetic CALorimeter (ECAL) has a thickness of around 30 radiation lengths ($X_{0}$) and provides together with the Hadron CALorimeter (HCAL) an overall calorimeter thickness of more than 10.5 nuclear interaction lengths ($\lambda$), to ensure 98\,\% containment of high energy hadron showers and to limit punch-through to the muon system. The ECAL is based on Liquid Argon (LAr) due to its intrinsic radiation hardness. The barrel HCAL is a scintillating tile calorimeter with steel and Pb absorbers, that uses wavelength shifting (WLS) fibres and Silicon Photomultipliers (SiPMs) for the readout. It is divided into a central barrel and two extended barrels. The HCALs for the endcap and forward regions are also based on LAr. The requirement of calorimetry acceptance up to $|\eta| \approx 6$ translates into an inner active radius of only 8\,cm at a z-distance of 16.6\,m from the IP. 
The transverse and longitudinal segmentation of both the electromagnetic and hadronic calorimeters is $\sim4$ times finer than the present ATLAS calorimeters.
%The ECAL is specified to have an energy resolution around 10\%/ $\sqrt{E}$ and the hadron calorimetry around 50\%/ $\sqrt{E}$ for single particles. The central beam pipe is assumed to be cylindrical with a radius of 20\,mm up to a z-distance of 8\,m. Between this point and the forward calorimeters there will be a conical beam pipe or straight beam pipe sections around a rapidity of $|\eta| = 6.7$, projecting towards the IP, which represents an angle of 2.5\,mrad.

\subsection{Calorimetry at the FCC-hh}

Calorimeters will play a crucial role to exploit the full physics potential of the FCC-hh. Proton collisions at unprecedented centre-of-mass energies will produce particles with energies up to the multi-TeV range. Due to the statistical nature of the energy measurement in calorimeters, the energy resolution improves with energy, which makes them the ideal candidates for the measurement of particles with energies in the multi-TeV range. In addition, calorimeters produce fast signals and can be read out at bunch-crossing frequency which makes them an ideal choice for a hardware trigger. In future collider experiments, the final 4-momentum measurement can be obtained by combining several sub-detectors. In particular, the tracker and the calorimeter measurements can be combined by using the \emph{particle flow} technique~\cite{Sirunyan:2017ulk}, that is already in use at the LHC experiments. This technique requires highly granular calorimeters such as the HGCAL, planned for the CMS upgrade~\cite{Collaboration:2293646} or  CALICE calorimeters~\cite{Repond:2018flg} for future linear collider experiments. Furthermore, calorimeters are the basis for the missing $E_\mathrm{T}$ measurement and play an important role on particle identification, background and pile-up rejection. The ability to resolve collimated decay products from highly boosted objects sets constraints on the position resolution pointing resolution, which will also help to identify the primary vertex for neutral particles such as photons. Last but not least, a time measurement with a resolution at the level of 30\,ps could be used to reject energy deposits from particles of many other collision vertices at the same bunch crossing (in-time pile-up).   
All these properties call for a high resolution, and finely granular calorimeter systems with adequate time-measurement capabilities whereas the harsh radiation environment limits the choice of available technologies.

The reference calorimeter system for the FCC-hh detector is composed of sampling calorimeters using liquid argon and scintillating tiles as active media. Liquid argon as an intrinsically radiation hard noble liquid is suitable for the FCC-hh radiation environment. The design of the scintillating tile calorimeter for the central hadronic calorimeter, allows for a very fine transverse segmentation and a good intrinsic energy resolution. A novel design of highly segmented liquid argon and tile calorimeters will be presented. Additionally, an alternative option with a digital electromagnetic calorimeter will be discussed.

%The calorimeter chapter is structured as follows: first an introduction is given in which physics benchmarks are presented to derive the requirements. Then a technical description of the calorimeter is given followed by its implementation in the FCC software. Finally, the simulated performance of the calorimeter system is presented and discussed in detail.

%% file: tex/introduction/requirements_general.tex
\subsection{Calorimeter Requirements}
\label{sec:intro:requirements}

\subsubsection{Benchmark Physics Channels and General Requirement Considerations}

\begin{figure}[h]
  \centering
  \begin{subfigure}[b]{0.49\textwidth}
    \includegraphics[width=\textwidth]{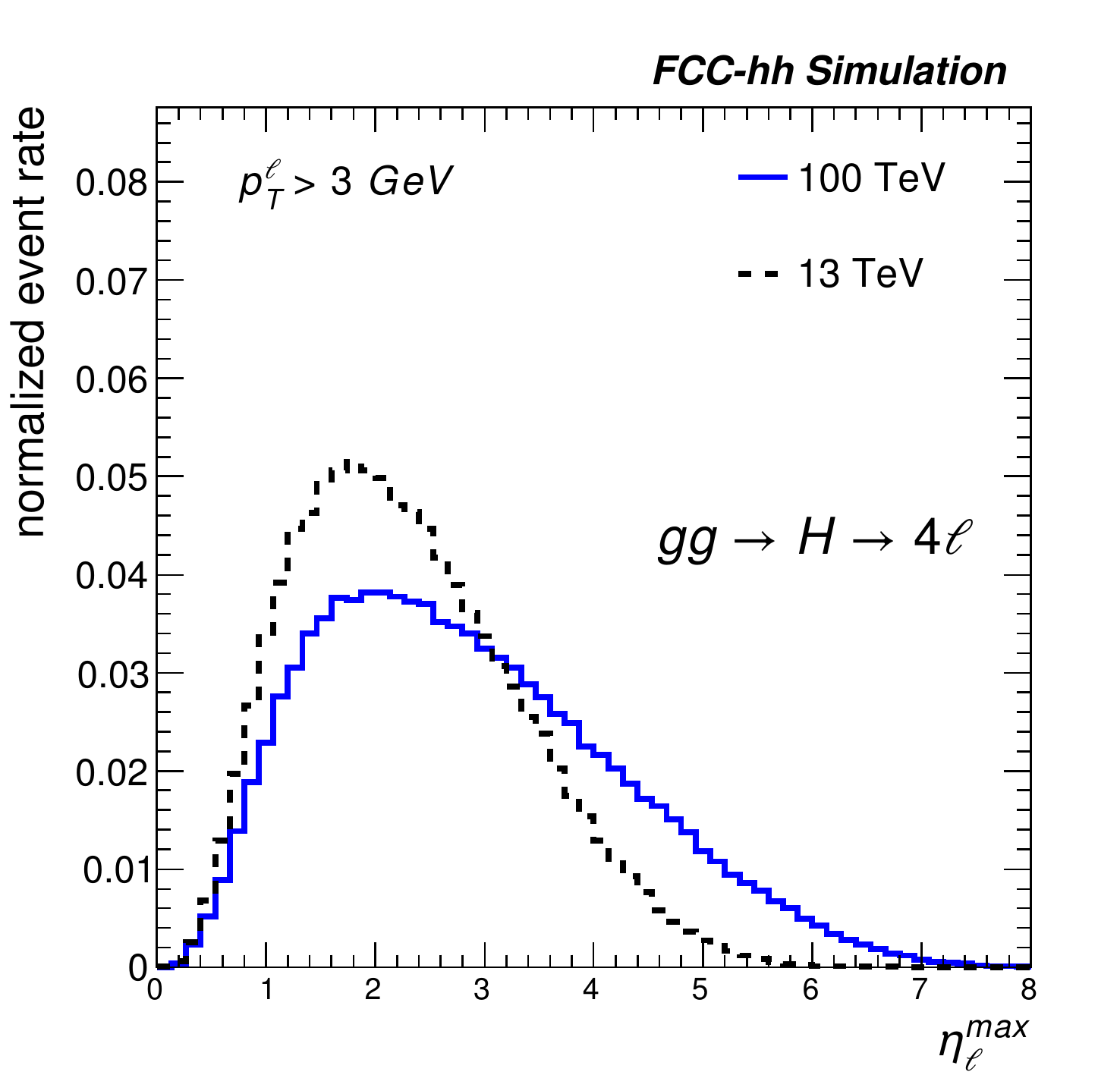}\caption{}
  \end{subfigure}
  \begin{subfigure}[b]{0.49\textwidth}
    \includegraphics[width=\textwidth]{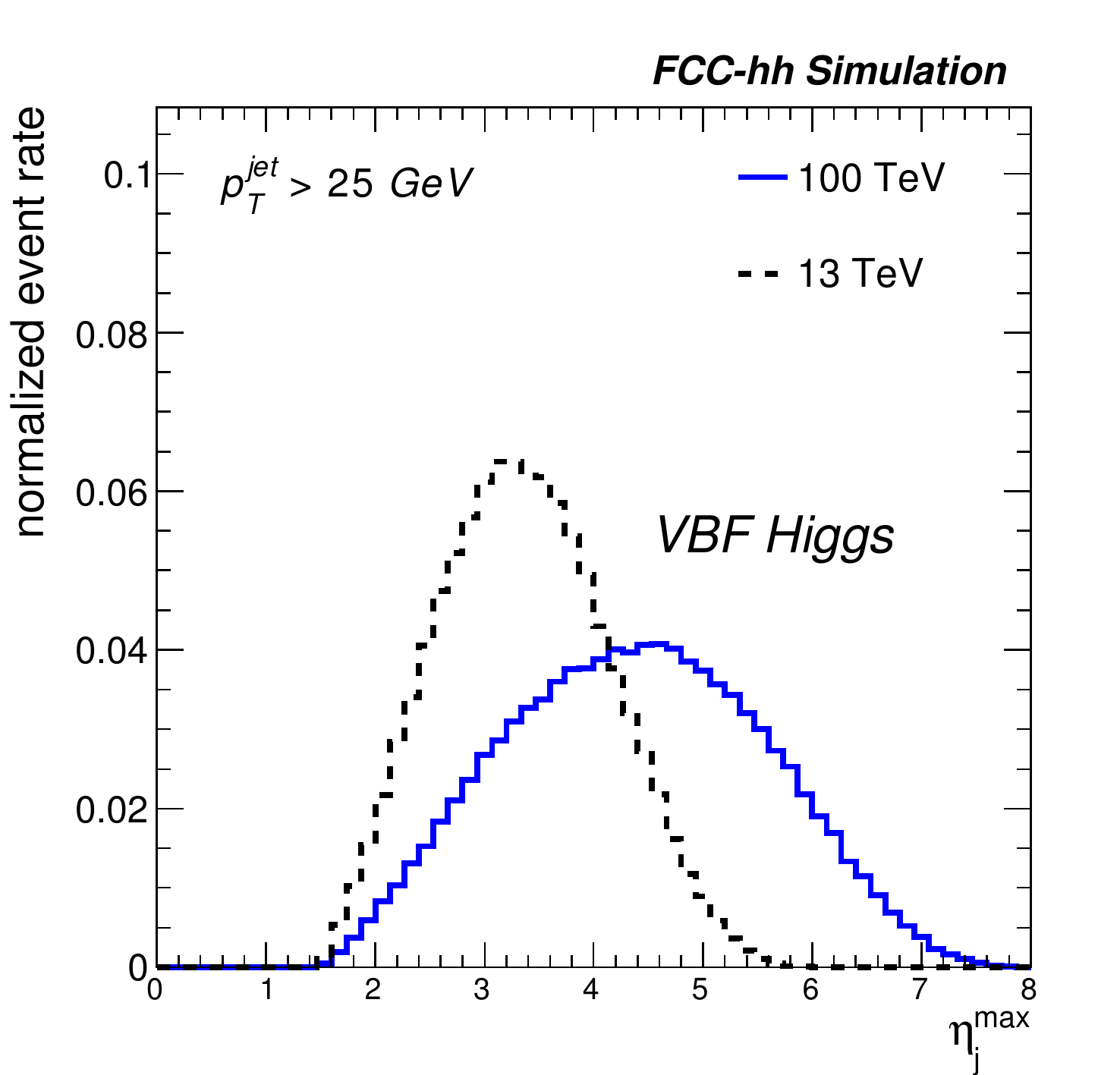}\caption{}
  \end{subfigure}
  \caption{highest lepton pseudo-rapidity for gluon-gluon fusion Higgs decaying to 4 leptons (a) and maximum jet pseudo-rapidity for vector-boson fusion Higgs (b)}
  \label{figure:intro:higgsgeneration}
\end{figure}

Calorimeters for the next generation of high energy machines like FCC, will have to operate efficiently in a very broad energy range. Final states produced at a given characteristic energy scale Q, will be produced on average at higher rapidities at \sqrtsfcc\ compared to \sqrtslhc. As an illustration Fig.~\ref{figure:intro:higgsgeneration}~(a) shows the highest lepton pseudo-rapidity $\eta^{max}$, for a gluon-gluon fusion produced Higgs decaying into four leptons for both 13 and 100\,TeV p-p collisions.
To reach 90\,\% fiducial acceptance in this channel, a detector coverage of $|\eta|<3.8(4.8)$ is needed for 13(100)\,TeV respectively. 
Fig.~\ref{figure:intro:higgsgeneration}~(b), shows the maximum jet pseudo-rapidity $|\eta^{max}_{j}|$, for a Vector-Boson-Fusion (VBF) produced Higgs. 
To reach 90\,\% fiducial acceptance for the forward jets, the pseudo-rapidity acceptance
will need to be extended from $|\eta|=4.5$ to $|\eta|=6$ . This increase has strong consequences on the detector design, as the very shallow polar angle of 0.28\,$^\circ$ for $|\eta|=6$ implies to have the calorimeters very far out of the interaction region and/or very close to the beam pipe. As an example, if the forward calorimetry is located at 16.5\,m in $z$, the calorimeter system must have an inner radius of 8.2\,cm in order to comply with the $|\eta|=6$ acceptance requirement. Needless to say, in the calorimeter endcaps the radiation levels will be extremely high, e.g. the 1 MeV neutron equivalent fluence will be $\approx 2 \; 10^{16} \mathrm{cm}^{-2}$ implying that radiation hardness will be a key requirement for such sub-detectors.

\begin{figure}[h]
  \centering
  \begin{subfigure}[b]{0.49\textwidth}
      \includegraphics[width=\textwidth]{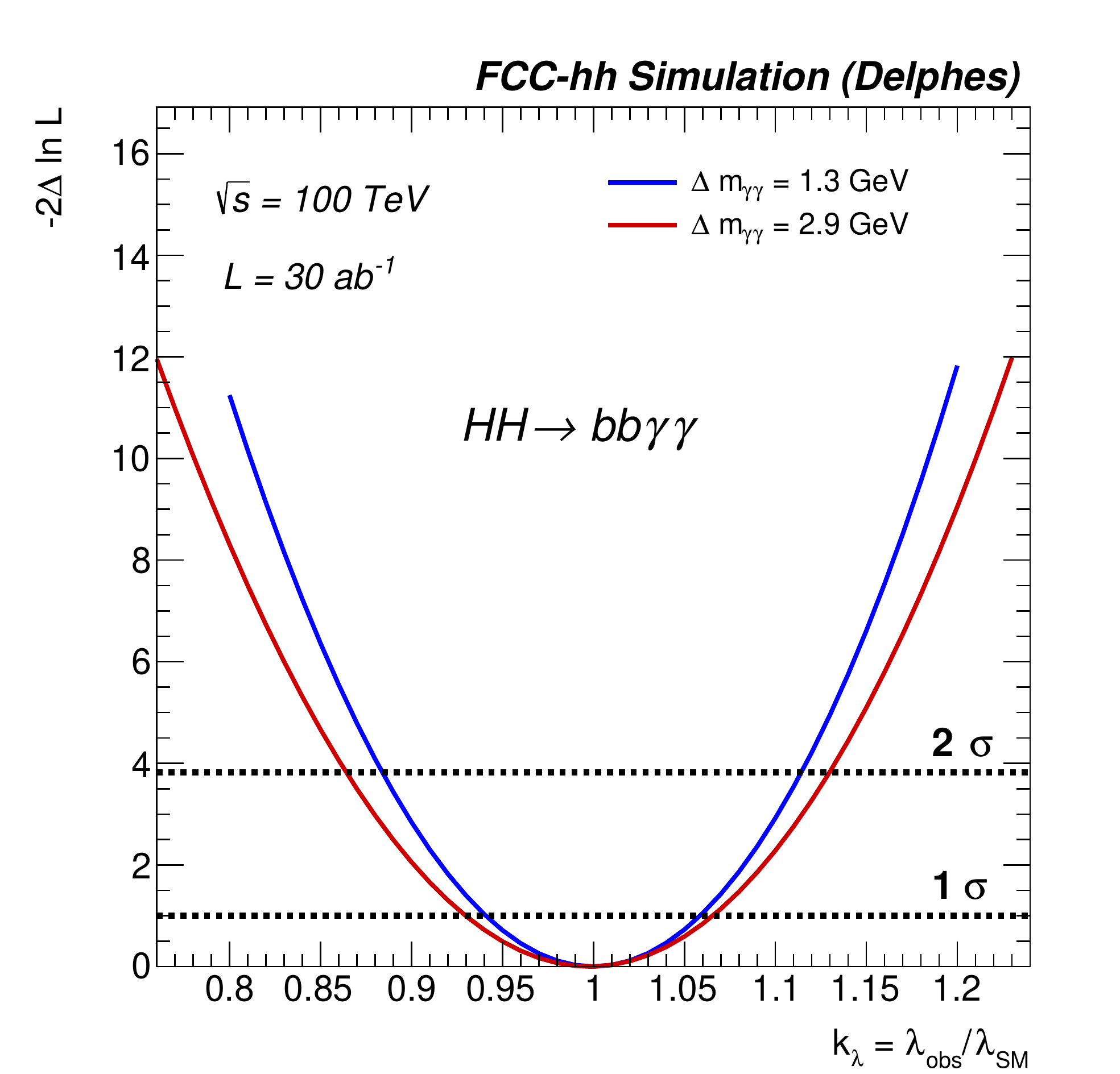}\caption{}
  \end{subfigure}
  \begin{subfigure}[b]{0.49\textwidth}
      \includegraphics[width=\textwidth]{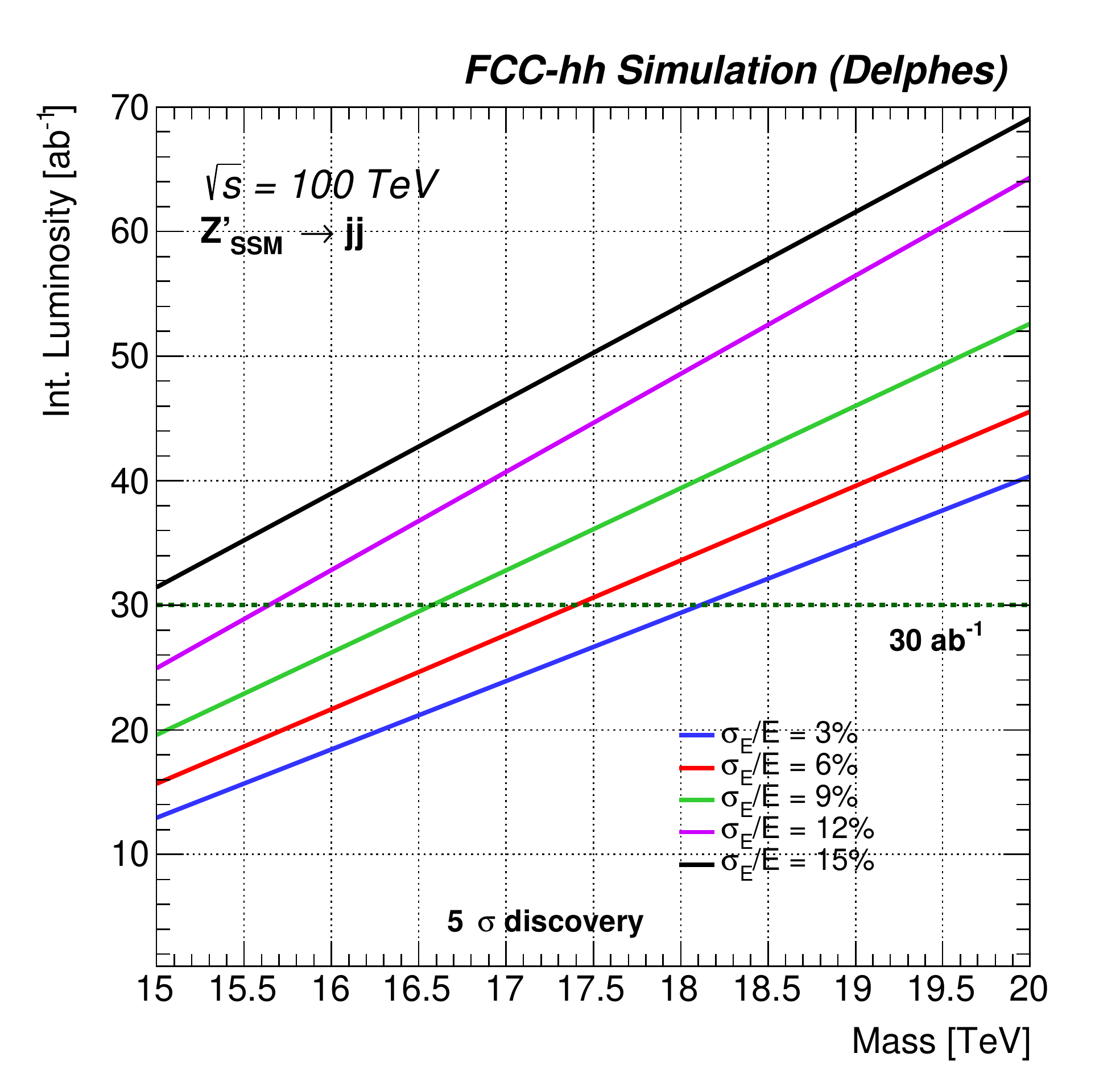}\caption{}
  \end{subfigure}
  \caption{Precision on the Higgs self coupling measurement in the bb$\gamma\gamma$ channel (a), di-jet mass reach in the case of a narrow resonance (b).}
  \label{figure:intro:dijet}
\end{figure}

FCC-hh is possibly the only machine that can allow for a  few percent level precision on the Higgs self coupling~\cite{L.Borgonovi:2642471}.
Since this process is very rare even at 100\,TeV, the full integrated luminosity and an excellent calorimetry will be needed to achieve the few percent accuracy. One of the most promising channels for double Higgs production is $\mathrm{HH \rightarrow bb\gamma\gamma}$ - which heavily relies on the precise measurement of the photon energy and position with electromagnetic (EM) calorimeter - and the two b-quarks. Figure~\ref{figure:intro:dijet}~(a) shows the precision at which the Higgs self coupling can be measured for Higgs mass resolutions of 1.3 and 2.9\,GeV respectively. At the one sigma level, the error on the self coupling increases from 5 to 6\,\%, thus having an excellent di-photon mass resolution is absolutely essential.

For new high mass particles that would eventually decay to high energetic objects in the central part of detector the requirements are different. The energy resolution of the calorimeter can be parameterised according to
\begin{equation}
\frac{\sigma_{E}}{E} \approx \frac{a}{\sqrt{E}} \oplus \frac{b}{E} \oplus c
\label{equation:intro:eneResolution}
\end{equation}

where $a$ is the stochastic term\footnote{The stochastic term is also called sampling term in sampling calorimeters.} due to shower fluctuations and sampling, $b$ is called the noise term due to electronic noise and pile-up and $c$ stands for the constant term due to various effects including shower leakage, construction non-uniformities and cell to cell calibration variations and is the dominant term for highly energetic objects.

It has been shown already~\cite{Carli:2016iuf} that with a total interaction length of about 11\,$\lambda$, hadronic showers are sufficiently contained to reach the desired 3\,\% constant term. 
As seen in  Fig.~\ref{figure:intro:dijet}~(b), if the mass resolution degrades, the discovery reach of a narrow resonance decaying to jets is strongly reduced, thus keeping the constant term at the 3\,\% level is important. 
Additionally, the calorimeter response has to be linear at the per-cent level over many orders of magnitude to extrapolate the absolute energy calibration with known resonances (e.g. Z or Higgs boson decays) to the multi-TeV range. This is important to limit systematic uncertainties on measured masses of high-mass resonances. 
In addition to good energy resolution and linearity, high granularity is relevant for efficiently reconstructing the high energetic unstable particles. For instance, decay products from a high $p_T$ particle which decays are collimated, with a typical angular distance $\Delta R \approx \frac{2m}{p_T}$. Thus, in order to have the ability to disentangle the sub-structure inside such boosted objects, the granularity of the calorimeters (both, lateral and longitudinal) should be increased significantly with respect to the LHC experiments. 

The high pile-up environment finally necessitates robust pile-up rejection. While the tracker will be the key instrument to assign particles to the correct primary vertex and hence allow to reject tracks from pile-up vertices, the ability to connect tracks to the correct primary vertex without using any timing information decreases heavily for rapidity of $|\eta|\ge 3$. It will therefore be essential to integrate a time measurement into the tracker and also into the calorimeters. Experience from simulations at the HL-LHC show that a time resolution of $\cal{O}$(30\,ps) can reduce pile-up effectively by a factor 6 (assuming a time distribution of primary vertices of 180\,ns)~\cite{Atlas:2019qfx}. 

In summary, key ingredients to be taken into account for the design of the calorimeter system are:
\begin{itemize}
\item excellent resolution and linearity of the response at the per-cent level from few GeVs up to multi-TeV particles
\item acceptance up to $|\eta|\leq 6$ 
\item time measurement of showers of $\cal{O}$(30\,ps).
\item high longitudinal and lateral segmentation
\end{itemize}

%% file: tex/introduction/requirements_ecal.tex
\subsubsection{Requirements for Electromagnetic Calorimetry}
\label{sec:intro:ecal}

\noindent
\textbf{Energy resolution over the energy range 10-500\,GeV:}
An excellent energy resolution is necessary to achieve a mass resolution close to 1\,\% 
for $H\rightarrow\gamma\gamma$ and $H\rightarrow4e$ decays. This can be achieved only if 
the stochastic term of the electromagnetic energy resolution stays at a level of $~\sim 10\,\%\sqrt{\mathrm{GeV}}/\sqrt{E}$ and the noise term is kept under control. The constant term should be smaller than 1\,\% in order to have a better mass resolution than the intrinsic width of heavy $Z'$ that occur in many models.
The goal for the energy resolution in the region $|\eta|\le4$ is 
\begin{equation}\label{eq:em-resolution}
\frac{\sigma_E}{E}=\frac{10\,\%\sqrt{\mathrm{GeV}}}{\sqrt{E}}\oplus\frac{0.3\,\mathrm{GeV}}{E}\oplus0.7\,\%~,
\end{equation}
neglecting the effect of pile-up. The expected average number of pile-up interactions $\langle\mu\rangle=200$ and $\langle\mu\rangle=1000$ for the FCC-hh baseline and ultimate scenario, respectively, will lead to energy deposits from pile-up collisions on top of the hard scatter of interest. Due to the bipolar read-out of the calorimeters, in long bunch trains these energy deposits will cancel on average, however, due to fluctuations of the exact number of collisions in each bunch crossing and the statistical nature of their energy deposits this pile-up will lead to additional noise in the calorimeter, referred to as pile-up noise in the following. Without any pile-up rejection procedure, the noise term could increase at $\langle\mu\rangle=200$ and depending on the size of the cluster, by a factor 2 to 6. It is therefore obvious that the tracker and timing information will be needed to reduce the impact of pile-up. An expected azimuthal non-uniformity due to the detector geometry will contribute to a global constant term not larger than $\sim 0.7\,\%$.

\noindent
\textbf{Rapidity coverage:}
As previously mentioned, the very large increase in the centre-of-mass energy with respect to LHC leads to decay products at higher rapidity,
thus an acceptance of up to $|\eta|=6$ is needed. High acceptance will be beneficial for both, detecting rare processes like double 
Higgs production, but also for tagging vector boson fusion or scattering induced processes as well as single Higgs production 
where forward jets and event tagging capabilities are needed. 
At pseudorapidity $4<|\eta|<6$, the EM calorimeter will mainly be needed for jet reconstruction and forward jet tagging, missing transverse energy (\etmiss) measurement, 
and pile-up rejection. 

\noindent
\textbf{Dynamic range:} 
The dynamic range for each read-out cell is defined by the range between the lowest and the highest energy that the calorimeter should be able to measure. The lower limit is typically set to a value close to the electronics noise per cell (see Sec.~\ref{sec:software:noise:electronics}), since such a choice allows to measure the noise and its auto-correlation by the calorimeter read-out. On top of that, the possibility of measuring the energy deposits of minimum ionising particles (MIPs) per cell or at least per longitudinal layer will be very beneficial for the layer calibration.   
The upper limit should be set close to the expected energy deposit per cell of electrons or photons from heavy 
resonances such as $Z'$, $W'$, or Gravitons with masses up to 50\,TeV. 
Taking into account these considerations a dynamic range of $\sim 2$\,MeV to $\sim 100$\,GeV ($\sim 16$\,bits) per cell will be necessary, depending on the exact cell position.  Detailed simulation studies are required to understand the maximum energy deposit in one read-out cell or per layer.

\noindent
\textbf{High segmentation and granularity:}
High granularity in the calorimeters will be necessary for particle identification, background rejection, position measurement of showers, photon pointing, and the correct connection of tracks with calorimeter clusters, which is crucial for both, pile-up rejection and particle flow reconstruction techniques. 
Many of these aspects will be further developed in the performance chapter, Sec.~\ref{sec:performance:egamma}, while in this section we introduce some of the most important aspects.
At FCC, the boost of relatively light SM particles will be large, leading to very collimated decay products. The calorimeters need to be 
able to resolve and reconstruct such highly boosted objects. To resolve boosted objects a cell size of a fraction of a Moli\`ere radius of about $\sim 2$\,cm ($R_M=5.7$\,cm in the EM calorimeter proposed in Sec.~\ref{sec:layout:lar}) is probably optimal. Simulations showed that below a certain cell size the separation of partial showers doesn't improve anymore.
Efficient $\gamma/\pi^0$ separation will require a layer (so-called strip layer) with very fine segmentation at the beginning of the shower. In addition, a large number of longitudinal layers, producing 3D images of the shower together with sophisticated analysis techniques, will allow for an efficient particle identification.

\begin{figure}[h]
  \centering
  \includegraphics[width=0.8\textwidth]{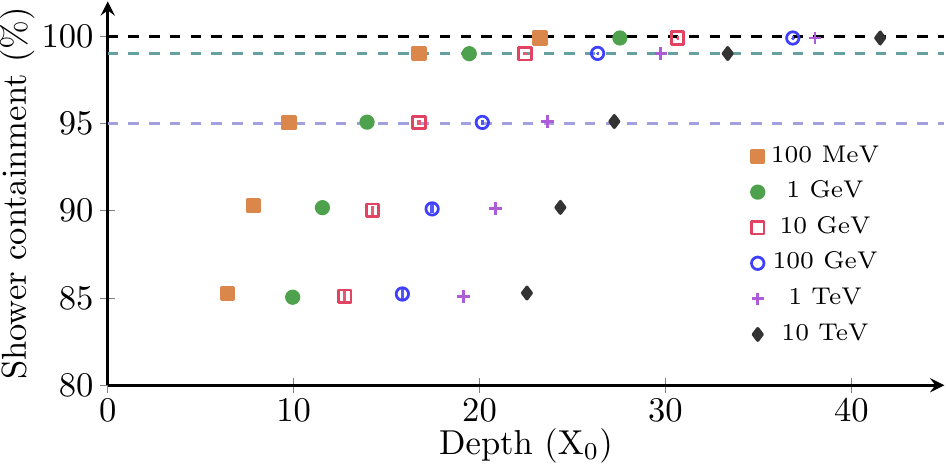}
  \caption{Dependence of the electron shower containment on the calorimeter depth expressed in the radiation lengths. The horizontal lines correspond to the shower containment of 95\%, 99\% and 100\% respectively.}
  \label{fig:intro:electronContainment}
\end{figure}

\noindent
\textbf{Total thickness of at least 30 radiation lengths at $\eta=0$:} The shower depths of electromagnetic showers increase with $\propto\ln(E)$. Longitudinal leakage of electromagnetic showers leads to a loss of resolution and also a deterioration of particle identification capabilities. Due to the higher particle energies with respect to LHC, the showers become longer and the calorimeter needs to be deeper to achieve $\cal{O}$(99\,\%) containment. Figure~\ref{fig:intro:electronContainment} shows that with a calorimeter depth of 30\,$\mathrm{X}_0$, a containment of $>99$\,\% can be achieved for particles $\le1$\,TeV.  

\noindent
\textbf{Others:} In addition to what is already listed above, excellent photon/jet, electron/jet and $\tau$/jet separation needs to be achieved.
This is extremely important as very rare signals that will decay to electron/photon/$\tau$ should be distinguishable from SM processes with jets in the final state. Moreover, some compressed SUSY models would benefit from identifying electrons with energies of few GeVs only, long lived or higly ionizing particles could give rise to peculiar signature in the detector which are relevant for the design.
Finally, it is also worth mentioning the need for an excellent angular resolution.

%% file: tex/introduction/requirements_hcal.tex
\subsubsection{Requirements for Hadronic Calorimetry}

\noindent
\textbf{Energy resolution over the energy range 20\,GeV - 10\,TeV:} For hadronic calorimetry, the energy resolution requirements are set by the required jet energy resolution for the different $\eta$ regions, 
\begin{eqnarray}
\frac{\sigma_{p_T}}{p_T} &=&\frac{50-60\,\%\sqrt{\mathrm{GeV}}}{\sqrt{p_T}} \oplus 3\,\%\quad \mathrm{for}\quad |\eta|\le 4~,\\ 
\frac{\sigma_{p_T}}{p_T} &=&\frac{100\,\%\sqrt{\mathrm{GeV}}}{\sqrt{p_T}} \oplus 10\,\%\quad\mathrm{for}\quad 4<|\eta|<6~.
\end{eqnarray}
Such resolutions have been found adequate for providing jet and di-jet mass reconstruction as well as \etmiss. A strong motivation for these performance goals is the discovery reach of heavy narrow resonances like a $Z'$, shown in Fig.~\ref{figure:intro:dijet}, and tested for different jet energy resolutions ranging from 3-20\,\%. At these very high energies, the constant term of the calorimeter resolution is dominating, thus has to be kept under control and $<3\,\%$.

\noindent
\textbf{Rapidity coverage:} As discussed for the electromagnetic calorimeters, and shown in the examples in Fig.~\ref{figure:intro:higgsgeneration}, a coverage for up to $\abseta\le6$ is essential to enable jet measurements and tagging. 

\noindent
\textbf{High segmentation and granularity:} The two main criteria to take into account for the granularity and segmentation 
are boosted high $p_T$ bosons or top quarks and pile-up mitigation. At FCC-hh, objects produced with momenta $p_T$ of up to 15\,TeV will have to be distinguishable from QCD jets. For example, the two quarks from a 5\,TeV $Z$ boson decay will be separated only by $\Delta R \sim 0.03$. Although the more granular electromagnetic calorimeter in front will help, the granularity of the hadronic calorimeter should also be of this order. 
For pile-up mitigation and particle-flow techniques, the same arguments apply as for the electromagnetic calorimeter.

\begin{figure}[htbp]
\begin{center}
  \begin{subfigure}[b]{0.45\textwidth}
	\includegraphics[width=\textwidth]{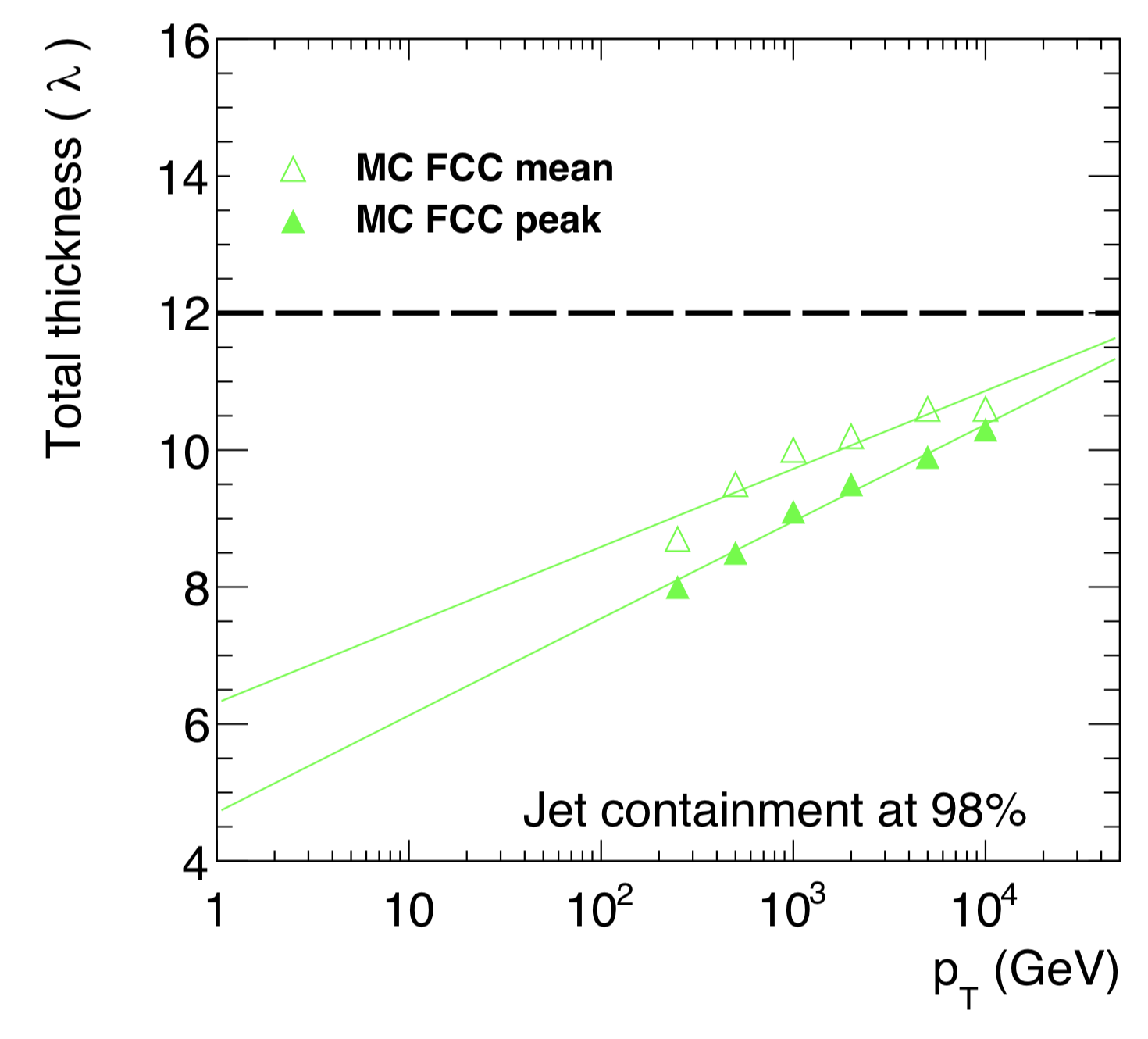}\caption{}
	\label{fig:intro:jetContainmentDepth}
  \end{subfigure}
  \begin{subfigure}[b]{0.48\textwidth}
	\includegraphics[width=\textwidth]{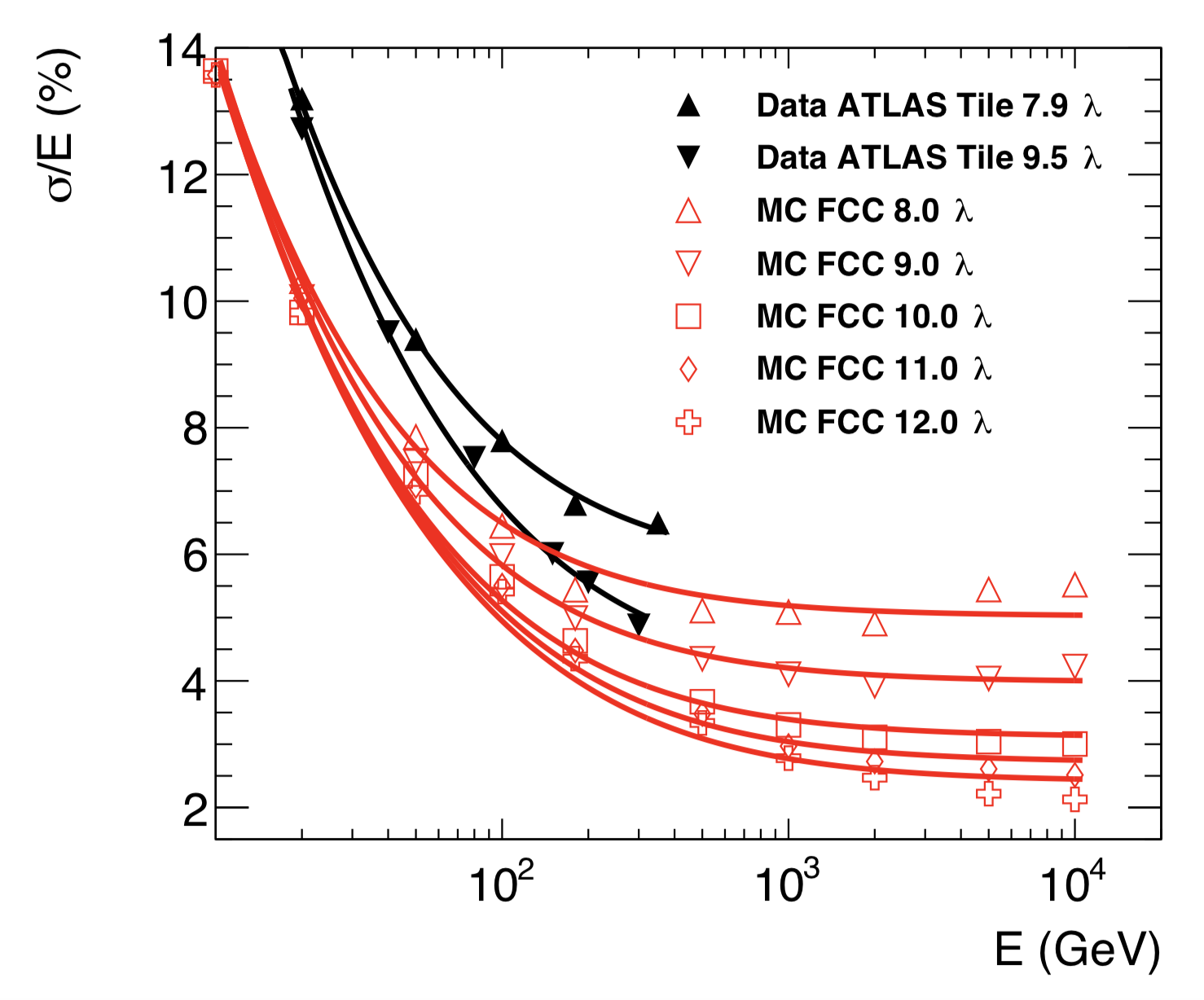}\caption{}
  	\label{fig:intro:jetContainmentReso}
\end{subfigure}
\caption{$\text{(a)}$ Calorimeter depth in interaction lengths $\lambda$ for 98\,\% jet containment as a function of jet $p_{\text{T}}$. Jets come from simulated $Z'\rightarrow q\bar{q}$ events. Two methods, so-called 'FCC mean' (mean of the distribution of the total thicknesses)  and 'FCC peak' (a mean of the Gaussian fit in the range of $\pm 2\sigma$ around the maximum), are used for the evaluation, more details can be found in~\cite{Carli:2016iuf}. $\text{(b)}$ 
Energy resolution for single pions as a function of the particle energy for the FCC-hh simulations compared to ATLAS data~\cite{Carli:2016iuf}.
}
\end{center}
\end{figure}

\noindent
\textbf{Total thickness of at least 11 interaction lengths at $\boldsymbol{\eta =0}$:} As shown in Fig.~\ref{fig:intro:jetContainmentDepth} and~\ref{fig:intro:jetContainmentReso}, a total (EM and hadronic) depth of about 11 interaction lengths ($\lambda$) is required for a sufficient shower containment compatible with a constant term of the energy resolution of about 3\,\%. Another important aspect is the leakage into the muon system, that needs to be kept at a minimum to avoid fake muon triggers. Additional studies are required to estimate this fake trigger rate, but the additional material of the solenoid between the hadronic calorimeter and the muon system will mostly eliminate this effect in the central region. 

\begin{figure}[h]
\begin{center}
  \begin{subfigure}[b]{0.49\textwidth}
	\includegraphics[width=\textwidth]{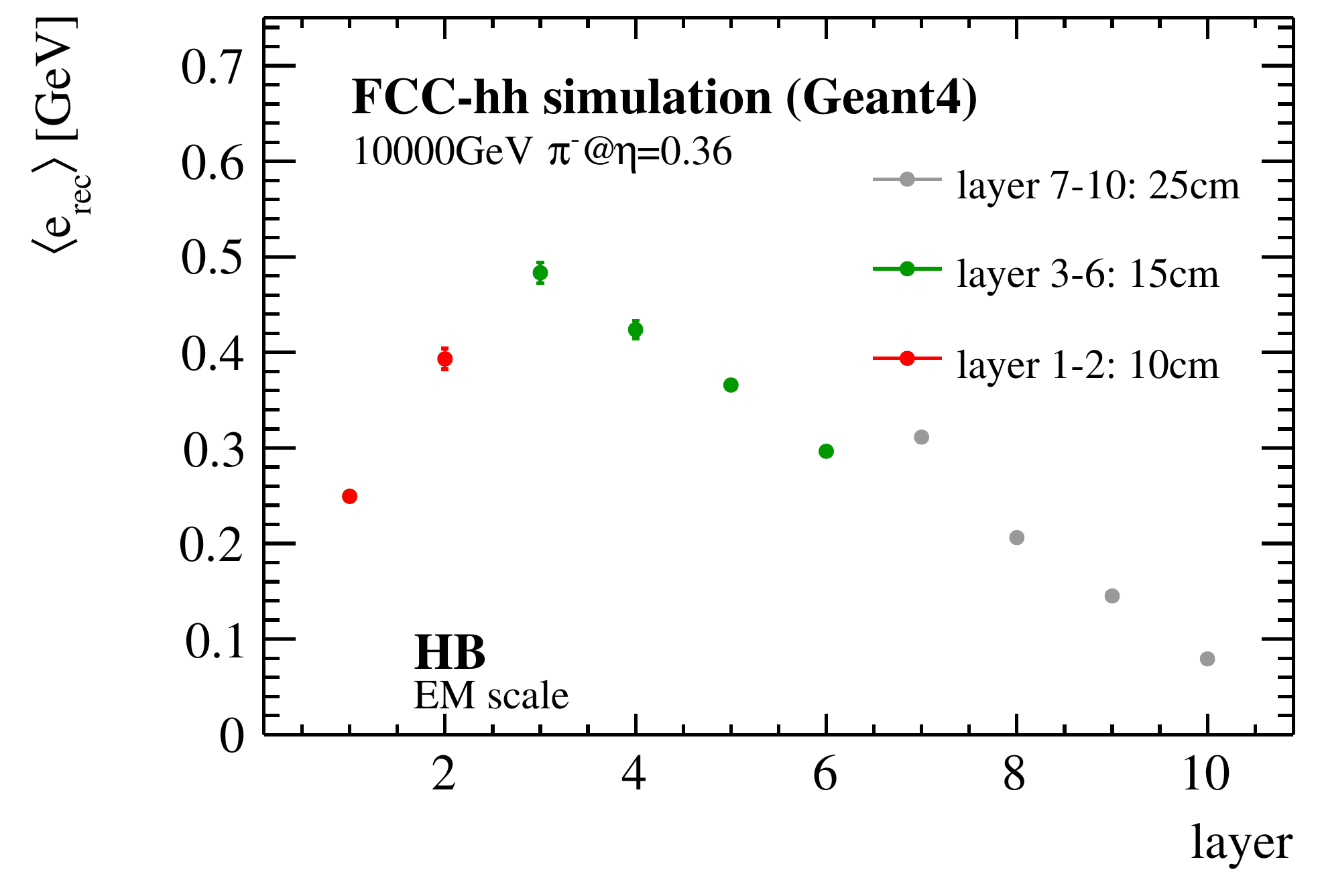}\caption{}
	\label{fig:intro:dynamicRangeProf}
  \end{subfigure}
  \begin{subfigure}[b]{0.49\textwidth}
	\includegraphics[width=\textwidth]{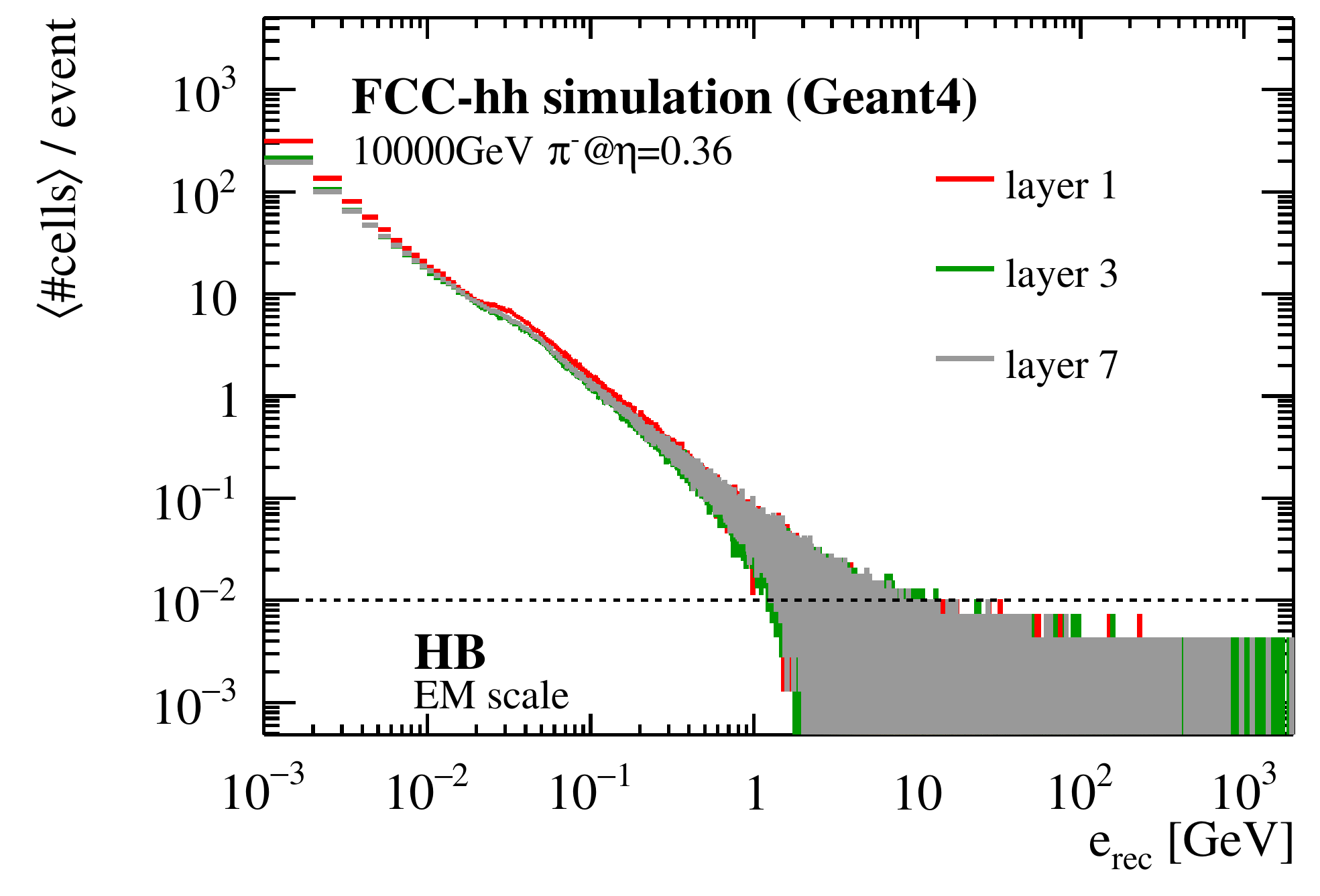}\caption{}
  \label{fig:intro:dynamicRange}
  \end{subfigure}
  \end{center}
\caption{$\text{(a)}$ Average energy deposited by 10\,TeV hadron in the hadronic Barrel (HB) calorimeter cells. $\text{(b)}$ The cell energy distributions per HB layer.}
\end{figure}

\noindent
\textbf{Dynamic range:} The necessary dynamic range has been studied for the Barrel HCAL (HB),  as proposed in Sec.~\ref{sec:layout:hcal}, in full simulations of the FCC-hh detector. It has been found that a 10\,TeV pion at $\eta=0.36$ deposits on average $100-500$\,MeV per cell in the hadronic calorimeter, with tails up to 2\,TeV. The variations with the calorimeter depth, in terms of layers, is shown in Fig.~\ref{fig:intro:dynamicRangeProf}, where the different colours correspond to the radial layer sizes of 10, 15, and 25\,cm. The radial layer size determines the cell size due to the perpendicular orientation of the scintillating tiles in the HB. The response to MIPs has been determined from a Landau-Gauss convoluted fit to the cell energy distribution, see details in Sec.~\ref{sec:layout:hcal:opti}, and features a most probable value (MPV) of 56\,MeV per HCAL cell in the Barrel region. These numbers are given for the HB in full granularity configuration, and prove the sensitivity to MIPs for an estimated electronic noise per cell of $\approx 10$\,MeV, see Sec.~\ref{sec:software:noise:electronics}. 
The required dynamic range of the HB cells is determined to 10\,MeV to 10\,GeV, for a minimum hit cell rate $>1$\,\%, see Fig.~\ref{fig:intro:dynamicRange}. However, this estimate from single MIPs and 10\,TeV pions only ensures the performance results shown for hadrons and jets, and will need further evaluation from future studies including e.g. rare decays. 
The dynamic range necessary for the performances presented in this document, corresponds to a 10\,bits readout. Similar studies have to be performed for the endcap and forward hadronic calorimeters, where the hit rate are higher, but cell sizes and sampling fractions are much smaller.

\noindent
\textbf{Others:}
Another important aspect for the hadronic calorimetry is the energy calibration (see Sec.~\ref{sec:layout:hcal:calibration}) and a good linearity of the response.
As already mentioned in Sec.~\ref{sec:intro:requirements}, good timing resolution is expected to help dealing with high pile-up.
Excellent jet identification and measurement in the full acceptance, very good di-jet mass resolution, forward jet tagging capabilities and \etmiss reconstruction will therefore be required and are mostly addressed in Sec.~\ref{sec:performance:jets}.

%% file: tex/layout/overview.tex
\subsection{Overview and Reminder of FCC-hh Detector Environment}

The layout of the calorimeter system of the reference FCC-hh detector has been driven by the following requirements:
\begin{itemize}
\item Use of technologies that withstand the high radiation environment.
\item Under these constraints best possible conventional calorimetry to ensure the best possible standalone energy measurement.
\item Use of technologies that can achieve timing resolution of $< 100$\,ps. 
\item High transversal and longitudinal granularity to optimise the combination with the tracker to enable particle flow techniques and use of 4D imaging for sophisticated particle ID and pile-up rejection algorithms.
\end{itemize}
  
In this section we will introduce the calorimeter system of the FCC-hh reference detector, which represents a possible implementation aimed at demonstrating that the performance requirements can be achieved. In Sec.~\ref{sec:performance} we will then show its performance and discuss the optimisations that have led to this design and which further improvements could be done. 

\begin{figure}[h]
  \centering
  \includegraphics[width=1\textwidth]{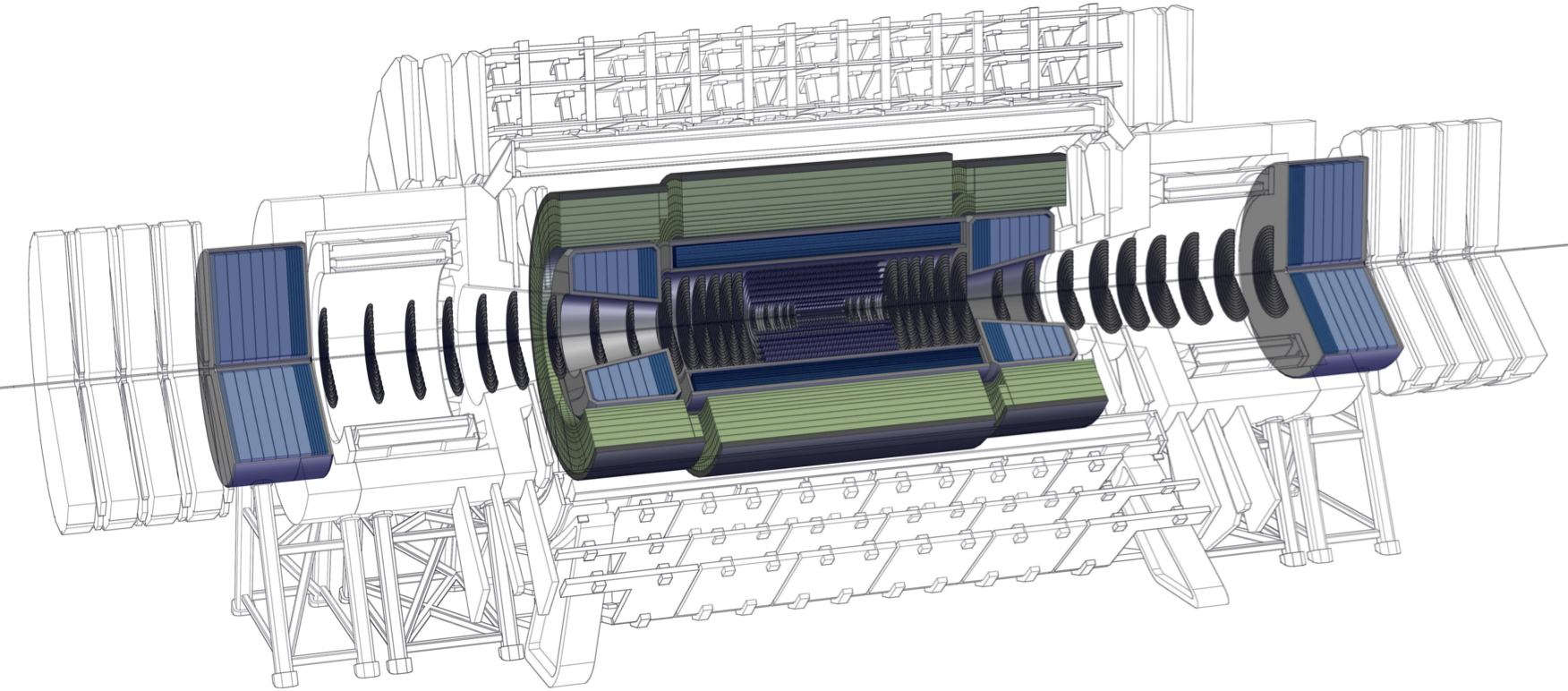}
  \caption{Calorimetry of the reference FCC-hh detector.}
  \label{fig:layout:calorimetry}
\end{figure}

\begin{figure}[h]
  \includegraphics[width=.9\textwidth]{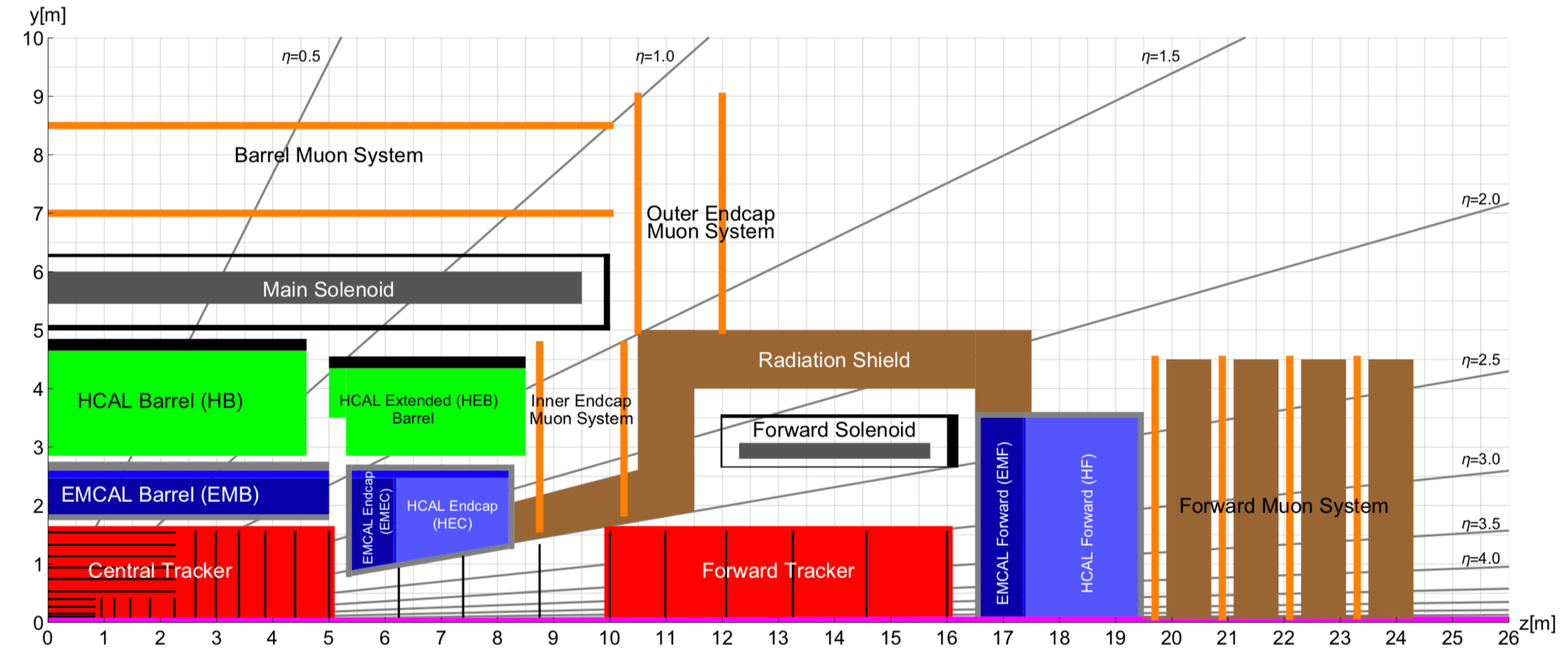}
  \caption{Longitudinal cross-section of the FCC-hh reference detector~\cite{Abada2019}.}
  \label{fig:layout:crossSec}
\end{figure}

\begin{figure}[h]
  \centering
  \begin{subfigure}[b]{0.49\textwidth}
    \includegraphics[width=\textwidth]{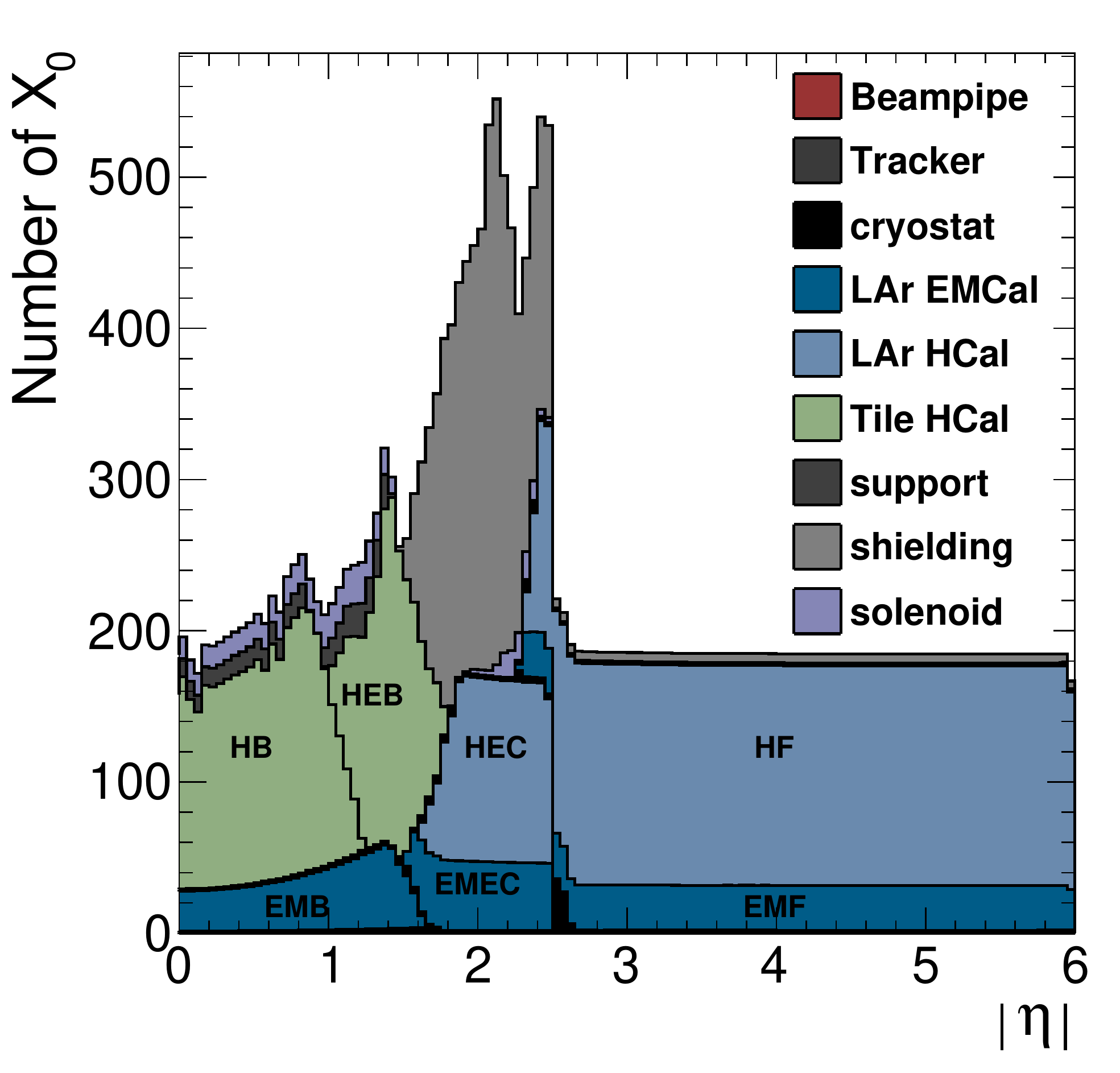}\caption{}
    \label{fig:layout:matScanX0}
  \end{subfigure}
  \begin{subfigure}[b]{0.49\textwidth}
    \includegraphics[width=\textwidth]{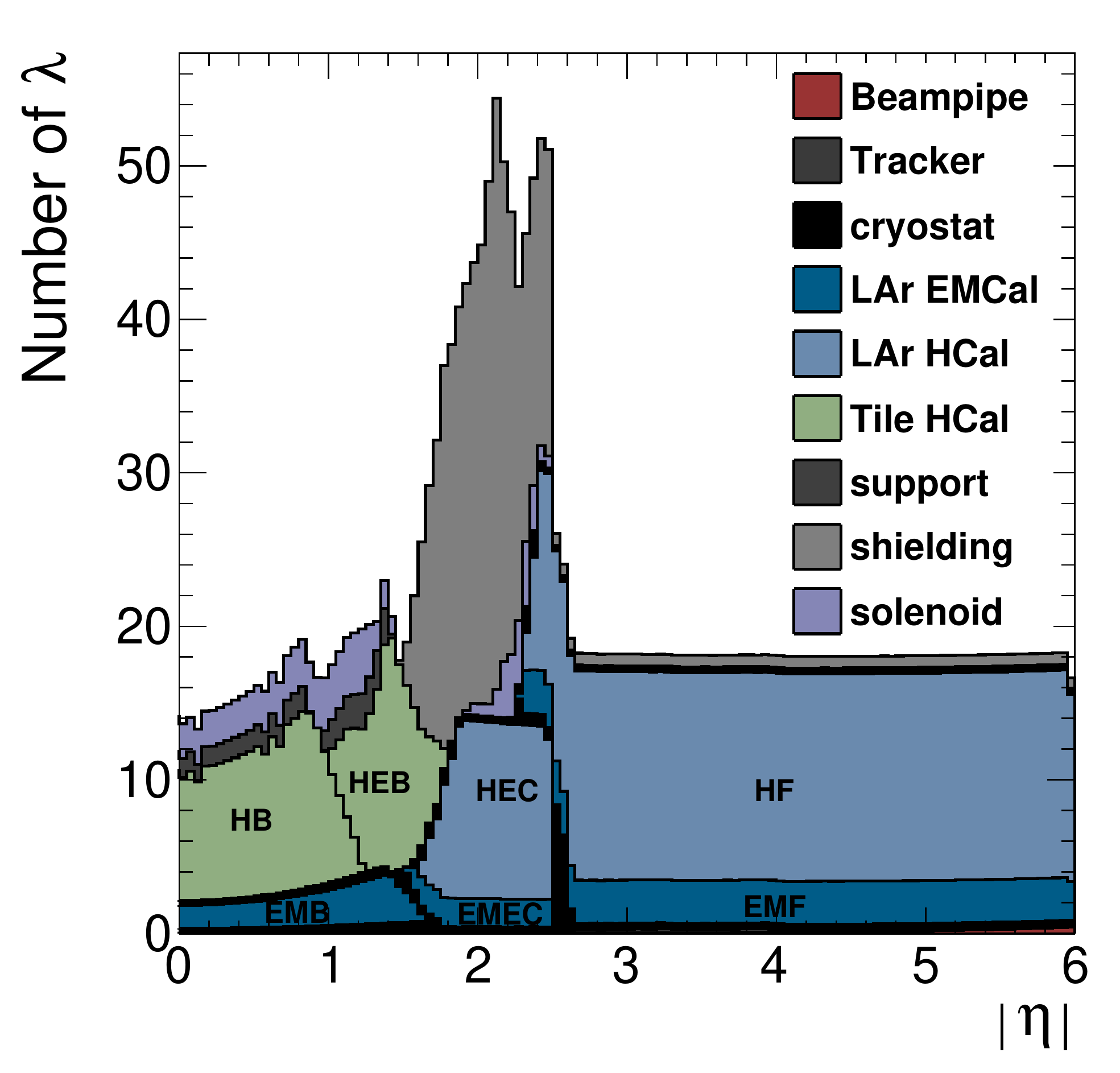}\caption{}
    \label{fig:layout:matScanLambda}
  \end{subfigure}
  \caption{Material budget of the reference detector expressed in units of \textbf{(a)} radiation length and \textbf{(b)} nuclear interaction length. The amount of material is measured from the interaction region (centre of the detector) to the outer boundary of the calorimeter. The spike at $\abseta=2.5$ corresponds to the inner wall of the cryostat of the endcap. The colour code is the same as in Fig.~\protect\ref{fig:layout:calorimetry}. }
\end{figure}

The overview of the calorimeter system of the reference FCC-hh detector is shown in Fig.~\ref{fig:layout:calorimetry}. It consists of a central and extended barrel, two endcaps, and two forward calorimeters, with the dimensions as given in Tab.~\ref{tab:layout:dimensions}. The sub-systems and the corresponding acronyms shown in Fig.~\ref{fig:layout:crossSec}. The central barrels and the endcaps are immersed in the magnetic field of the main solenoid of $\sim 4$\,T. Due to the high integrated luminosity goals and high collision rates at the FCC-hh, the radiation environment in the detector is very challenging. The expected radiation dose and 1\,MeV neutron equivalent fluence (NIEL) is presented in Tab.~\ref{tab:layout:dimensions} for an integrated luminosity of $30\,\mathrm{ab}^{-1}$. These unprecedented radiation requirements especially in the forward region call for radiation hard technologies and front-end electronics that can be placed in the back of the calorimeters at areas of reduced radiation exposure. Figs.~\ref{fig:layout:matScanX0} and~\ref{fig:layout:matScanLambda} show graphically the material budget of the reference detector which is described in the following sections.

\begin{table}[]
\centering
{\footnotesize
\begin{tabular}{|c|cccccc|}
\hline
         & $R_{min}$       & $R_{max}$        & $z$ coverage          & $\eta$ coverage        & Dose        & $1\UMeV~\text{n}_\text{{eq}}$ fluence      \\\hline
Unit     & m              & m              & m                     &                         & MGy         & $\times10^{15}~\mathrm{cm}^{-2}$           \\\hline
EMB      & 1.75           & 2.75           & \absz < 5             & $\abseta<1.67$          & 0.1         & 5                                        \\\hline
EMEC     & 0.82--0.96           & 2.7           & 5.3 <\absz<6.05       & $1.48<\abseta<2.50$     & 1           & 30                                       \\\hline
EMF      & 0.062--0.065          & 3.6          & 16.5 < \absz< 17.15   & $2.26<\abseta<6.0$      & 5000        & 5000                                     \\\hline
HB       & 2.85           & 4.89           & \absz < 4.6           & $\abseta<1.26$          & 0.006       & 0.3                                      \\\hline
HEB      & 2.85            & 4.59              & 4.5  <\absz < 8.3     & $0.94<\abseta<1.81$     & 0.008       & 0.3                                      \\\hline
HEC      & 0.96-1.32           & 2.7           & 6.05<\absz< 8.3       & $1.59<\abseta<2.50$     & 1           & 20                                       \\\hline
HF       & 0.065--0.077          & 3.6         & 17.15 <\absz < 19.5   & $2.29<\abseta<6.0$      & 5000        & 5000                                     \\\hline
\end{tabular}}
\caption{Dimensions of the envelopes for the calorimeter sub-systems (including some space for services) and the maximum radiation load at inner radii (total ionising dose is estimated for 30$\Uab^{-1}$). The abbreviations used in the first column are explained in the text.}
\label{tab:layout:dimensions}
\end{table}

For the electromagnetic calorimetry liquid argon was chosen as active material due to its intrinsic radiation hardness (provided LAr impurities can be kept low)\footnote{Experience of the ATLAS LAr calorimeter shows that purity levels down to 0.3\,ppm oxygen equivalent are possible to achieve if all materials used inside the cryostat are carefully chosen. }. As the radiation levels increase with pseudo-rapidity and in the vicinity of the beam-pipe, also the hadronic calorimeter in the endcaps and in the forward region uses the liquid argon technology. Lead/steel absorbers (thickness $=$~2.0\,mm) have been foreseen for the barrel (EMB) and endcap (EMEC) electromagnetic calorimeters. Tungsten absorbers are an interesting option due to the resulting smaller Moli\`ere radius and hence smaller clusters, which could reduce the impact of pile-up by up to a factor $\sim 1.5$. The LAr based hadronic endcap (HEC) and forward calorimeter (EMF, HF) foresee copper absorbers, following the example of the ATLAS forward calorimeters~\cite{CERN-LHCC-96-041}. The hadronic calorimeters in the central part of the detector (hadronic barrel, HB, hadronic extended barrel HEB) are based on scintillating plastic tiles within an absorber structure consisting of steel and lead. The baseline detectors are described in detail in Sec.~\ref{sec:layout:lar} for the liquid argon calorimeters and in Sec.~\ref{sec:layout:hcal} for the hadronic scintillator calorimeter. %The alternative options that are being studied for the central electromagnetic calorimeter, based on silicon and tungsten, are discussed in Sec.~\ref{sec:alternative}.

The proposed longitudinal and transversal granularity for the reference system is summarised in Tab.~\ref{tab:layout:mainParameters}. The strip layer is introduced to allow an efficient $\gamma/\pi^0$ separation. The segmentation in the strip layer is $\Delta\eta \approx 0.0025$ (two photons originating from a decay of a 50\,GeV $\pi^0$ are separated by $\Delta R\approx0.005$.
Overall, the granularity is $2-4 \times$ higher in each dimension ($\eta-\phi-layer$) compared to the calorimeters in the ATLAS experiment.

\begin{table}[]
\resizebox{\textwidth}{!}{%
\begin{tabular}{|c|ccccccc|}
\hline
                       & \multirow{2}{*}{material}   & \multicolumn{2}{c}{minimal depth}            & \multicolumn{3}{c}{granularity}                                                    & \# channels  \\
                       &                             &  $\#X_{0}$            &  $\#\lambda$          & $\Delta\eta$                     & $\Delta\varphi$          &layers                & $\left(10^6\right)$   \\ \hline
\multirow{2}{*}{EMB}   & \multirow{2}{*}{LAr/Pb}     & \multirow{2}{*}{26.5}& \multirow{2}{*}{1.5}  & 0.01                             & 0.009                    & \multirow{2}{*}{8}   & \multirow{2}{*}{$\sim$1.7}\\
                       &                             &                      &                       & (0.0025 in strip layer)          & (0.018 in some layers)   &                      &  \\\hline
\multirow{2}{*}{EMEC}  & \multirow{2}{*}{LAr/Pb}     & \multirow{2}{*}{45}  & \multirow{2}{*}{1.8}    & 0.01                             & 0.009                    & \multirow{2}{*}{6}   &  \multirow{2}{*}{$\sim$0.6} \\
                       &                             &                      &                       & (0.0025 in strip layer)          & (0.018 in some layers)   &                      &   \\\hline
EMF                    & LAr/Cu                      & 30                   & 2.8                   & 0.025                            & 0.025                    & 6                    & $\sim$0.1 \\\hline
HB                     & Sci/Pb/steel                & 136                  & 9.4                   & 0.025                            & 0.025                    & 10                   & $\sim$0.2 \\\hline
HEB                    & Sci/Pb/steel                & 141                  & 9.8                   & 0.025                            & 0.025                    & 8                    & $\sim$0.1\\\hline
HEC                    & LAr/Cu                      & 119                  & 11.3                  & 0.025                            & 0.025                    & 6                    & $\sim$0.5\\\hline
HF                     & LAr/Cu                      & 145                  & 13.5                   & 0.025                            & 0.025                    & 6                    & $\sim$0.1 \\ \hline
\end{tabular}}
\caption{Depth and proposed granularity of the FCC-hh reference calorimeter. 
}
\label{tab:layout:mainParameters}
\end{table}

%%\todo[inline]{JF: Moved this paragraph from end of 7.2.2 to here}
As explained in Sec.~\ref{sec:intro:ecal}, the increase of the centre-of-mass collision energy results in higher transverse momenta of the produced particles. Therefore, the required depth of the electromagnetic calorimeter is $\sim$30\,$X_0$, and of  $\sim$11\,$\lambda$ for the full EM + hadronic calorimeters. The thickness of the reference detector as a function of pseudo-rapidity in units of radiation length and interaction length is shown in Fig.~\ref{fig:layout:matScanX0} and \ref{fig:layout:matScanLambda} respectively. The thickness is measured including all inactive materials of the detector, as well as the tracker and the beam-pipe. %% Thickness of material in front of the calorimeter is shown in Fig.~\ref{fig:layout:lar:ecal:matScanPreEcal}.
At $\eta=0$ the total depth of the EMB calorimeter is $\sim29.5\,X_0$. It increases with pseudorapidity (up to $\abseta=1.5$) as particles traverse the detector with a smaller angle with respect to the beam-pipe (and cryostat). Material in front of the electromagnetic calorimeter is presented in Fig.~\ref{fig:layout:lar:upstreamEnergy} and discussed in Sec.~\ref{sec:layout:lar:corrections:upstream}. For the endcap (EMEC) and the forward electromagnetic calorimeter (EMF) the depth is above 30\,$X_0$. Including the hadronic calorimeters (HB, HEB, HEC and HF), a total depth in terms of interaction lengths of $>11\lambda$ over the full rapidity range is achieved.

%% file: tex/layout/electromagnetic.tex
\subsection{Liquid Argon Calorimeters}
\label{sec:layout:lar}

Liquid Argon (LAr) calorimetry has proven to provide excellent electromagnetic energy measurements, with high resolution, linearity and uniformity of the response, high stability and ease of calibration. Additionally, it is an intrinsically radiation hard material that can be used in the detectors with high particle fluence rates and ionisation doses. LAr-based calorimetry is successfully operating in ATLAS experiment~\cite{CERN-LHCC-96-041, Aad:2008zzm}. It has been chosen for the FCC-hh reference detector as a baseline technology for the electromagnetic calorimetry, but also, due to the expected high radiation dose, for the hadronic calorimeters at pseudo-rapidities of $>1.4$.

The electromagnetic barrel calorimeter is located within a 10\,m long and 1\,m thick double-vessel cylindrical cryostat, with an inner radius of 1.75\,m, covering the pseudorapidity range $\abseta<1.5$. The length is dictated by the length of the central tracker, which provides full lever arm for momentum spectroscopy for pseudorapidity up to $\abseta<2$. The calorimeter endcaps are located next to the barrel, starting at $\left|z\right|=5.3$\,m, and are $3$\,m thick (along the beam axis). They are positioned closer to the beam axis, with the active volume spanning up to $\abseta=2.5$, housed in two double-vessel cryostats. The forward detector is localised far from the centre of the detector, from $\left|z\right|=16.5$\,m to $\left|z\right|=19.5$\,m. In order to cover pseudorapidity up to $\abseta=6$, the inner radius of the forward calorimeter must be $r=8.2$\,cm, which leaves little space for the cryostat and the beam pipe but is regarded as feasible\footnote{In ATLAS the forward calorimeter's inner radius is 7.2\,cm~\protect\cite{CERN-LHCC-96-041}}.

%\begin{table}[ht]
%  \centering
%  \begin{tabular}{|c| c c c  c c|}
%    \hline
%    	&$\left<X_0\right>$& $\left<\lambda\right>$&	\multicolumn{3}{c|}{depth } \\
%    \hline
%    Unit	&cm&cm&	m & $\#X_0$& $\#\lambda$ \\
%    \hline
%    B. ECAL	&2.43	&&0.65 & 27 & 2\\
%    B. ECAL + HCAL	& &	&0.65+ & 27 & 2+\\
%    E. ECAL	&1.71&	&0.58 & 34 + X  & 1.5\\
%    E. ECAL	+ HCAL&1.71&	&0.58+ & 34 + X  & >11\\
%    F. ECAL	&1.62	&&0.48 & 30 + X  & 8.5\\
%    F. ECAL + HCAL	&1.62&	&0.48 & 30 + X  & >11\\
%    \hline
%  \end{tabular}
%  \caption{Radiation lengths $X_0$ and nuclear interaction lengths $\lambda$ of LAr calorimeters. The depth of the calorimeter is given for barrel at $\eta=0$, for endcaps and forward detector along the beam axis. \protect\todo[inline]{calculate for hadronic part as well}}
%  \label{tab:layout:lar:depth}
%\end{table}

\subsubsection{Barrel}\label{layout:barrel}
\subsubsubsection{Geometry Layout}

\begin{figure}[ht]
\centering
  \includegraphics[width=0.75\textwidth]{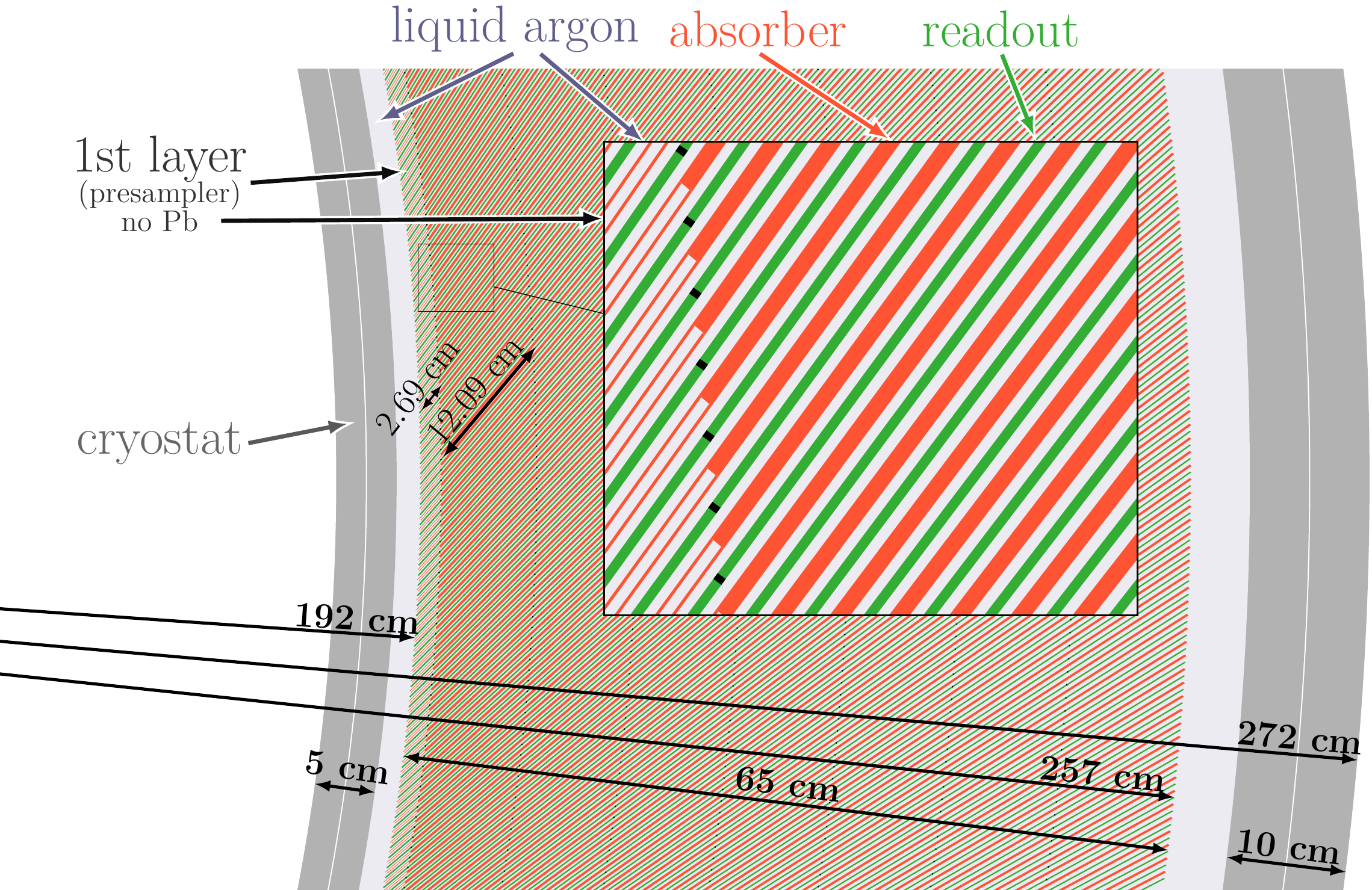}
  \caption{The cross section of the electromagnetic barrel calorimeter.}
  \label{fig:layout:lar:barrel}
\end{figure}

As explained above, it has been attempted to adapt LAr calorimetry to high granularity read-out. For the moment cell sizes as described in Tab.~\ref{tab:layout:mainParameters} are foreseen for the different parts of the calorimeter. To achieve this granularity a design as depicted in Fig.~\ref{fig:layout:lar:barrel} has been chosen. Straight lead/steel absorbers are interleaved with LAr gaps and straight electrodes with HV and read-out pads forming a cylinder of 192\,cm (257\,cm) inner (outer) radius respectively. The increase of the longitudinal segmentation (compared to the accordion ATLAS LAr calorimeter~\cite{CERN-LHCC-96-041} with three-layer kapton electrodes) is possible thanks to the use of  multi-layer electrodes realised as straight printed circuit boards (PCB). Therefore the electrodes as well as the absorber plates are arranged radially, but azimuthally inclined by 50\,degrees with respect to the radial direction as shown in Fig.~\ref{fig:layout:lar:barrel}. This ensures readout capabilities of the electrodes via cables arranged at the inner and outer wall fo the cryostat together with a high sampling frequency. The inclination of the plates also
provide for incoming particle, a uniform response in $\varphi$ down to few GeV particles. 
Together with honeycomb spacers defining the exact width of the LAr gap, this relatively simple structure should lead to a high mechanical precision and hence small impact on the energy resolution and uniformity. 
However, LAr gaps in this design are radially increasing (opening angle with the calorimeter depth), leading to a sampling fraction changing with depth. Due to the longitudinal layers, the shower profile will be measured for each shower, and the energy calibration will correctly handle the non uniform sampling fraction. As described in Sec.~\ref{sec:layout:lar:barrel:calibration}, with eight longitudinal layers or more, the effect on the energy resolution is negligible. In the current simulation, the electrodes and absorbers are assumed to be single piece. However, due to mechanical constraints and producibility, both absorbers and electrodes will have to be divided into pieces (in $z$ or projectively in $\eta$), forming distinct detector wheels, which could be manufactured separately. Strong outer rings\footnote{To avoid additional material in front of the active calorimeter, the main mechanical structure must sit at the outside radius.} together with spacers at the inner and outer radius will hold the electrodes and absorbers in place with high precision. A detailed engineering design is needed to realise these wheels with the required precision\footnote{Note, that for a uniformity of $\sim 0.7\,\%$ the absorber width and also the LAr gap width needs to be controlled at that level (at least for the sum of all absorbers and gaps in one cluster). A precision of $\cal{O}$(10\,$\mu$m) will therefore be necessary. Such a precision was achieved for the ATLAS LAr calorimeter (see~\protect\cite{CERN-LHCC-96-041}). Non-uniformities can also be corrected using in-situ calibration with $\mathrm{Z}\rightarrow\mathrm{e}^+\mathrm{e}^-$ events.}. 

\begin{figure}[ht]
  \centering
    \begin{subfigure}[b]{0.8\textwidth}
    \includegraphics[width=\textwidth]{lar/readoutWithAxes}
    \caption{}\label{fig:layout:lar:readoutEta}
  \end{subfigure}
  \begin{subfigure}[b]{0.19\textwidth}
    \includegraphics[width=\textwidth]{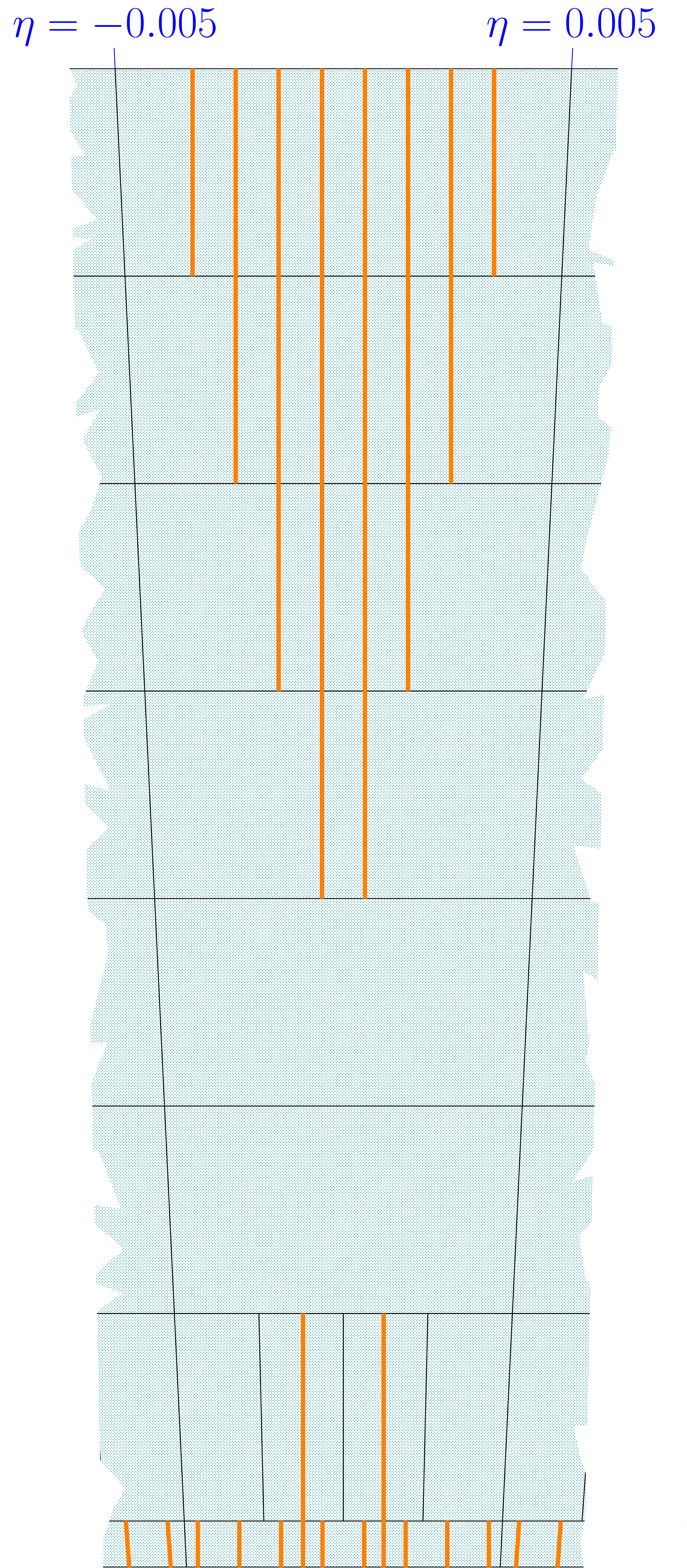}
    \caption{}\label{fig:pcb}
  \end{subfigure}
  \caption{Read-out PCB: ({\bf a}): Read-out electrode and cell segmentation on one electrode longitudinally (8 layers) and in pseudorapidity. The first layer is 4.5 times smaller than the rest for the correction described in Sec.~\ref{sec:layout:lar:corrections:upstream}. The cell size in pseudorapidity is $\Delta\eta=0.01$ in all layers except for the second (strip layer), where it is equal to $\Delta\eta=0.0025$. ({\bf b}): Top view of the signal pads and the signal traces and shields from each layer. The signals are extracted to the front for the first three layers, and to the back of the detector for the  remaining five layers.
    %Assuming the traces to be connected close to the edge of the signal pad, there are no traces passing through the third and fourth layer.
  }
\end{figure}

The thickness of the active detector is 650\,mm, composed of 1408 absorber plates, which are inclined from the radial direction by $50^\circ$. Each absorber is 980\,mm wide, in total 9.9\,m long\footnote{As mentioned above, it has to be studied how to best divide the detector in $z$ or $\eta$.} and 2\,mm thick. The absorbers are composed of a sandwich of lead (thickness 1.4\,mm) with steel sheets on both sides (thickness 0.2\,mm) to yield a flat and conductive surface glued with 0.1\,mm epoxy impregnated fabric (prepreg). Steel increases the mechanical strength, ensures the uniformity of the surface, and serves as a second HV electrode for the electric field in the liquid argon gap needed for the ionisation charge drift. Between two absorbers there are two liquid argon gaps of $2\times1.15$\,mm thick at the inner radius and $2\times 3.09$\,mm at the outer detector radius. The gaps are separated by a 1.2\,mm thick electrode. Two of those double gaps are read-out together, forming a $\varphi$ cell. That gives $\varphi$ granularity of $2\pi/704=0.009$. The segmentation in $\eta$ and depth (layers) is formed by cells on the readout electrode. The granularity in pseudorapidity is $0.0025$ in the second (strip) layer, and $0.01$ in the remaining 7 layers\footnote{Some of the simulations in the performance section are done for a granularity of $0.01$ in all layers.}. The fine segmentation in the strip layer is needed for a good $\gamma$/$\pi^{0}$ separation. The thickness of the first layer is 4.5 times smaller (Fig.~\ref{fig:layout:lar:readoutEta}) as the signal from this layer is used to correct for the energy deposited in the upstream material, described in Sec.~\ref{sec:layout:lar:corrections:upstream}. To achieve a $\varphi$-uniform response of this first layer, the absorbers do not contain lead in the middle to form a ``LAr-only'' presampler layer. 

\begin{figure}[ht]
  \centering
  \begin{subfigure}[b]{0.4\textwidth}
    \includegraphics[width=\textwidth]{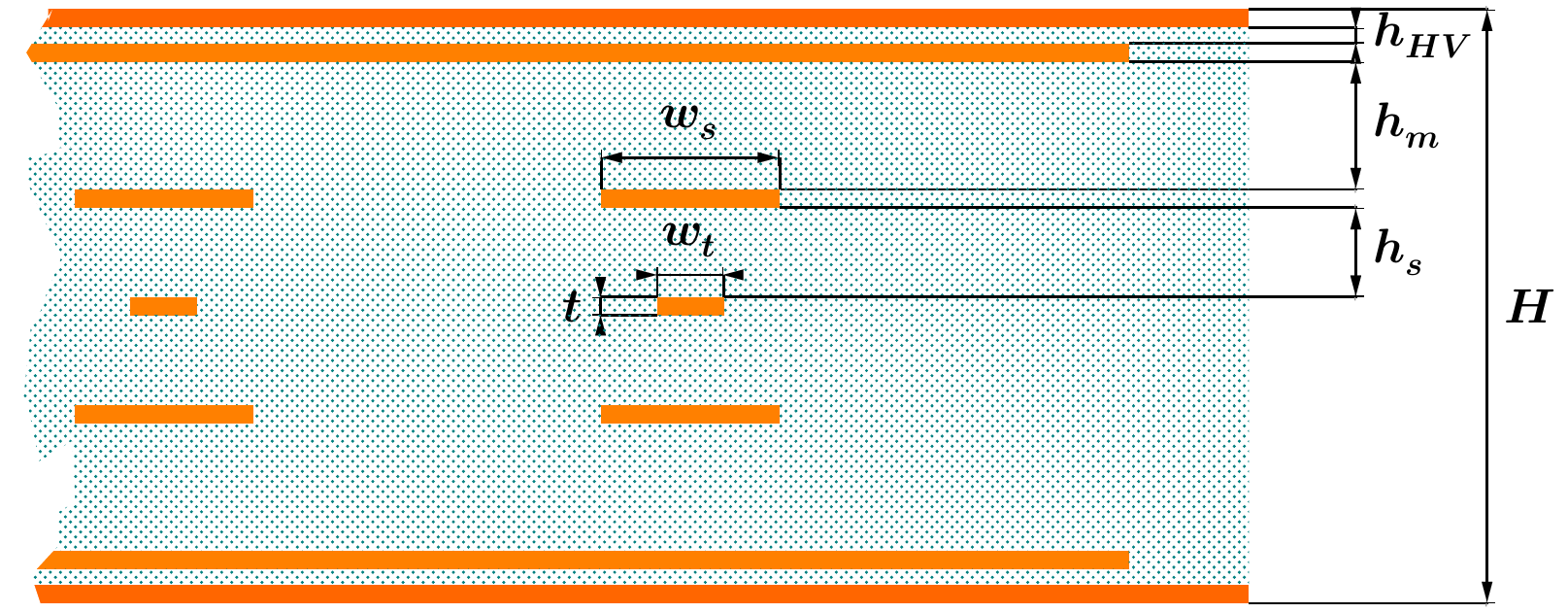}
    \caption{}\label{fig:pcb:traces}
  \end{subfigure}
  \begin{subfigure}[b]{0.59\textwidth}
    \includegraphics[width=\textwidth]{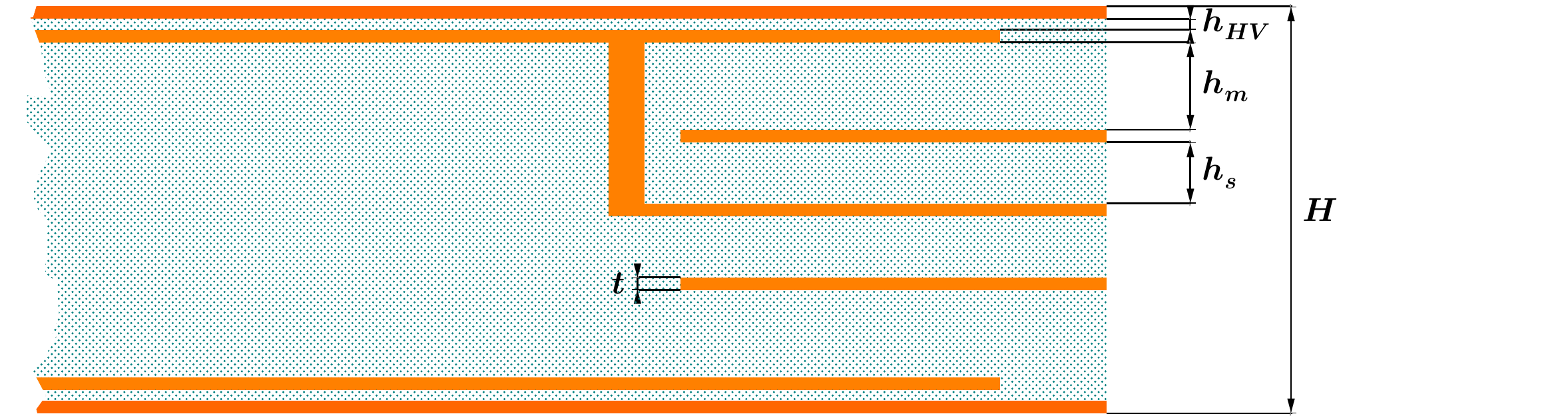}
    \caption{}\label{fig:pcb:traces:long}
  \end{subfigure}
  \caption{Cross-sections of the read-out PCB: ({\bf a}): Cross-section perpendicular to the signal traces: High voltage layers create an electric field in the liquid argon gap of the detector. The signal pads, located in the layer below, collect the ionisation signal which is extracted to either back or front of the electrode by the signal traces. In between the pads and the traces there is a layer of shields that minimise the cross talk from the pads to the signal traces connected to different pads. ({\bf b}): Cross-section parallel to the signal traces: One signal pad is connected to a signal trace by vias inside the PCB.}
\end{figure}

The electrodes will be realised as multi-layer PCBs ($\varepsilon_r=4$) with the following seven layers described here from outside to the inside:
\begin{itemize}
\item Two outside HV layers that produce a $\sim 1$\,kV/mm electric field in the LAr gaps. Due to the changing LAr-gap width several HV channels in depth will be foreseen. In order to limit the current and possible damage during discharges and to decouple these layers from the read-out, they need to be protected by $\cal{O}$(10\,k$\Omega$) HV resistors.
\item Two read-out layers with printed signal pads of the size of the desired read-out channels at a distance of $h_{HV}=100\,\mu\mathrm{m}$ from the HV layers. A schematic view of the read-out layer of the electrodes is depicted in Fig.~\ref{fig:layout:lar:readoutEta}.  The radial depth of the layers is the same for the whole barrel (i.e. for pseudorapidity ranging from $0$ to $1.5$): a first layer of 20\,mm and seven layers of 90\,mm in depth. This creates large difference in the thickness of layers expressed in the units of a radiation length (for particles originating near the interaction point). It may be addressed in the future by decreasing the thickness of layers as the pseudorapidity increases, so that measurements of the shower evolution are more uniform for different pseudorapidity values.
\item Two shielding layers on ground $h_m=285\,\mu\mathrm{m}$ inside the read-out layer to shield the signal pads from the signal traces. The width of these shields has been assumed to be $w_s=250\,\mu\mathrm{m}$ for the noise calculations in Sec.~\ref{sec:software:noise:electronics}, but will need to be optimised to keep cross-talk from the signal pads to the signal traces low ($< 0.1\,\%$). Larger shields, however, will translate into larger cell capacitance to ground and hence larger noise. 
\item One layer with $w=127\,\mu\mathrm{m}$ wide and $t=35\,\mu\mathrm{m}$ thick signal traces that are connected with vias to each of the signal pads. The signal traces together with the shielding layers should form transmission lines with an impedance in the range of $25\,\Omega\le Z\le 50\,\Omega$. For $Z=50\,\Omega$ the distance of the signal traces and the shields has to be $h_s\sim 170\,\mu\mathrm{m}$\footnote{The following approximation for a strip line between two ground shields has been used (see~\protect\cite{Zumbahlen:1520867}):
\begin{equation*}
Z[\Omega]=\frac{60}{\sqrt{\varepsilon_r}}\mathrm{log}\frac{1.9(2h_s+t)}{0.8w_t+t}
\end{equation*}
}.
\end{itemize}
Figure~\ref{fig:pcb} shows a cross section of such a read-out electrode for an impedance of $Z=50\,\Omega$. The middle layer used for the extraction of the signal to the front or back of the detector is sketched respectively in Fig.~\ref{fig:pcb:traces} and~\ref{fig:pcb:traces:long}. 

Since the electronic noise of a calorimeter cell is proportional to its capacitance, it was checked by how much the cell capacitance increases due to the shields inside the PCBs. The capacitance of  the signal pads to the shields $C_s$ are in parallel to the capacitance of the LAr gap $C_d=\varepsilon_{LAr}\varepsilon_0 A/d$ of width $d$ and area $A$ ($\varepsilon_{LAr}=1.5$), and therefore needs to be added to the total cell capacitance. An approximation for the capacitance of a microstrip line\footnote{The following approximation for a microstrip line on top of one ground shield has been used (see~\protect\cite{Zumbahlen:1520867}):
\begin{equation*}
C_s[\mathrm{pF/cm}]=\frac{0.26(\varepsilon_r+1.41)}{\mathrm{log}\frac{5.98h_m}{0.8w_s+t}}
\end{equation*}
} 
summed to the gap capacitance $C_d$ yields read-out cell capacitance (4 LAr gaps per read-out cell) of $C_{cell}$ ranging from 100\,pF to 500\,pF at $\eta=0$ and up to 1000\,pF at $\eta=1.5$. Such cell capacitance is similar to cells of the ATLAS LAr calorimeter, although their size is much smaller. In Sec.~\ref{sec:software:noise:electronics} the capacitance for the barrel detector is calculated and it is explained how realistic electronic noise values for each cell are estimated and used for the performance simulation.

\subsubsubsection{Optimisation of Longitudinal Layers}
\label{sec:layout:lar:barrel:calibration}

\begin{figure}[h]
  \centering
  \includegraphics[width=0.85\textwidth]{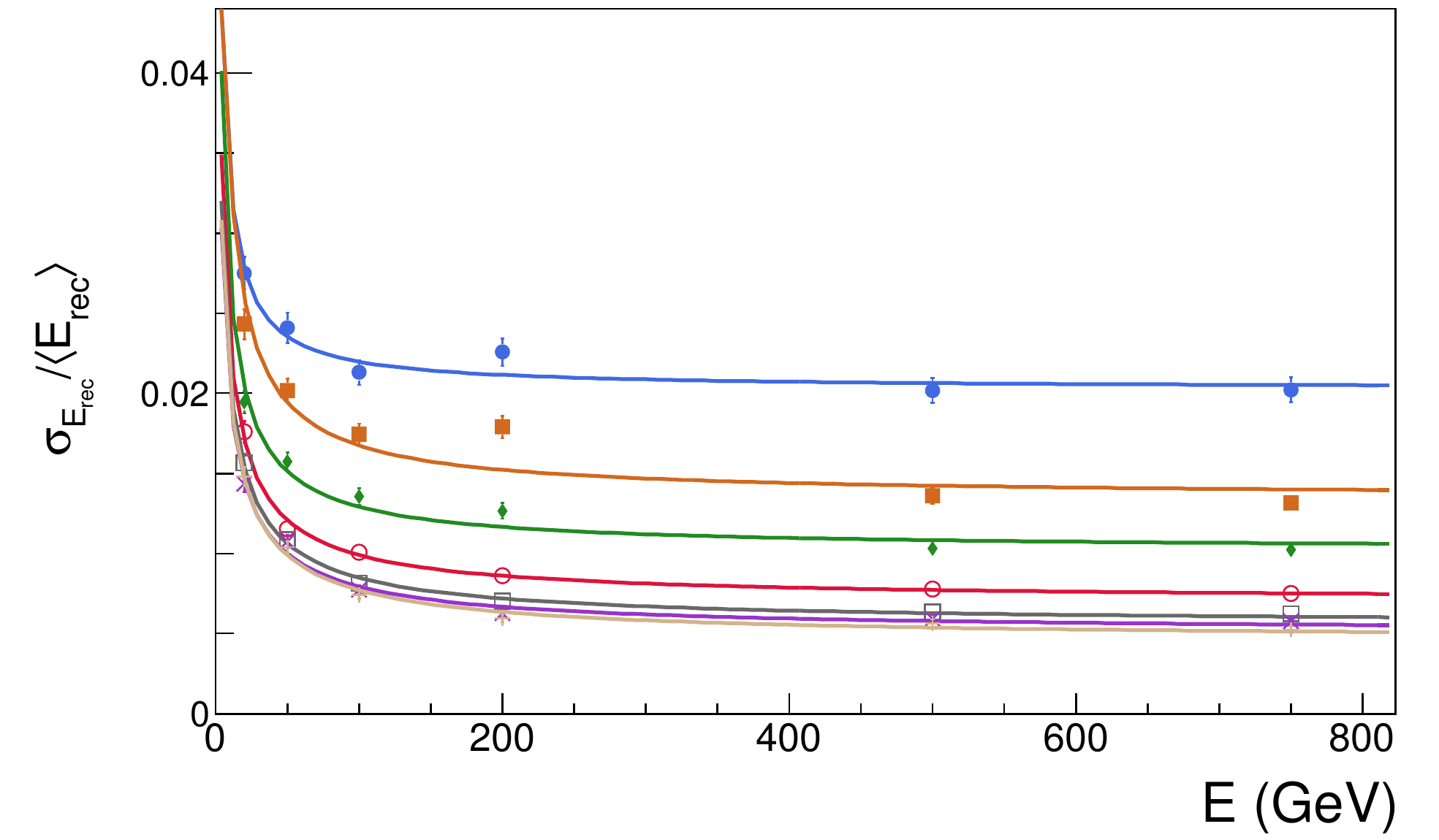}
  \begin{tikzpicture}[overlay]
    \node at (-5,5.8) {\includegraphics[width=0.35\textwidth,trim={9cm 6.5cm 1cm 0.6cm},clip]{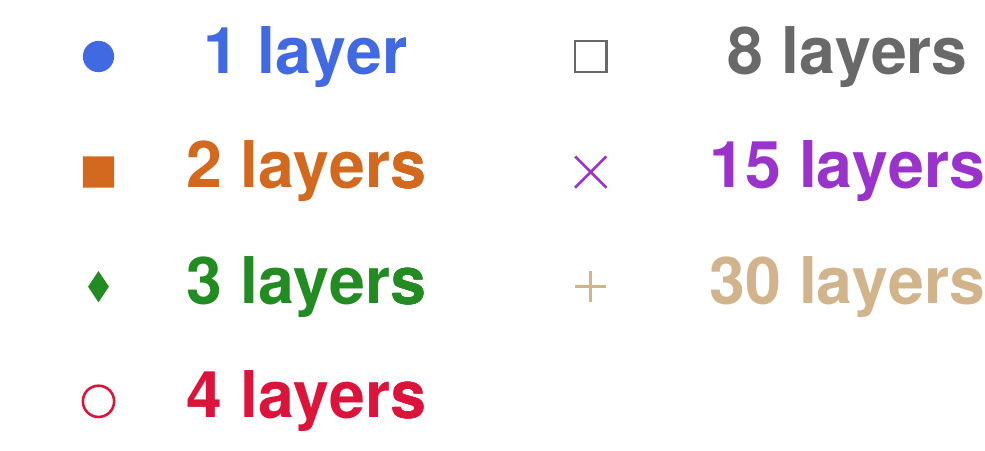}};
  \end{tikzpicture}
  \caption{Energy resolution of electrons for different number of longitudinal layers used for cell energy calibration. The geometry layout used in this study assumed $30^{\circ}$ inclination angle of absorber and readout plates from the radial direction (at the inner radius). Furthermore, it was performed without the tracker in front and the cryostat in order to compare only the effect of calibration on the energy resolution. The barrel was divided into layers of identical thickness. Sampling and constant terms are listed in Tab.~\ref{tab:layout:lar:enRes:DiffCalibLayers}.
}
  \label{fig:layout:lar:enRes:DiffCalibLayers}
\end{figure}

Due to the increasing LAr gap with radius, the sampling fraction changes with depth. Fluctuations of the shower depth would therefore immediately translate into different reconstructed energies and hence a degraded resolution. The energy reconstruction must therefore take this into account and apply a corrected sampling fraction to energy deposits at different depths. Figure~\ref{fig:layout:lar:enRes:DiffCalibLayers} and Tab.~\ref{tab:layout:lar:enRes:DiffCalibLayers} show the energy resolution of electrons using different number of longitudinal layers with different sampling fractions. The values of the sampling fraction are extracted from the simulation, by comparing the deposits in the active and the passive materials. Calibration of the shower energy using only one sampling fraction value leads to a high constant term $c=2\,\%$. Using at least 8 longitudinal layers significantly improves the resolution ($c=0.6\,\%$). Therefore 8 layers are chosen for the FCC-hh EMB calorimeter. Division to more layers does not improve significantly the resolution, while it would increase the number of read-out channels significantly.

\begin{table}[ht]
  \centering
 \begin{tabular}{|c| c c|}
   \hline
   number of &  &  \\
    layers & $a~(\sqrt{\UGeV})$ & $c$ \\
    \hline
   1&	8.4\,\% $\pm0.7$\,\%&	2.03\,\%$\pm0.05$\,\%\\
   2&	9.9\,\%$\pm0.5$\,\%&	1.35\,\%$\pm0.03$\,\%\\
   3&         7.9\,\%$\pm0.3$\,\%&	1.02\,\%$\pm0.03$\,\%\\
   4&	6.9\,\%$\pm0.3$\,\%&	0.71\,\%$\pm0.02$\,\%\\
   8&	6.4\,\%$\pm0.2$\,\%&	0.56\,\%$\pm0.02$\,\%\\
  15&	6.1\,\%$\pm0.2$\,\%&	0.51\,\%$\pm0.02$\,\%\\
  30&	6.2\,\%$\pm0.2$\,\%&	0.46\,\%$\pm0.02$\,\%\\
   \hline
 \end{tabular}
\caption{Energy resolution of electrons for different number of layers of equal thickness used for the cell energy calibration. The geometry layout used in this study assumed $30^{\circ}$ inclination angle of absorber and readout plates from the radial direction (at the inner radius).}
\label{tab:layout:lar:enRes:DiffCalibLayers}
\end{table}

\begin{figure}[ht]
  \centering
    \includegraphics[width=0.7\textwidth]{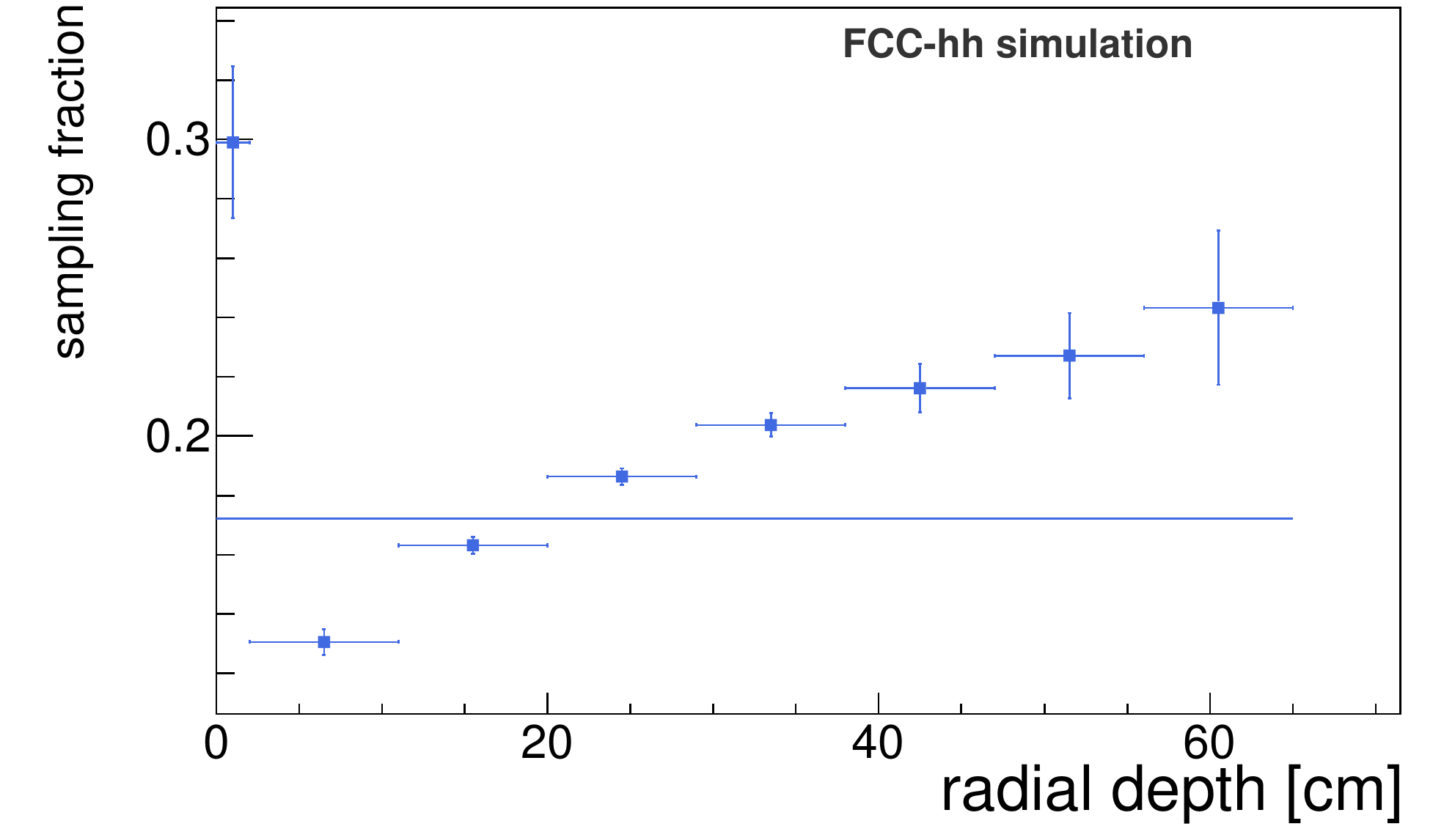}
  \caption{Average sampling fraction ($E=50$-200\,GeV) calculated from the energy deposited by electrons in each of the 8 layers of the detector. Horizontal line represents the average sampling fraction, obtained without longitudinal segmentation.}\label{fig:layout:lar:samplFraction:avg}
\end{figure}

The obtained values of the sampling fractions for each of the 8 detector layers in the current detector setup are presented in Fig.~\ref{fig:layout:lar:samplFraction:avg}. Thickness of the first layer (presampler) is smaller by a factor of 4.5 from other layers in order to provide an input to the correction for the energy deposited in the material in front of the detector, as explained in Sec.~\ref{sec:layout:lar:corrections:upstream}. Results for 50, 100 and 200\,GeV electrons have been found to be similar and have been averaged. These values have been used for all performance simulations shown in this report. However, as also performed in ATLAS, a MVA based recalibration using shower depth and shower shape variables could be considered on the reconstructed clusters in order to further improve the energy measurement.

\subsubsection{Endcap Calorimeters}
\subsubsubsection{Geometry Layout}

\begin{figure}[h]
  \centering
  \includegraphics[width=0.75\textwidth]{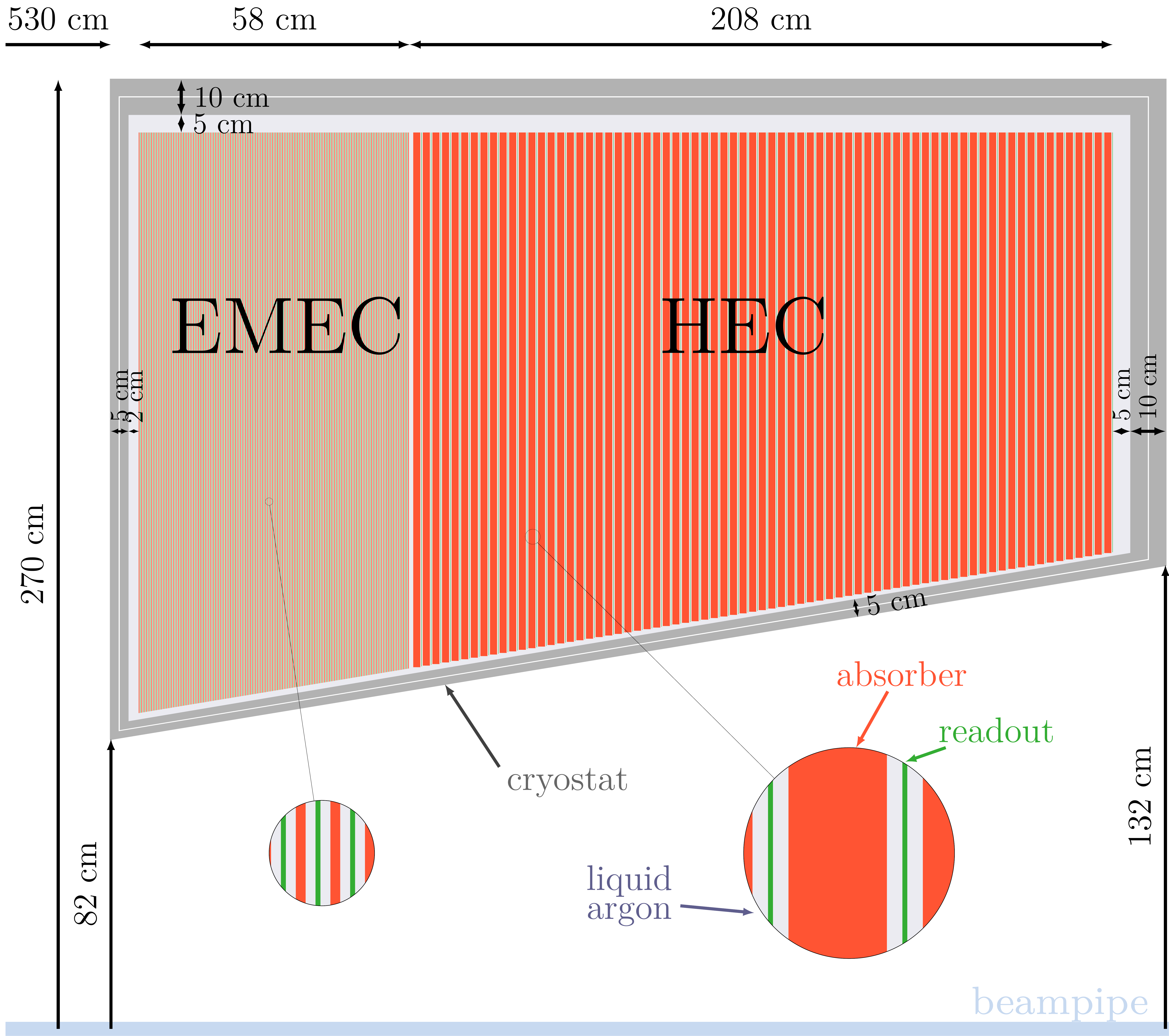}
  \caption{Cross section of the calorimeter endcap. EM indicates the electromagnetic calorimeter, H the hadronic part. They differ in terms of the thickness and material of the absorber plates.}
  \label{fig:layout:lar:endcap}
\end{figure}

Due to the radiation load, both the electromagnetic (EMEC) and the hadronic calorimeters (HEC) in the endcaps use liquid argon as the active medium. Both calorimeters share one endcap cryostat. The endcaps are located on both sides of the central barrel, from $\left|z\right|=5.3$\,m to $\left|z\right|=8.3$\,m. The outer radius of the endcaps is $r=2.7$\,m, similar to the electromagnetic barrel calorimeter. The inner radius changes with $\left|z\right|$ leading to a conical inner bore of the cryostat, allowing that the active detector volume covers a pseudorapidity region up to $\abseta=2.5$. The cross section through the upper half of one endcap is presented in Fig.~\ref{fig:layout:lar:endcap}. The segmentation used in the simulation is summarised in Tab.~\ref{tab:layout:mainParameters} and follows the segmentation of the barrel region.

The layout of the detector is again inspired by the ATLAS calorimeters, but uses also in the electromagnetic endcaps parallel discs of absorbers and readout electrodes instead of accordion-shaped electrodes and absorbers. Material and thickness of the absorbers differ for the electromagnetic and the hadronic part. The electromagnetic calorimeter is made of $1.5$\,mm thick lead discs, glued inside steel sleeves as described in detail for the barrel in Sec.~\ref{layout:barrel}. Between two absorbers there are two liquid argon gaps, $2\times0.5$\,mm thick. The readout electrode, realised as a seven-layer PCB as described in Sec.~\ref{layout:barrel}, is positioned in between the LAr gaps, providing the HV for the drift field inside the gap, and housing the read-out pads and signal traces for the read-out. The thickness of the drift gap is decreased compared to the ATLAS detector (which has a $2.16$\,mm gap) because of the larger particle densities expected at FCC-hh which would lead to space-charge inside the drift gaps. The hadronic part of the detector uses copper as passive material, with $40$\,mm thick discs and $2\times1.5$\,mm liquid argon gaps. The thickness of the read-out electrode PCB is $1.2$\,mm, as in the barrel detector. The read-out electrodes are rather large disk-shaped panels in that design. A re-partitioning into smaller size electrodes will need to be studied as well as the exact layout of the signal traces inside the PCBs, avoiding too long traces which could lead to an attenuation of the signal. 

%\subsubsubsection{Energy Calibration}

The ratio of active to passive material in this detector is constant, therefore the energy calibration could be performed using one calibration constant. The sampling fraction is equal to $f_{\mathrm{sampl}}=0.072$ for the electromagnetic endcap and $f_{\mathrm{sampl}}=0.030$ for the hadronic endcap.

\subsubsection{Forward Calorimeters}
\subsubsubsection{Geometry Layout}

\begin{figure}[h]
  \centering
  \includegraphics[width=0.7\textwidth]{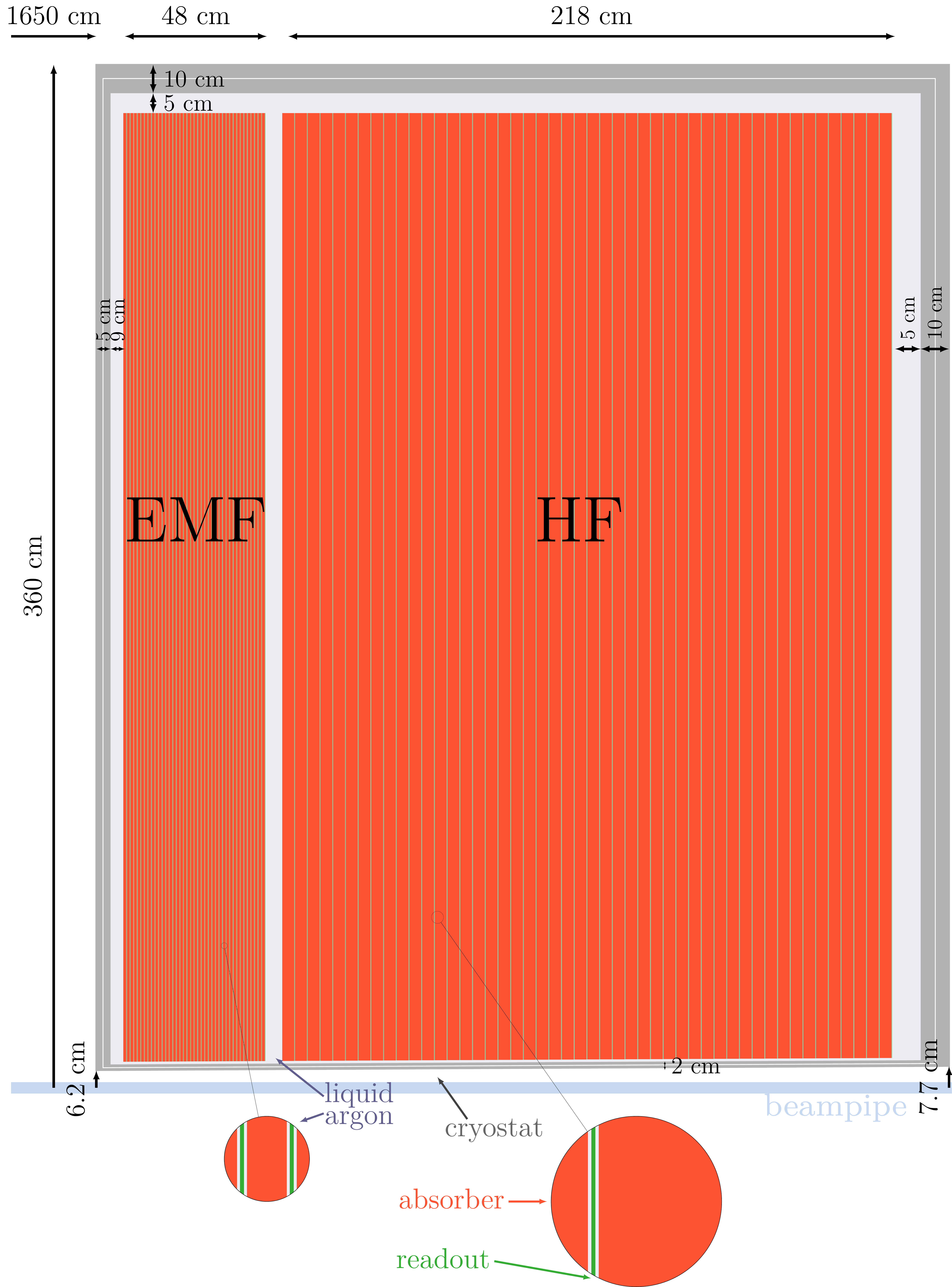}
  \caption{The cross section view of the forward calorimeter. EMF indicates the electromagnetic calorimeter, HF - the hadronic part. They differ in terms of the thickness of the absorber.}
  \label{fig:layout:lar:forward}
\end{figure}

As shown in Tab.~\ref{tab:layout:dimensions} the forward calorimeters (EMF and HF) will have to withstand an unprecedented integrated ionisation dose of up to 5000\,MGy and a 1\,MeV neutron equivalent fluence of $5\times10^{18}$\,cm$^{-2}$. This goes far beyond the specifications of any detector system that is operating nowadays, e.g. the forward calorimetry at HL-LHC will experience 1\,MeV neutron equivalent fluence of up to $3\times10^{17}$\,cm$^{-2}$. Some ATLAS calorimeter components were tested up to such fluence as summarised in Sec. 2.5 of~\cite{Collaboration:2285582}. 
However, it is rather difficult to extrapolate by an additional factor of 15 from existing experience. Very careful irradiation studies will therefore be needed to qualify all the materials used for these detectors. 
As LAr calorimetry is based on a liquid active material, we believe that it has the best chance to withstand this hostile radiation environment. 
The proposed layout of the forward detector is similar to  the layout of the calorimeter endcaps, however adapting the dimensions as presented in Fig.~\ref{fig:layout:lar:forward}. The forward calorimeter is positioned far from the centre of the detector, from $\left|z\right|=16.5$\,m to $\left|z\right|=19.5$\,m. The outer radius is $r=3.6$\,m, while the inner radius of the cryostat is just outside the beam-pipe ($r\approx5$\,cm), so that the active volume of the detector covers the region up to $\abseta=6$ ($r\approx8.2$\,cm for $\left|z\right|=16.5$).  The segmentation used in the simulation is summarised in Tab.~\ref{tab:layout:mainParameters} and for both the electromagnetic and hadronic detectors it follows the granularity of the hadronic barrel.

Absorbers used in forward region are proposed to be copper, in both parts of the calorimeter (EMF and HF). In order to avoid ion build-up due to large energy densities, the thickness of LAr gap is reduced to $0.1$\,mm. The thickness of the absorber discs in the electromagnetic part is $0.9$\,mm in order to keep similar sampling fraction as in the ATLAS forward detector. In the hadronic part the copper discs are $40$\,mm thick. Such a design of parallel plates could turn our to be difficult to realise due to the large discs. An alternative design could - inspired by the ATLAS forward calorimeter - consist of copper rods inside a copper matrix forming a 100$\,\mu\mathrm{m}$ drift gap in between. The values of the sampling fraction for the electromagnetic and hadronic forward calorimeters are equal to $f_{\mathrm{sampl}}=0.0033$ and $f_{\mathrm{sampl}}=0.00083$, respectively.

\subsubsection{Cryostats}

These calorimeters will be housed in five different cryostats that can be seen in Fig.~\ref{fig:layout:calorimetry}. The barrel EM calorimeter sits inside a 10\,m long cylindrical barrel cryostat (inside volume of $\sim 110$\,m$^3$), the EM endcaps and hadronic endcaps are housed in two cylindrical endcap cryostats with a conical inner bore (inside volume of $\sim 50$\,m$^3$ each) and the forward calorimeters will be located inside two cylindrical forward cryostats (inside volume of $\sim 110$\,m$^3$ each).
Very similar, though slightly smaller cryostats were designed for the ATLAS LAr calorimeters. They are realised as double-vessel aluminium cryostats (see~\cite{CERN-LHCC-96-041}). It is very likely that a similar design, scaled to the new dimensions could be used for the FCC-hh LAr calorimeters, but R\&D has started within the CERN EP R\&D program, to develop a cryostat with the necessary mechanical properties and lowest possible material budget of the inner bore. The outer vessels of the barrel and endcap cryostats will be situated between the EM calorimeter and the hadronic calorimeter, and therefore should also be kept thin to ensure that even lower energetic particle will reach the hadronic calorimeter despite the strong magnetic field. %\todo[inline]{AZ: we can add reference now?}

The barrel calorimeter is immersed in a 10\,m long cylindrical cryostat with an inner bore of 185\,cm radius and an outer radius of 272\,cm, that has to support the $\sim 350$\,tons of the barrel calorimeter (including $\sim 275$\,tons of absorber material) immersed in a LAr bath of 75\,m$^3$ weighing $\sim 100$\,tons. To minimise the material upstream of the EM calorimeter, the inner bore of this cryostat needs to be as thin as possible in terms of radiation lengths. 
In the simulation the two vessels of the cryostat have been assumed to be $2\times 25$\,mm thick aluminium in the inner bore, and $2\times 50$\,mm aluminium at the outer detector radius. Space between the active detector and the cryostat is filled with liquid argon and is reserved for the necessary services. 

The two endcap cryostats (see Fig.~\ref{fig:layout:lar:endcap}) have also been assumed to have $2\times 25$\,mm thick aluminium front walls in inner bore walls and thicker walls behind the calorimeters. The cryostats for the forward calorimeters have been assumed to have a very thin inner bore to allow space for the beam pipe (see Fig.~\ref{fig:layout:lar:forward}). A detailed engineering design needs to be performed for all of these cryostats taking into account the huge load they have to support. 

%\clearpage
\subsubsection{Cryostat Feedthroughs}

The number of read-out channels of the above described LAr calorimeters is not yet fixed, but needs to be optimised after detailed simulations evaluating the needed granularity for pile-up rejection, particle identification and particle flow techniques. However, we anticipate that the number of channels will strongly increase in comparison with noble liquid detectors nowadays (e.g. ATLAS LAr calorimeter, 183000 channels). Assuming a granularity as summarised in Table~\ref{tab:layout:mainParameters}, signals of $\sim 2$\,million channels will have to be fed out of the five cryostats. Whereas e.g. in ATLAS the density of signal cables at the feed-through flange is about 6--7 per cm$^2$ (ATLAS is using glass sealed gold pin carriers), values of up to 20--50 signals per cm$^2$ should be achieved to accommodate the higher number of read-out channels. New ways of sealing these cables have to be studied. Epoxy based sealing technologies exist, also seals of strip lines using solder can be realised, or feeding the signals through sealed PCBs. All these technologies will need to be developed further to achieve the required cable density and required reliability for 20 years of operations. An R\&D project has been started to survey existing techniques, to design and construct test feed-throughs with selected promising techniques and to further optimise these techniques. Close collaboration with industry and other interested laboratories will be very important. Cold tests and electrical tests of these test-feed-throughs have to be carried out to test their cryogenic reliability and electrical properties.
The signal feed-throughs of the barrel cryostat will sit at both ends of the cryostat (highest $|z|$) on the outer warm vessel and will lead the signals into read-out boxes with read-out electronics that will be located in the foreseen gap between the hadronic barrel (HB) and extended barrel (HEB). The feed-throughs on the endcap cryostats could be located on the forward wall of the endcap cryostats at largest possible radius. The feedthroughs of the forward cryostats could be located on the outer radius.

On top of the signal feedthroughs, there will be at least two HV feedthroughs per cryostat, bringing the HV for the drift gaps into the cryostats. The proximity of the cryogenic system will also use several cryogenics feedthroughs per cryostat for controlling and monitoring the cryostat operation. 
%\todo[inline]{AZ:Do we mean again EP RD? if so - ref}

\subsubsection{Read-Out Electronics}
\label{sec:layout:electronics}

Particles crossing the LAr filled drift gap will ionise the Argon atoms. Due to the high electric field ({$\sim1$\,kV/mm}) the electrons will immediately be separated from the Ar ions and both will start to drift inside the electric field. This drift of charges will induce triangular current signals in the read-out pads of the electrodes dominated by drift of the electrons (typical drift time of $\sim 200$\,ns/mm\footnote{The exact drift time will depend on the LAr temperature, the exact field and the gap width.}). %\todo[inline]{AZ: Isn't it better if we give it per mm? e.g. 200ns/mm ? Especially that in our detector the thickness increases (with radius) by 100\%.}

Based on various considerations, in particular maintainability and long-term reliability in the strong radiation environment of FCC-hh, it is foreseen to have all active read-out electronics located outside the cryostats. ATLAS has chosen this approach for the EM calorimeter and has proven that excellent noise performance can be reached despite the long cable connections from the detector cell outside the cryostat to the preamplifier. In ATLAS the signal to noise ratio of a muon (MIP) in the second (first) calorimeter layer in ATLAS is $\sim 7 (\sim 3)$~\cite{Aharrouche:2009zzh}, respectively. It should be noted that measurement of the response to muons per layer is extremely useful to inter-calibrate the different longitudinal layers independently from the exact material knowledge in front of the calorimeter. In the proposed calorimeter design the signal-to-noise ratio of muons  per layer will degrade due to longer cables (stronger attenuation), and smaller longitudinal layer dimensions, however due to PCB electrodes, comparable cell capacitances as ATLAS could be achieved. The read-out electronics will be located in boxes mounted directly on the read-out feedthroughs. The signals will be guided on transmission lines through the read-out electrode PCBs and then on coaxial cables from the detector to the feedthroughs. The impedances of the transmission lines $Z$ must accurately match the preamplifier input impedance $Z_{pa}$ which defines the preamplifier time constant $\tau_{pa}=C_{cell}Z_{pa}$, with $C_{cell}$ being the cell capacitance. As described in Sec.~\ref{layout:barrel} transmission lines in the range of $25\,\Omega\le Z\le 50\,\Omega$ seem adapted for the expected cell capacitance.
%\todo[inline]{AZ: Can someone reread the sentence above to make sure it makes sense? (I do not seem to understand it)}

Similar to ATLAS~\cite{CERN-LHCC-96-041}, bipolar shaping seems to be the optimal choice. Due to the signal shape that has a zero net area, the average signal in any read-out cell is also zero except for settling effects at the beginning of bunch trains. Pile-up signals from the same bunch crossing and pile-up from previous bunch crossings will therefore cancel to zero on average. However, due to the statistical nature of the proton collisions, the created particles and their energy deposits inside the calorimeter, pile-up will induce fluctuations of the baseline that can best be described as pile-up noise. Section~\ref{sec:software:noise:pileup} will describe how this pile-up noise was estimated for the FCC-hh simulations. 
 
The shaping time will need to be optimised taking into consideration the electronics noise, decreasing with higher shaping time, and the pile-up noise, increasing with higher shaping time. Also, the series noise contribution from the additional capacitance of the long transmission lines of impedance $Z$ can only be neglected if $C_{cell}Z\gg\tau_{sh}$. For ATLAS and a pile-up of $\left<\mu\right>=25$ an optimum around $\tau_{sh}=45$\,ns was found~\cite{CERN-LHCC-96-041}, but due to the much higher peak pile-up expected at FCC-hh and the constraint mentioned above, the best choice will likely be at lower values. For the simulation results presented, we assumed that a similar shaping as in ATLAS could be achieved. The shaped signals will then be digitally sampled with bunch crossing frequency (40\,MHz) or twice this frequency, within a dynamic range of 16\,bits and sent via optical links into the counting room. There these data can be used as input to the hardware trigger and will, after a positive trigger decision, be written to disk.
%\todo[inline]{AZ: Actually, we never used any of that (shaping) in simulation. Do we want to add out-of-time pile-up noise (as factor 1.4--1.6 wider noise than in-time pu)? I can rerun it and add to performance results.}

With this architecture, which has also been chosen for the HL-LHC upgrade of the ATLAS LAr calorimeters~\cite{Collaboration:2285582}, the full history of energy deposits is available in the counting room and therefore could be used to actively subtract the impact of out-of-time pile-up from preceeding bunch crossings. Signal reconstruction algorithms based on this idea are currently being developed and tested for HL-LHC~\cite{Collaboration:2285582}.

Since a precise time measurement of energy deposits inside the calorimeter will be essential to reduce impact of pile-up, the design of the read-out electronics will need to take into account the precise timing requirements at the 30\,ps level. In comparison the ATLAS LAr calorimeter achieves timing resolution of $\cal{O}$(65\,ps) for high energetic clusters. The timing resolution is discussed in more details in Sec.~\ref{sec:performance:egamma:timing}.

%\subsubsection{Cryogenics System}
%\todo[inline]{To be written.}

%\subsubsection{Cryostat}

%5 cryostats, design aiming to reduce as much as possible inactive material. RD needed.

%LAr volume:\\
%Barrel: 70 m3 (ATLAS: 40 m3)\\
%Endcaps: 14 m3 x 2 (ATLAS: 19 m3 x 2) \\
%Forward Cryostats: 12 m3 x 2\\
%Sum: 122 m3 (ATLAS: 78 m3).

%\subsubsubsection{Barrel}
%Volumes for Barrel\\
%Cryostat volume: 24.67 m3\\
%Inside cryostat volume: 108.2 m3\\
%Modules (1408 absorber + PCB): 38.6 m3\\
%LAr (including 'placeholders for services'): 69.7 m3
%\subsubsubsection{Endcaps}
%Volumes for endcap (one side!):
%Total: 58.6 m3\\
%Inside-cryo: 49.1 m3\\
%cryo (aluminium): 9.5 m3 (but assumes 10cm inner cryo!)\\
%discs: 35.3 m3\\
%LAr: 13.9 m3
%\subsubsubsection{Forward}
%Volumes for forward (one side!):
%Total: 122.1 m3\\
%Inside-cryo: 109.6 m3\\
%cryo (aluminium): 12.4 m3\\
%discs: 97.6 m3\\
%LAr: 12.0 m3

%% file: tex/layout/hadronic.tex
\subsection{Scintillator Tile Calorimeters}
\label{sec:layout:hcal}
\subsubsection{Hadronic Barrel and Extended Barrel}
Because of the reduced radiation requirements behind the EM barrel calorimeters, cost and performance considerations, a hadronic calorimeter based on scintillating tiles is proposed for the barrel (HB) and extended barrels (HEB) of the FCC-hh reference detector. The calorimeter design has been inspired by the ATLAS Tile Calorimeter~\cite{CERN-LHCC-96-042}.

\begin{figure}[htbp]
	\begin{center}
		\includegraphics[width = .85\textwidth]{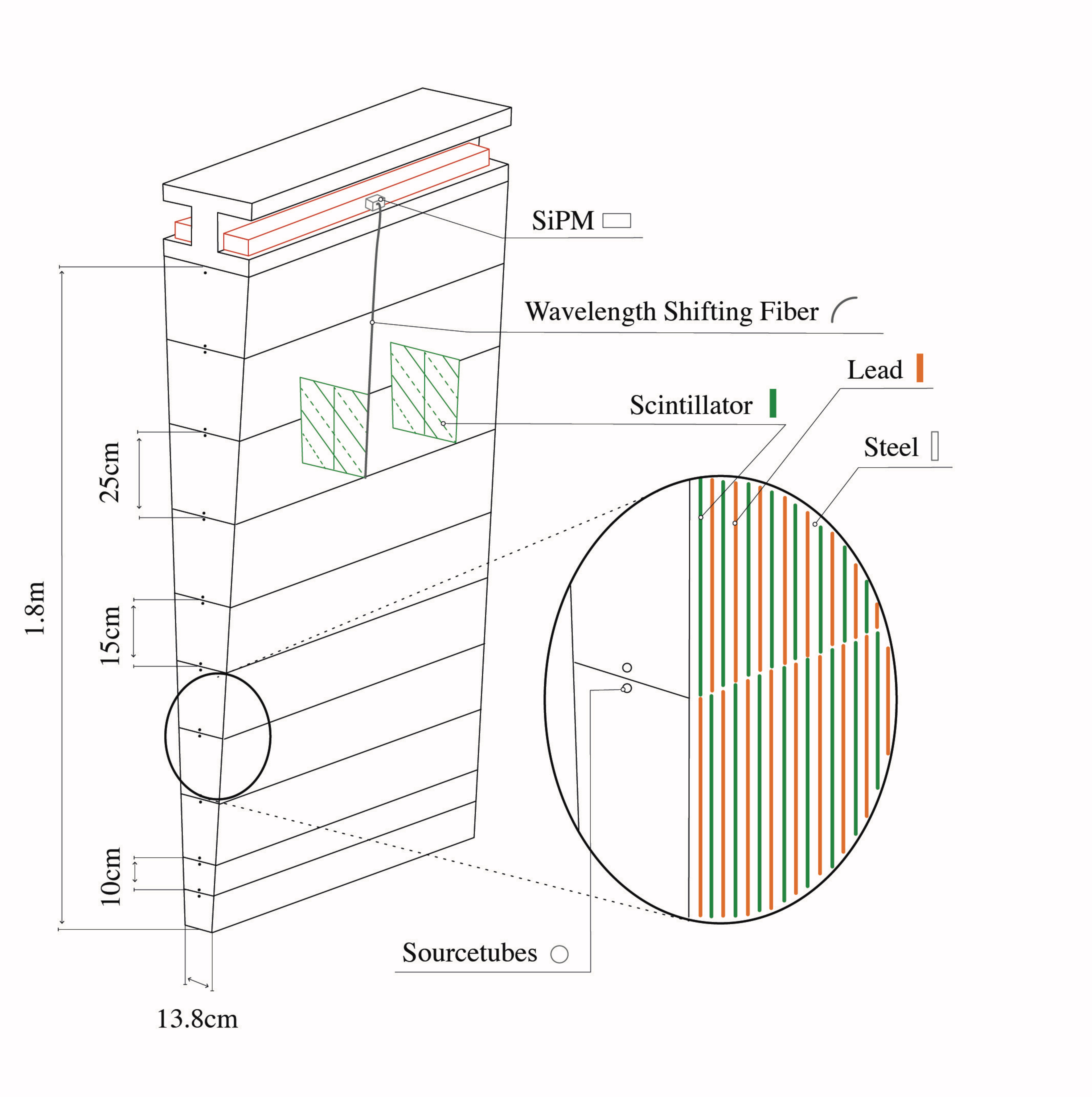}
		\caption{Schematic of one module of the hadronic barrel Tile calorimeter. The optical components (the two scintillating tiles per layer, wavelengh shifting fiber and the SiPM) are shown. The tubes designed for a movable radiation source (for details about the calibration system see Sec.~\ref{sec:layout:hcal:calibration}) are also sketched.}
		\label{fig:layout:hcal:geometry}
	\end{center}
\end{figure}

\begin{figure}[htbp]
	\begin{center}
    \begin{subfigure}[b]{0.5\textwidth}
		\includegraphics[width = \textwidth]{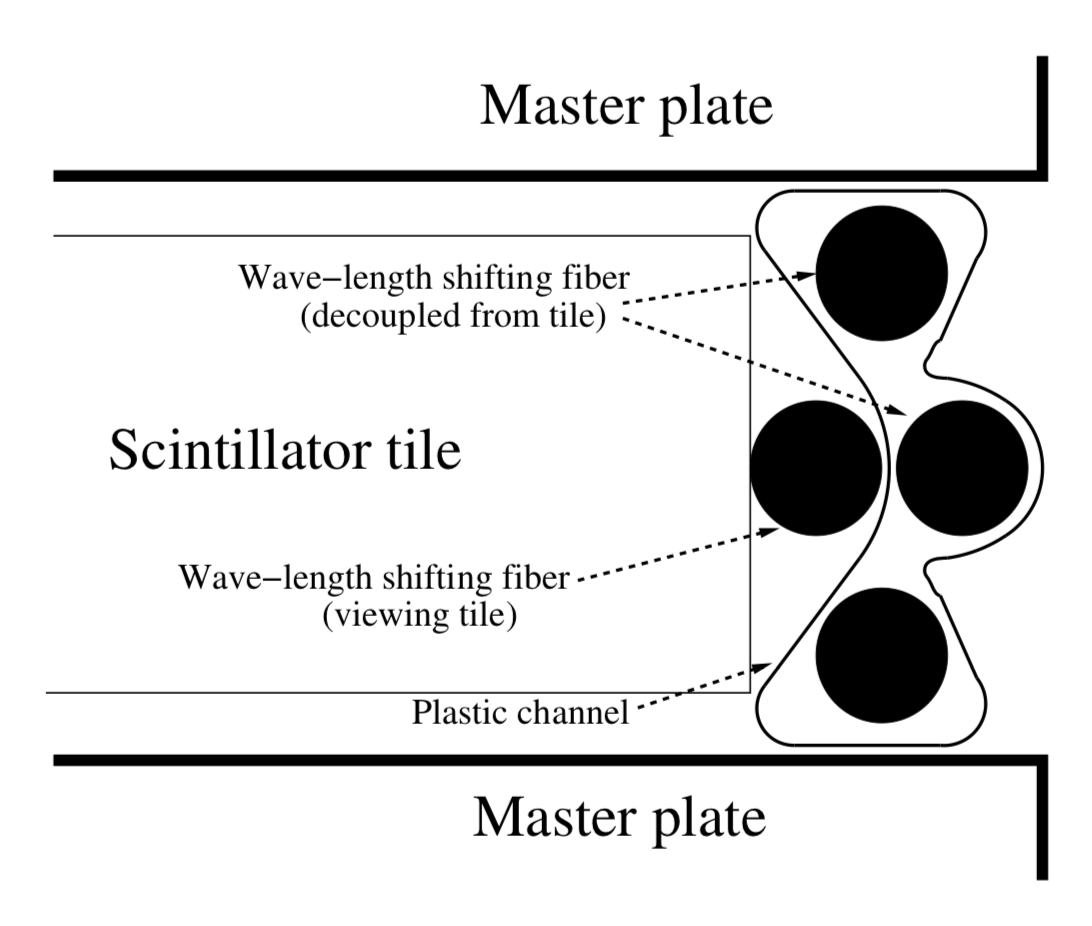}\caption{}
		\label{fig:layout:hcal:profile}
    \end{subfigure}
    \hfill
    \begin{subfigure}[b]{0.4\textwidth}
		\includegraphics[width = \textwidth]{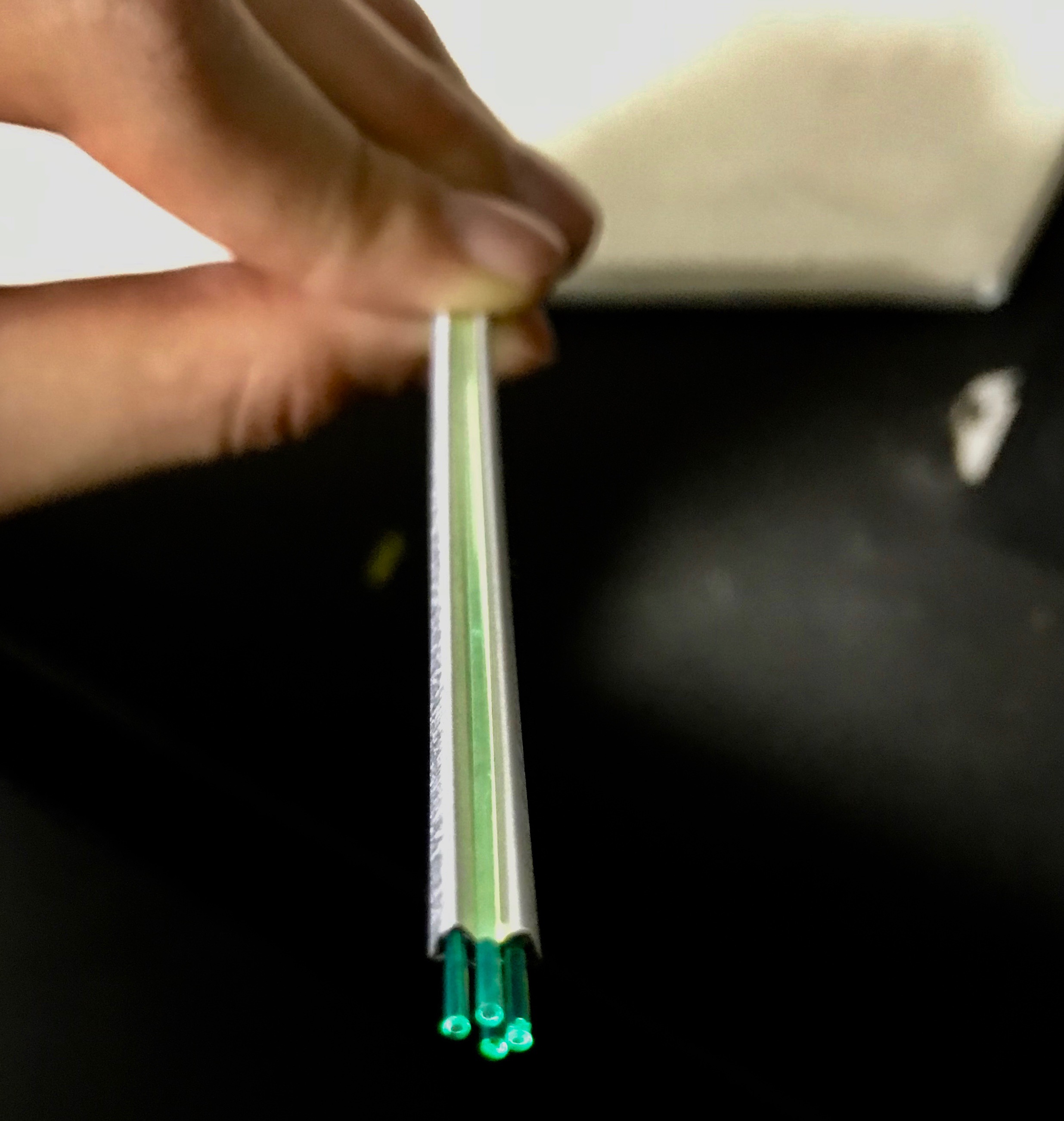}\caption{}
		\label{fig:layout:hcal:profilePic}
    \end{subfigure}
		\caption{(a) Schematic of profile to connect WLS fibres with scintillating plastic tiles and transport towards outer radius.~\cite{CERN-LHCC-96-042} (b) Picture of fibre filled plastic profile as used in the ATLAS TileCal.}
	\end{center}
\end{figure}

The hadronic ``Tile'' calorimeter is a sampling calorimeter using stainless steel, lead and scintillating plastic tiles, with a ratio between volumes of 3.3:1.3:1. The choice of mixing different absorber materials will be further discussed in Sec.~\ref{sec:layout:hcal:opti}. The central barrel and two extended barrels are divided into 128 modules in the $\phi$ direction. Each module has 10 and 8 longitudinal layers in the central barrel and extended barrels respectively. The geometry of the barrel module is sketched in Fig.~\ref{fig:layout:hcal:geometry} and a summary of the main dimensions and parameters is given in Tables~\ref{tab:layout:dimensions}, \ref{tab:layout:mainParameters} and~\ref{tab:layout:hcal:summary}. Each module contains 2 scintillating tiles per longitudinal layer, which will be separated via a reflective material and read out by wavelength shifting (WLS) fibres of 1mm diameter into two separate silicon photomultipliers (SiPMs). This increases the granularity in $\phi$ by a factor of two to $\Delta\phi = 2\pi/256 \approx 0.025$. The orientation of the scintillating tiles perpendicular to the beam line, in combination with wavelength-shifting fibre readout, allows for almost seamless azimuthal calorimeter coverage. For fibre transport and cross-talk suppression between tiles, plastic profiles similar to those shown in Fig.~\ref{fig:layout:hcal:profile} and \ref{fig:layout:hcal:profilePic} will be integrated along the outer sides of each module. The absorber structure consists of 0.5\,cm thick master Stainless Steel plates and lead spacers of 0.4\,cm thickness, as illustrated in Fig.~\ref{fig:layout:hcal:geometry}. The scintillating tiles of 0.3\,cm thickness are filled in the empty gaps. These tiles have a size of 6.9\,cm to 11.3\,cm in length, and 10, to 15, to 25\,cm in height, increasing with the radius and layer. The sequence of scintillator and Pb tiles iterates with the layer in radius, see the zoom in Fig.~\ref{fig:layout:hcal:geometry}. The granularity provided by a one-to-one readout of scintillator tile and SiPM results in a $\eta$ granularity is smaller than 0.006. However, a granularity of 0.025 is expected to be sufficient and thus the default choice and the merging of the SiPM individual signals that form one cell will be merged at the read-out level. Simulation studies of the angular resolution support this choice, and are discussed in detail in Sec.~\ref{sec:performance:hadronic:angular}.

The extended barrel consists of two parts, with only the second part covering the full radial space of 1.74\,m. The first part is only 30\,cm long (along beam direction), which ensures enough space for the supports of the cryogenics system needed for the LAr calorimeter in front. The gap between the barrel and the extended barrel could be additionally instrumented with thin scintillator counters to recover partially the performance in a region occupied by services and electronics from the electromagnetic calorimeter. The WLS fibre readout not only ensures an optimal space usage of active and absorber material only thus very homogenous calorimeter response, but allows for the readout electronics to sit at the outer radius, in an area of moderate radiation levels. These reduced radiation levels at the outer radius lie within the tolerances for current technologies of readout electronics and SiPMs. An additional advantage is the easy access to the electronics, which allows for upgrades in maintenance periods.

%%% Physics of TileCal 
\begin{table}[htp]
\begin{center}
\begin{tabular}{|c|c|c|c|c|}
\hline
granularity & long. layers HB (HEB) & $\left<\lambda\right>$\,[cm] & \#$\lambda$ ($\eta=0$) \\
\hline
default: $\Delta\eta=0.025$, $\Delta\phi=0.025$ 			&  \multirow{2}{*}{10 (8)} 	&  \multirow{2}{*}{21.68} 	&  \multirow{2}{*}{8.3} \\
full:  \hspace*{.3cm} $\Delta\eta<0.006$, $\Delta\phi=0.025$ 	&  					&  					&  \\
\hline
\end{tabular}
\end{center}
\caption{Summary of Tile calorimeter specifications: granularity, longitudinal layers in barrel (HB) and extended barrel (HEB), and nuclear interaction length.
}
\label{tab:layout:hcal:summary}
\end{table}%

\subsubsection{Mechanics}
\begin{figure}[htbp]
    \begin{subfigure}[b]{0.48\textwidth}
	\begin{center}
		  \includegraphics[width = \textwidth]{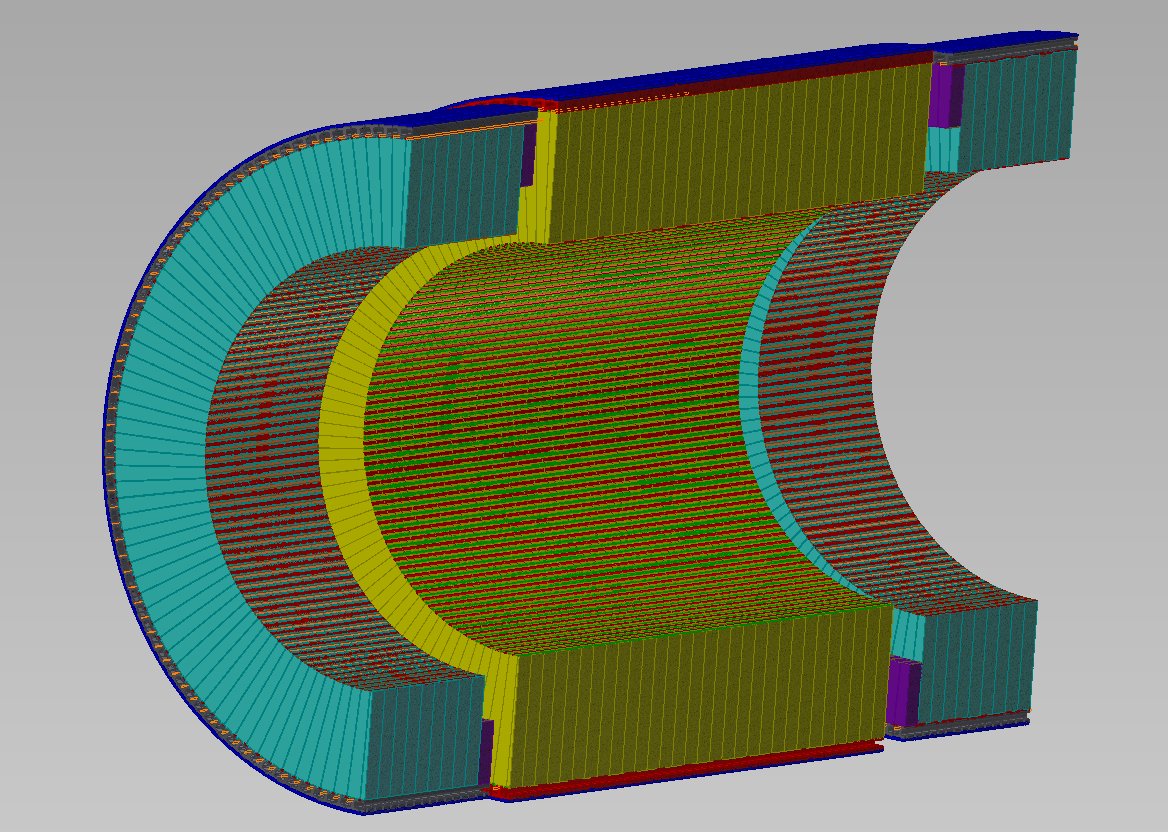}\caption{}
		  \label{fig:layout:hcal:techDrawingBarrel}
	\end{center}
    \end{subfigure}
    \begin{subfigure}[b]{0.48\textwidth}
	\begin{center}
		  \includegraphics[width = \textwidth]{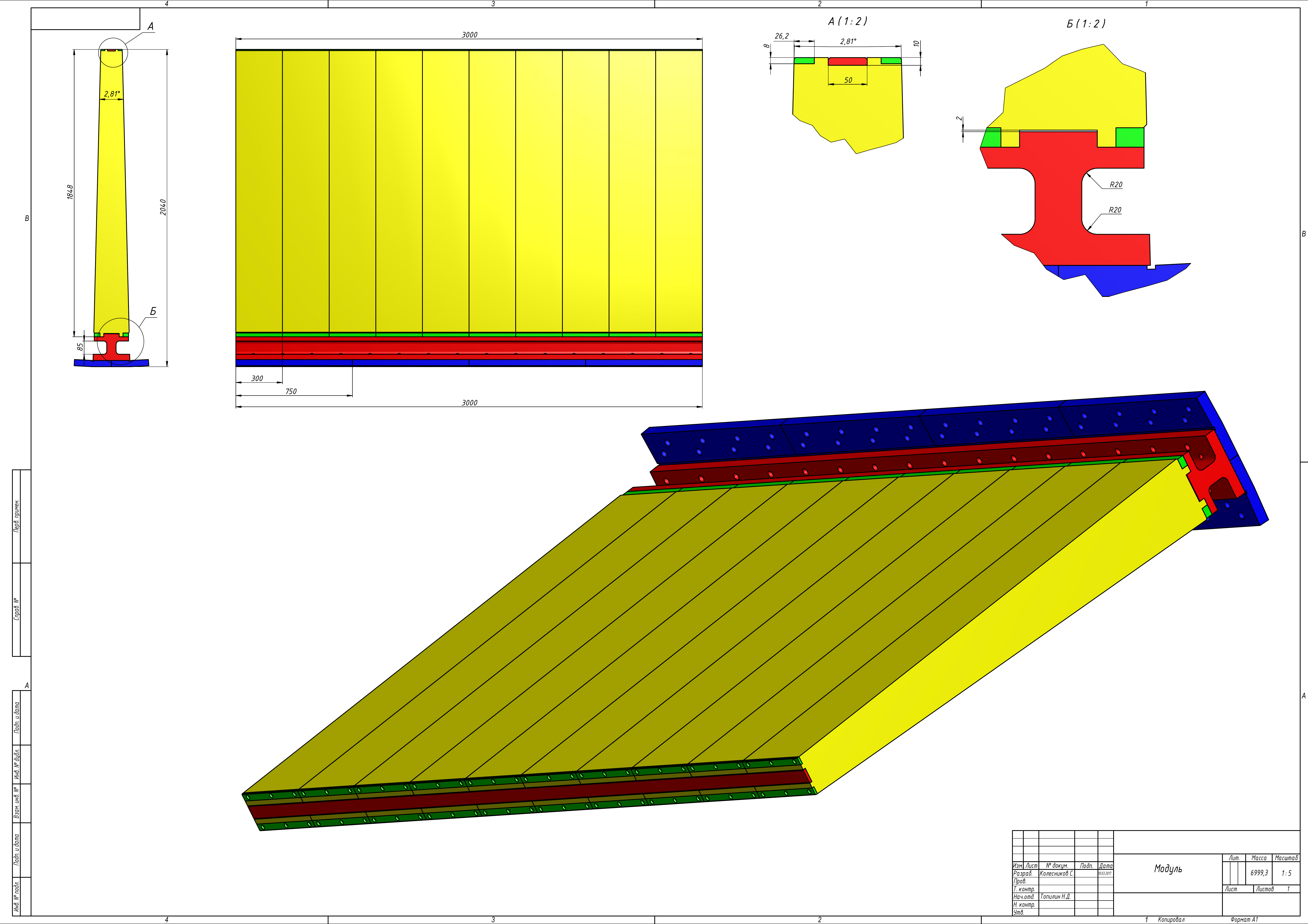}\caption{}
		  \label{fig:layout:hcal:techDrawingModule}
	\end{center}
    \end{subfigure}
		\caption{Technical drawings of the hadronic barrel (HB) and extended barrel (HEB). (a) Half of full calorimeter, with the division of the HEB in part 1 (purple) and part 2(turquoise). (b) Detailed view of one HB module.}
\end{figure}

The mechanical structure has been designed and found to be mechanically feasible to construct. A cut through of the barrel and extended barrel structure is displayed in Fig.~\ref{fig:layout:hcal:techDrawingBarrel}. The foreseen outer steel structure housing the readout electronics and yielding mechanical support is shown in red in Fig.~\ref{fig:layout:hcal:techDrawingModule}. These studies include an estimate of the total weight of the whole calorimeter, which includes the scintillating tiles as well as the outer steel support structures. In total the HB and HEB will weight approximately 4.4\,kt, see Table~\ref{tab:layout:hcal:mechanics}. The central barrel consists of 128 modules with 21\,t each, weighing in total 2.7\,kt.

%%% Specs of TileCal
% old estimates, Steel:Sci
%\begin{table}[htp]
%\begin{center}
%\begin{tabular}{|c|c|c|c|}
%\hline
%			 &  radial depth (active) [m] & length [m] & weight [t] \\
%\hline
%HB 		 & 2.04 (1.8) & 9.2 & $2,732$ \\
%\hline
%HEB	part 1& 1.09 (0.85) & 0.3 & \\
%HEB	part 2& 1.74 (1.5) & 3.2 & $837$ \\
%\hline
%total 			& & 15.8 & $4,406$\\
%\hline
%\end{tabular}
%\end{center}
%\caption{Summary of dimensions and total weight of the HB and HEB; one HB module weighing $\thicksim21$\,t.}
%\label{tab:layout:hcal:mechanics}
%\end{table}%

% new estimates by Helder Steel:Pb:Sci
\begin{table}[htp]
\begin{center}
\begin{tabular}{|c|c|c|c|}
\hline
	 &  material & volume [m$^3$] & weight [t] \\
\hline
\hline
	& Pb & 76.5 & 845 \\
	& Scintillator & 57.4 & 59 \\
	& Steel plates & 193.8 & 1521 \\
	& Steel support &  39.1 & 307 \\
\hline
HB	& & &              2,732 \\
\hline
	& Pb & 22.8 & 258 \\
	& Scintillator & 17.0 & 18 \\
	& Steel plates & 57.8 & 453 \\
	& Steel support &  13.8  & 108 \\
\hline
$1\times$ HEB		& &  & 837\\
\hline
total 		& & & 4,406\\
\hline
\end{tabular}
\end{center}
\caption{Summary of dimensions and total weight of the HB and HEB; one HB module weighing $\sim21$\,t.}
\label{tab:layout:hcal:mechanics}
\end{table}%

\subsubsection{Light Collection, Readout and Electronics}
\label{sec:layout:hcal:readout}
The incredibly challenging environment of 100\,TeV centre-of-mass proton collisions every 25\,ns with up to $\left<\mu\right> = 1000$ collisions per bunch-crossing, sets stringent requirements on the sensitive material as well as the signal readout devices and electronics. The maximum radiation dose to be expected in the HB region is 8\,kGy for the scintillating plastic tiles and WLS fibres, see Tab.~\ref{tab:layout:dimensions}. Ongoing R\&D on radiation hard scintillator for the upgrades of the LHC experiments show promising results and prove that these technologies will be able to withstand the radiation levels expected at the FCC-hh~\cite{Jivan:2015sua}. 

The scintillation light guided through the WLS fibre is read out by Silicon Photomultipliers (SiPMs), which are matrices of single-photon avalanche diodes operated in Geiger mode. These devices allow single photon detection, and achieve photon detection efficiencies (PDEs) between 20 and 60\,\%. Each scintillating tile will be connected via a WLS fiber to one SiPM. The single SiPMs will be arranged within arrays on PCBs, digitised, summed and sent via optical links to the counting room where they can be used for the hardware trigger and, after a positive trigger decision, will be written to disk.

At the outer radius of the hadronic barrel, the radiation levels to be expected are of the order of $10^{11}/\mathrm{cm}^2$ 1\,MeV neutron equivalent fluence. The resulting damage of the silicon substrate and the effect on the dark count rate, leakage current, over voltage, and PDE has been studied in the context of the CMS hadronic calorimeter upgrade for the HL-LHC and proven to function up to  $2.2\times10^{14}$\,n/cm$^{2}$~\cite{Heering:2016lmu}. Nonetheless, the strong temperature dependence of the devices will require either temperature control or cooling and precise temperature monitoring. 

First tests have started on single channel level, focusing so far on the response of the optical components used in the ATLAS Tile calorimeter: scintillating tiles made of polystyrene doped with 1.5\,\% pTp and 0.04\,\% POPOP and double cladding Y11 wavelength-shifting fibres from Kuraray~\cite{Abdallah_2013}.

\begin{figure}[htbp]
	\begin{center}
    \begin{subfigure}[b]{0.48\textwidth}
		\includegraphics[width = 1.\textwidth]{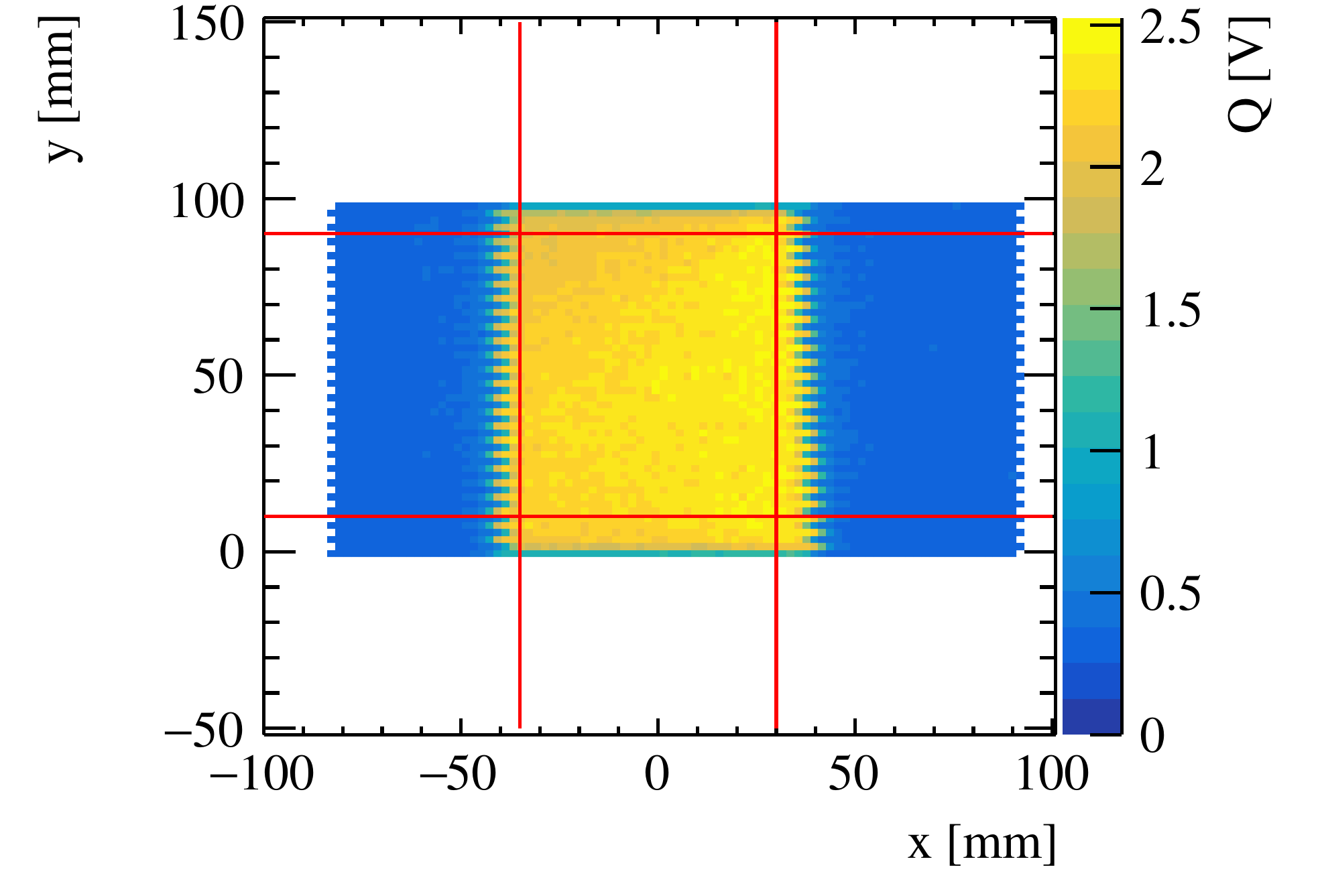} \caption{}
		\label{fig:technical:hcal:SiPMTileScan}
\end{subfigure}
    \begin{subfigure}[b]{0.48\textwidth}
		\includegraphics[width = 1.\textwidth]{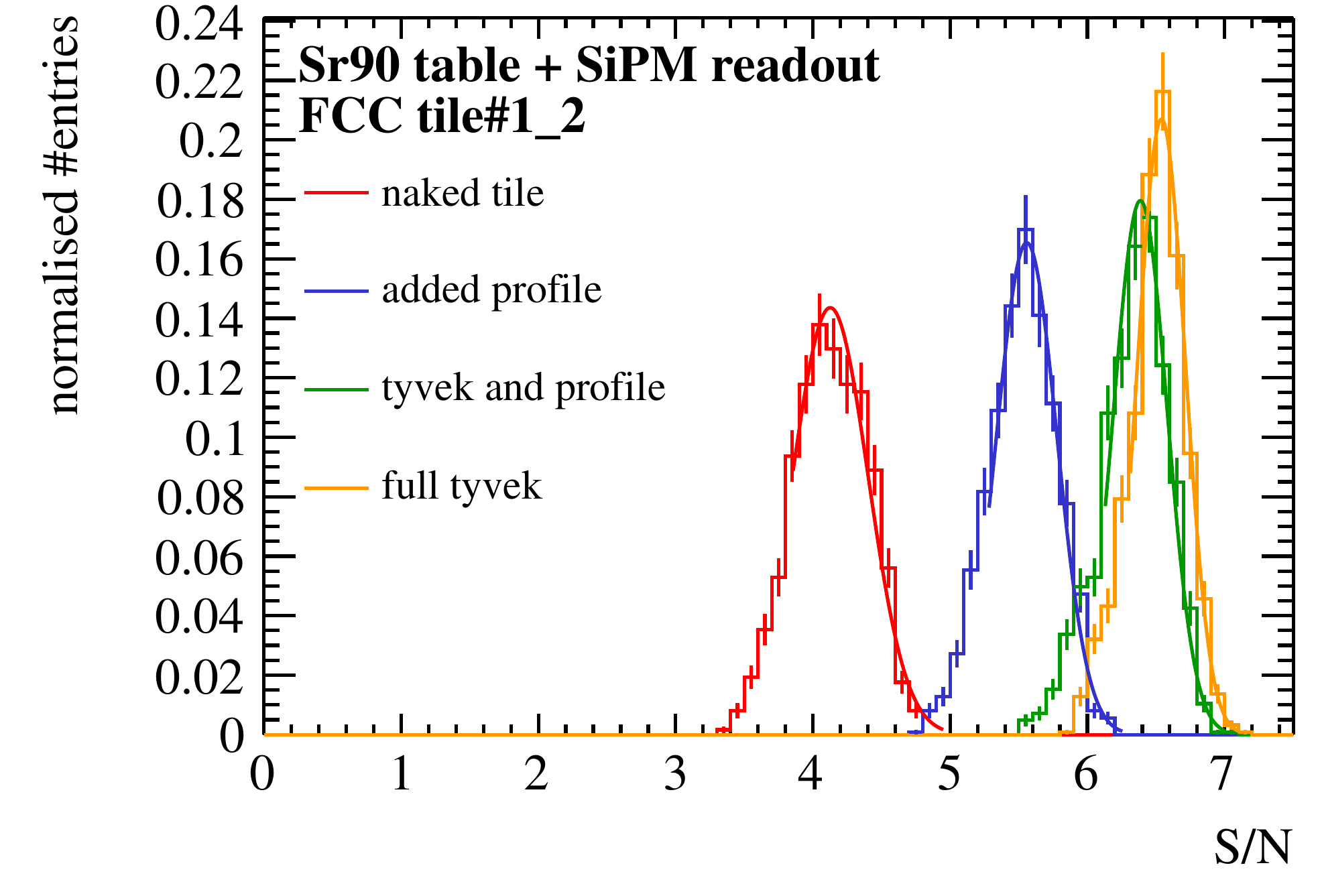} \caption{}
		\label{fig:technical:hcal:SiPMTileScanDistr}
\end{subfigure}
    \caption{(a) 2D scan of FCC tile at the inner most radius, wrapped in tyvek, and read out with a WLS fibre on the right side. The red lines indicate the area cut used for (b) the response distributions for different tile configurations. The curves correspond to Gaussian fits within a range of $-1/+2\,\sigma$. }
	\end{center}
\end{figure}

\begin{figure}[htbp]
	\begin{center}
    \begin{subfigure}[b]{0.48\textwidth}
		\includegraphics[width = 1.\textwidth]{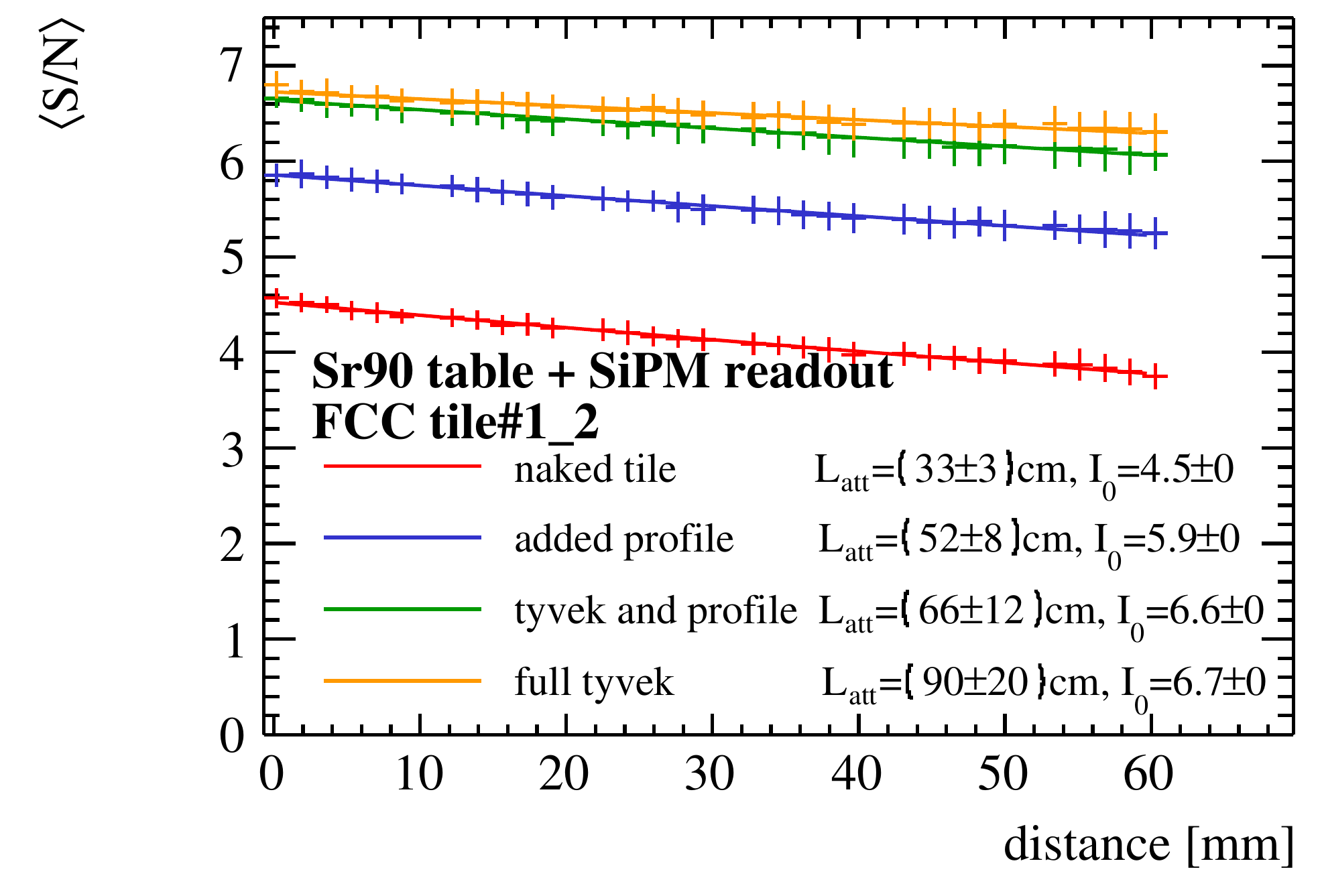} \caption{}
\end{subfigure}
    \begin{subfigure}[b]{0.48\textwidth}
		\includegraphics[width = 1.\textwidth]{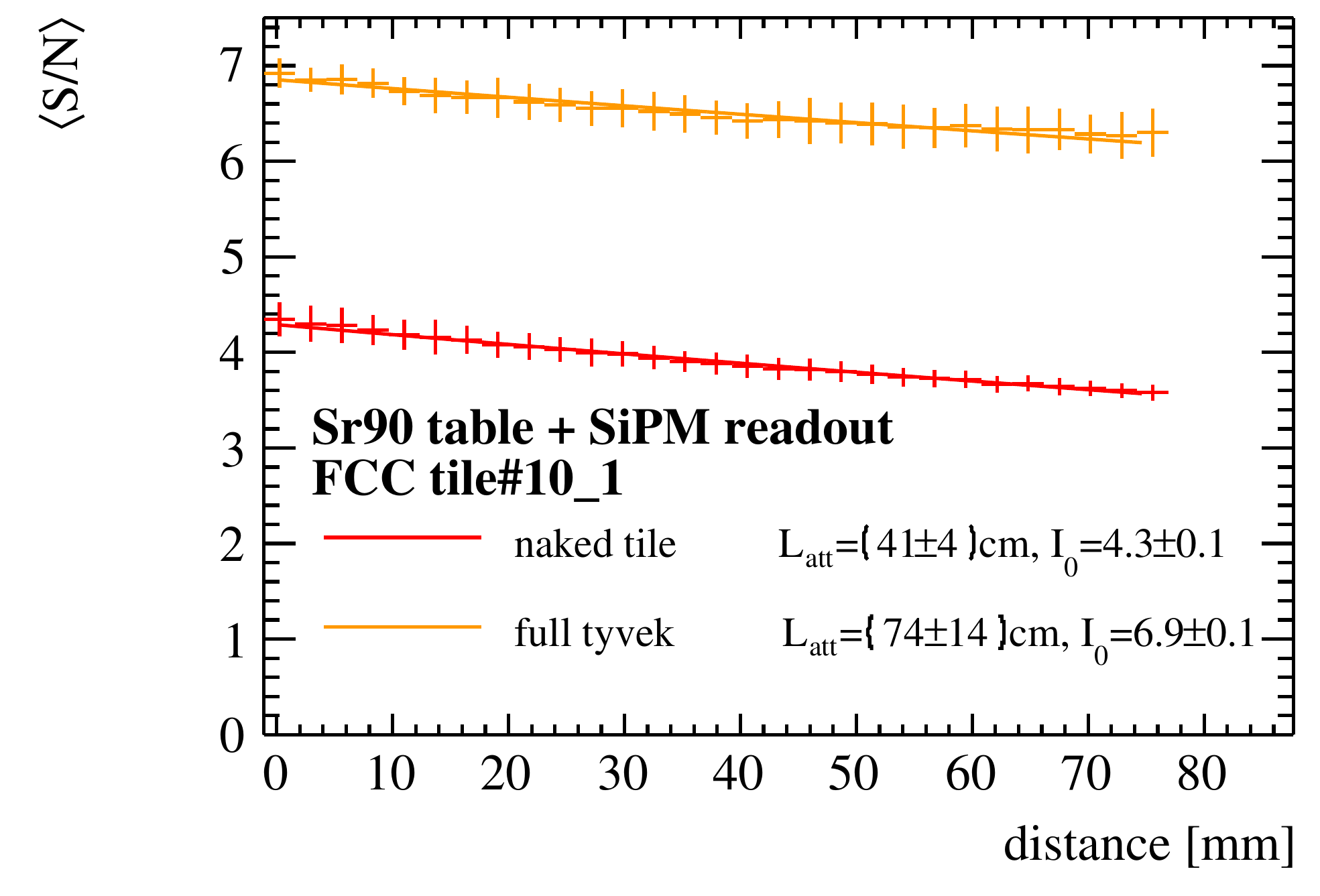}\caption{}
\end{subfigure}
		\caption{The mean S/N ratio as a function of the distance to the readout tile edge of (a) a tile sized for the inner radius, and (b) a tile at the most-outer radius. The tile attenuation length was obtained through an exponential fit in each case. Error bars correspond to the rms of the response measured on a grid covering the surface of the tile.}
		\label{fig:technical:hcal:SiPMTileScanAtt}
	\end{center}
\end{figure}

Tiles were cut to the dimensions of the first and tenth FCC-hh HB layer. These tiles are then coupled to WLS fibres of required length by contact with one tile edge, and wrapped in tyvek\textregistered{}~\cite{tyvek}, to enhance light collection efficiency. The right-angled trapezoid surface of the tested tiles is scanned using a $Sr^{90}$ source mounted on a 2D stage. %Several configurations were tested by measuring the light yield relative to a fixed reference tile. 
These tests focused on the response uniformity, fibre coupling, fibre length and wrapping options.
The attenuation length of the WLS fibres is $>2$\,m, thus acceptable for the transport of the light produced in the first HB layer tiles at the inner-most radius.
With the scintillation light collected by the WLS fibre of 1\,mm diameter on one side of the tile, as foreseen in the FCC-hh TileCal modules, the WLS fibres are connected by simple contact to a $1\times1\,$mm$^2$ Multi-Pixel Photon Counter (MPPC)\footnote{from Hamamatsu, type S12571-015C,  \url{www.hamamatsu.com}}. The SiPM output signals are integrated over $\thicksim$1\,ms, and read out with a multimeter. Different wrapping materials and configurations have been studied, using simple back-reflection on a WLS fibre plastic profile (shown in Fig.~\ref{fig:layout:hcal:profilePic}) on the opposite tile edge, up to full tyvek wrapping. Figure~\ref{fig:technical:hcal:SiPMTileScan} shows the measured response over the full tile area with a spread in response of <5\,\% (including a reflective material on the opposite tile edge to the readout fibre). 
The attenuation length $L_{att}$ of the tile is determined from the fit of the response $\left<S/N\right>$ to an exponential function $I_{0}\cdot\exp{\left(-\Delta x/L_{att}\right)}$, as a function of the distance from the readout tile edge $\Delta x$, see Fig.~\ref{fig:technical:hcal:SiPMTileScanAtt}. The response height is measured in a quantity related to the signal to noise ratio, by the normalisation of the measured charges to the measurement points outside the tile volume. The attenuation length in case of a layer \#1 and fully wrapped tile reaches 90\,cm, with a S/N ratio of about 6.7. These results are comparable with previous measurements using the standard ATLAS Tile PMT readout. The distributions of the response for the different tile configurations are displayed in Fig.~\ref{fig:technical:hcal:SiPMTileScanDistr}, and the mean values, as well as the widths, are summaries for both tile types in Table~\ref{tab:technical:hcal:SIPMmeasurementSummary}.

\begin{table}[htp]
\begin{center}
  \begin{tabular}{|c|c|c|c|c|c|c|c|c|}
    \hline
				& \multicolumn{4}{c|}{FCC tile \#1} 						& \multicolumn{4}{c|}{FCC tile \#10} \\
    & rms/mean 	& $\sigma/\mu$ 	&	$L_{att}$	& $I_{0}$	& rms/mean 	& $\sigma/\mu$ 	&	$L_{att}$	& $I_{0}$	\\
    \hline
unit				&		\%	&	\%			&	cm 		& 		&		\%	&	\%			&	cm 		& 		\\
\hline
naked tile			&	6.5		&	6.7			& 33 			& 4.5 	& 7.6	 & 6.3 & 41& 4.3\\
naked tile + profile	&	4.5		& 	4			& 52			& 5.9 	& - & - & - & - \\
tyvek + profile		& 	3.9		& 	3.1			& 66 			& 6.6 	& - & - & - & -  \\
full tyvek			&	3.2		& 	2.7			& 90 			& 6.7 	& 4.4 & 3.8 & 74 & 6.9 \\
\hline
\end{tabular}
\end{center}
\caption{Summary of uniformity tests for different tile configurations for tile sizes in the first (\#1) and last (\#10) HB layer. Profile stands for back-reflection on a WLS fibre plastic profile on the opposite tile edge.
  \label{tab:technical:hcal:SIPMmeasurementSummary}
  }
\end{table}%

%%%\cite{Abdallah:2013}

While it is not yet possible to estimate the exact characteristics of the optical materials and SiPMs for the final detector, these preliminary tests are very  encouraging and clearly indicate the usefulness of a campaign to optimise the design of various aspects of the optics system.

\subsubsection{Calibration Systems}
\label{sec:layout:hcal:calibration}

\subsubsubsection{Caesium Calibration}
The Caesium calibration system could be based on a movable $^{137}$Cs $\gamma$ source ($E_\gamma=661.7$\,keV) that is moved through the calorimeter body via source tubes penetrating all scintillators of a module (see source tubes in Fig.~\ref{fig:layout:hcal:geometry}). The individual channel response to the energy deposits is used to equalise the global response and calibrate the calorimeter to the electromagnetic scale. 
ATLAS is successfully using this technique and achieves a precision of 0.5\,\%~\cite{Aad:2010af}.

\subsubsubsection{SiPM Characterisation and Calibration} 
The SiPMs will be characterised before being connected to the WLS fibres to determine the breakdown voltage, gain, and response-temperature coefficients. At the start, the operating voltage will be adjusted to equalise the response of all the cells. The cells inter-calibration will be done with the caesium calibration system, while a fraction of the modules should be tested both with the caesium source and in testbeams (with electrons and muon beams) to settle the absolute electromagnetic scale. The variations over time, to account for temperature variations, ageing and radiation damage will be monitored with the caesium calibration system.  To monitor and calibrate the stability of the SiPMs a calibration system using LEDs or lasers injecting light into some fibres will be implemented. Together with the caesium calibration this will allow to disentangle variations of SiPMs from the optics system (tile and fibres). 

\begin{figure}[htbp]
	\begin{center}
		  \includegraphics[width = .8\textwidth]{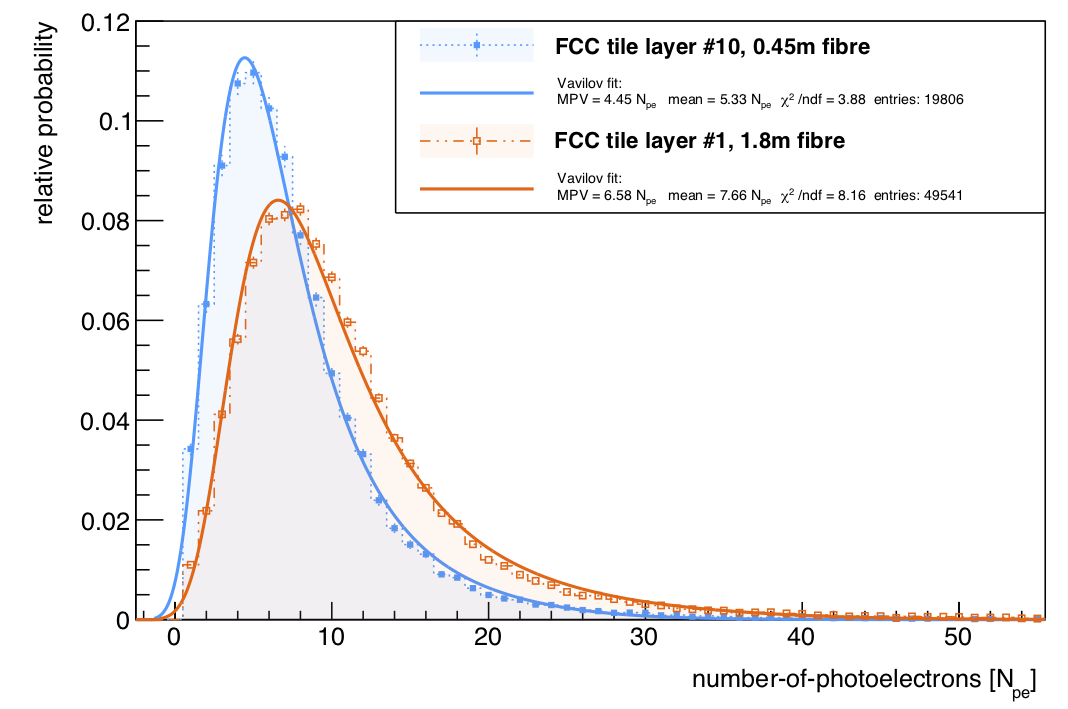}
		\caption{Energy loss distribution of cosmic muons in the smallest and largest FCC-hh tile corresponding to inner-most and outer-most layer radius. The tiles were wrapped in tyvek and connected through 0.45 and 1.80 m long WLS fibres. \label{fig:layout:hcal:pe_per_mip}}
	\end{center}
\end{figure}

\begin{figure}[htbp]
	\begin{center}
		  \includegraphics[width = .9\textwidth]{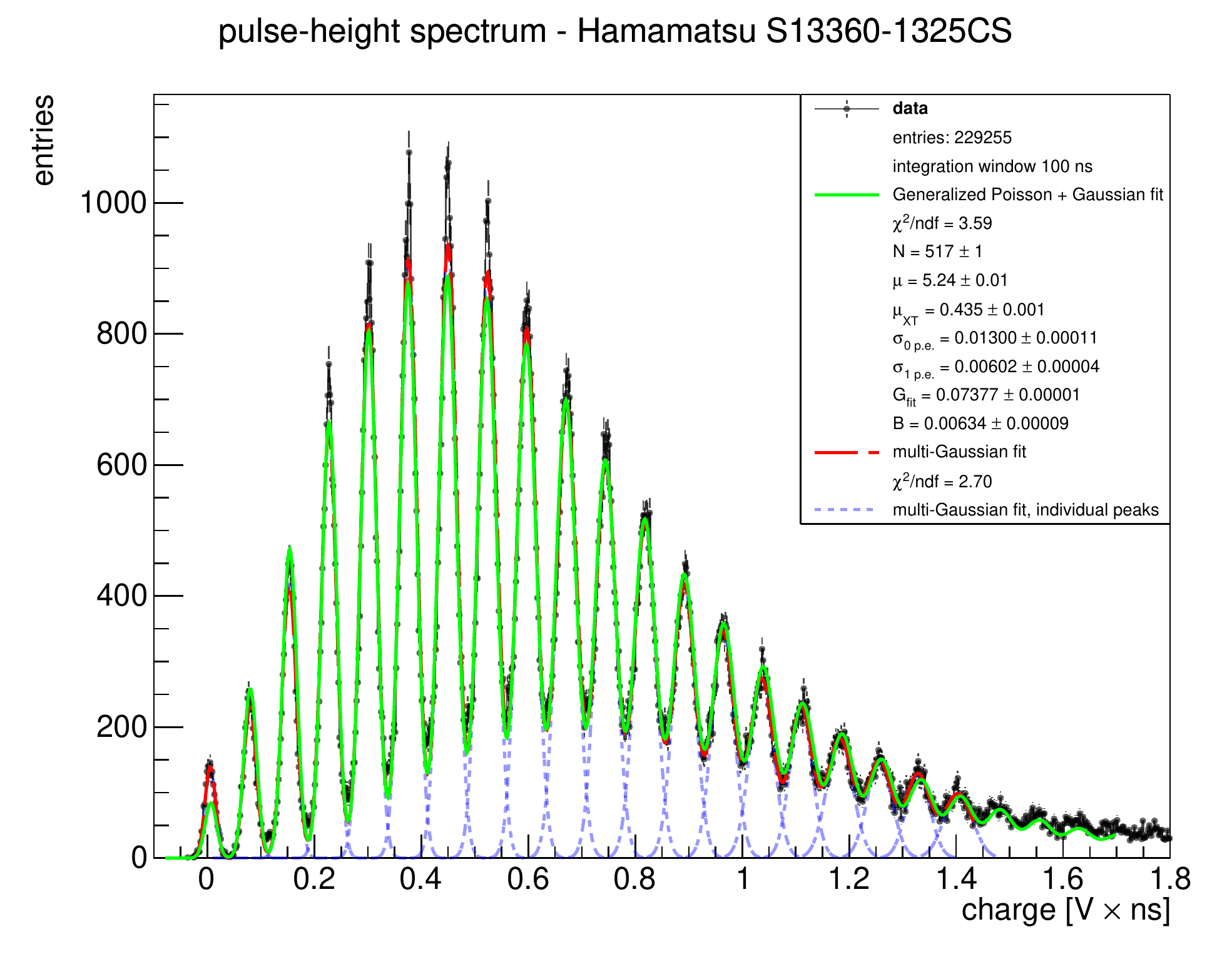} 
\caption{Single photon spectrum of 230 thousand cosmic muon events in FCC-hh tiles in different configurations~\cite{Schliwinski:2687718}. The green line corresponds to a generalised poisson fit.  \label{fig:layout:hcal:gain}}
	\end{center}
\end{figure}

First measurements of the light yield for cosmic muons have proven the sensitivity and determined the expected response of the FCC hadronic barrel calorimeter to MIPs. Figure~\ref{fig:layout:hcal:pe_per_mip} shows the response of the smallest and largest FCC tile in first and tenth layer attached to 0.45 and 1.8\,m long WLS fibre and read out by a SiPM (as described in Section~\ref{sec:layout:hcal:readout}). The measured light yield results in 5 to 7\,photo-electrons per MIP. Additionally, it could be shown that cosmic muon runs can be used for calibrations of the SiPM gain due to the small responses within the range of a few photo-electrons. The single photon spectrum for $\thicksim230,000$ events is shown in Figure~\ref{fig:layout:hcal:gain}. The spectrum is fitted with a generalised poisson function and enables the extraction of the gain from the distances between the peaks~\cite{Schliwinski:2687718}.

\subsubsection{Optimisation of Absorber Materials}
\label{sec:layout:hcal:opti}
\begin{figure}[htbp]
	\begin{center}
    \begin{subfigure}[b]{0.48\textwidth}
		  \includegraphics[width = \textwidth]{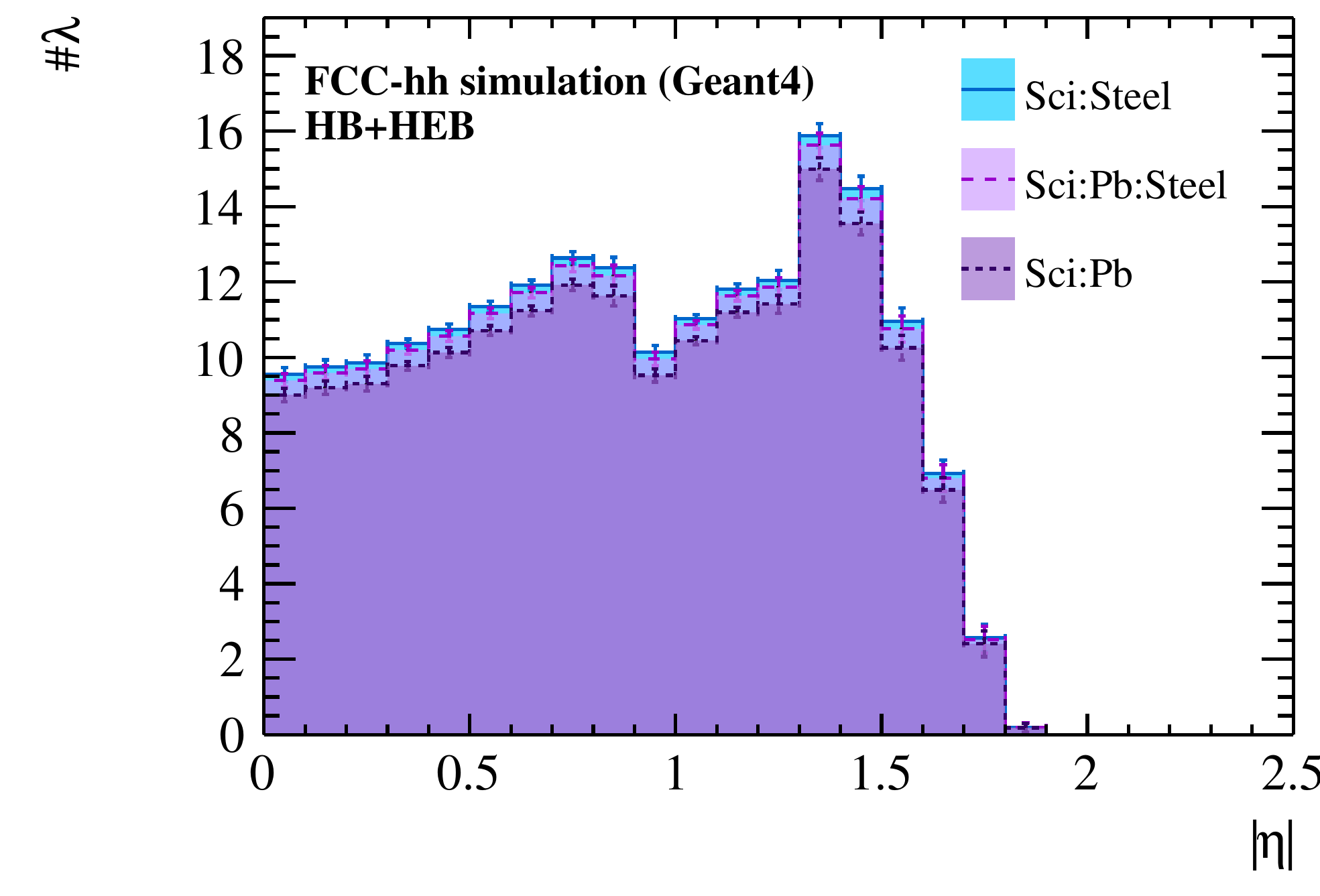} \caption{}
		  \label{fig:layout:hcal:depthsLambda}
    \end{subfigure}
    \begin{subfigure}[b]{0.48\textwidth}
		  \includegraphics[width = \textwidth]{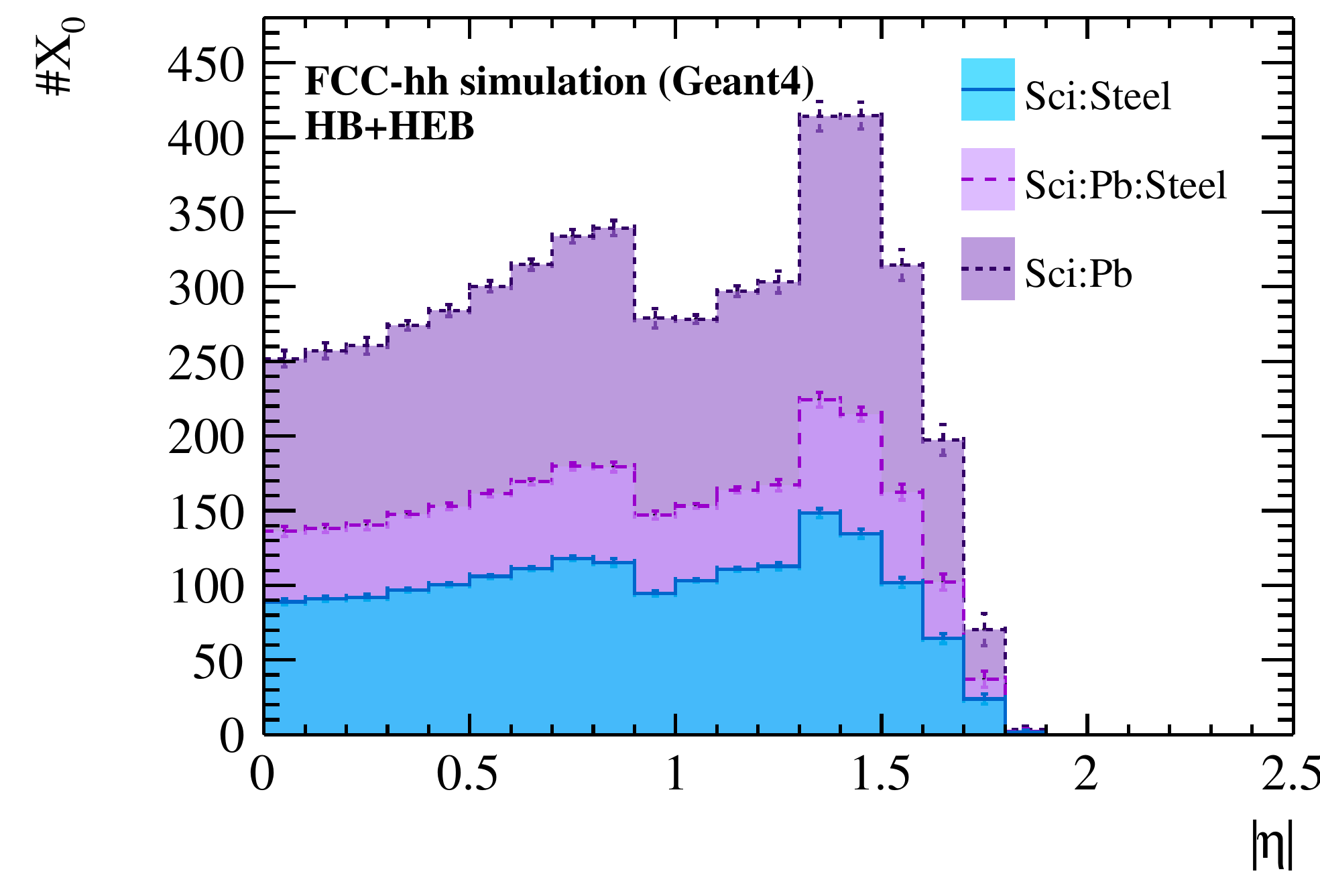} \caption{}
		  \label{fig:layout:hcal:depthsX0}
    \end{subfigure}
		\caption{Depth of the hadronic scintillator tile calorimeter as a function of $\eta$ in nuclear interaction lengths (a) and radiation lengths (b). Shown is the impact of the material choice for full Steel:Sci, Steel:Pb:Sci mix and full Pb:Sci option. The material of the outer support structure is included.}
	\end{center}
\end{figure}

Even though the overall design follows the ATLAS Scintillator-Steel calorimeter, the FCC-hh HB and HEB uses an absorber structure consisting of a major Stainless Steel structure (\textit{masters}) with additional lead tiles (\textit{spacers}), while keeping the absorber dimensions thus volume fractions the same. 
The partial replacement of Stainless Steel with lead absorbers, resulting into a calorimeter closer to compensation, aims to improve the hadronic performance in terms of linearity and resolution. But the expected slight decrease of total calorimeter thickness in terms of nuclear interaction lengths $\lambda$ and consequently poor containement of hadronic showers has been evaluated. The impact on the calorimeter depth (in units of $\lambda$ and $X_0$)  has been studied for three absorber scenarios: full Steel (Sci:Steel with a ratio of 1:4.7), Pb mixture (Sci:Pb:Steel with a ratio of 1:1.3:3.3) and full Pb (Sci:Pb with a ratio of 1:4.7), see Fig.~\ref{fig:layout:hcal:depthsLambda} and~\ref{fig:layout:hcal:depthsX0}. Whereas the decrease of depth in terms of interaction lengths is small, the increase in terms of radiation lengths is rather dramatic when adding more Pb. While this is not a problem for the calorimetry performance, this is a source of an increased amount of multiple scattering of muons before reaching the muon system.

\begin{figure}[htbp]
	\begin{center}
    \begin{subfigure}[b]{0.48\textwidth}
		\includegraphics[width = \textwidth]{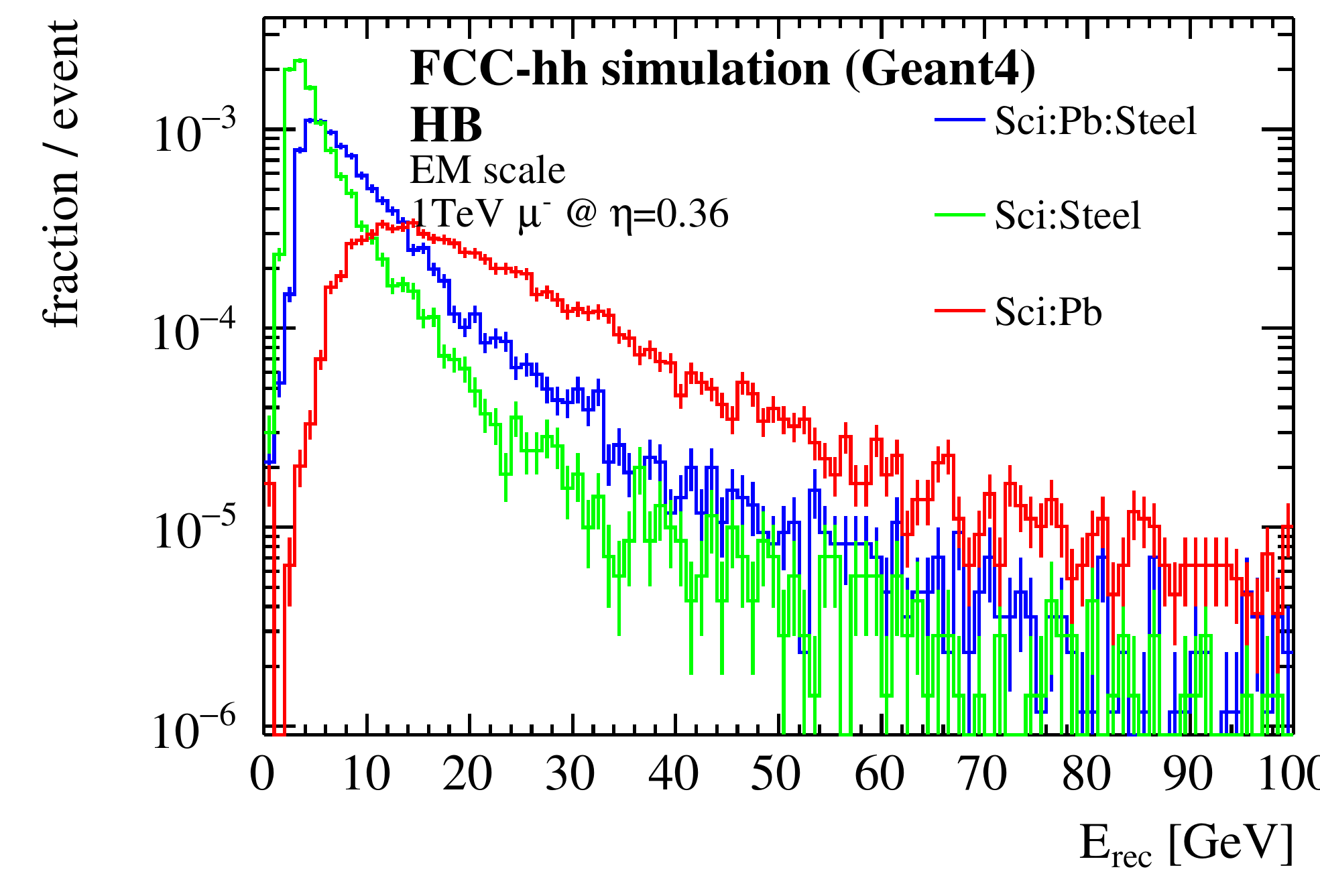}\caption{}
		\label{fig:layout:hcal:muonsEtot}
    \end{subfigure}
    \begin{subfigure}[b]{0.48\textwidth}
		\includegraphics[width = \textwidth]{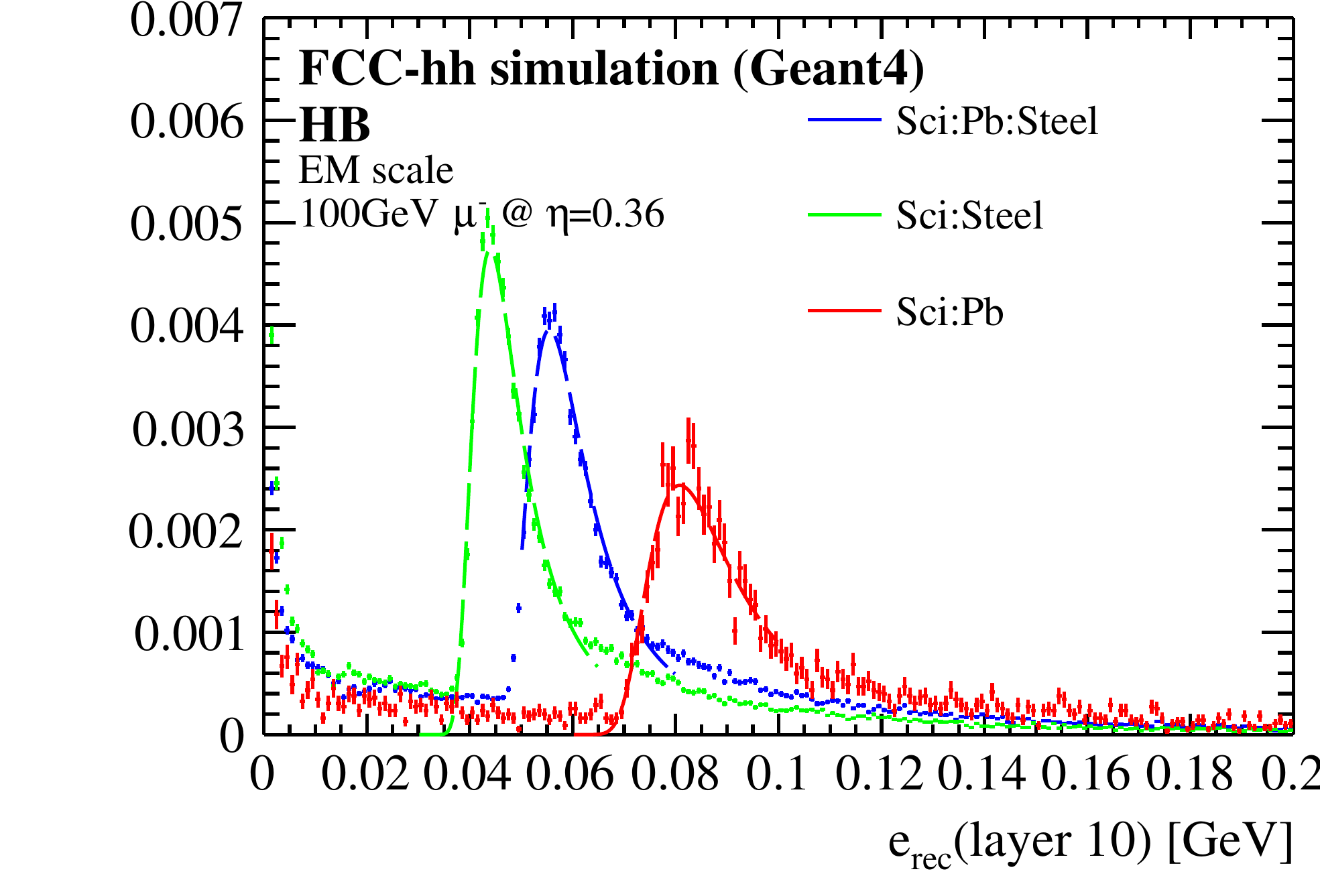}\caption{}
		\label{fig:layout:hcal:muonsEcell}
    \end{subfigure}
    \begin{subfigure}[b]{0.48\textwidth}
		\includegraphics[width = \textwidth]{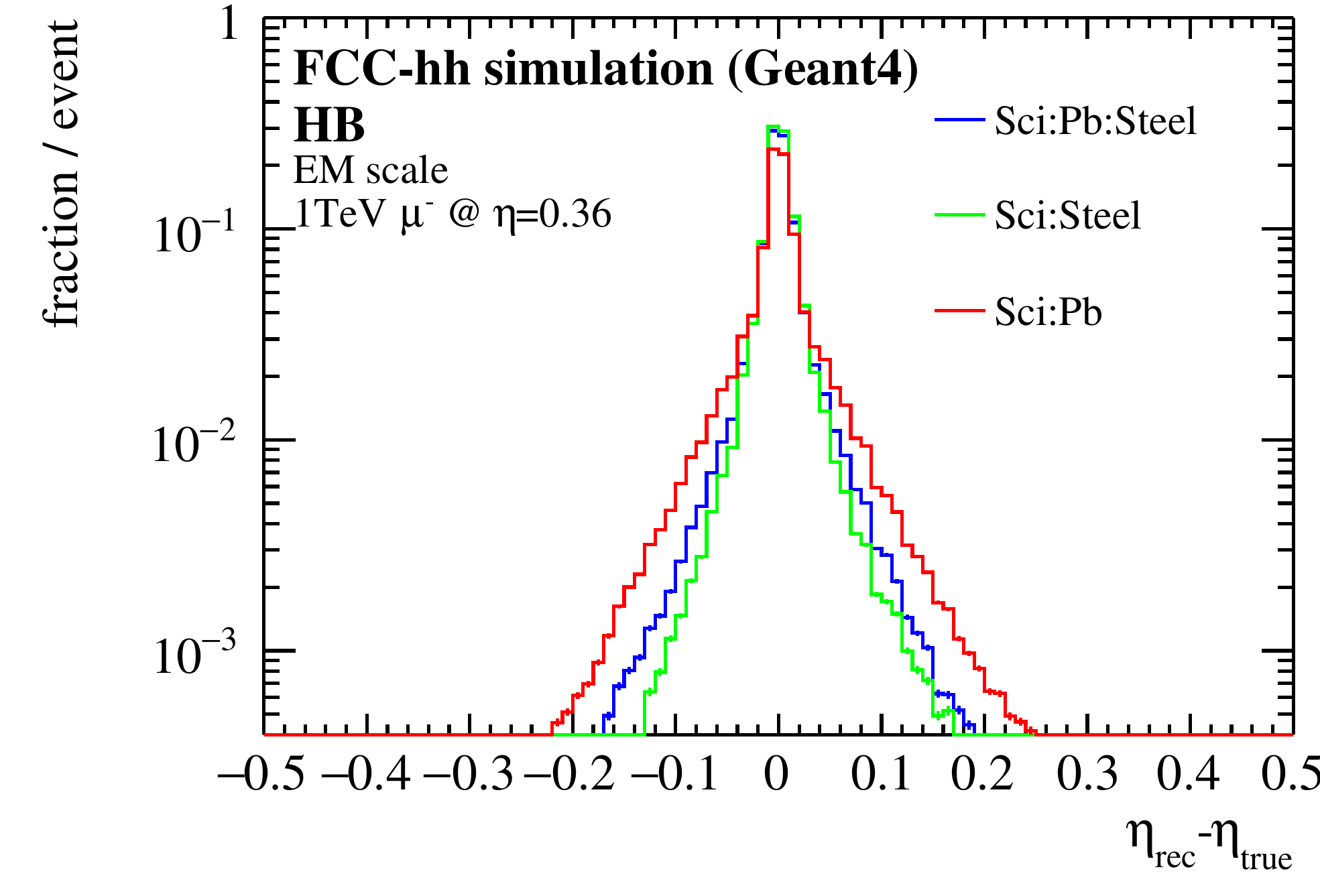}\caption{}
			\label{fig:layout:hcal:muonsEta}
    \end{subfigure}
		\caption{Total energy loss (a), energy deposit per tile in last HB layer 10 (b), and angular distribution (c) of 100\,GeV and 1\,TeV muons in three HB absorber options.}
	\end{center}
\end{figure}

While the needed depth in nuclear interaction length ($\lambda$) to contain jets and hadronic showers up to 98\,\% can be parameterised with a function similar to the one shown in Fig.~\ref{fig:intro:jetContainmentDepth}, the impact of the absorber choice on the contained jet and pion energies can be estimated by 
\begin{equation}
	\text{E}^{98\%} = \exp{\left(\frac{\lambda-b}{a}\right)},
\end{equation}
with parameters $a_{jet} = 0.495$,  $b_{jet} = 6.3$ for jets and $a_{\pi^{+}}=0.64$, $b_{\pi^{+}} = 5.4$ for single pions, from~\cite{Carli:2016iuf}. Furthermore, the energy is estimated with additional 2\,$\lambda$ of the EMB in front (see Sec.~\ref{sec:layout:lar}) and the resulting values are summarised for the three absorber scenarios in Table~\ref{tab:layout:hcal:depths}. 
As expected, it shows the strong increase of calorimeter depth in terms of radiation lengths with an increased amount of Pb absorbers (second column). This increase however, does not only affect hadronic showers, but strongly impacts traversing muons which will experience increased multiple scattering, deteriorating the accuracy of the momentum measurement. The resulting effect on the muons' energy loss and angular distributions is shown in Fig.~\ref{fig:layout:hcal:muonsEtot}, ~\ref{fig:layout:hcal:muonsEcell}, and ~\ref{fig:layout:hcal:muonsEta}. The peak total energy loss of muons in the three scenarios are summaries in the fourth column of Tab.~\ref{tab:layout:hcal:depths}. Figure~\ref{fig:layout:hcal:muonsEcell} shows the expected energy per cell for 100\,GeV muons in the last HB layer (layer 10). The most probable value of the measured energy ranges from 40 to 80\,MeV which is well above the expected electronics noise per cell of $10$\,MeV and shows the calorimeters sensitivity to MIPs.

\begin{figure}[htbp]
	\begin{center}
    \begin{subfigure}[b]{0.48\textwidth}
		  \includegraphics[width = \textwidth]{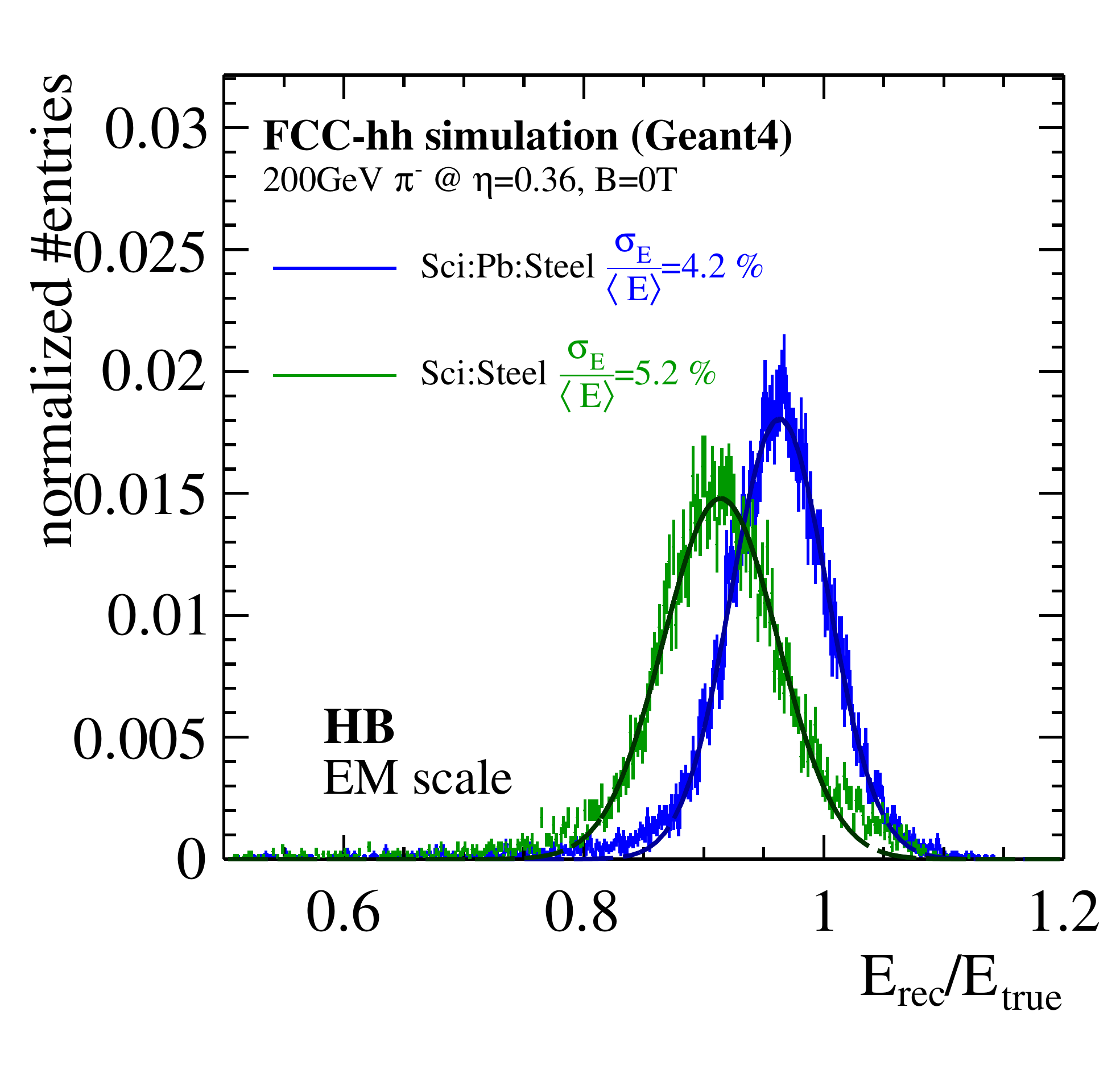}\caption{}
		  \label{fig:layout:hcal:singlePionEnergyDistr}
    \end{subfigure}
    \begin{subfigure}[b]{0.48\textwidth}
		  \includegraphics[width = \textwidth]{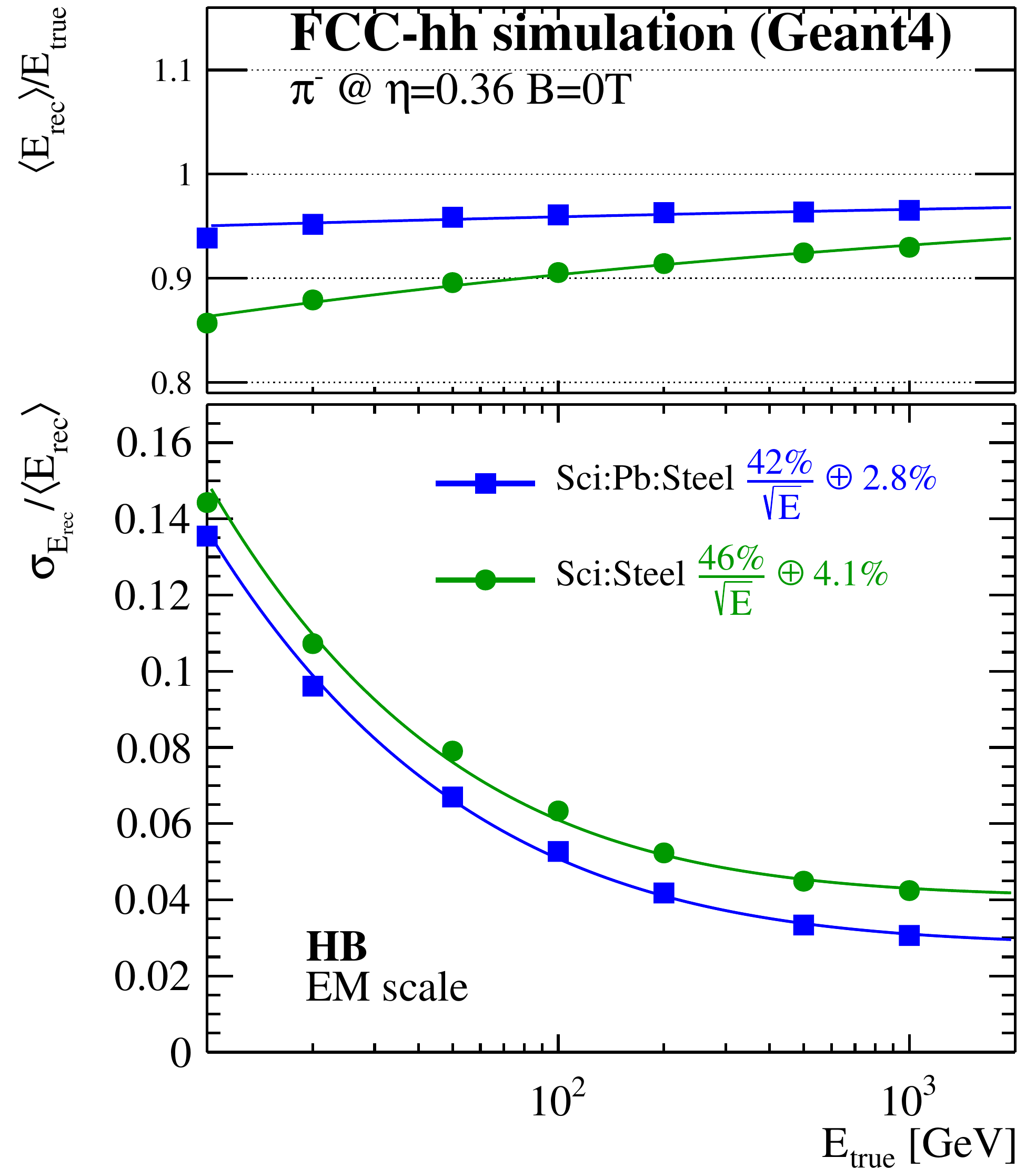}\caption{}
  		\label{fig:layout:hcal:singlePionEnergyResolution}
    \end{subfigure}
		\caption{(a) Reconstructed energy distributions for 200\,GeV pion showers, where the curves shows the Gaussian fit. (b) Single pion energy resolution (bottom) and linearity (top) for the FCC-hh HB for particles at $\eta=0.36$ in an energy range of 10\,GeV to 1\,TeV. The resolution is compared to the ATLAS type design with a Sci:Steel ratio of 1:4.7. 
    }
	\end{center}
\end{figure}

The jet and hadronic shower containment is less affected by the lead fraction due to the rather moderate decrease in depth in terms of nuclear interaction lengths with higher Pb content (third column). However, the detector response to hadronic showers is strongly affected by the higher atomic number of lead compared to iron, which suppresses the response to the electromagnetic component of the shower. Thus the intrinsic non-compensation $e/h>1$, due to the partially invisible deposits of the hadronic shower components, can be brought closer to compensation.
The effect of an $e/h$ ratio closer to 1 (closer to compensation) can be seen in an improved energy resolution and linearity as shown in Fig.~\ref{fig:layout:hcal:singlePionEnergyResolution}. It should be pointed out that the constant term is better for the Sci:Pb:Steel option despite the reduced shower containment.

\begin{figure}[htbp]
	\begin{center}
    \begin{subfigure}[b]{0.49\textwidth}
		  \includegraphics[width = \textwidth]{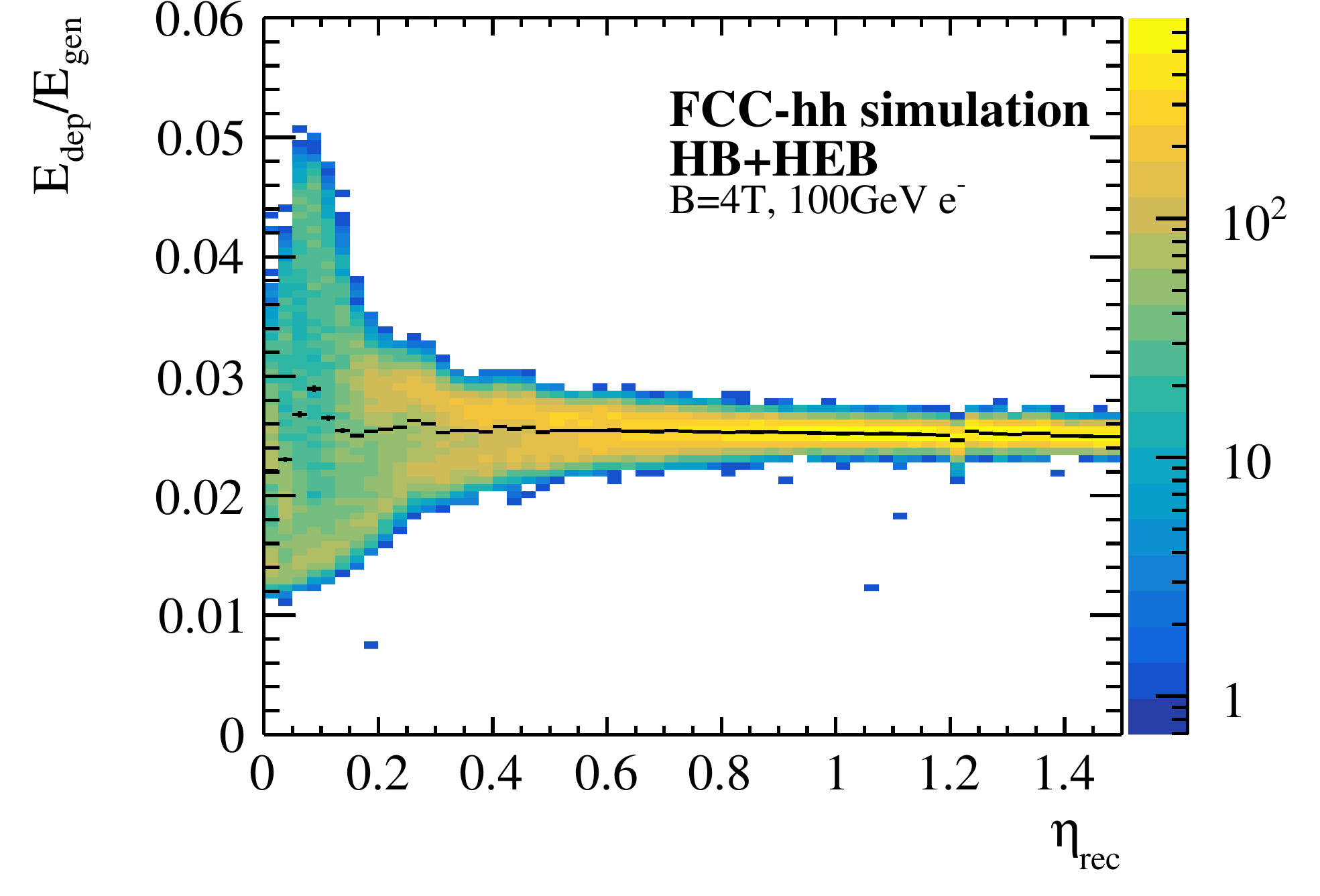}\caption{}
		   \label{fig:layout:hcal:etaResponseFull}
    \end{subfigure}
    \begin{subfigure}[b]{0.49\textwidth}
		  \includegraphics[width = \textwidth]{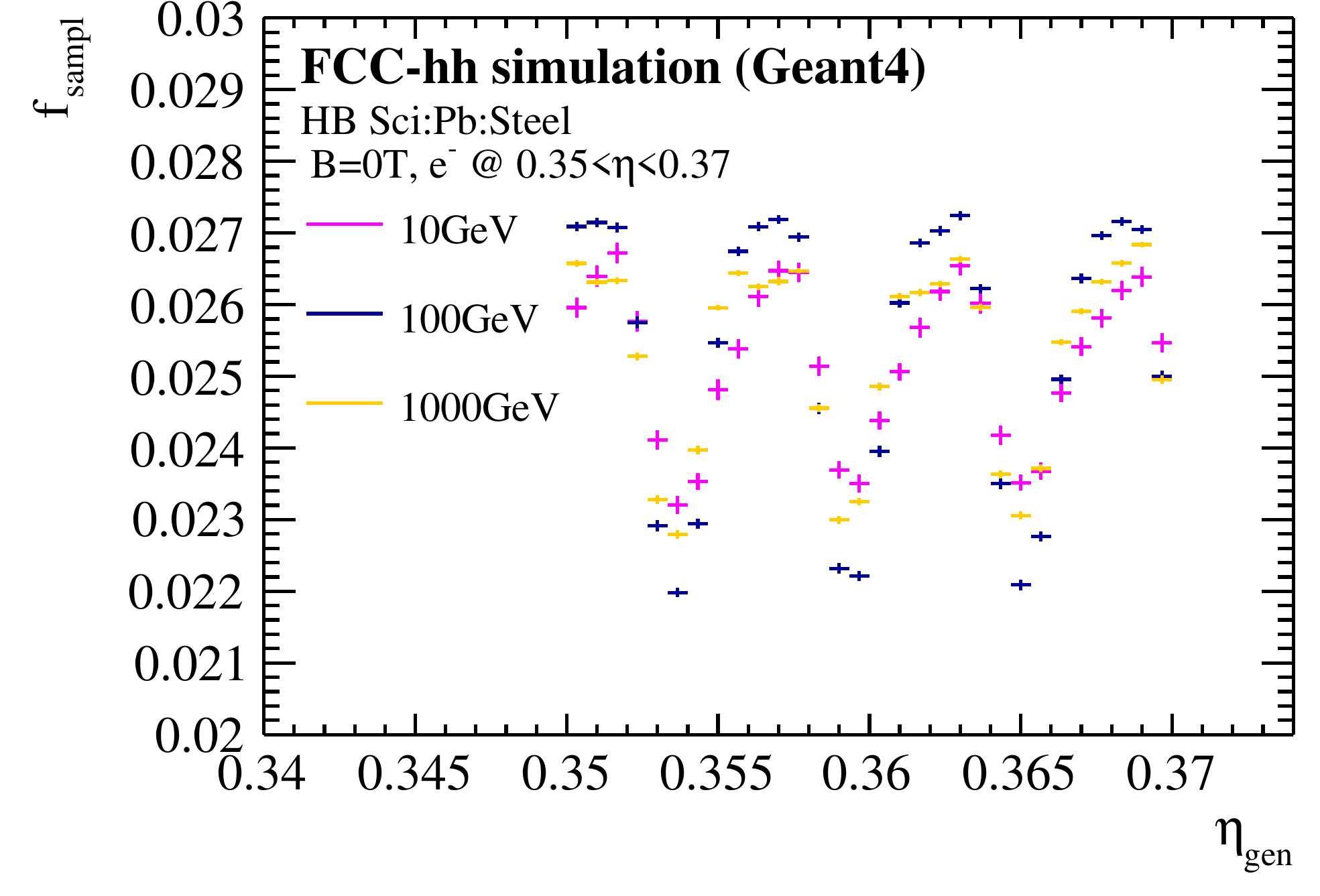}\caption{}
		   \label{fig:layout:hcal:etaResponse}
    \end{subfigure}
    \caption{(a) Response to 100\,GeV electrons over full eta range of the HB. (b) Response modulations in the scintillating tiles as a function of $\eta$, in absence of magnetic field.}
	\end{center}
\end{figure}

An example of the deposited true energy for electrons of 100\,GeV is presented in Fig.~\ref{fig:layout:hcal:etaResponseFull} as a function of pseudo-rapidity. The deep around $\eta=1.2$ originates from the gap between hadronic barrel and extended barrel. The dependence of the sampling fraction on the incident angle $\eta$ is given in Fig.~\ref{fig:layout:hcal:etaResponse}, and determined for electron energies ranging from 10 to 1000\,GeV. The modulation occurs due to the perpendicular tile orientation to the beam axis, with a distance between two adjacent tiles of  18 mm, but this is of no concern because  most of the hadronic particles will start showering in the electromagnetic calorimeter. The response for singles pions in the HB standalone has been tested for the same $\eta$ range, and the modulation has been found to be negligibly small. 
The single hadron energy resolutions are in the following always shown for an incident angle of $\eta=0.36$ to ensure the response to be unaffected by the geometry of the calorimeter. As discussed above, fluctuations in the response at $\eta=0$ are expected due to the perpendicular orientation of the absorber-scintillator structure.

\begin{table}[htp]
\begin{center}
\begin{tabular}{|c|c|c|c|c|c|c|}
\hline
			& \#$X_{0}$ (active) & \#$\lambda$ (active) & $\text{E}^{\mu}_{\text{peak}}$ & $f_{\text{sampl}}$ & $\text{E}^{\text{98\%}}_{\text{jets}}$ & $\text{E}^{\text{98\%}}_{\pi^{+}}$\\
& $\eta=0$ & $\eta=0$ 	& $\eta=0.36$  	& 	& 		& \\
\hline
unit			&    		&	& GeV	& \% & TeV	& TeV \\
			\hline
Sci:Steel 		& \multirow{2}{*}{89 (78)} 	&  \multirow{2}{*}{9.5 (8.4)}&	& $3.14\pm0.01$ 	& 4.2 	& 2.5\\
(B=4\,T)		&					&				& 4 		& $3.22\pm0.01$	& 		&	\\ \hline
Sci:Pb:Steel 	&  \multirow{2}{*}{136 (123)}& \multirow{2}{*}{9.4 (8.3)}&  	& $2.49\pm0.01$	& 3.1	   	& 2.1	 \\
(B=4\,T)		&					&				& 5 		& $2.55\pm0.01$	&		&	\\ \hline
Sci:Pb	 	&  \multirow{2}{*}{252 (242)}& \multirow{2}{*}{9.0 (7.9)}&  	& $1.75\pm0.01$	& 1.4 	& 1.1 \\
(B=4\,T)	 	& 					&  				& 14 &					& 		&  \\
			\hline
\end{tabular}
\end{center}
\caption{ Summary of major parameters for 3 different absorber scenarios of the Tile HB/HEB.
%at $\eta=0$: Total depth in $X_{0}$ and $\lambda$; the maximum energy of jets and hadrons contained up to 98\,\%, calculated adding 2\,$\lambda$ for the ECAL in front (see Sec.~\ref{sec:layout:lar}), with parameters $a_{jet} = 0.495$,  $b_{jet} = 6.3$ and $a_{\pi^{+}}=0.64$,  $b_{\pi^{+}} = 5.4$ from~\cite{Carli:2016iuf}; the determined $e/h$ ration; the energy resolution parameters, following Equation~\ref{sec:intro:energyResolution}. 
}
\label{tab:layout:hcal:depths}
\end{table}%

%To ensure a sufficient energy reconstruction, 
The deposited energy $\text{E}_{\text{dep}}$ is calibrated to the electromagnetic (EM) scale:
\begin{equation}
	\text{E}_{\text{rec}}=\frac{\text{E}_{\text{dep}}}{f_{\text{sampl}}}
\end{equation}
with $f_{\text{sampl}}$ = 2.55\,\% (3.22\,\% for Sci:Steel) in a magnetic field of 4\,T as shown in Tab.~\ref{tab:layout:hcal:depths}. The sampling fractions have been determined from electron simulations with particle energies of 20\,GeV to 1\,TeV within a pseudo-rapidity range of $0.35\leq\eta\leq0.37$. The uncertainties on $f_{\text{sampl}}$ are determined as the standard deviation over the full energy range. For determining the calorimeters mean response and resolution the $E_{\text{rec}}$ distributions are fitted with a Gaussian in the range of $\pm 2\,\sigma$ around the mean of the Gaussian. An example is shown in Fig.~\ref{fig:layout:hcal:singlePionEnergyDistr}. 

The $e/h$ ratio has been determined with a fit to  the ratio of the mean calorimeter response and the true particle energy $\left<E_{\text{rec}}\right>/E_{\text{true}}$ as a function of energy, see the top plot in Fig.~\ref{fig:layout:hcal:singlePionEnergyResolution}. A perfect linear and compensated calorimeter would make this curve flat and around 1 over the full energy range.  The response shows values between 0.85 and 0.95 (0.95 and 0.98) for Sci:Steel and Sci:Pb:Steel respectively. This change in the response depends on the electromagnetic fraction, increasing with the pion energy. The linearity can be  described with the formula~\cite{Adragna:2009zz}: 
\begin{equation}   
\label{eq:EoverH}
\frac{\left<E_{\text{rec}}\right>}{E_{\text{true}}}=\left(1-F_{\text{h}}\right)+F_{\text{h}}\times\left(\frac{e}{h}\right)^{-1}~,
\end{equation}
where the energy dependent hadronic fraction in a hadronic shower is written as
\begin{equation}
F_{\text{h}}=\frac{E}{E_{0}}^{k-1}~,
\end{equation}
with a fixed value of $E_{0}=1$\,GeV. The fits to the linearity are shown as straight lines in Fig.~\ref{fig:layout:hcal:singlePionEnergyResolution}, and the extracted $e/h$ ratios are listed in Table~\ref{tab:layout:hcal:perf}.

\begin{table}[htp]
\begin{center}
\begin{tabular}{|c|c|c|c|}
\hline
			& $e/h$ 			& k 				& resolution \\
			& 				& 				& $\pi^- @ \eta=0.36$ \\
\hline
unit			&    				&				& \%  \\
			\hline
Sci:Steel 		& $1.24\pm0.01$ 	& $0.849\pm0.002$	& $46\%/\sqrt{\text{E}}\oplus 4.1$\,\% \\
Sci:Pb:Steel 	& $1.06\pm0.01$ 	& $0.917\pm0.004$	& $42\%/\sqrt{\text{E}}\oplus 2.8$\,\%  \\
			\hline
\end{tabular}
\end{center}
\caption{Summary of the resolution and linearity fit parameters in Fig.~\ref{fig:layout:hcal:singlePionEnergyResolution}.}
\label{tab:layout:hcal:perf}
\end{table}%

The excellent single pion energy resolution, as well as the improved linearity due to the $e/h$ ratio close to 1, motivates the choice of the Sci:Pb:Steel mixture as the reference design for the hadronic scintillator calorimeters.

%% file: tex/software/general.tex
The detector simulation studies presented in this and the following section are performed within FCC-hh software framework (FCCSW)\cite{fccsw}. %, described in Sec.~\ref{chapter:software}. 
The geometry description is implemented using the DD4hep toolkit~\cite{dd4hep:website}. Geant4~\cite{Agostinelli:2002hh} is used for the simulation of particle transport through the detectors. In this section we describe the reconstruction methods of particle energies in the detectors, introduced in Sec.~\ref{sec:layout}. Additionally, the handling of electronics and pileup noise will be discussed. 

%% file: tex/software/reconstruction.tex
\subsection{Digitisation and Reconstruction}
\label{sec:software:reco}

Particles traversing the detector material produce particle showers and deposit their energy. Calorimeters proposed for FCC-hh experiments are sampling calorimeters, where only a fraction of the energy ($f_\mathrm{sampl}$) is deposited inside the active material, and only these deposits are used for the energy reconstruction. 
%Energy deposited in the detector is accumulated for each cell of the calorimeter.  
In order to account for the energy deposited in the passive material, a calibration is made using this equation

\begin{equation}
E_{\text{cell}} = \frac{E_{\text{deposited}}}{f_{\text{sampl}}}
\label{eq:layout:lar:corrections:calib}
\end{equation}

Values of the sampling fraction depend on the calorimeter. For the EMB the sampling fraction depends on the layer of the detector and is described in Sec.~\ref{sec:layout:lar:barrel:calibration}. For the other calorimeter parts only one sampling fraction value can be used as the ratio of the passive to active material is constant. The sampling fraction is estimated using a simulation registering energy deposits in both the active and the passive material and then calculating the fraction of energy inside the active material. 
These energy deposits then lead to a current signal in the LAr-based calorimeter or a light signal in the Tile calorimeter which gets amplified, shaped and digitised. However, these steps are neglected in the energy reconstruction at the moment and will need to be implemented later on\footnote{Note that the widening LAr gap with depth in the EMB will lead to different drift fields across the LAr gaps and hence different current responses for energy deposits depending on their depth. This effect decreases the effect of an increasing sampling fraction with depth. We therefore believe that neglecting this effect does not artificially improve the simulated energy resolution.}. 
The digitisation does not include any signal modelling of the readout systems. However, the saturation of the light output of scintillator materials used in the hadronic calorimeter in the barrel is included following Birk's law~\cite{CRAUN1970239}.

On top of the simulated energy deposit, the readout from each calorimeter cell will be affected by electronic noise. We assume uncorrelated Gaussian noise for each read-out channel, with a mean centred around zero and a standard deviation estimated from prior experience, see Sec.~\ref{sec:software:noise} for a description how the electronic noise is implemented.

\subsubsection{Clustering}

In order to reconstruct the energy deposited by single particles in the calorimeters, cluster of read-out cells are created and summed. There are two types of reconstructions implemented in FCCSW, the main difference being the resulting final cluster shapes: The sliding window algorithm produces cluster of a fixed size (in $\Delta\eta\times\Delta\varphi$) and a constant size in radius $r$. Instead, the topological clustering starts with a seed cell and then adds adjacent cells according to their energy deposits to form a cluster. As a result each reconstructed cluster has a different shape. The sliding window algorithm can be used for both analogue and digital calorimeters. In the latter instead of energy, number of hits in a read-out pad are used. The sliding window algorithm is used for the reconstruction of photons and electrons, whereas the topological clustering is optimised for the reconstruction of hadrons and jets. Both clustering algorithms are based on the standard calorimeter reconstruction algorithms used at the ATLAS experiment~\cite{Lampl:1099735, Aad:2016upy}.

\subsubsubsection{Sliding window algorithm}
\label{sec:software:reco:slidingWindow}

The sliding window algorithm considers the calorimeter as a two-dimensional grid in $\eta$-$\varphi$ space, neglecting the longitudinal segmentation of the calorimeter. There are $N_\eta\times N_\varphi$ elements building this space, each of size $\Delta\eta^{\mathrm{tower}}\times \Delta\varphi^{\mathrm{tower}}$. The energy of each tower is the sum of energies deposited in all cells within the tower.

\begin{figure}[ht]
  \centering
  \includegraphics[width=0.75\textwidth]{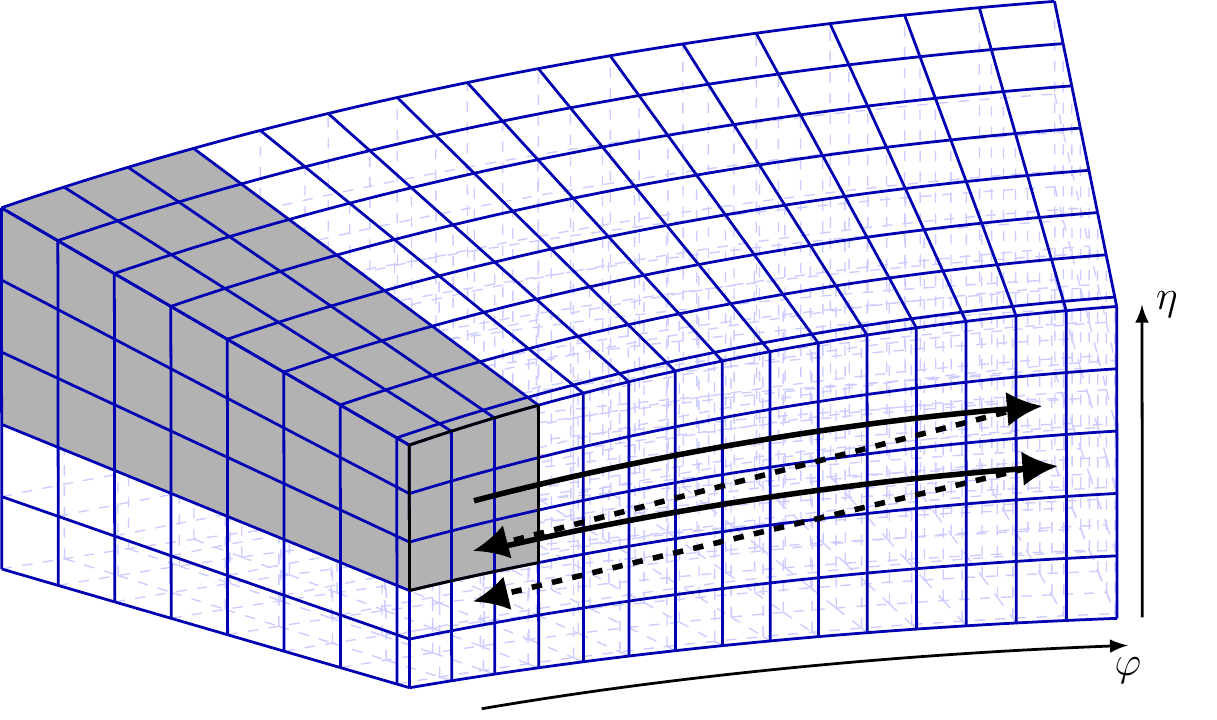}
  \caption{An illustration of the basic concept of the sliding window algorithm. A window of fixed size (here $N_\eta^{\mathrm{seed}}\times N_\varphi^{\mathrm{seed}} = 3\times 3$) is moved across the tower grid.}
  \label{fig:software:reco:sliWin:map}
\end{figure}

First, the grid of towers is scanned for local maxima: A window of a fixed size $N_\eta^{\mathrm{seed}}\times N_\varphi^{\mathrm{seed}}$ (in units of $\Delta\eta^{\mathrm{tower}}\times \Delta\varphi^{\mathrm{tower}}$) is moved across the grid, as depicted in Fig.~\ref{fig:software:reco:sliWin:map}, so that each tower is once in the middle of the window. Since the windows are symmetric around the central tower, their sizes are expressed with an odd number of towers in each direction. If the sum of the transverse energy of towers within the window is a local maximum and is larger than the threshold $E_{\mathrm{T}}^{\mathrm{threshold}}$, a pre-cluster is created. The size of the seeding window and the threshold energy are optimised to achieve the best efficiency of finding pre-clusters while reducing the fake rate.

The position of a pre-cluster is calculated as the energy-weighted average of the $\eta$ and $\varphi$ positions of the centre of cells within the fixed-sized window. The window for the position calculation may have different (smaller) size $N_\eta^{\mathrm{pos}}\times N_\varphi^{\mathrm{pos}}$ than the seeding window in order to mitigate the effect of noise. The exact position in pseudo-rapidity is corrected afterwards, as described in Sec.~\ref{sec:layout:lar:corrections:eta}.

\begin{figure}[ht]
  \centering
  \includegraphics[width=0.75\textwidth]{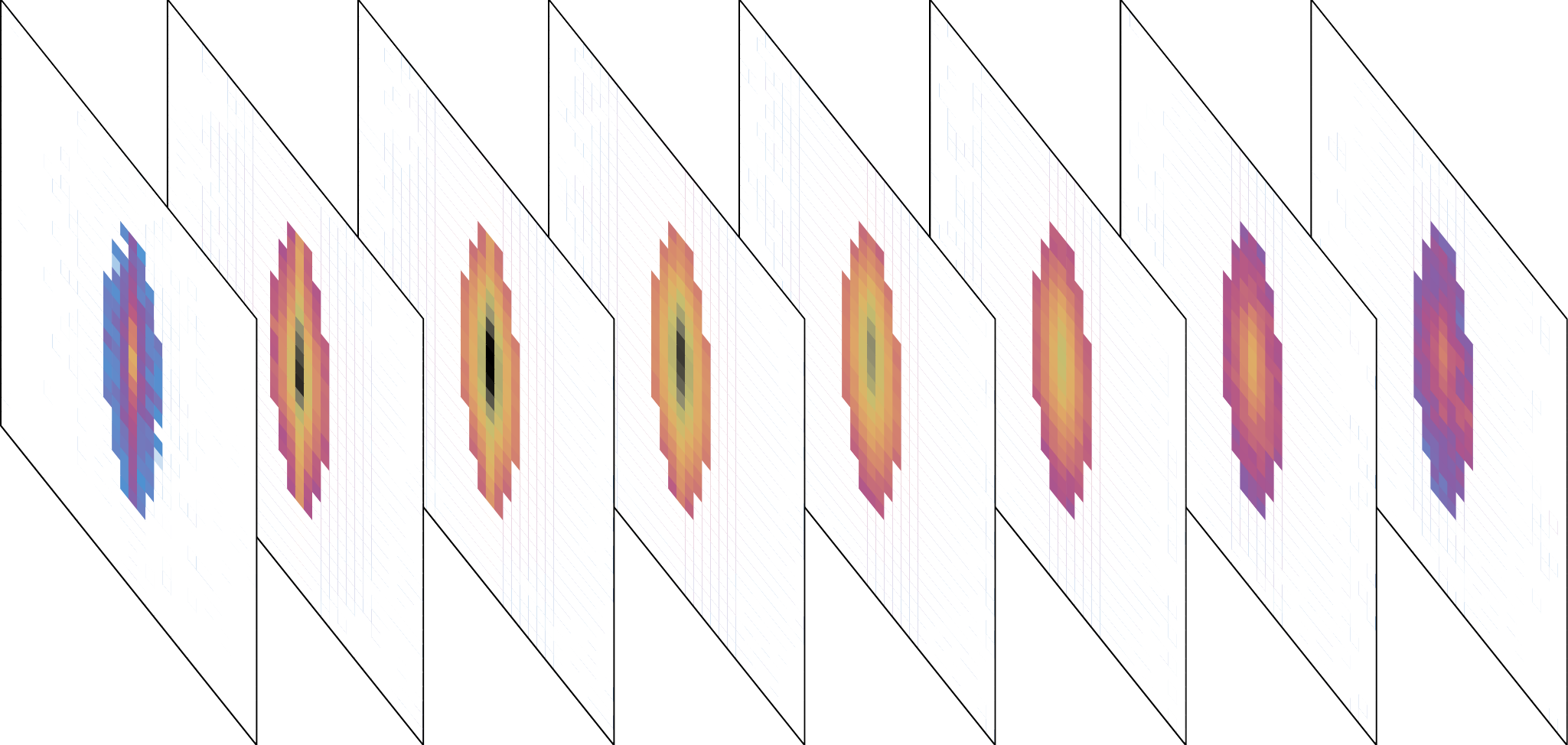}
  \caption{Shape of a reconstructed shower created by a 50\,GeV photon in the EMB at $\eta=0$. Each projection represents one calorimeter layer. Energy is collected for all layers from the cells within an ellipse which axes are defined by the final reconstruction window ($N_\eta^{\mathrm{fin}}\times N_\varphi^{\mathrm{fin}} = 7\times19$ for no pile-up).
}
  \label{fig:software:reco:sliWin:3d}
\end{figure}

In the next step the overlapping pre-clusters are removed and only the more energetic one is kept. 
A final cluster is built of all cells located within a fixed size window $N_\eta^{\mathrm{fin}}\times N_\varphi^{\mathrm{fin}}$ around the tower containing the calculated position of the pre-cluster. This window needs to be large enough to contain most of the shower, thus limiting the effect of the lateral shower leakage. However, the more cells contained in the cluster the higher the noise contribution. Therefore, the final shape of the cluster is elliptic, reducing the number of cells containing mostly noise contribution. The final cluster shape is shown in Fig.~\ref{fig:software:reco:sliWin:3d}.

\begin{figure}[ht]
  \centering
  \includegraphics[width=0.5\textwidth]{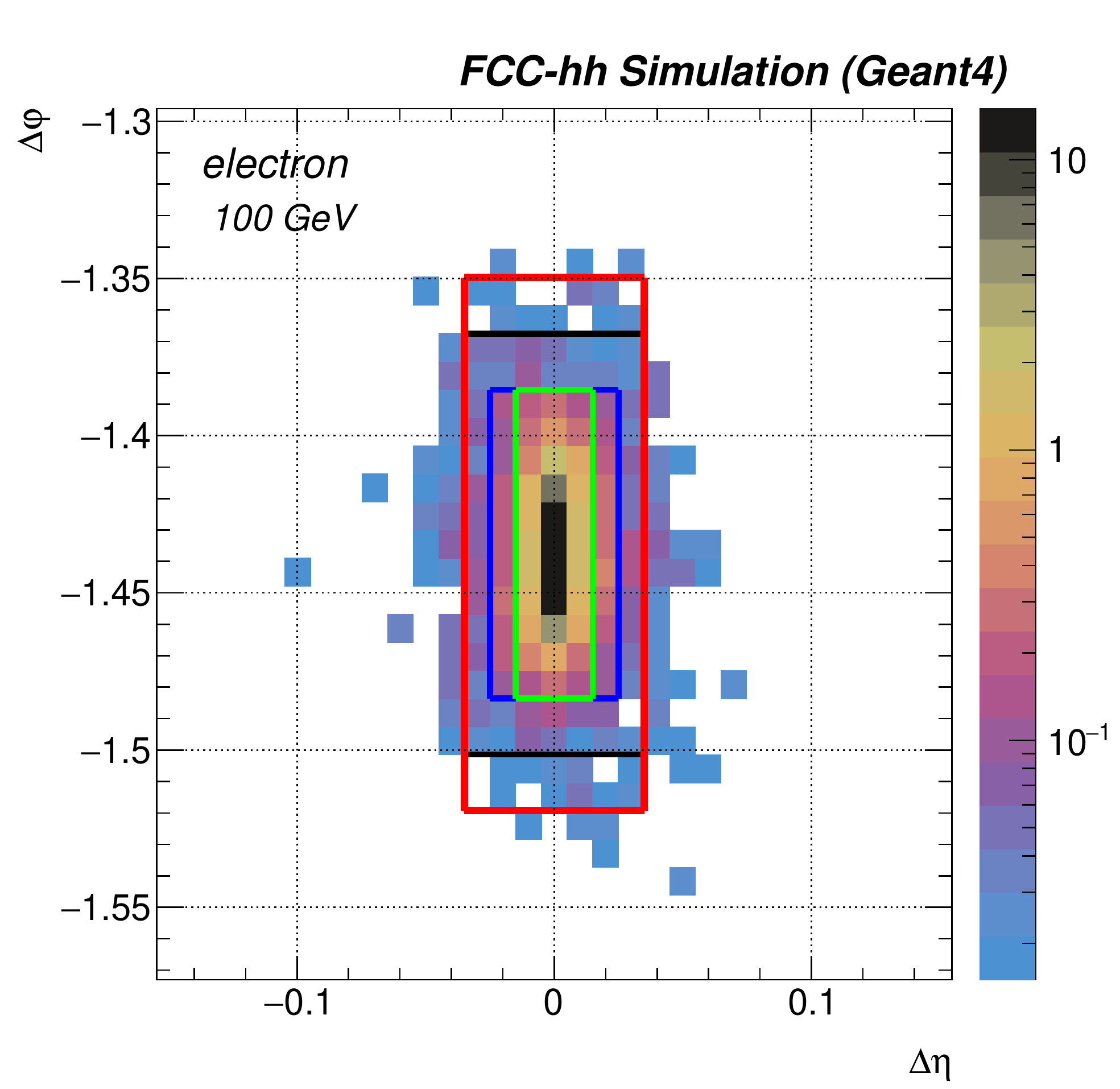}
  \caption{Transverse profile of the shower created by a 100\,GeV electron. All towers in which an electron deposited energy are included (without the detector noise). Reconstructed cluster is formed from the cells inside the red window ($N_\eta^{\mathrm{fin}}\times N_\varphi^{\mathrm{fin}}=7\times 19$). Black window corresponds to the seeding window ($N_\eta^{\mathrm{seed}}\times N_\varphi^{\mathrm{seed}}=7\times 15$), and green to the window used to calculate the position ($N_\eta^{\mathrm{pos}}\times N_\varphi^{\mathrm{pos}} = 3\times 11$). Blue window represents the area from where the overlapping pre-clusters are removed ($N_\eta^{\mathrm{seed}}\times N_\varphi^{\mathrm{seed}}=5\times 11$),
}
  \label{fig:software:reco:sliWin:profile}
\end{figure}

Results for window sizes as indicated in Tab.~\ref{tab:software:reco:sliWin:sizes} are presented in Sec.~\ref{sec:performance:egamma}. The transverse profile of a shower of a 100\,GeV electron can be seen in Fig.~\ref{fig:software:reco:sliWin:profile}. An example of the sliding-window size ($\Delta\eta^\text{fin}\times\Delta\varphi^\text{fin}=0.07\times0.17 \Rightarrow N_\eta^{\mathrm{fin}}\times N_\varphi^{\mathrm{fin}} = 7\times 19$) is indicated with a red line. In order to mitigate the effect of noise the size of that window is decreased in the presence of pile-up.

\begin{table}[ht]
  \centering
    \begin{tabular}{|c |c c c c|}
      \hline
     $\left<\mu\right>$ & $E_{\mathrm{T}}^{\mathrm{threshold}}$&  $N_\eta^{\mathrm{seed}}\times N_\varphi^{\mathrm{seed}}$ & $N_\eta^{\mathrm{pos}}\times N_\varphi^{\mathrm{pos}}$ & $N_\eta^{\mathrm{fin}}\times N_\varphi^{\mathrm{fin}}$ \\
      \hline
     0 &  \multirow{3}*{3\,GeV} & \multirow{3}*{$7\times15$} & \multirow{3}*{$3\times11$} & $7\times19$ \\
     200 &   &  &  & \multirow{2}*{$3\times9$} \\
     1000 &   &  & &  \\
      \hline
    \end{tabular}
  \caption{Parameters used in the sliding window reconstruction for different pile-up scenarios. The final cluster is of an elliptic shape and the size represents the axes of an ellipse.}
  \label{tab:software:reco:sliWin:sizes}
\end{table}

\begin{figure}[ht]
  \centering
  \begin{subfigure}[b]{0.48\textwidth}
  \includegraphics[width=1\textwidth]{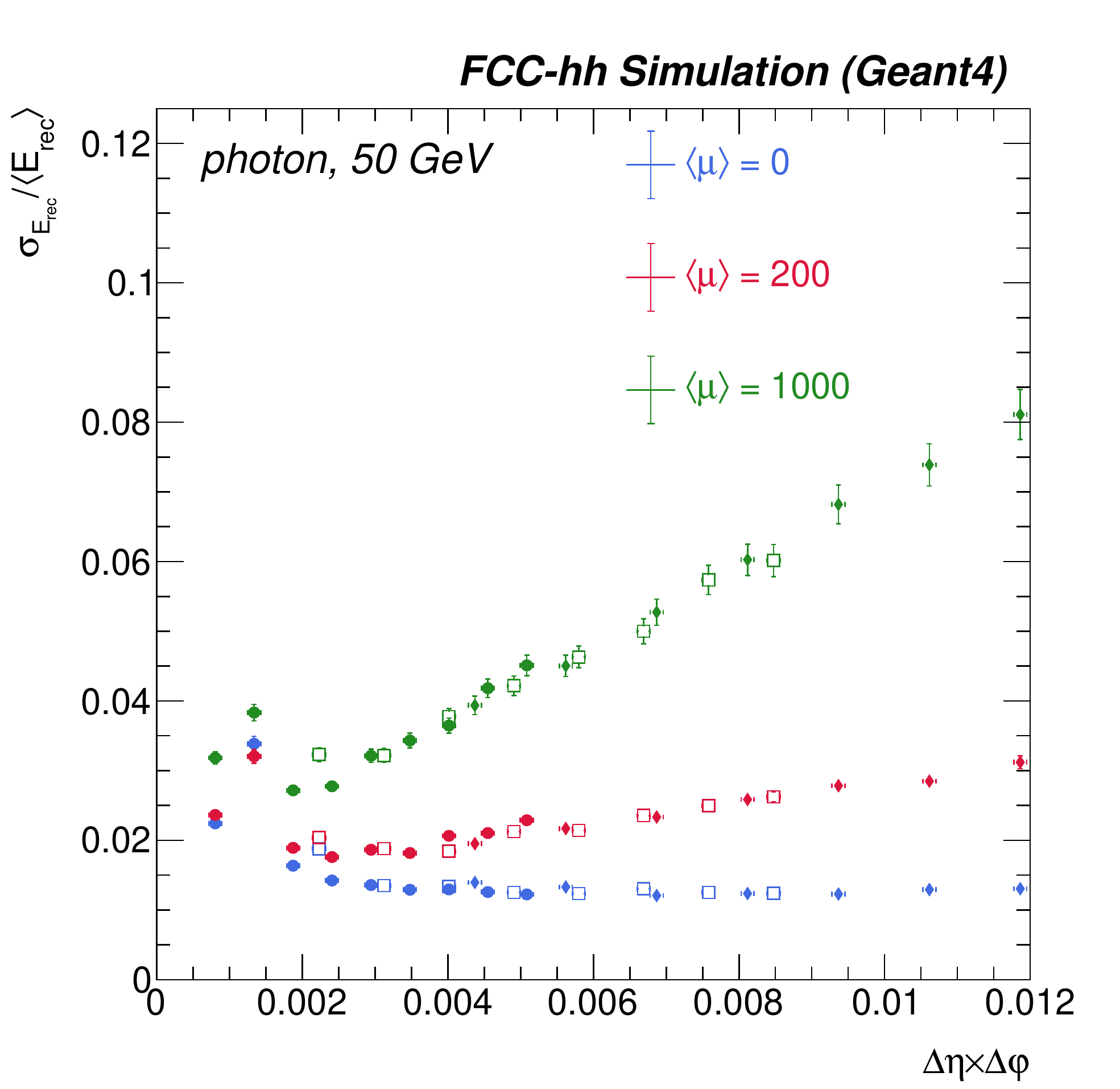}
  \caption{}
  \label{fig:software:reco:sliWin:clusterSizeOptimisation}
  \end{subfigure}
  \hspace{5pt}
  \begin{subfigure}[b]{0.48\textwidth}
  \includegraphics[width=1\textwidth]{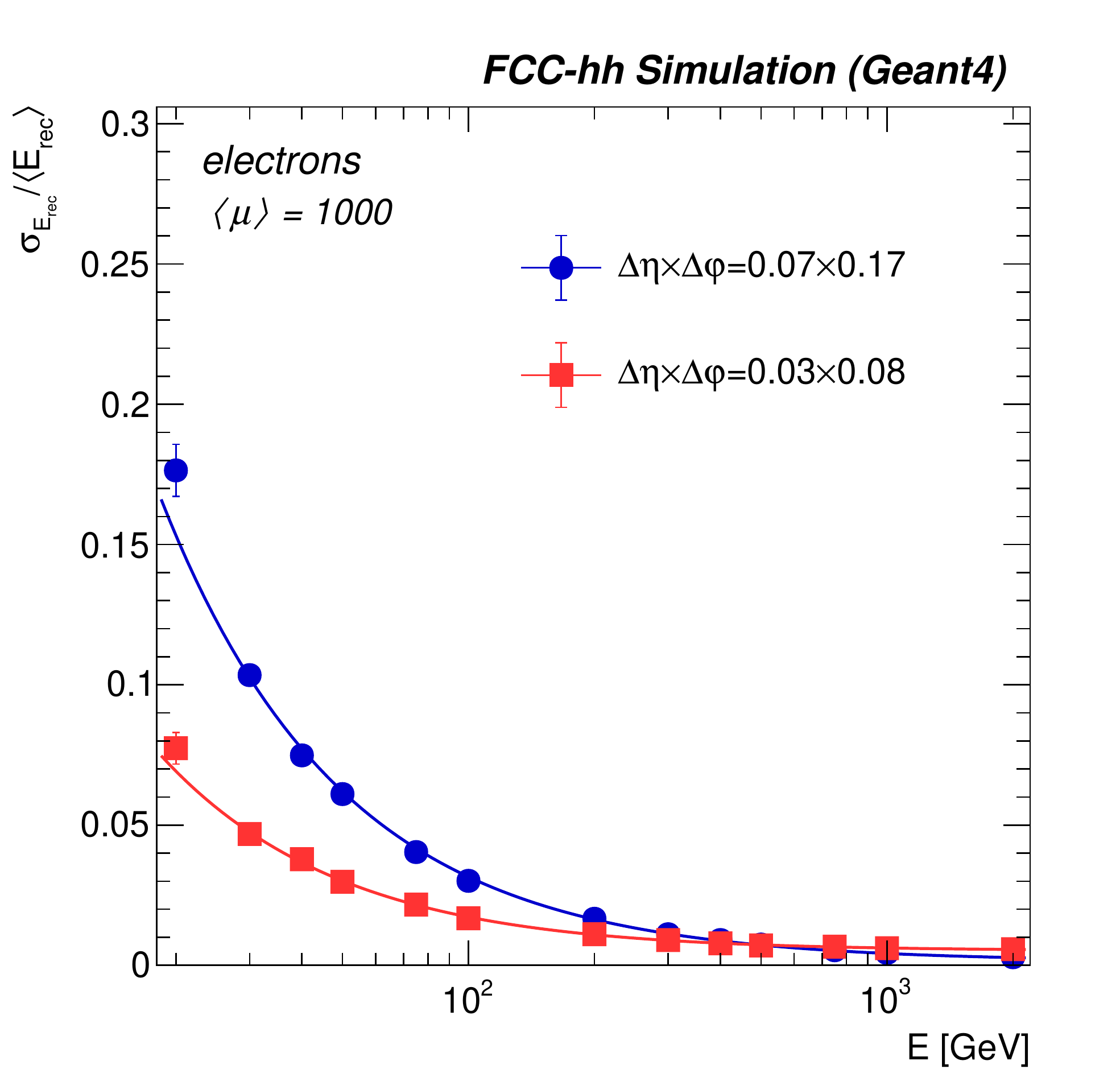}
  \caption{}
  \label{fig:software:reco:sliWin:clusterSizeImprovement}
  \end{subfigure}
  \caption{Energy resolution \textbf{(a)} for 50\,GeV photons as a function of the size of the reconstructed cluster. No pile-up environment (blue symbols) and pile-up $\left<\mu\right>=200,~1000$ (red, green) are presented. Different markers describe different width of cluster in $\Delta\eta$: $N^{\mathrm{fin}}{\eta}=3$ (full circles),  $N^{\mathrm{fin}}{\eta}=5$ (hollow squares), $N^{\mathrm{fin}}_{\eta}=7$ (full diamonds).; \textbf{(b)} Energy resolution for electrons in the presence of pile-up $\left<\mu\right>=1000$. Blue circles represent larger window ($\Delta\eta\times\Delta\varphi=0.012$), and red squares a smaller window ($\Delta\eta\times\Delta\varphi=0.0024$). Resolution can be fitted to $\frac{\sigma_E}{E}=\frac{10\%}{\sqrt{E(GeV)}} \oplus\frac{b}{E} \oplus 0.5\%$, where $b=3$\,GeV and $b=1.4$\,GeV for larger and smaller windows, respectively.}
\end{figure}

Fig.~\ref{fig:software:reco:sliWin:clusterSizeOptimisation} presents the energy resolution for 50\,GeV photons in the EMB as a function of the transverse size of the reconstructed cluster for three pile-up scenarios: $\left<\mu\right>=0, 200, 1000$. Without the presence of pile-up, the energy resolution is saturating for clusters larger than $\Delta\eta\times\Delta\varphi=0.004$. For smaller clusters not enough energy is collected hence the degradation due to larger sampling term. For pile-up of $\left<\mu\right> = 200$ the noise originating from simultaneous collisions is degrading the energy resolution for large clusters. This effect is even more prominent for $\left<\mu\right> = 1000$. The minimum between the two degrading effects (increased sampling term for small clusters and noise term for large clusters) is located around 0.0025 (0.002) for $\left<\mu\right>=200 (1000)$. Therefore the final cluster size used for reconstruction in pile-up environment has been chosen to be a window of $\Delta\eta\times\Delta\varphi=0.0023$ which corresponds to $N_\eta^{\mathrm{fin}}\times N_\varphi^{\mathrm{fin}}=3\times 9$ in units of tower size. The improvement of the energy resolution for $\left<\mu\right> = 1000$  can be seen in Fig.~\ref{fig:software:reco:sliWin:clusterSizeImprovement}, where the noise term has been reduced by more than $50\%$ from $b=3$\,GeV to $b=1.4$\,GeV.

\subsubsubsection{Topological clustering}
\label{sec:software:reco:topo}
This clustering algorithm builds so-called topo-clusters from topologically connected calorimeter cells. The algorithm explores the spatial distribution of cell signals in all three dimensions to connect neighbouring cells, thus reconstructing the energies and directions of the incoming particles. The collection of cells into topologically connected cell signals is an attempt to extract significant energy deposits by particles and reject signals coming from electronic noise or fluctuations due to pile-up. The logic of this algorithm follows the topo-clustering of ATLAS~\cite{Aad:2016upy}.

\begin{figure}[htbp]
	\begin{center}
		\includegraphics[width = .98\textwidth]{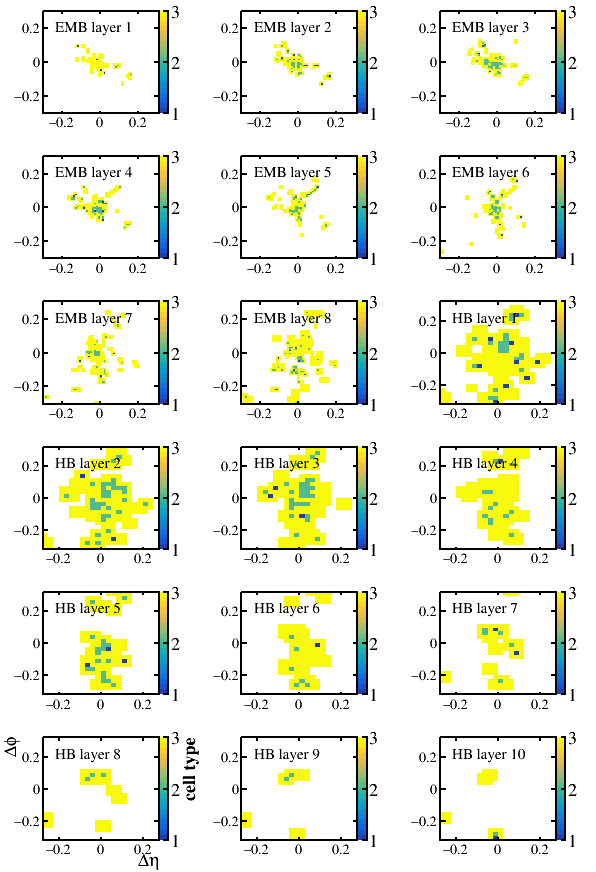}
		\caption{Cell types of seeds (blue), neighbours (green), and last iteration cell (yellow) shown in $\Delta\eta=\eta_{\text{cell}}-\eta_{\text{gen}}$ and $\Delta\phi=\phi_{\text{cell}}-\phi_{\text{gen}}$, for topo-cluster of a 100\,GeV $\pi^{-}$ shower per layer in the combined EMB+HB system at $\eta_{\text{gen}}=0.36$ with electronics noise.}
		\label{fig:software:reco:clusteredCellTypes}
	\end{center}
\end{figure}

Topo-clusters created in the FCC-hh calorimeters are not expected to contain all cells with signals created by a single particle, but rather fractional responses of particle (sub-)showers dependent on the spatial separation.
%% Algorithm logic
The main observable controlling the topo-cluster building is the cell significance $\xi_{\text{cell}}$ which is defined as the absolute value of the ratio of the cell signal to the expected noise in this cell,
\begin{equation}
\xi_{\text{cell}} = \left|\frac{E_{\text{cell}}}{\sigma^{\text{noise}}_{\text{cell}}}\right|~.
\end{equation}
To avoid positive biases all thresholds are applied on absolute values. 
The cluster formation starts with a highly significant seed cells that has a significance of $\xi_{\text{cell}}\ge S$, $S$ being a tunable parameter. In the next step, the seed cells are ordered by energy and a proto-cluster per seed is created. Starting with the highest energy proto-cluster, the next cell neighbours are added if their cell significances $\xi_{\text{cell}}$ are larger than a parameter $N$, while the newly added cells become the next seed cells of that cluster. This step is repeated until no more neighbouring cells pass the required criterion $\xi_{\text{cell}}\ge N$. In this way the cluster growth is controlled by the threshold $N$. Finally, the cell collection is finalised by adding all neighbours (of cells collected up to this point) that display a cell significance $\xi_{\text{cell}}$ larger $P$. 
In the FCC-hh calorimeters the neighbours are defined in 3D. Hence, all cells sharing a border or a corner within their own or a nearby layer are neighbours. Additionally, neighbours across sub-calorimeters are defined. This is done for the EMB and HB by adding the cells in the first HB layer with a distance in $\varphi$ of $\le\frac{1}{2}\left(\Delta\phi_{\text{EMB}}+\Delta\phi_{\text{HB}}\right)$ and of $\le\frac{1}{2}\left(\Delta\eta_{\text{EMB}}+\Delta\eta_{\text{HB}}\right)$ in $\eta$ to the list of neighbours to the last EMB layer cells.
The cluster are characterised by a core of cells with highly significant signals, surrounded by an envelope of cells with less significant signals. The types of clustered cells are shown as 1, 2 and 3 corresponding to their significances above thresholds $S$, $N$ and $P$ in Fig.~\ref{fig:software:reco:clusteredCellTypes}. 

\begin{figure}[htbp]
	\begin{center}
		\includegraphics[width = .98\textwidth]{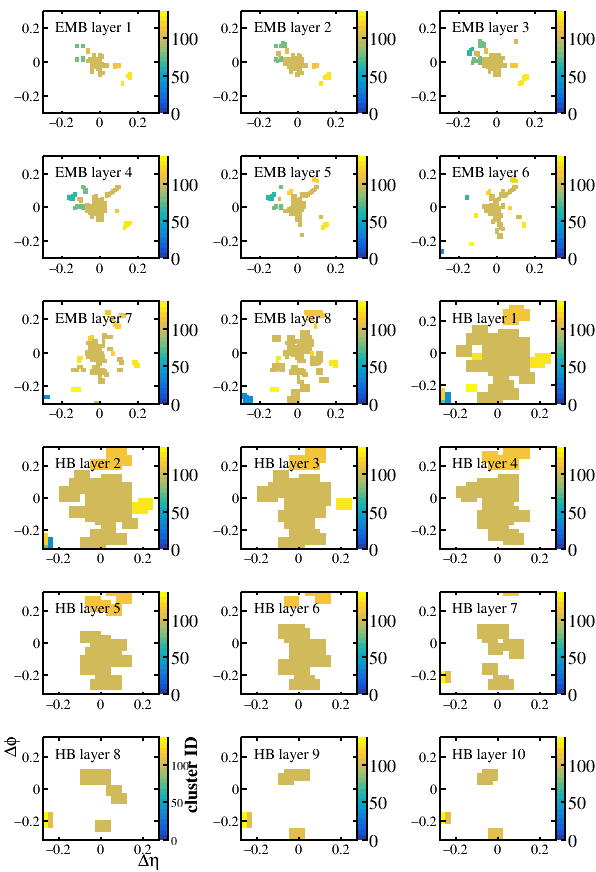}
		\caption{Cluster ID of cells, shown in $\Delta\eta=\eta_{\text{cell}}-\eta_{\text{gen}}$ and $\Delta\phi=\phi_{\text{cell}}-\phi_{\text{gen}}$, for topo-cluster of the same 100\,GeV $\pi^{-}$ shower as in Fig.~\ref{fig:software:reco:clusteredCellTypes}.}
		\label{fig:software:reco:clusteredCellClusterID}
	\end{center}
\end{figure}

In each event, cluster IDs are assigned, counting from 0 for each proto-cluster. These cluster IDs allow for a clear classification of clustered cells. Before assigning a cell to a proto-cluster, it is checked wether the cell already belongs to another cluster with a different ID. In case a cell already belongs to another cluster, and its significance is above $N$ or $P$, the two cluster are merged. For the particular case of $N$ or $P$ equal 0, the cluster stay separate and the cell is assigned to the cluster with higher energy. The mechanism of cluster merging allows for cluster with more than one seed cell. The cluster ID of the cells in all EMB and HB layers is shown in Fig.~\ref{fig:software:reco:clusteredCellClusterID} for an event of a 100\,GeV $\pi^-$ shower at $\eta=0.36$.

\begin{figure}[htbp]
	\begin{center}
  \begin{subfigure}[b]{0.48\textwidth}
		\includegraphics[width = \textwidth]{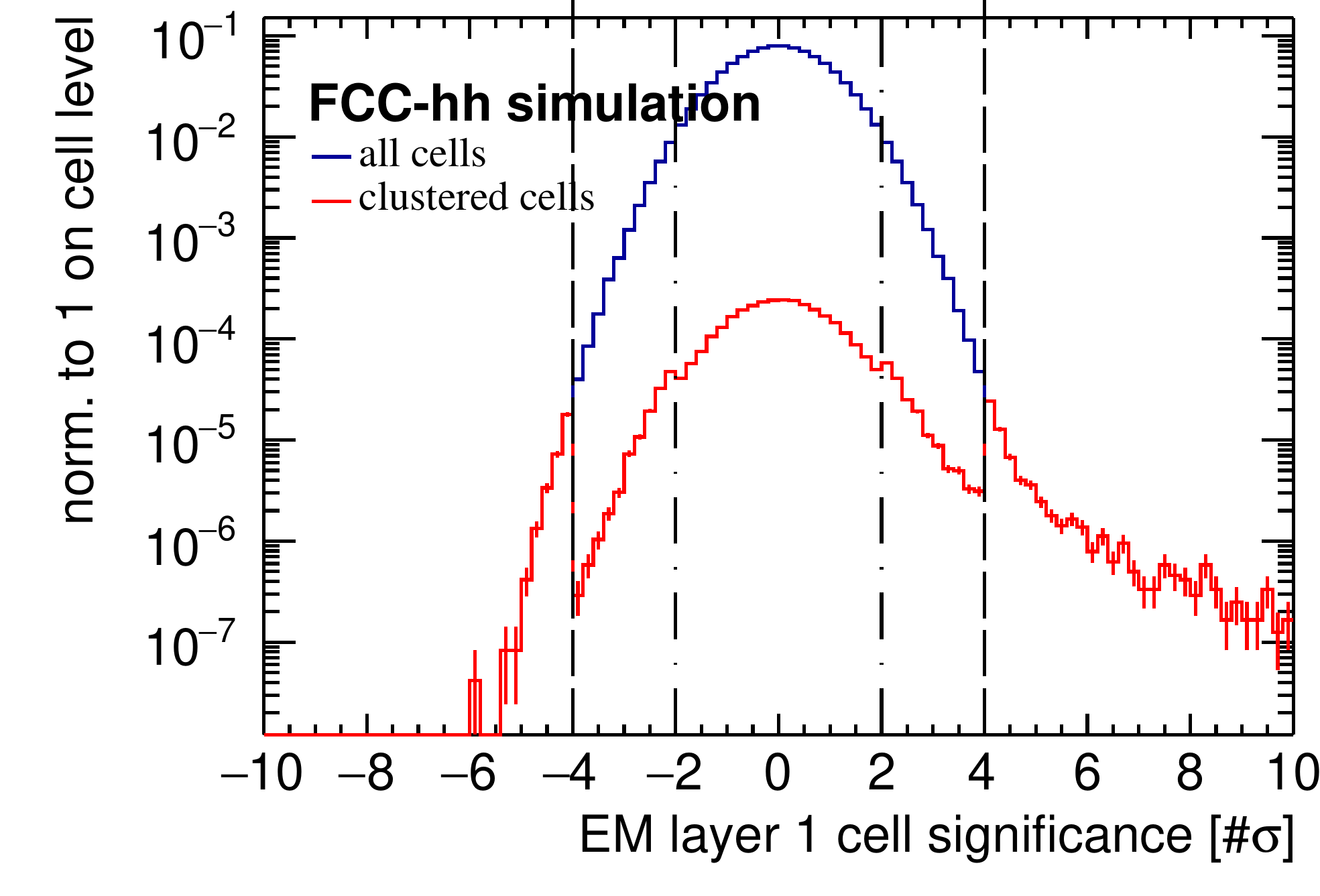}\caption{}
  \end{subfigure}
  \begin{subfigure}[b]{0.48\textwidth}
		\includegraphics[width = \textwidth]{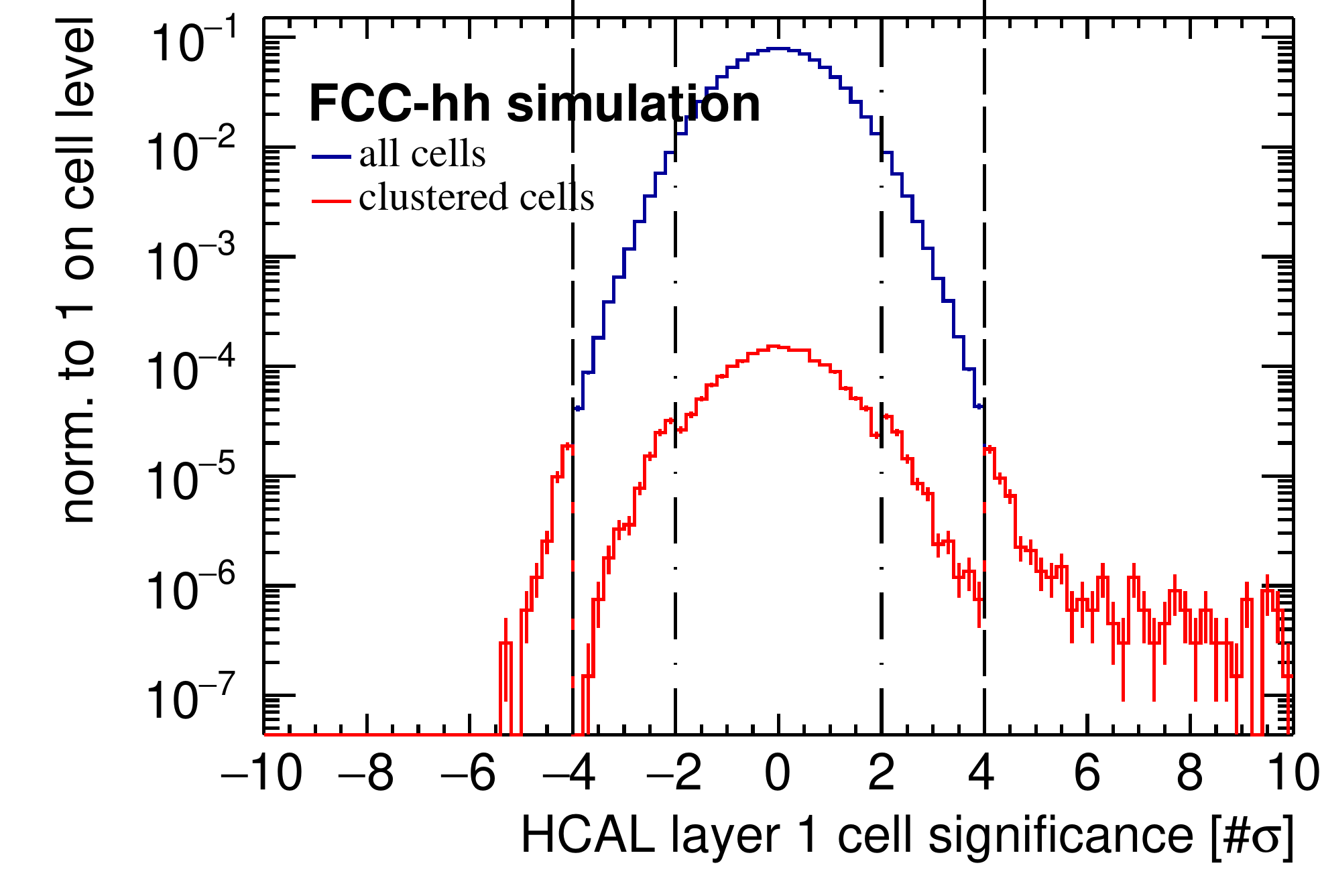}\caption{}
  \end{subfigure}
		\caption{Cell significances of 100 minimum bias events before (blue) and after (red) topo-clustering in the first EMB layer (a) and the first HB layer (b).}
		\label{fig:software:reco:MinBiasTopo}
	\end{center}
\end{figure}

The default configuration of $S=4$, $N=2$, and $P=0$, has been optimised for single charged hadrons on test-beam data of ATLAS calorimeter prototypes, and has proven good performance in LHC Run 1 data~\cite{Aad:2016upy}. Additionally, the thresholds have been tested in FCCSW and optimised for different pile-up scenarios, see Sec.~\ref{sec:performance:hadronic:topoClusterEMscale}. An example of the noise suppression power of the algorithm is shown in Fig.~\ref{fig:software:reco:MinBiasTopo}, which illustrates the cell selection that reduced the number of cells by three orders of magnitude after topo-clustering for minimum bias events with electronics noise in the EMB and HB calorimeters (see more on the noise modelling in FCCSW in Sec.~\ref{sec:software:noise}). 

\subsubsection{Jet reconstruction}
\label{sec:software:reco:jets}
The jet reconstruction is based on the anti-k$_{\text{T}}$ algorithm using the \textsc{FastJet} software package~\cite{Cacciari:2011ma,antikt}. This algorithm is based on the distances $d_{i,j}$ between entities, which in our case are calorimeter cells or cluster
\begin{equation}
d_{i,j} = \mathrm{min}\left(1/p^{2}_{Ti},1/p^{2}_{Tj}\right)\frac{\Delta R^{2}_{ij}}{R^{2}},
\end{equation}
with $\Delta R^{2}_{ij}=\left(\eta_{i} - \eta_{j}\right)^{2} + \left(\phi_{i} -\phi_{j}\right)^{2}$ and p$_{T}$, $\eta_{i}$ and $\phi_{i}$ being the transverse momentum, rapidity and azimuth of particle $i$, respectively, $R$ is the parameter of the algorithm.  
The algorithm proceeds by identifying the smallest of the distances $d_{i,j}$ and recombining entities $i$ and $j$, while calling $i$ a jet and removing $B$ from the list of entities if it is $d_{i,j}=d_{i,B}$ with 
\begin{equation}
d_{i,B} = p^{2p}_{Ti}~,
\end{equation}
where $p=-1$ for the anti-k$_\mathrm{T}$ algorithm. The distances are recalculated and the procedure repeated until no entities are left. Following the example of ATLAS, the default jet parameter R is set to $R = 0.4$, if not stated differently.

%% file: tex/software/noise.tex
\subsection{Noise}
\label{sec:software:noise}
Each read-out channel will be affected by electronic noise due to series and parallel noise of the read-out electronics. On top of that, energy deposits from particles coming from pile-up collisions will add to the energy deposits of the collision of interest, the hard scatter\footnote{The following terminology will be used in this section: The proton collision of interest is called the hard scatter emerging from the primary vertex, whereas other (minimum bias) collisions occurring during the same (prior) bunch crossing(s) are called in-time (out-of-time) pile-up collisions, respectively.}. In-time pile-up will increase the energy deposits, whereas - depending on the read-out electronics - out-of-time pile-up from prior bunch crossings might reduce the cell signals due to negative signal undershoots (see also discussion in Sec.~\ref{sec:layout:electronics}). In case of bipolar shaping the in-time and out-of-time pile-up contribution will cancel in average (for infinite bunch trains). However, due to the stochastic nature of pile-up there will be fluctuations that will be refered to pile-up noise in the following.  
If one neglects significant correlations between different cells, pile-up noise can be treated very similarly to electronic noise. 
\subsubsection{Electronic Noise}
\label{sec:software:noise:electronics}

In order to obtain realistic simulation results, especially for topological clustering, it is necessary to assume values of electronic noise for each read-out channel. In the following sections we will explain how these realistic noise estimates were obtained for the FCC-hh LAr calorimeter and the hadronic Tile calorimeter. 
%The electronics noise is estimated assuming the readout electronics similar to ATLAS.

\subsubsubsection{LAr Calorimeter}
\label{sec:software:noise:electronics:lar}
In order to estimate realistic noise levels for the FCC-hh LAr calorimeters without a detailed design of the read-out electrodes, the transmission lines, the signal feed-throughs and the read-out electronics, it is assumed that an extrapolation from the ATLAS calorimeter middle layer can be made. Empirical formulas are used to calculate the capacitances of the FCC read-out cells. The conversion factor which translates the capacitance into the electronic noise is estimated using measurements performed with the ATLAS LAr calorimeter. A correction due to different sampling fractions in ATLAS and FCC-hh LAr calorimeters has to be considered. The correction factor between a cell electronic noise and its capacitance ($\sigma_{noise}/C_{cell}$) is extracted using values from ATLAS ~\cite{Baffioni:2006eza}, yielding $0.04\,\mathrm{MeV}/\mathrm{pF}\times f_{sampl}^\mathrm{ATLAS} / f_{sampl}^\mathrm{FCC-hh}$, where $f_{sampl}^\mathrm{ATLAS} = 0.18$ is the sampling fraction of the ATLAS LAr calorimeter and $f_{sampl}^\mathrm{FCC-hh}$ the depth dependent sampling fraction of the FCC-hh LAr calorimeter (See Fig.~\ref{fig:layout:lar:samplFraction:avg}). The extrapolation from ATLAS neglects the fact that an optimisation of the read-out electronics in terms of noise will have to be performed for FCC-hh which will likely lead to shorter preamplifier rise times and shaping times than those used in ATLAS. It is also neglected that the signal attenuation inside the read-out PCBs and along the (longer) transmission lines might lead to slightly higher noise values at FCC-hh. However, we believe that the following estimates predict the actual electronic noise with an accuracy of about a factor 2. 

As described in Sec.~\ref{layout:barrel}, the capacitance of the read-out cells is composed of the LAr-gap capacitance $C_d=\varepsilon_{LAr}\varepsilon_0 A/d$ for a LAr gap of width $d$ and an area $A$ ($\varepsilon_{LAr}=1.5$) and the capacitance $C_s$ of the read-out pads to the signal shields of width $w_s=250\,\mu\mathrm{m}$ traversing below the pads inside the PCB (distance $h_m=285\,\mu\mathrm{m}$) and shielding the signal traces from the read-out pads (see Fig.~\ref{fig:pcb}, \ref{fig:pcb:traces} and \ref{fig:pcb:traces:long}).

\begin{figure}[ht]
  \centering
  \includegraphics[width=1\textwidth]{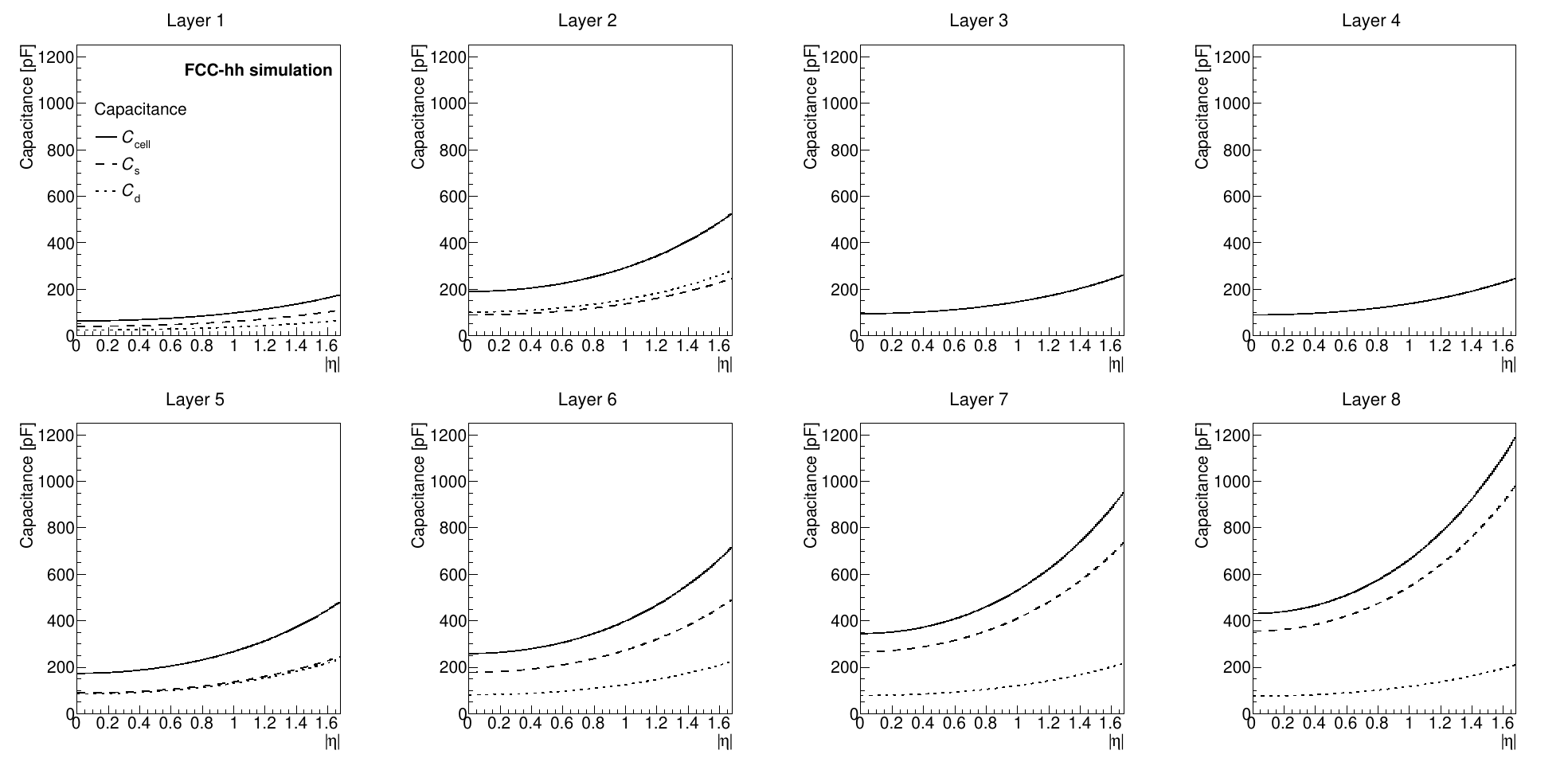}
  \caption{Capacitance $C_{cell}$, $C_s$ and $C_d$ calculated for each longitudinal layer of the EMB as a function of pseudorapidity. The signal pads of the first three layers are read out via traces leading to the inner radius of the calorimeter and all other layers via the outer radius, as depicted in Fig.~\protect\ref{fig:pcb}. Here a granularity of $\Delta\eta\times\Delta\varphi = 0.01\times 0.009$ is assumed for all layers. 
}
  \label{fig:software:noise:capacitance:layers}
\end{figure}

For each cell, the capacitance between the signal shields and the read-out pads $C_s$ depends on the length of the shield and the number of signal traces that are passing below each read-out pad. In order to minimise that number, cells from the first and second detector layer will be read out via the front of the detector, while the rest will be read out via the back (see Fig.~\ref{fig:pcb}). 
The resulting capacitance of read-out cells $C_{cell}$ in all layers as a function of pseudorapidity is presented in Fig.~\ref{fig:software:noise:capacitance:layers}. 

\begin{figure}[ht]
  \centering
  \includegraphics[width=0.75\textwidth]{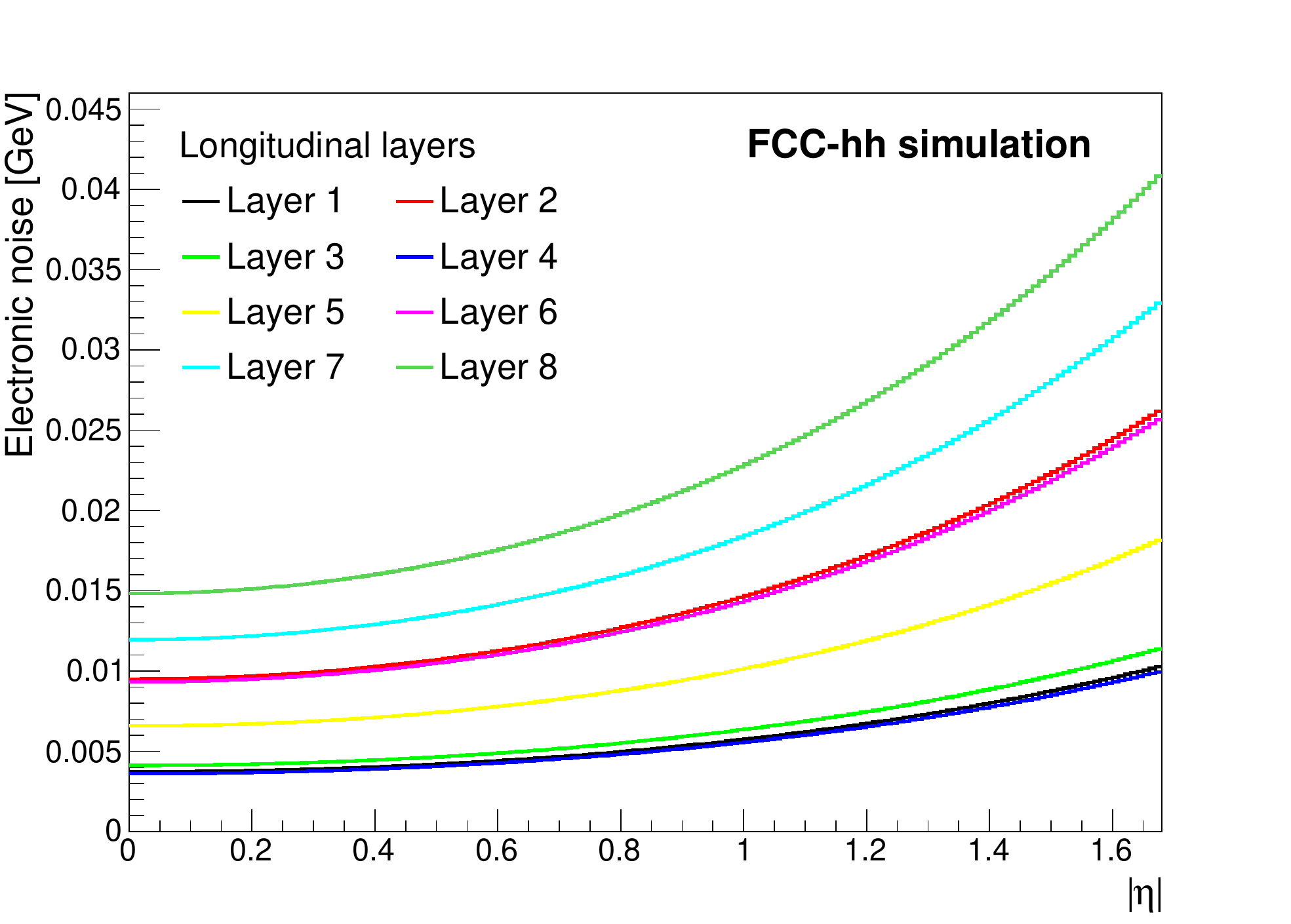}
  \caption{Estimated electronic noise per cell for each longitudinal layer of the EMB as a function of pseudorapidity.}
  \label{fig:software:noise:capacitance:summary}
\end{figure}

Assuming the above mentioned conversion factor $\sigma_{noise}/C_{cell}$, we obtain values of electronic noise per cell as presented in Fig.~\ref{fig:software:noise:capacitance:summary}. This electronic noise is assumed to be uncorrelated between cells.\footnote{During the design of the read-out electrodes, feed-throughs and read-out electronics the correlated noise contribution and cross-talk needs to be simulated in detail and kept at an absolute minimum. A cross-talk between neighbouring strip cells of $\le 7\,\%$ has been achieved in ATLAS.} The electronic noise of clusters can therefore be calculated by the quadratic sum of the noise in individual cells. For a cluster of size $\Delta\eta\times\Delta\varphi=0.07\times0.17$ this yields $\sigma_{noise} \approx 300$\,MeV at $\eta = 0$.

As discussed, this estimation of the electronic noise in the LAr calorimeter  is very preliminary and is based on many approximations and assumptions. A detailed design and simulation of the full read-out chain will have to be carried out in order to obtain more reliable results.

\subsubsubsection{Tile Calorimeter}
The electronic noise in the Tile calorimeter will be dominated by electronic noise of the SiPMs read-out electronics. Due to the strong dependence on the missing design details in the SiPM readout chain, a conservative assumption is used in the following based on estimations for the HL-LHC upgrade of the read-out electronics of the ATLAS Tile calorimeter, that assumes 10-15\,MeV per read-out channel on EM scale. We therefore add uncorrelated random Gaussian noise of 10\,MeV on EM scale for each read-out cell. Additional effects due to the dark count rate, cross talk, after pulses, and saturation of pixels of the SiPM are not included in the simulation because all these effects vary strongly between SiPM types and manufacturers, and have shown great improvements over the past years.  
%~\cite{privateTileElectronicsNoise}

\subsubsection{Pile-up Noise}
\label{sec:software:noise:pileup}

Energy deposits of particles from simultaneous collisions in the same bunch crossing will create a background to the energy deposits of the hard scatter collision of interest. These pile-up energy deposits will create a bias to the energy measurement. This positive bias can be reduced by using bi-polar shaping, as discussed in Sec.~\ref{sec:layout:electronics} and proposed for the calorimeters of the reference detector, whereas fluctuations of these pile-up energy deposits will remain.  We call these fluctuations pile-up noise. This noise is usually composed of the energy deposits from the same bunch crossing (the in-time pile-up) and those from prior bunch crossings (the out-of-time pile-up) creating negative contributions through the signal undershoots. In case of signal collection and shaping times smaller than the time between two bunch crossings ($<25$\,ns) - as is the case for Si sensors, out-of-time pile-up will not influence the energy measurement, but the bias and fluctuations of the in-time pile-up will remain. For our case of longer signal collection and shaping times it is also possible to minimise the impact of out-of-time pile-up. Due to the known history of energy deposits from prior bunch-crossings, the out-of-time pile-up can be unfolded from the measurement of the current bunch crossing. Such an approach has been studied in detail and demonstrated for the ATLAS LAr Calorimeter HL-LHC upgrade~\cite{Collaboration:2285582}. 

In-time pile-up can be reduced by using timing information of energy deposits and rejecting those deposits which are not consistent with the time-of-flight of particles from the primary vertex. It is also possible to reject pile-up deposits if they can be attributed to a charged particle track reconstructed with the tracker that does not originate at the primary vertex. Such techniques will need to be developed and studied to exploit the full physics potential of an FCC-hh experiment. 

In the following we assume that the out-of-time pile-up can be corrected for, but estimate the full in-time pile-up noise contribution (without any above mentioned corrections). 

\subsubsubsection{LAr Calorimeter}

%Both, the in-time and the out-of-time pile-up play an important role in the energy measurement of the LAr calorimeter. 
The ATLAS detector with its planned updates for the HL-LHC proves that LAr calorimetry can provide excellent energy measurement even in a high pile-up environment~\cite{Collaboration:2285582}.

%Ionisation electrons drift in the electric field applied aver the LAr gap inducing a triangular signal, which can be further amplified and shaped into a bipolar shape. It is chosen so that the integral of the area is 0. This way, on the average, energy deposited by particles in the collisions from the in-time pile-up is compensated by the signal from the out-of-time pile-up. Naturally, none two events are identical, therefore fluctuations in the energy deposits give rise to the noise in the detector, called the pile-up noise. Moreover, reading the negative values from the detector indicate that no zero-suppresion technique could be used to reduce the amount of data.

%Ideally, pile-up noise should be estimated by overlaying the signal event with the background events from the same collision. However, in LAr calorimeters, where mitigation of the shift in the deposited energy is due to the signal shaping, many tools need to be still implemented to fully simulate the signal read-out chain. Therefore, the pile-up noise is estimated looking at the event-to-event fluctuations in the deposited energy (RMS of energy distributions).

\begin{figure}[ht]
  \centering
  \includegraphics[width=0.75\textwidth]{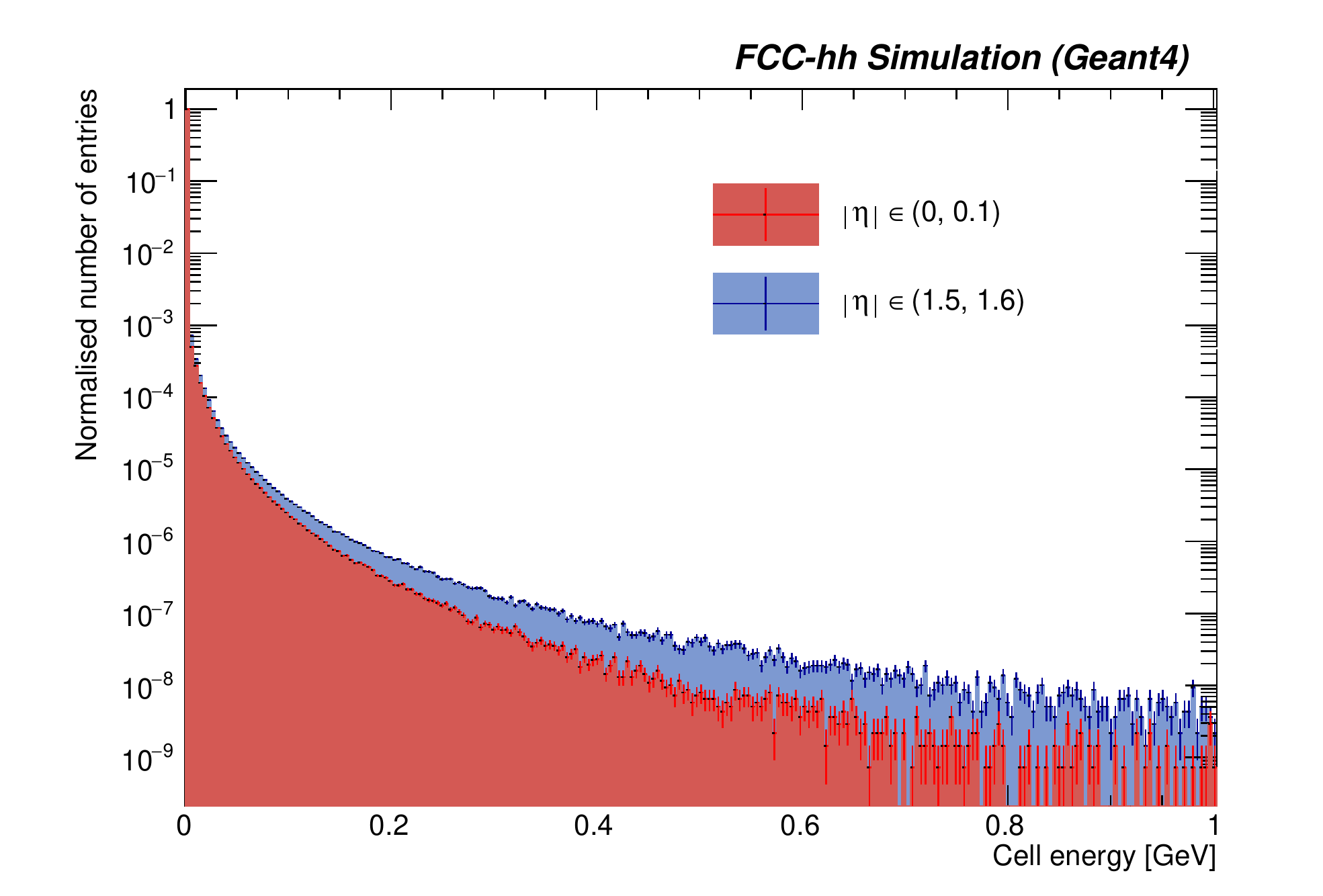}
  \caption{Energy distribution in calorimeter cells of size $\Delta\eta\times\Delta\varphi=0.01\times0.009$ in the third calorimeter layer within $\left|\eta\right|$ range between 0 and 0.1 (red) and for  $\left|\eta\right|\in (1.5,1.6)$ (blue).}
  \label{fig:software:noise:pileup:energy}
\end{figure}

In order to estimate the in-time pile-up, a sample of minimum bias events has been simulated with the Pythia8~\cite{Sjostrand:2014zea} event generator. The electronic noise was switched off for these special simulations with single pile-up collision per event. The energy distribution for a single calorimeter cell is presented in Fig.~\ref{fig:software:noise:pileup:energy}. The standard deviation of the energy distribution is used to quantify the fluctuations caused by the pile-up collisions, i.e. the pile-up noise.

The size of the sample was not large enough to overlay hundreds of pile-up collisions per event. Therefore the standard deviation obtained with one minimum bias collision per event has been scaled with $\sqrt{\left<\mu\right>}$, where $\left<\mu\right>$ is the desired average number of simultaneous collisions per bunch crossing. It has been checked with simulation that this scaling yields the correct values. 
%Fig.~\ref{fig:software:noise:pileup:scaling} shows a validation of such scaling. RMS of energy distribution for one minimum bias event scaled by $\sqrt{\left<\mu\right>}=\sqrt{200}$ (blue markers) matches the RMS of energy distributions obtained by overlaying 200 minimum bias events (red markers). 
In order to increase the sample size, azimuthal symmetry is assumed. 

%In-time pile-up noise for the EMB estimated for all cells as a function of pseudorapidity and detector layer, is presented in Fig.~\ref{}. One can see the largest noise is in the second and third layer, which correspond to the shower maxima created by the low energetic particles from the background events.
%
%\begin{figure}[ht]
%  \centering
%  \includegraphics[width=0.5\textwidth]{lar/pileup_noise_scaling_validation}
%  \caption{RMS of energy distribution for one minimum bias event scaled by $\sqrt{\left<\mu\right>}=\sqrt{200}$ (blue markers) matches the RMS of energy distributions obtained by overlaying 200 minimum bias events (red markers). \protect\todo[inline]{redo plot, add ratio}}
%  \label{fig:software:noise:pileup:scaling}
%\end{figure}

\begin{figure}[ht]
  \centering
  \begin{subfigure}[b]{0.48\textwidth}
  \centering
    \includegraphics[width=\textwidth]{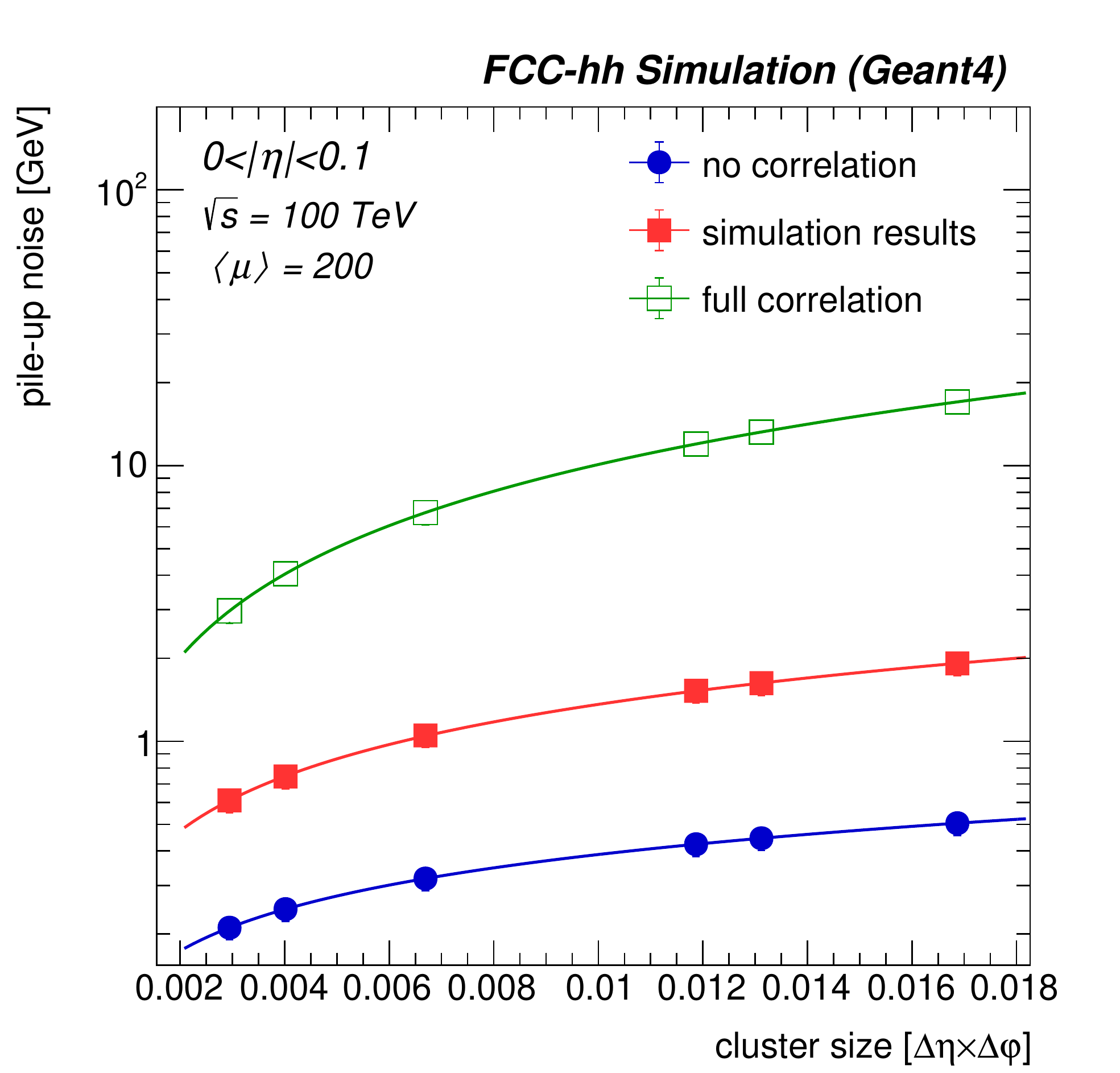}
    \caption{}\label{fig:software:noise:pileup:cluster:size}
  \end{subfigure}
  \begin{subfigure}[b]{0.48\textwidth}
  \centering
    \includegraphics[width=\textwidth]{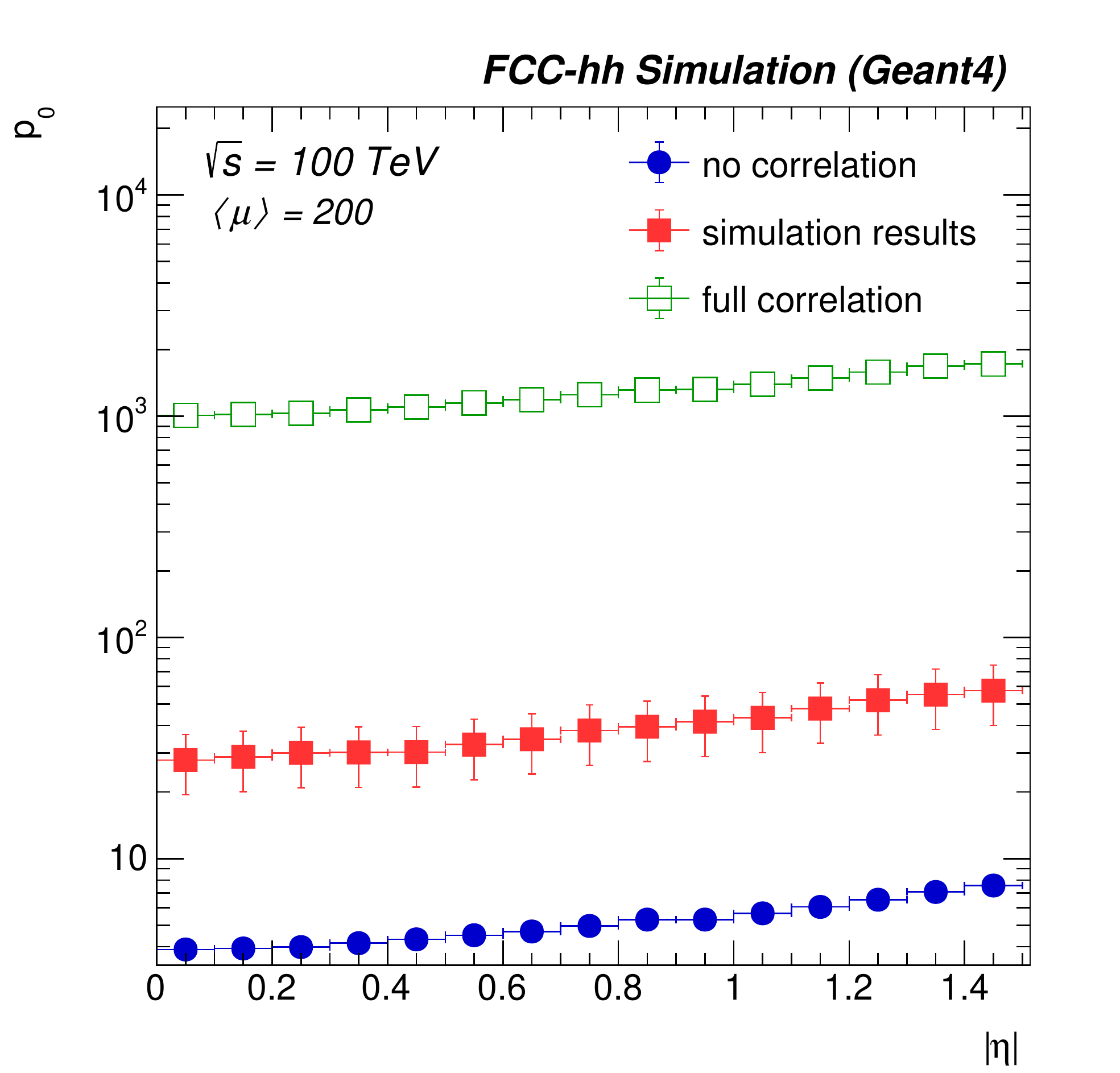}
    \caption{}\label{fig:software:noise:pileup:cluster:parameters}
  \end{subfigure}
  \caption{(a): Cluster noise as a function of cluster size in terms of $\Delta\eta\times\Delta\varphi$ for clusters in the EM calorimeters within $\abseta<0.1$. The full squares show the minimum bias simulation result, whereas the filled circles show the quadratic sum of pile-up noise of individual cells (no correlation) and the open squares show the linear sum of individual cells (full correlation). (b): Parameter $p_0$ as a function of $\eta$ using the parametrisation of Eq.~\eqref{eq:software:noise:pileup:parametrisation} for all three cases shown in (a).
  }
\end{figure}

Pile-up noise is correlated between neighbouring cells, and therefore cannot be treated in the same way as the electronic noise which is modelled on a cell-by-cell basis. The impact of pile-up has been studied for the two clustering algorithms separately. The first approach is used for the EM calorimeters and is relatively straight forward for fixed-size clusters as used in the sliding window algorithm, where all clusters consist of the same number of cells. Instead of looking at the individual cell, the noise in a cluster of dimensions $\Delta\eta\times\Delta\varphi$ is studied, assuming same size in all longitudinal layers. As expected, a clear dependence on the cluster size is found, which is presented in Fig.~\ref{fig:software:noise:pileup:cluster:size} (squares) for $\left<\mu\right>=200$. This dependence on cluster size can be parameterised in bins of pseudorapidity using 

\begin{equation}
  \sigma = p_0 \cdot (\Delta\eta\times\Delta\varphi)^{p_1}
  \label{eq:software:noise:pileup:parametrisation}
\end{equation}

where $p_0$ and $p_1$ are the two fit parameters and  $\Delta\eta$ and $\Delta\varphi$ the cluster dimensions in $\eta$ and $\varphi$, respectively. Figure~\ref{fig:software:noise:pileup:cluster:size} also shows the result obtained (filled circles) if the pile-up noise in cells of $\Delta\eta\times\Delta\varphi=0.01\times0.009$ is summed up quadratically (no correlation assumed, $p_1=1/2$) as well as the case (open circles) of summing the pile-up noise linearly (assuming fully correlated pile-up noise, $p_1=1$). As presented in Fig.~\ref{fig:software:noise:pileup:cluster:parameters} $p_0$ depends on pseudorapidity, while parameter $p_1$ is constant, $p_1=0.66\pm 0.01$ in our case. Using these noise parametrisations, pile-up studies with different window sizes can be performed. 
%In Sec.~\ref{sec:performance:egamma} it is shown, that for $\left<\mu\right>=1000$, it will be crucial to further optimize cluster sizes and especially make them smaller in the first calorimeter layers. 

\begin{figure}[ht]
  \centering
  \begin{subfigure}[b]{0.48\textwidth}
  \centering
  \includegraphics[width=1\textwidth]{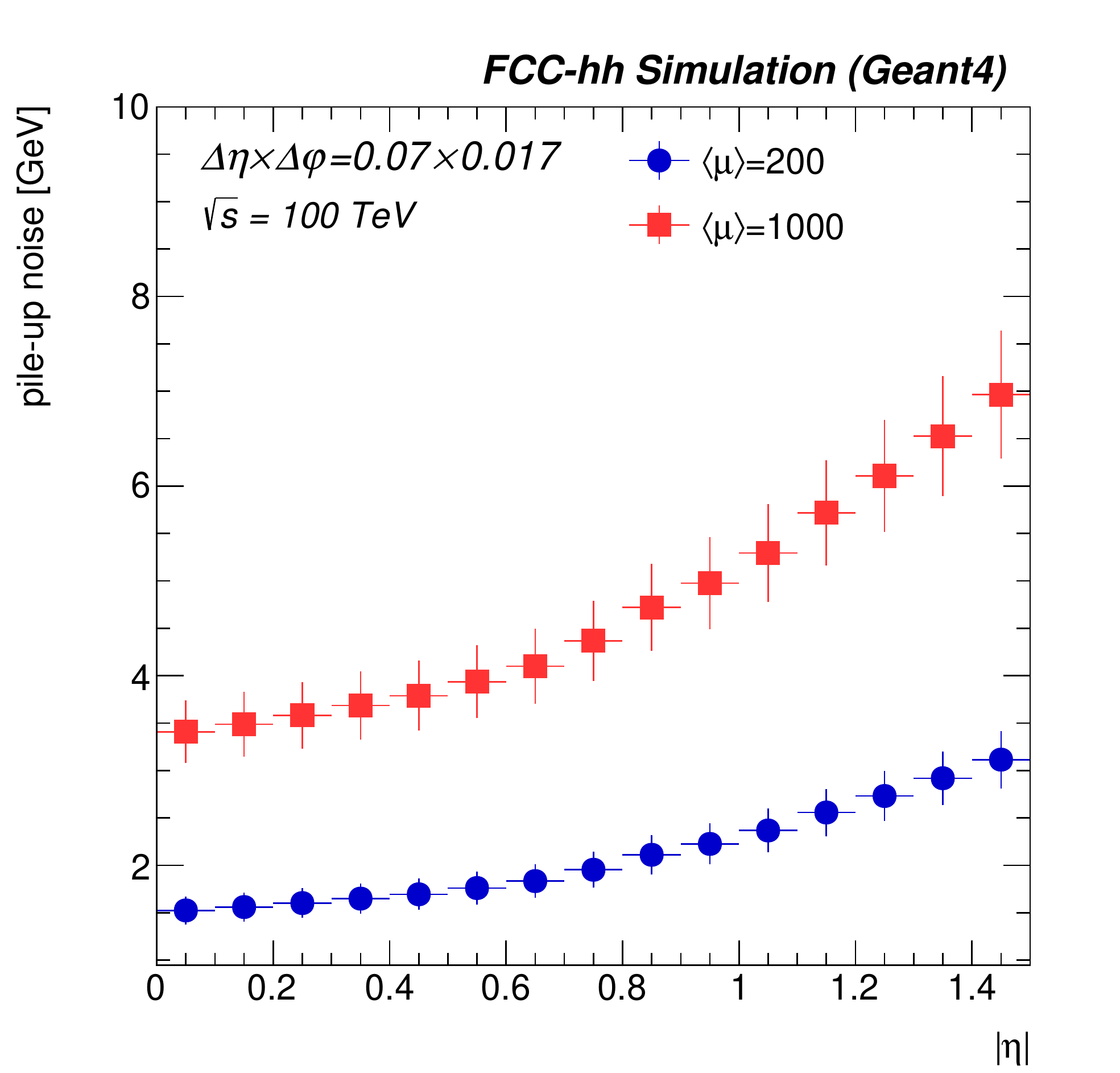}\caption{}
  \end{subfigure}
  \begin{subfigure}[b]{0.48\textwidth}
  \centering
  \includegraphics[width=1\textwidth]{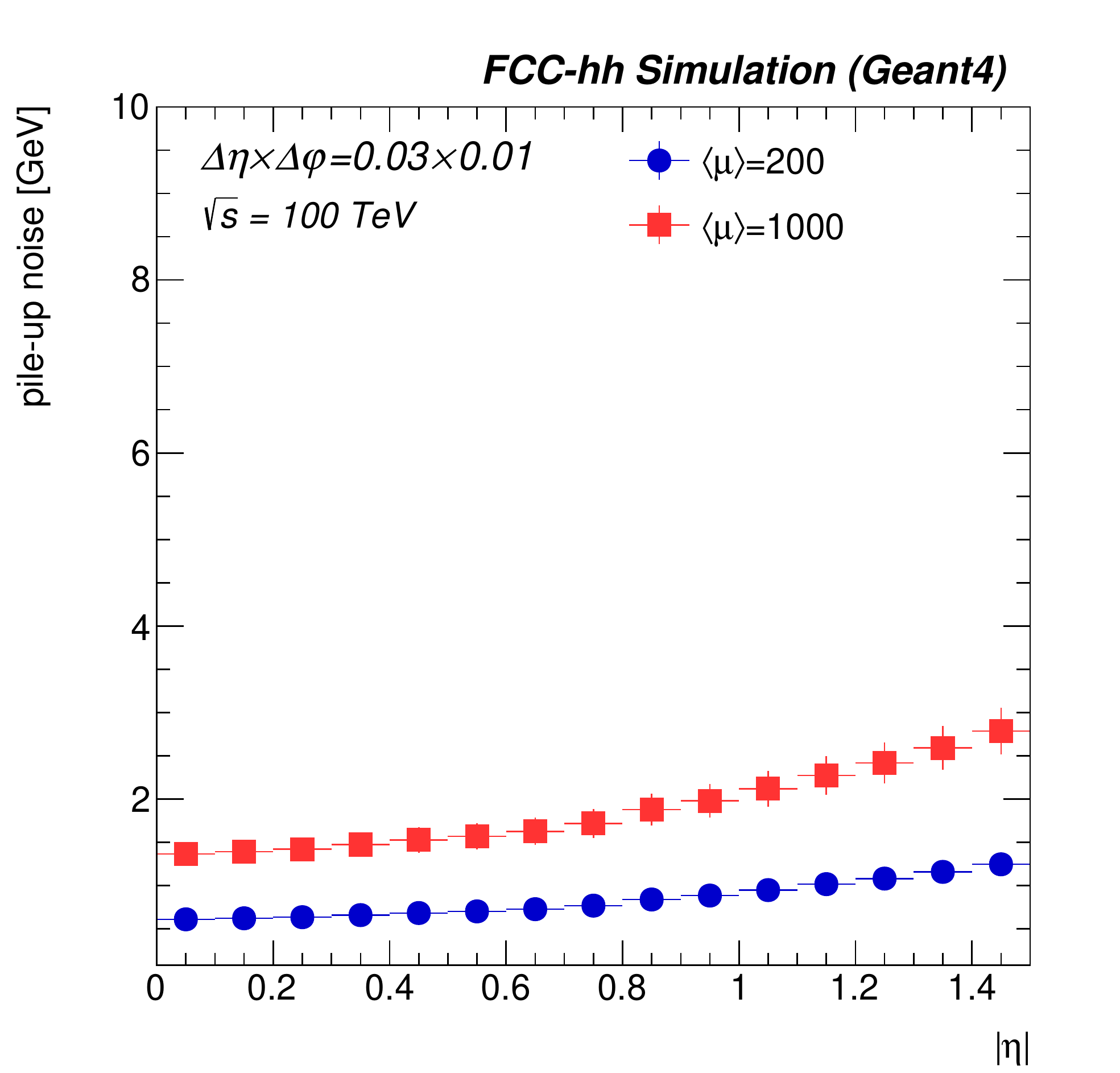}\caption{}
  \end{subfigure}
  \caption{Pile-up noise estimated for electromagnetic barrel (EMB) for two scenarios: $\left<\mu\right>=200$ (blue circles) and $\left<\mu\right>=1000$ (red squares) for cluster of size (a) $\Delta\eta\times\Delta\varphi=0.07\times0.17$ and (b) $\Delta\eta\times\Delta\varphi=0.03\times0.1$.
}
  \label{fig:software:noise:pileup:cluster}
\end{figure}

The pile-up noise dependence on pseudo-rapidity for $\left<\mu\right>=200$~and~$1000$ for clusters with a sliding window size of $\Delta\eta\times\Delta\varphi=0.07\times0.17$ is presented in Fig.~\ref{fig:software:noise:pileup:cluster}. As can be seen, a large pile-up noise rising in $\eta$ from 3.5\,GeV to 6.5\,GeV for $\left<\mu\right>=1000$ is obtained. This result suggests that it will be crucial to further optimise cluster sizes and especially make them smaller in the first calorimeter layers and to use timing information and the tracker measurement to further reject pile-up energy deposits. Nevertheless, these values have been used in the performance section to smear the cluster energies of sliding window clusters. 

\subsubsubsection{Combined LAr and Tile Calorimeter}

\begin{figure}[ht]
  \centering
  \begin{subfigure}[b]{0.49\textwidth}
    \includegraphics[width=\textwidth]{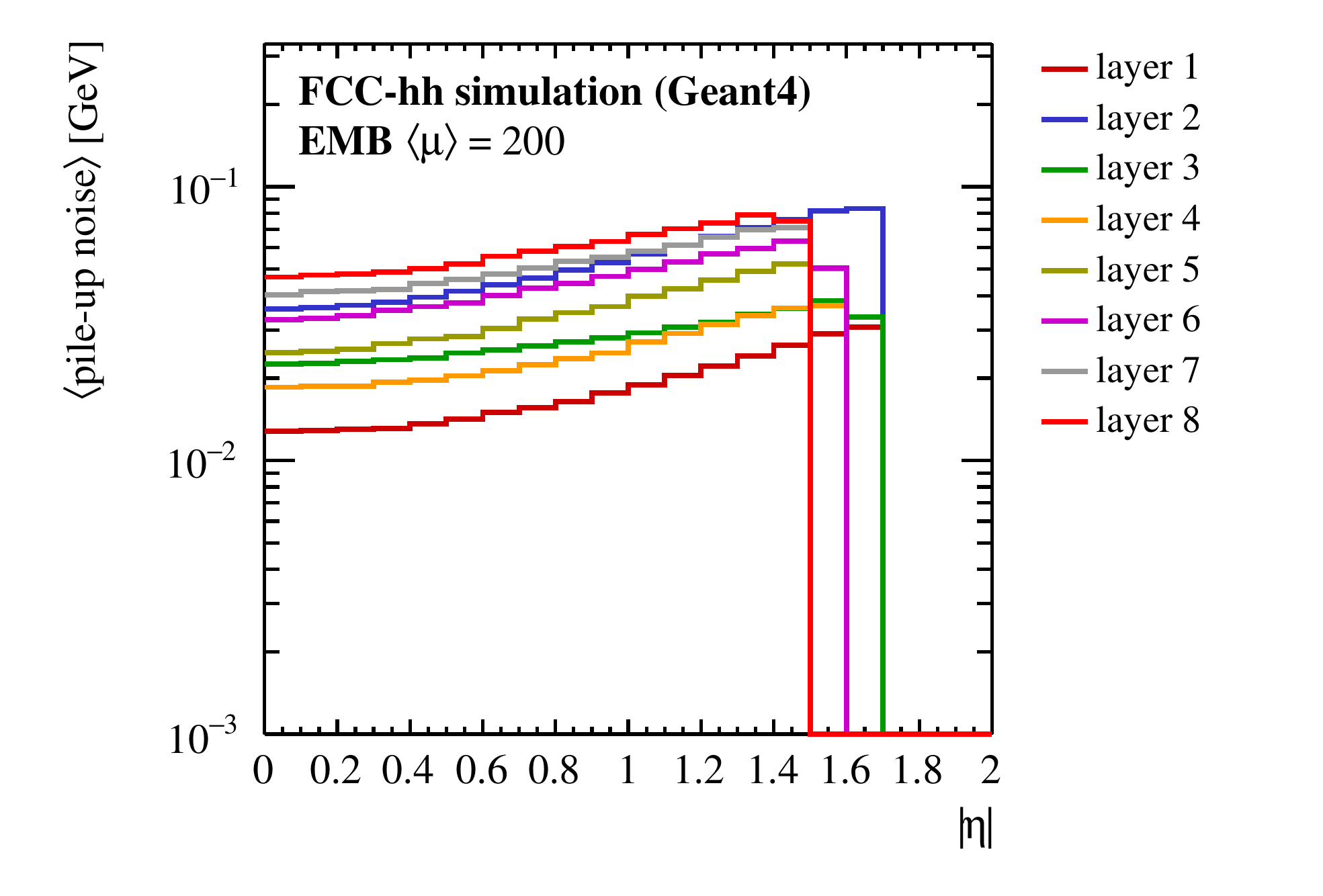}\caption{}\label{fig:software:noise:pileup:cell_EMB}
  \end{subfigure}
  \begin{subfigure}[b]{0.49\textwidth}
    \includegraphics[width=\textwidth]{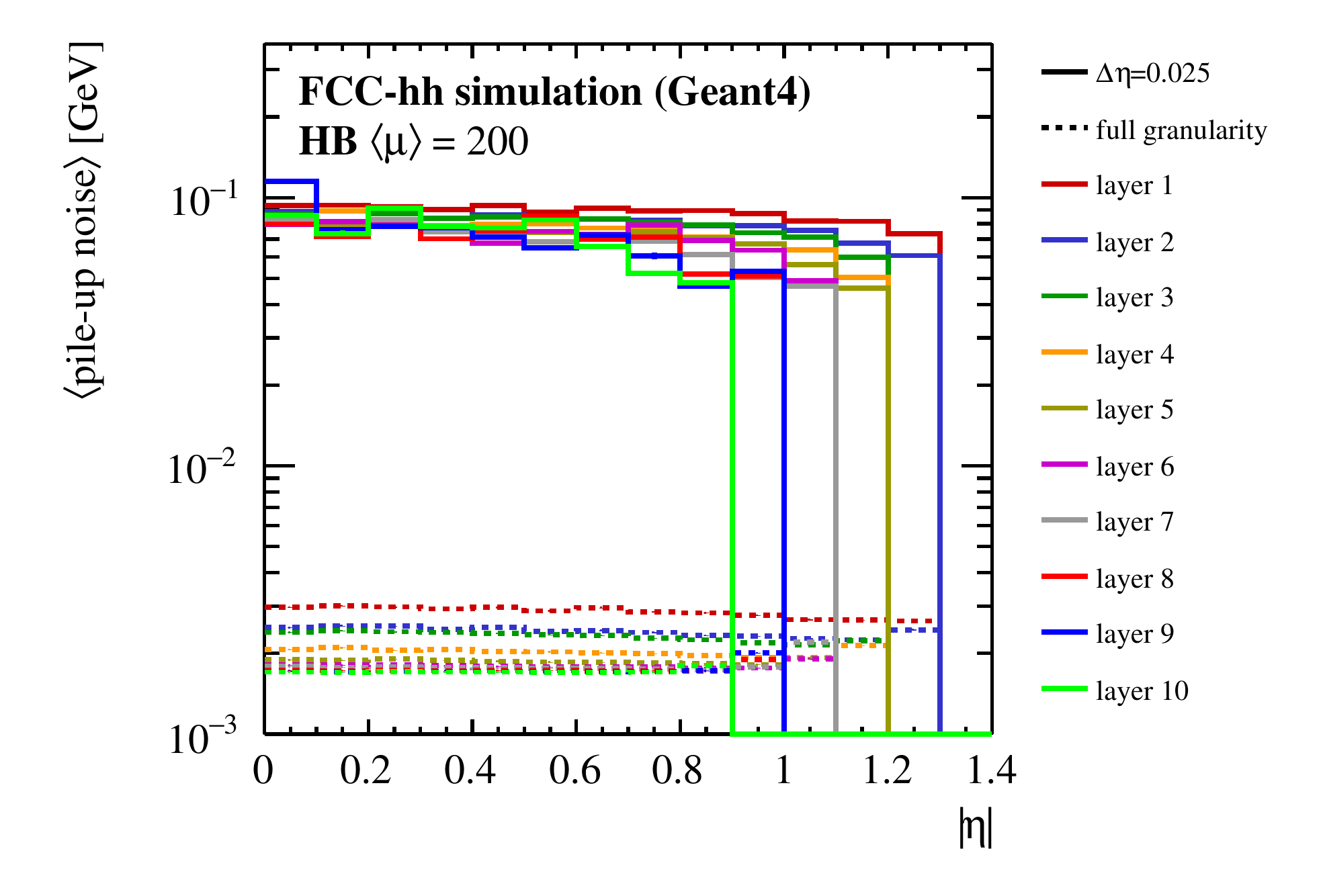}\caption{}\label{fig:software:noise:pileup:cell_HB}
  \end{subfigure}
  \caption{Mean energy deposit per cell for $\left<\mu\right>=200$ in (a) the EMB, and (b) the HB. The values per scintillating tile in the HB is shown in dashed lines.}
\end{figure}

\begin{figure}[ht]
  \centering
  \begin{subfigure}[b]{0.49\textwidth}
  \includegraphics[width=\textwidth]{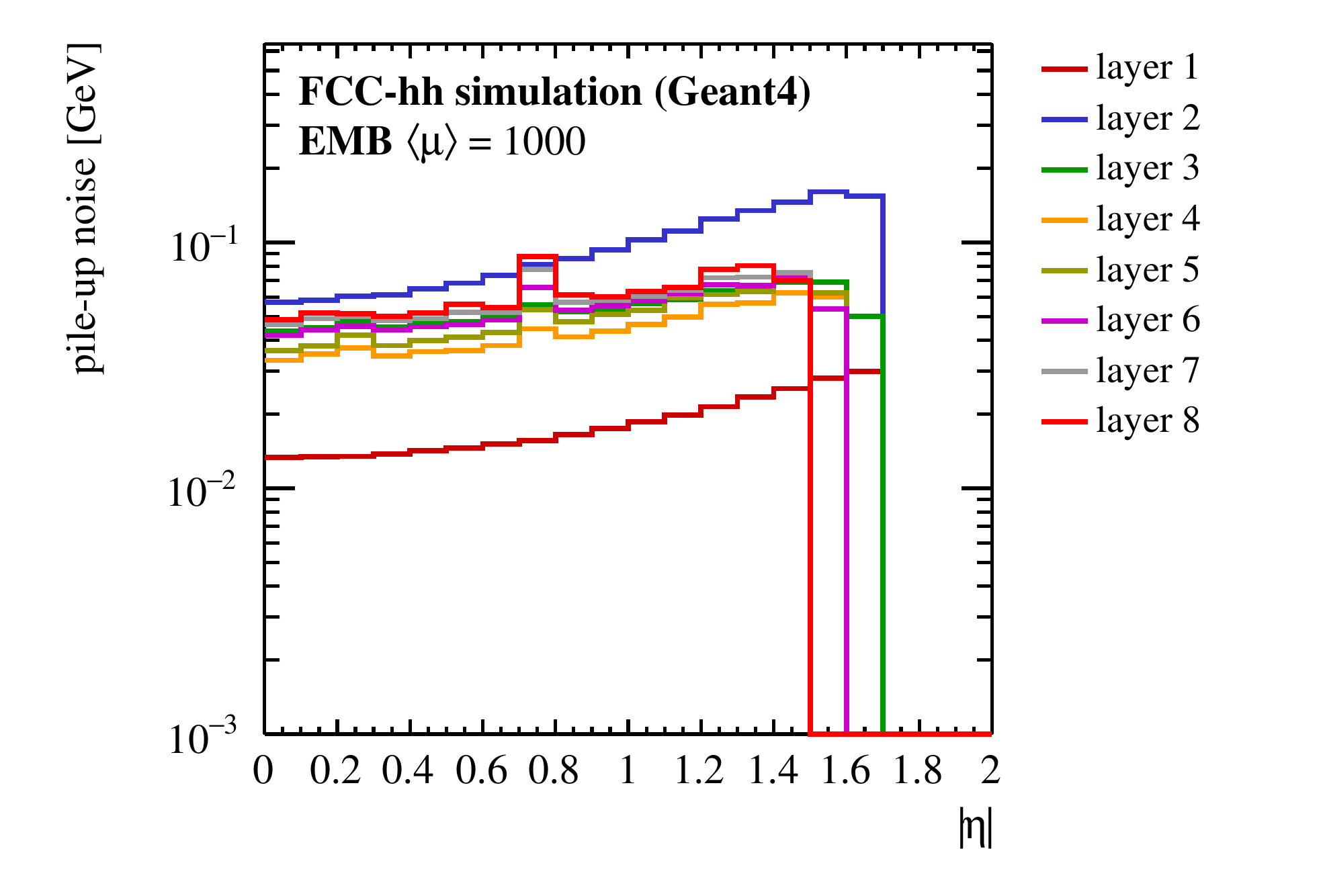}
  \end{subfigure}
  \begin{subfigure}[b]{0.49\textwidth}
  \includegraphics[width=\textwidth]{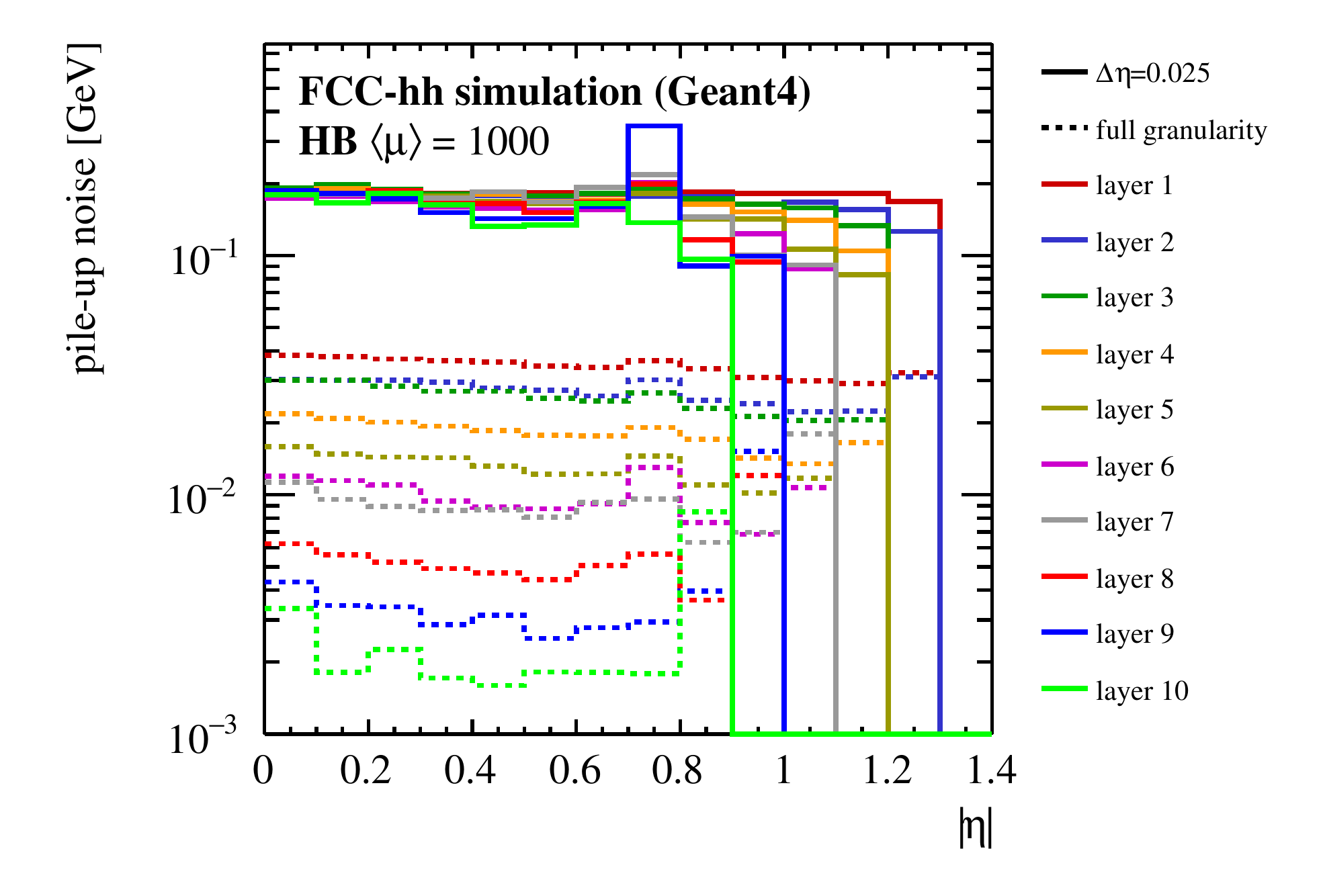}
  \end{subfigure}
  \caption{Pile-up noise level per cell for $\left<\mu\right>=1000$ in (a) the EMB, and (b) the HB. The noise per scintillating tile in the HB is shown in dashed lines.}
\label{fig:software:noise:pileup:rms_PU1000}
\end{figure}

The situation becomes more difficult for clusters of variable size and dispersion like topo-cluster. Due to the different cluster volumes and centre of gravity in all three dimensions a simple scaling as presented above does not work. Instead, the performance within a realistic pile-up scenario is studied by the overlay of minimum bias events on top of the hard scatter event, with the topological clustering applying energy thresholds using the expected noise level per cell (electronic noise and pile-up noise quadratically summed) within the cluster formation. However, before the topo-clustering is run, the mean cell energy is corrected for the mean noise per cell, determined from the merged, corresponding number of minimum bias events. The mean cell energies are shown for the EMB and hadronic calorimeter HB per layer and as a function of $\eta$ in Fig.~\ref{fig:software:noise:pileup:cell_EMB} and~\ref{fig:software:noise:pileup:cell_HB}. The cell noise for each tile of the HB shows the expected decreasing noise levels with increasing $\eta$, while the HB with $\Delta\eta=0.025$ segmentation increases the noise due to the merging of up to 7 tiles per cell for higher pseudo-rapidity. Within the topo-cluster algorithm, the cell significance is determined from the expected noise levels. The pile-up noise is estimated by the standard deviation of the cell energy distributions, one example is shown in Fig.~\ref{fig:software:noise:pileup:energy}. The noise levels, determined from merged minimum bias events, are shown for the pile-up scenario of 1000 collisions per bunch crossing in Fig.~\ref{fig:software:noise:pileup:rms_PU1000}. The impact of pile-up noise on the energy reconstruction of hadronic showers is presented and discussed in Sec.~\ref{sec:performance:hadronic:pileup}. 

%% file: tex/software/dnn.tex
\subsection{Reconstruction and Identification using Deep Neural Networks}
\label{sec:software:dnn}
On top of the more conventional reconstruction algorithms presented above, it was also tried to use deep neural networks (DNNs) for particle identification and energy reconstruction in an attempt to make use of the high granularity of the proposed calorimeter system. In this section the implementation of these DNNs is described, the results will be presented in Sec.~\ref{sec:performance:hadronic:compDNN}.

In the last decade, significant advances have ben made with respect to the design and application of DNNs. These were enabled by new algorithms, but also by developments in computing hardware, such as the capabilities of graphics processor units (GPUs) to compute thousands of operations in parallel. Their architecture, in principle consisting of a set of matrix multiplications, and the dedicated hardware can make these networks very fast, such that they are well suited also for triggering applications.

In contrast to boosted decision trees (BDTs), which are widely used in high energy physics and can be interpreted as shallow neural networks, DNNs allow to exploit the structure and symmetries of the input data and can therefore process a large input dimensionality. 
In this context, the sensor signals from the calorimeters can be interpreted as 3-dimensional energy images. Based on these images, DNN-based energy reconstruction and identification of electrons, photons, muons, charged and neutral pions is studied. 

The particles are generated at $\phi=0$ and $\eta=0.36$ with a flat energy spectrum between 10 and 1000\,GeV.
The image is centred using the mean energy deposits of the particle in the barrel calorimeters. For the EMB, $34\times 34 \times 8$ pixels are defined in $\eta$, $\phi$, and layer number, which corresponds to about $\Delta \eta=0.34$ and $\Delta \phi=0.31$ given the EMB granularity. For the HB, the energy deposits of the sensors are considered in $17\times 17 \times 10$ pixels, corresponding to a similar area with $\Delta \eta=0.43$ and $\Delta \phi=0.42$. These energy deposits are superimposed with energy deposits from 0, 200, or 1000 pile-up interactions. The total sample for the particle identification contains about 1\,M events, with equal contributions from all particles. For the energy reconstruction, only charged pions are considered with a total sample size of 1.2\,M events. For performance evaluation, a separate dataset is used in both cases to avoid possible biases from overtraining. 

For both energy reconstruction and identification, the DNN architecture is mostly based on convolutional neural networks\cite{DBLP:journals/computer/Fukushima88, DBLP:journals/nn/Fukushima11, DBLP:journals/tsmc/FukushimaMI83, DBLP:journals/neco/LeCunBDHHHJ89, DBLP:conf/nips/CunBDHHHJ89, DBLP:conf/shape/CunHBB99}. 
For the identification network, the first convolutional layer consists of 8 filters with a kernel size of $5\times 5\times 5$ pixels. In case of the EMB image, strides of $2\times 2\times 1$ are used, such that the output corresponds to $17 \times 17\times 8$ pixels, which allows to merge the EMB and HB images along the layer dimension without information loss.
The combined calorimeter image is fed through two paths of convolutional layers. One path is dedicated to muons, only considering the innermost $7 \times 7$ pixels in $\eta$, $\phi$ which consists of two layers. The other path consists of 5 layers and covers the full image. Both paths are merged and then fed through two dense layers with 128 or 32 nodes, respectively. The final network output is configured as a multi-classifier with 5 output nodes, each corresponding to the predicted probability of the signature stemming from an electron, photon, muon, or charged or neutral pion. The network contains 250\,k free parameters in total.

The energy reconstruction network architecture is based on the concept that convolutional layers are used to locally determine  the sum of the cell energies and a topology based correction to it. The architecture is inspired by the ResNet~\cite{DBLP:journals/corr/HeZRS15} model: in total 4 blocks with convolutional layers are used before their output is passed on to dense layers. Each block consists of one convolutional layer with a kernel size that equals the strides size and parallel additional layers with a larger kernel size, in the following referred to as direct layer and correction layers, respectively. Within the direct layer, the first filter is set to sum the first feature per pixel, while the other filters contain trainable weights. The correction path consists of 3 sequential convolutional layers with larger kernel sizes and more filters. The corresponding weights are trainable and initialised with low values. The last of these layers has the same kernel size and strides as the direct layer, such that its output is added as a small correction. 
Before the 4 blocks, common for EMB and HB, both images are fed through one direct layer without an additional correction path. For the EMB part, this layer is used to reduce the size to $17 \times 17\times 8$ pixels such that it can be merged with the HB image along the calorimeter layer dimension for further processing.
Finally, the output is fed through 3 dense layers with 64, 12, and 1 nodes. An additional multiplicative offset correction is trained using this output fed into a wide dense layer. The typical number of filters for the direct and correction layers is between 16 and 32, kernel sizes do not exceed 30 pixels, such that the total number of free parameters in the model is 130\,k.

The technical implementation is done in Keras~\cite{chollet2015keras} using tensorflow~\cite{tensorflow2015-whitepaper} as backend. The minimisation is performed using the Adam~\cite{DBLP:journals/corr/KingmaB14} optimiser. For the identification, the cross entropy loss is minimised, for the energy determination, a Huber loss~\cite{huber1964} is applied. The loss is modified, such that it follows
\begin{equation}
L = \frac{(E_{p}-E)^2}{E - 8 \text{\,GeV}}
\end{equation}
for $L<0.2$ and grows linearly with the same slope as at $L=0.2$ with $E_{p}-E$. Here, $E$ is the energy of the generated particle and $E_{p}$ the predicted energy by the network. The additional constant term in the denominator introduces an additional focus on the reconstruction of low energetic particles.

%% file: tex/performance/egamma.tex
\subsection{Electrons and Photons}
\label{sec:performance:egamma}
\subsubsection{Reconstruction of e/$\gamma$-Objects}
As explained in Sec.~\ref{sec:software:reco}, the energy deposits of particles showering inside the calorimeter need to be summed into clusters to reflect a first, uncorrected estimate of the energy of those particles.
The primary algorithm used for electrons and photons (often also called e/$\gamma$-objects) is the sliding window algorithm (see Sec.~\ref{sec:software:reco:slidingWindow}). As described, it scans a two-dimensional grid of calorimeter cells in pseudorapidity and in azimuth, looking for local maxima. Around the local maximum a cluster of fixed size in $\Delta\eta$ and $\Delta\varphi$ is formed. The energy of this cluster is a sum of the energies of all cells within the cluster. The position in $\eta$ and $\varphi$ is calculated as the energy-weighted mean of the individual cell positions. 
To improve precision corrections outlined in the following section are applied to these quantities. The cluster energy is corrected for the energy lost in the material in front of the detector, mainly the cryostat. The so-called upstream material correction will be discussed in Sec.~\ref{sec:layout:lar:corrections:upstream}.
As explained in Sec.~\ref{sec:layout:lar:corrections:eta}, logarithmic weights of energy are used to improve the position resolution.

Reconstructed clusters take into account the effect of electronics noise in each cell of the detector and pile-up noise is added to the cluster energy. 

%\subsubsubsection{Corrections for electrons and photons}
%\label{sec:layout:lar:corrections}

\subsubsubsection{Upstream Material Correction}
\label{sec:layout:lar:corrections:upstream}

\begin{figure}[h]
  \centering
    \begin{subfigure}[b]{0.49\textwidth}
    \includegraphics[width=\textwidth]{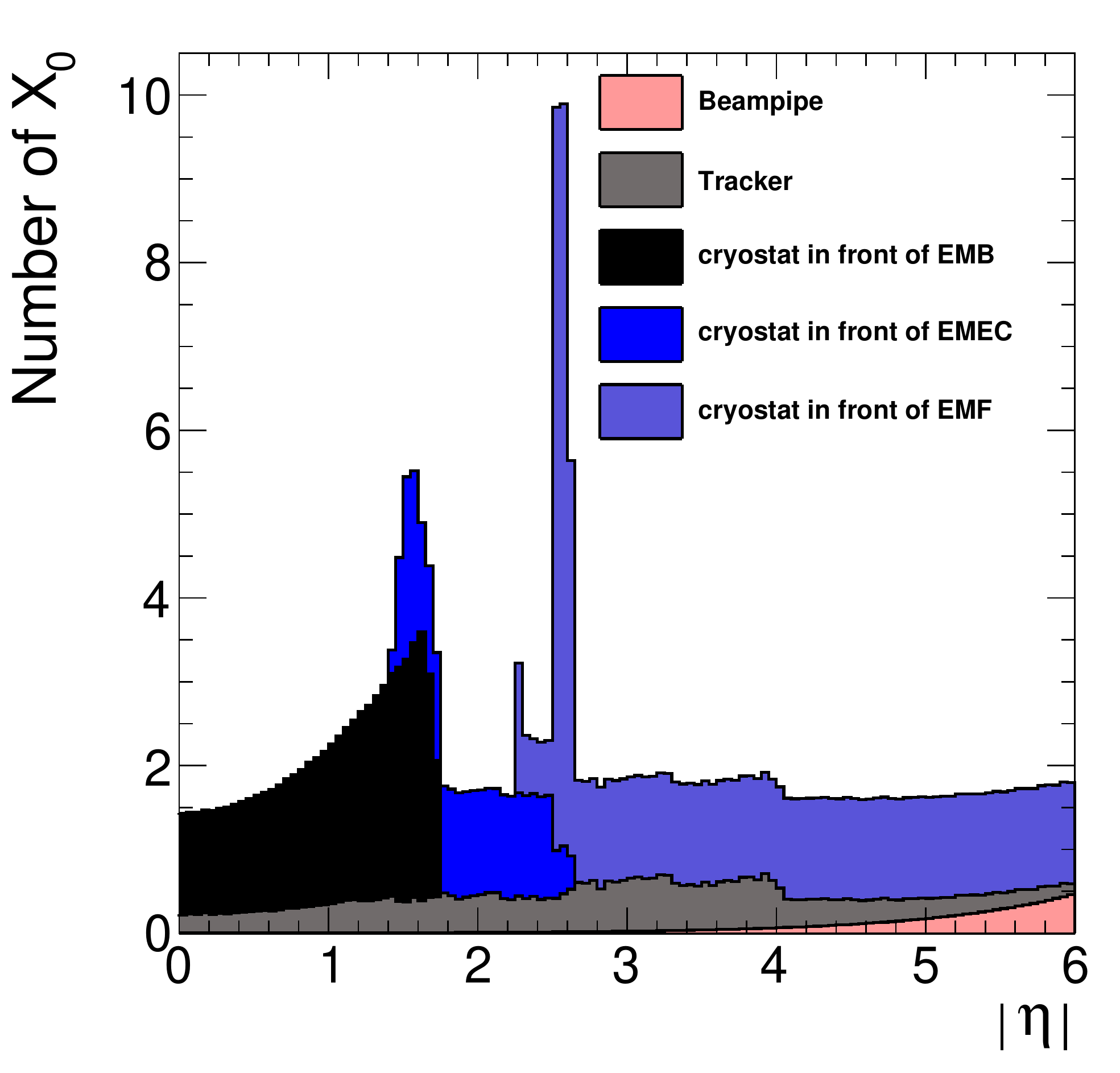}
  \begin{tikzpicture}[overlay]
    \node[anchor=south east] at (0.98\textwidth,7.5) {\textbf{\small{FCC-hh Simulation (Geant4)}}};
  \end{tikzpicture}
  \caption{}\label{fig:layout:lar:matScanLArPre}
  \end{subfigure}
  \begin{subfigure}[b]{0.49\textwidth}
    \includegraphics[width=\textwidth, trim={0 0 0 1.75cm},clip]{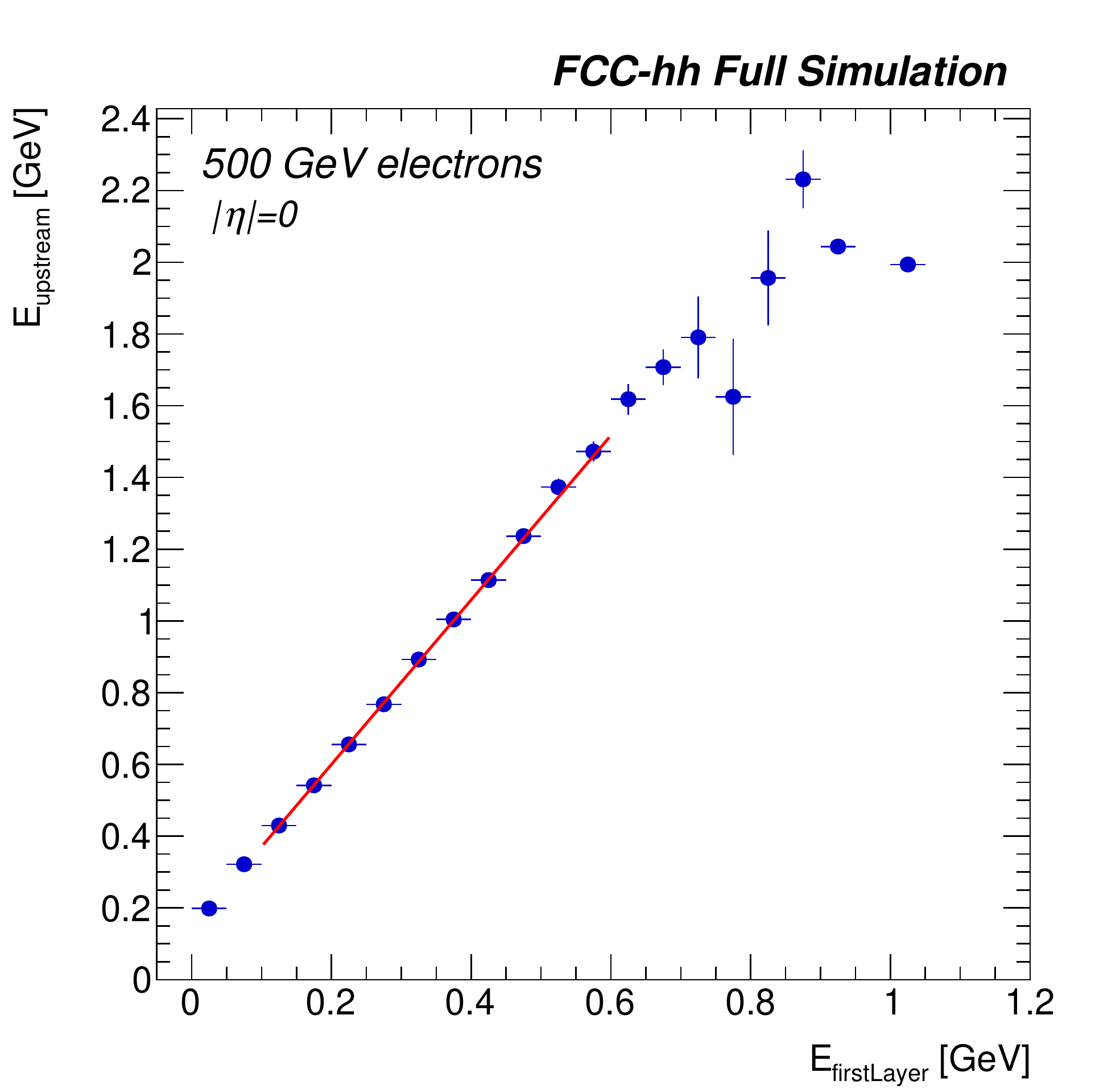}
  \begin{tikzpicture}[overlay]
    \node[anchor=south east] at (0.95\textwidth,7.3) {\textbf{\small{FCC-hh Simulation (Geant4)}}};
  \end{tikzpicture}
  \caption{}\label{fig:layout:lar:upstreamEnergy}
  \end{subfigure}

  \caption{Left (a): Material in front of the calorimeter expressed in units of radiation length, measured from the centre of the detector to the outer boundary of the LAr calorimeter. The ``spike'' at $\abseta=2.5$ corresponds to the wall at the inner bore of the endcap cryostat. Right (b): Linear correlation between the material upstream and the energy deposited in the first calorimeter layer, for 500\,GeV electrons and $\abseta=0.25$. The parametrisation shown in Eq.~\eqref{eq:layout:lar:upstreamFirst} is used.}
\end{figure}
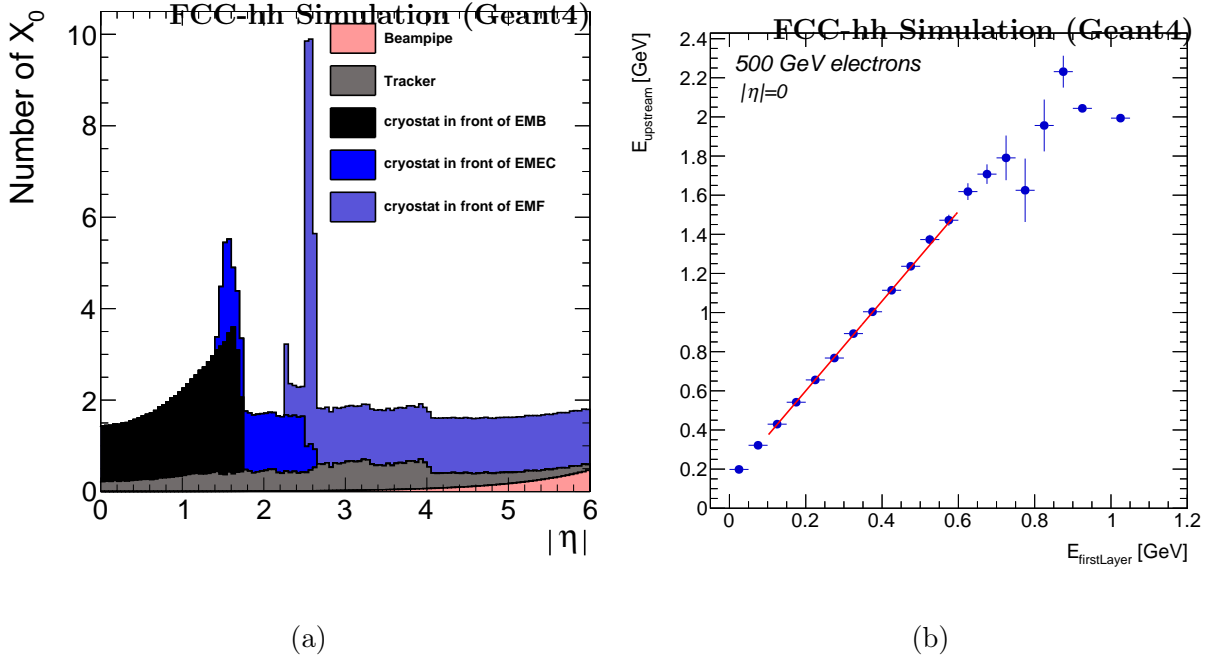

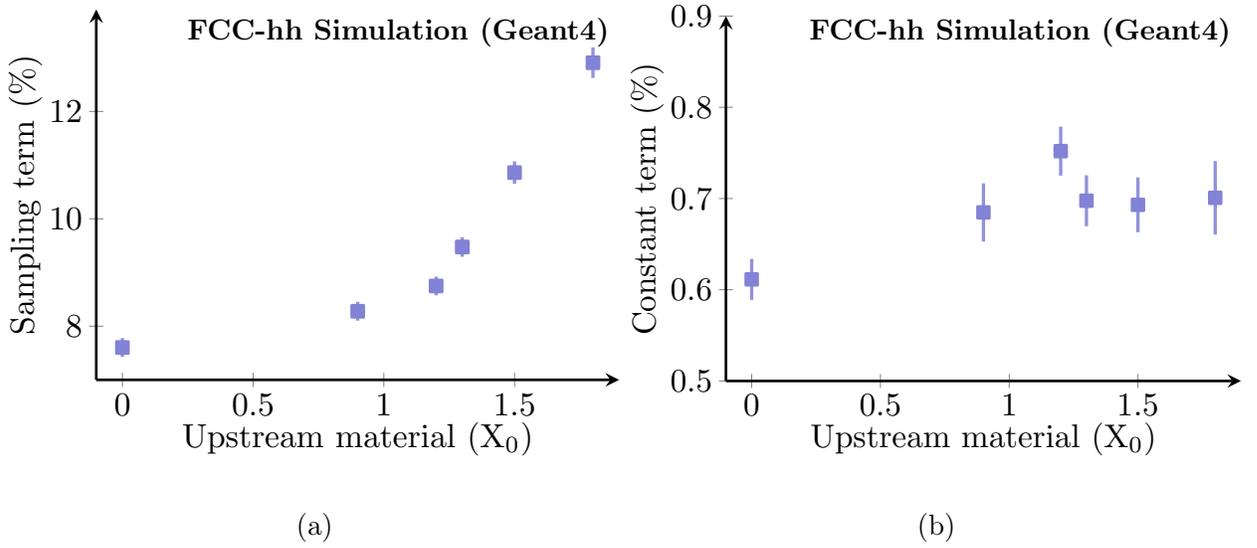
\begin{figure}[h]
  \centering
  \begin{subfigure}[b]{0.49\textwidth}
    \includegraphics[width=\textwidth]{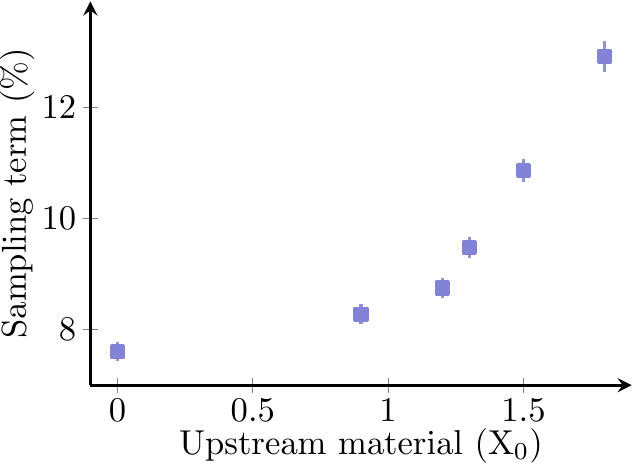}
  \begin{tikzpicture}[overlay]
    \node[anchor=south east] at (\textwidth,5.8) {\textbf{\small{FCC-hh Simulation (Geant4)}}};
  \end{tikzpicture}\caption{}
  \end{subfigure}
  \begin{subfigure}[b]{0.49\textwidth}
    \includegraphics[width=\textwidth]{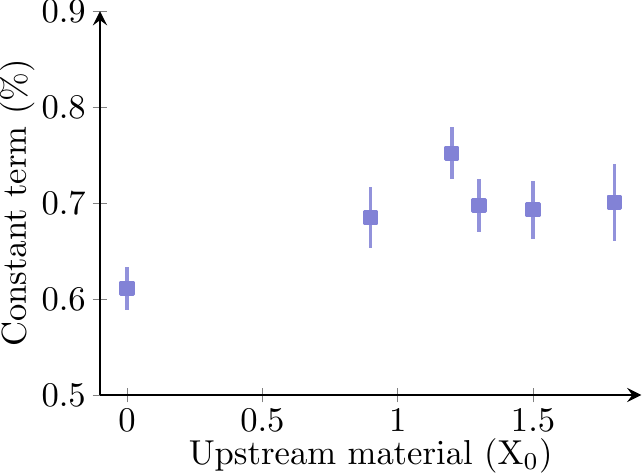}
  \begin{tikzpicture}[overlay]
    \node[anchor=south east] at (\textwidth,5.8) {\textbf{\small{FCC-hh Simulation (Geant4)}}};
  \end{tikzpicture}\caption{}
  \end{subfigure}
  \caption{Dependence of the electron energy resolution on the amount of upstream material expressed in terms of sampling term $a$ (left plot, (a)) and constant term $c$ (right plot, (b)) for the parametrisation of the energy resolution $\frac{\sigma_E}{E}=\frac{a}{\sqrt{E}} \oplus c$. This study was done for an absorber inclination of $30^\circ $.}
  \label{fig:layout:lar:upstreamMaterialInfluence}
\end{figure}

In high energy physics experiments particles coming from the interaction point have to traverse a significant amount of material before reaching the calorimeter (beam pipe, inner tracker, services,...). Upstream material in typical experiments ranges at $\eta=0$ from 0.5\,$X_0$, as realised in the CMS experiment to 3\,$X_0$ in the ATLAS experiment where the EM calorimeter sits inside a cryostat behind the solenoid coil. For the reference detector for FCC-hh we expect a value between these two, since the calorimeter is located inside the solenoid coil, but - due to the cryogenic temperatures necessary for LAr - will sit inside a cryostat. The expected amount of material in front of the active LAr calorimeter is presented in Fig.~\ref{fig:layout:lar:matScanLArPre}, showing values below or around 2\,$X_0$ for most of the pseudorapidity apart from the transition regions between EMB and EMEC and between EMEC and EMF. In the EMB the amount of material increases with $\eta$ due to flatter angle of incoming particles with respect to the cryostat walls.
Since particles will lose energy and start the electromagnetic cascade in this un-instrumented material in front of the active calorimeter, the energy measurement has to be corrected for this lost energy. 
If no correction is applied, the energy resolution, will degrade as a function of the traversed upstream material as can be seen in Fig.~\ref{fig:layout:lar:upstreamMaterialInfluence}. It presents the degradation for the $30^\circ$ inclination of the absorbers, but the same applies for a larger inclination (this optimisation brought the constant term down, leaving the sampling term unchanged). The constant term is almost independent of material, whereas the sampling term increases rapidly for upstream material thicknesses larger than one radiation length.

The energy deposited in the upstream material fluctuates and cannot be corrected for globally. However, for single particle showers, there is a strong correlation between the energy detected in the first calorimeter layer and the energy deposited upstream. This linear relationship is shown in Fig.~\ref{fig:layout:lar:upstreamEnergy} and can be used to estimate the energy lost upstream $E_\mathrm{upstream}$ as a function of the energy measured in the first calorimeter layer $E_\mathrm{firstLayer}$. Equation~\eqref{eq:layout:lar:upstreamFirst} shows the parametrisation that has been used, the two parameters $P_0$ and $P_1$ are functions of the cluster energy $E_\mathrm{cluster}$ and $|\eta|$.   
%As an example, Fig.~\ref{fig:layout:lar:upstreamEnergy} shows $P_0$ and $P_1$ as a function of $E_\mathrm{cluster}$ for 500\,GeV electrons at $|\eta|=0.25$.

\begin{equation}
E_{\mathrm{upstream}} = P_{0}(E_{\mathrm{cluster}},\abseta) + P_{1}(E_{\mathrm{cluster}},\abseta) \cdot E_{\mathrm{firstLayer}}~.
\label{eq:layout:lar:upstreamFirst}
\end{equation}

\begin{figure}[ht]
  \begin{subfigure}[t]{0.45\textwidth}
  \centering
  \includegraphics[width=\textwidth, trim={0 0 0 1.75cm},clip]{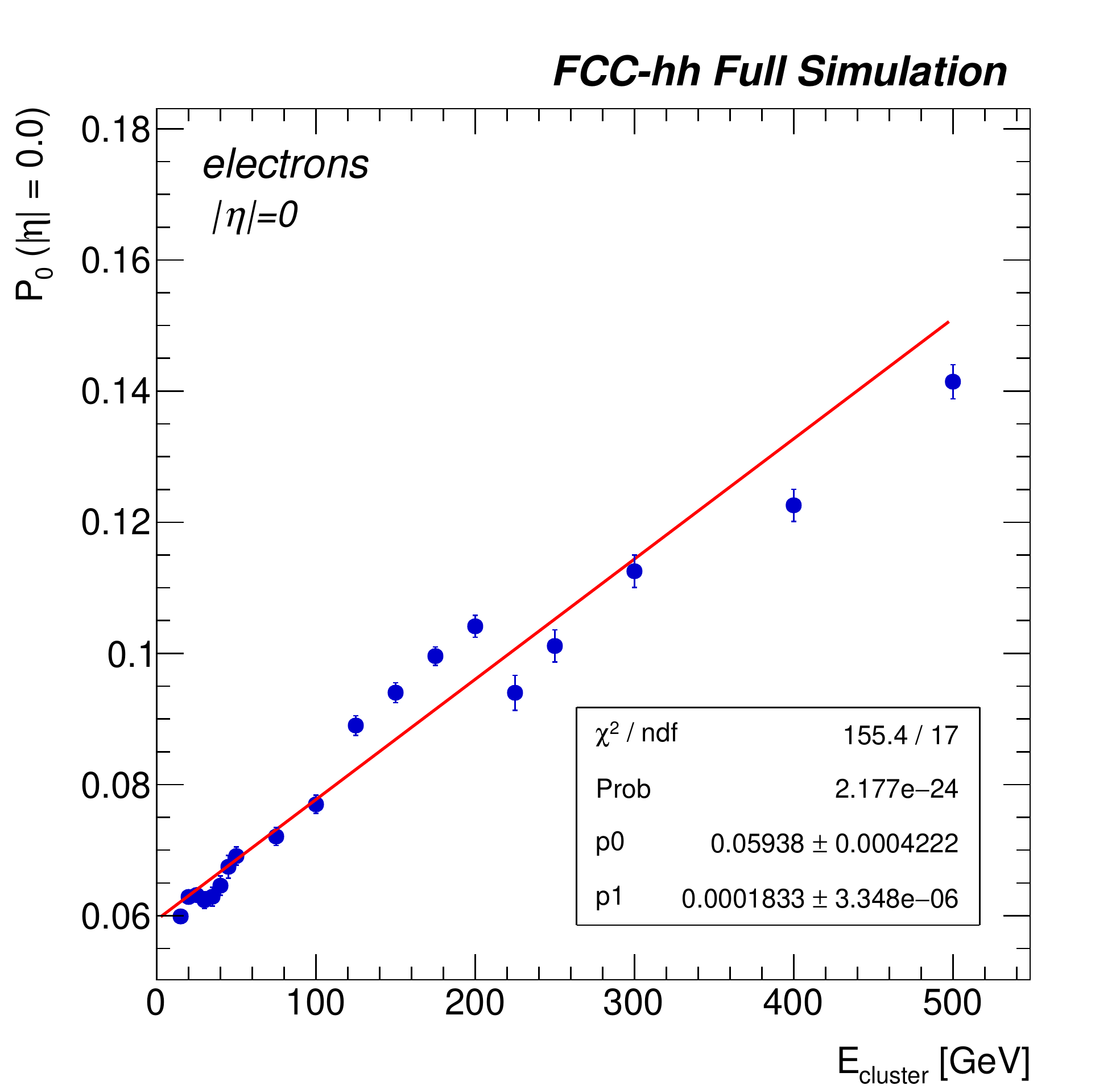}
  \begin{tikzpicture}[overlay]
    \node[anchor=south east] at (0.45\textwidth,6.7) {\textbf{\small{FCC-hh Simulation (Geant4)}}};
  \end{tikzpicture}
  \caption{}\label{fig:layout:lar:upstreamParametersEta0}
  \end{subfigure}
  \hspace{0.05\textwidth}
  \begin{subfigure}[t]{0.45\textwidth}
  \centering
    \includegraphics[width=\textwidth]{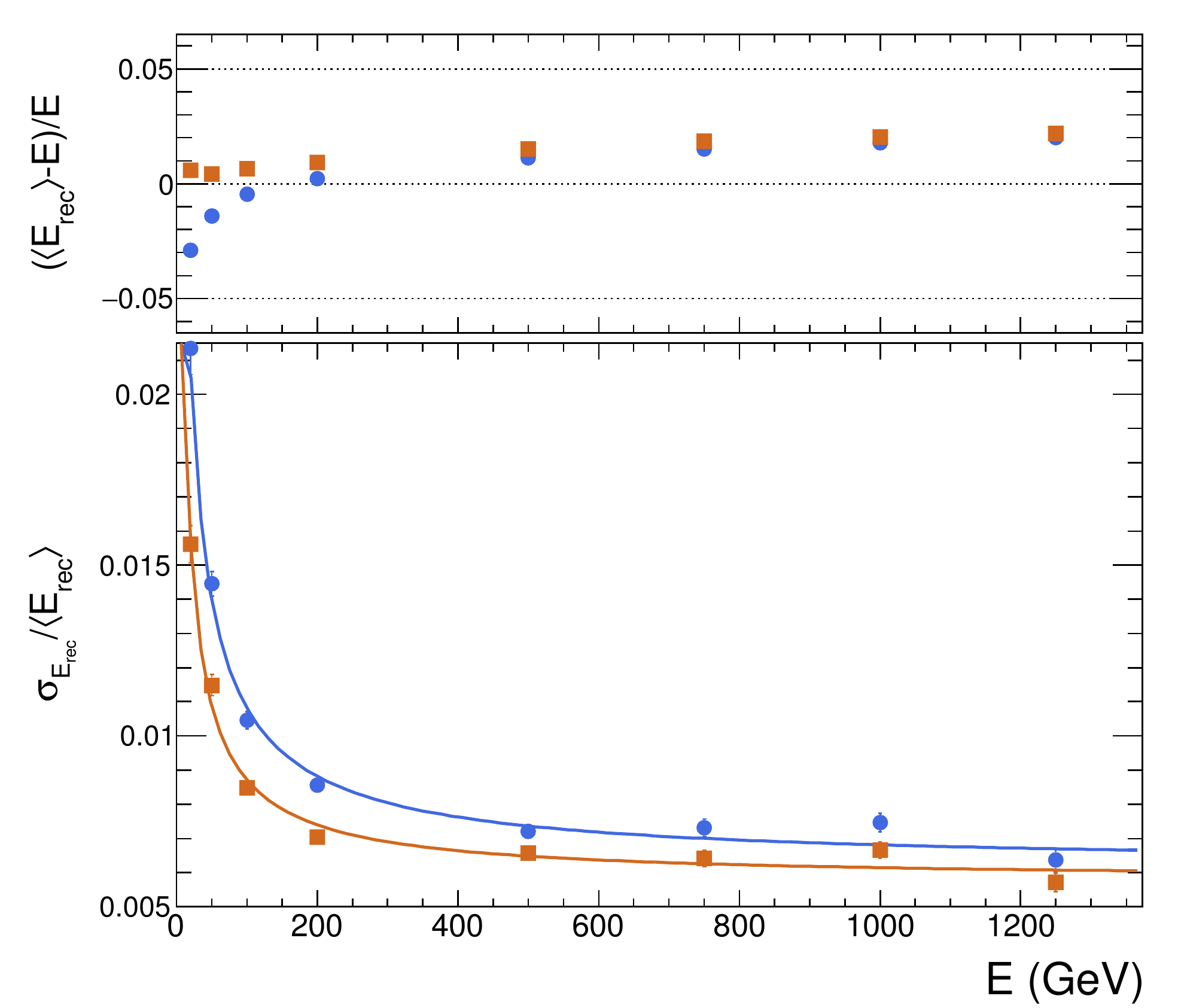}
  \begin{tikzpicture}[overlay]
    \node[anchor=south east] at (0.5\textwidth,6.5) {\textbf{\small{FCC-hh Simulation (Geant4)}}};
  \end{tikzpicture}
    \caption{}\label{fig:layout:lar:enRes:Corrections}
  \end{subfigure}
  \caption{(a) Parameter $P_0$ of Eq.~\ref{eq:layout:lar:upstreamFirst} as a function of initial particle energy for $\abseta = 0$. $P_0$ is described with Eq.~\eqref{eq:layout:lar:upstreamP0}. In the legend $p_0$ coresponds to $P_{00}$ and $p_1$ to $P_01$. (b) Energy resolution of electrons at $\abseta = 0$. The effect of the upstream material correction is presented (blue circles: no correction, orange squares: correction). Presented results refer to the geometry with an inclination angle of absorber and readout plates of $30^\circ$.
%\todo[inline]{JF: Add a legend in the plot (b).}
  }
\end{figure}
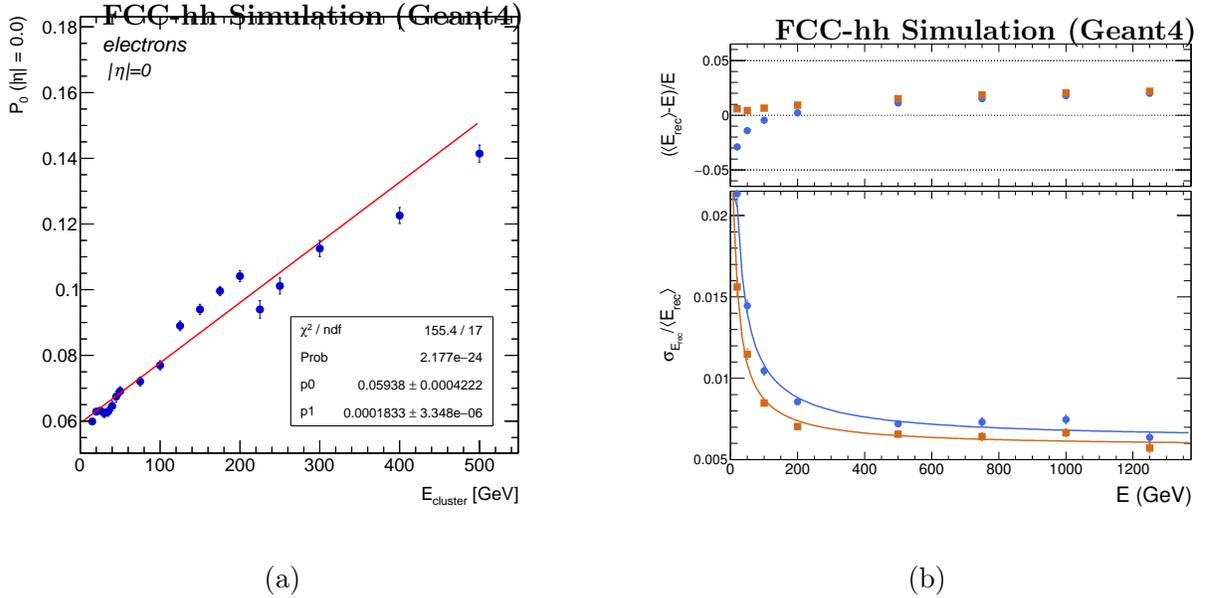

The energy dependence of parameters $P_0$ and $P_1$ for $\eta=0$ is presented in Fig.~\ref{fig:layout:lar:upstreamParametersEta0}. This dependence can again be parameterised using simple functions of the cluster energy $E_\mathrm{cluster}$ as shown in Eq.~\eqref{eq:layout:lar:upstreamP0} for $P_0$ and Eq.~\eqref{eq:layout:lar:upstreamP1} for $P_1$. Those functions were chosen to best fit the obtained data.

\begin{equation}
P_{0} = P_{00}(\abseta) + P_{01}(\abseta) \cdot E_{\text{cluster}}
\label{eq:layout:lar:upstreamP0}
\end{equation}
\begin{equation}
P_{1} =  P_{10}(\abseta) + \frac{P_{11}(\abseta)}{\sqrt{E_{\text{cluster}}}}
\label{eq:layout:lar:upstreamP1}
\end{equation}

Since the amount of upstream material strongly varies with pseudorapidity (see Fig.~\ref{fig:layout:lar:matScanLArPre}), the parameters $P_{00}$, $P_{01}$, $P_{10}$, and $P_{11}$ are all extracted for several $|\eta|$ values (every $\Delta\eta=0.25$). The energy of reconstructed electrons is then calculated as the sum of the energy deposited in the calorimeter $E_\mathrm{cluster}$ and the estimated energy lost upstream $E_\mathrm{upstream}$ according to Eq.~\eqref{eq:layout:lar:upstreamSecond}:

\begin{align}
\label{eq:layout:lar:upstreamSecond}
E = &E_{\text{upstream}} + E_{\text{cluster}}\\
E_{\text{cluster}} = &\sum_{\text{deposits}}^{} E_{\text{deposit}} \cdot f_{\text{sampl}}^{\text{layer}}\nonumber\\
E_{\text{upstream}} = &P_{00} + P_{01} \cdot E_{\text{cluster}} + (P_{10} + \frac{P_{11}}{\sqrt{E_{\text{cluster}}}}) \cdot E_{\text{firstLayer}}\nonumber
\end{align}

%\begin{figure}[ht]
%  \centering
%  \includegraphics[width=0.9\textwidth]{egamma/correctionParams_etaDep.eps}
%  \caption{The correction parameters as a function of pseudorapidity.}
%  \label{fig:layout:lar:upstreamParametersAll}
%\end{figure}

The effect of the corrections on the energy resolution for electrons at $\abseta=0$ is presented in Fig.~\ref{fig:layout:lar:enRes:Corrections}. The sampling term of the energy resolution improves significantly. Note also the improvement in response linearity. The corrected energy resolution for electrons impinging the calorimeter at different pseudorapidities is presented in Tab.~\ref{tab:layout:lar:upstreamMaterialAfter}.
As can be seen, the upstream energy correction achieves very good results and assures excellent electron resolution for the all studied pseudorapidities.

This upstream material correction is inspired by the corrections done for the ATLAS LAr calorimeter~\cite{Aad:2009wy}. The same procedure should be repeated for photons as a function of their conversion radius as well as extended into the other detector parts, the EMEC and the EMF.

Table~\ref{tab:layout:lar:upstreamMaterialBefore} shows the sampling term $a$ and constant term $c$ of the energy resolution for several values of pseudorapidity before (a) and after (b) correction. Especially at higher $\eta$ (more upstream material) a large improvement is obtained confirming the efficiency of the upstream material correction proposed above.

\begin{table}[ht]
  \begin{subfigure}[b]{0.49\textwidth}
    \begin{tabular}{c| c c}
      \hline
      $\eta$ & $a~(\sqrt{\UGeV})$ & $c$ \\
      \hline
      0&	8.7\,\% $\pm 0.1$\,\%& 0.64\,\%$\pm0.01$\,\%\\
      0.25& 8.9\,\%$\pm 0.2$\,\%&0.62\,\%$\pm0.01$\,\%\\
      0.5&	9.2\,\%$\pm 0.3$\,\%&0.59\,\%$\pm0.01$\,\%\\
      0.75 & 10.5\,\%$\pm 0.3$\,\%&0.53\,\%$\pm0.02$\,\%\\
      1&	12.3\,\%$\pm 0.3$\,\%&0.50\,\%$\pm0.03$\,\%\\
      \hline
    \end{tabular}
    \caption{} \label{tab:layout:lar:upstreamMaterialBefore}
  \end{subfigure}
  \begin{subfigure}[b]{0.49\textwidth}
    \begin{tabular}{c| c c}
      \hline
      $\eta$ & $a~(\sqrt{\UGeV})$ & $c$ \\
      \hline
      0&	 6.9\,\%$\pm0.1$\,\%&	0.60\,\%$\pm0.01$\,\%\\
      0.25& 6.5\,\%$\pm0.2$\,\%&	0.58\,\%$\pm0.01$\,\%\\
      0.5&	 6.6\,\%$\pm0.2$\,\%&	0.53\,\%$\pm0.01$\,\%\\
      0.75 & 7.5\,\%$\pm0.3$\,\%&	0.41\,\%$\pm0.01$\,\%\\
      1&	8.5\,\%$\pm0.3$\,\%&	0.24\,\%$\pm0.02$\,\%\\
      \hline
    \end{tabular}
    \caption{}\label{tab:layout:lar:upstreamMaterialAfter}
  \end{subfigure}
  \caption{Energy resolution of electrons for different pseudorapidity values. \textbf{(a)} No correction is applied. \textbf{(b)} The correction for upstream material improves the energy resolution. Presented results refer to the geometry with an inclination angle of absorber and readout plates of $30^\circ$.}
\end{table}

\subsubsubsection{Pseudorapidity Correction}
\label{sec:layout:lar:corrections:eta}

The most straightforward method to calculate the pseudorapidity of an incident particle showering in the calorimeter is to calculate the centre of gravity of the shower~\eqref{eq:layout:lar:corrections:etaGravity}, where $\eta_i$ is the pseudorapidity of cell $i$ and $E_i$ is energy deposited in that cell.

\begin{equation}
\eta_{\mathrm{rec}} = \frac{\sum_{i}E_i\eta_i}{\sum_{i}E_i}
\label{eq:layout:lar:corrections:etaGravity}
\end{equation}

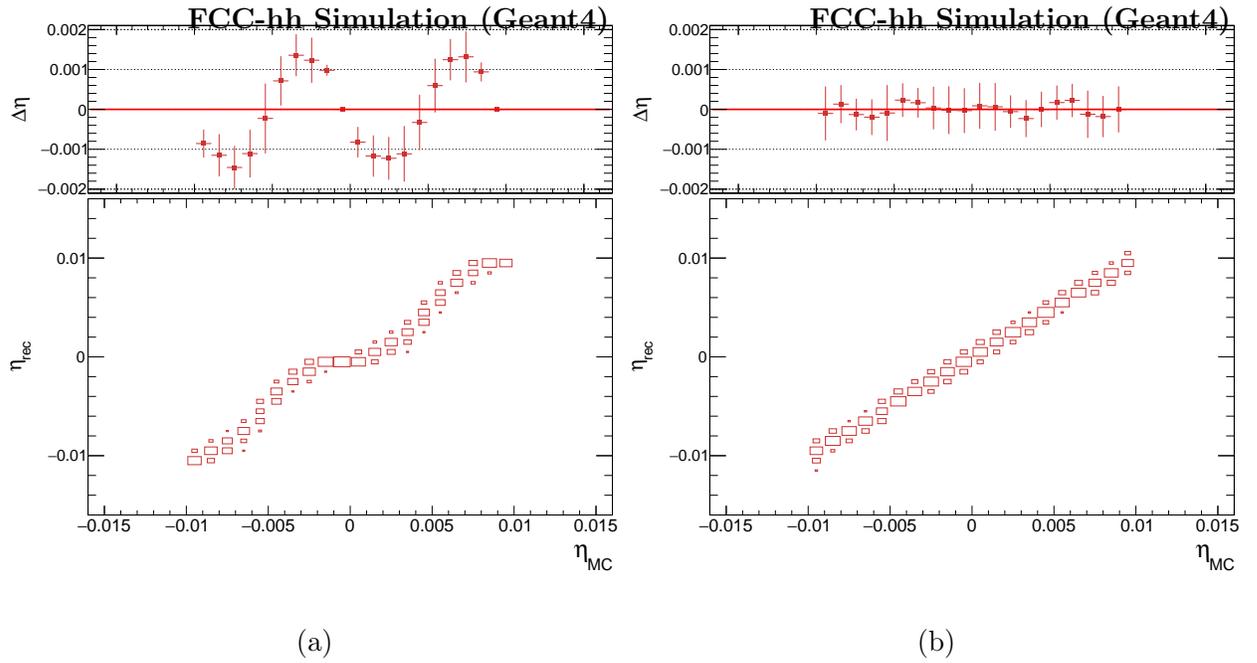
\begin{figure}[ht]
  \centering
  \begin{subfigure}[b]{0.49\textwidth}
  \includegraphics[width=\textwidth]{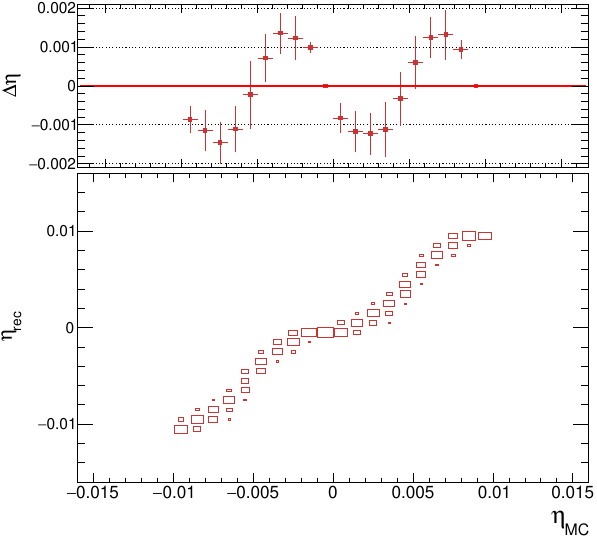}
  \begin{tikzpicture}[overlay]
    \node[anchor=south east] at (\textwidth,7.5) {\textbf{\small{FCC-hh Simulation (Geant4)}}};
  \end{tikzpicture}
  \caption{}\label{fig:layout:lar:corrections:sshape}
  \end{subfigure}
  \begin{subfigure}[b]{0.49\textwidth}
  \includegraphics[width=\textwidth]{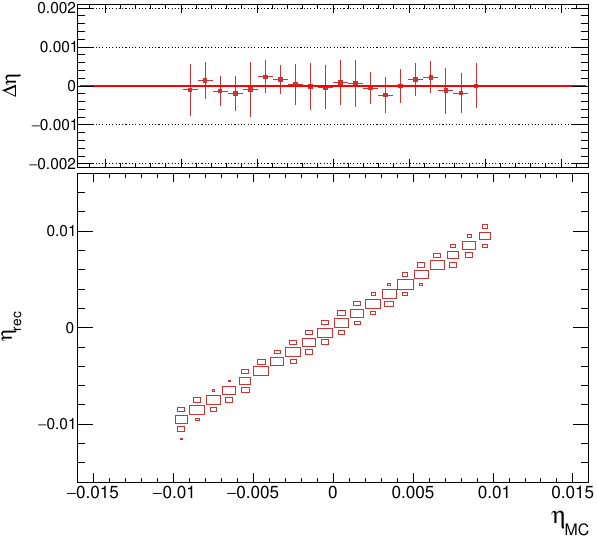}
  \begin{tikzpicture}[overlay]
    \node[anchor=south east] at (\textwidth,7.5) {\textbf{\small{FCC-hh Simulation (Geant4)}}};
  \end{tikzpicture}
  \caption{}\label{fig:layout:lar:corrections:sshapeFixed}
  \end{subfigure}
  \caption{Reconstructed pseudorapidity $\eta_\mathrm{rec}$ as a function of the incident pseudorapidity $\eta_\mathrm{MC}$. The top plots represent the difference $\Delta\eta$ between both values. The pseudorapidity is reconstructed \textbf{(a)} using energy weighting, Eq.~\eqref{eq:layout:lar:corrections:etaGravity}, and \textbf{(b)} logarithmic weighting according to Eq.~\eqref{eq:layout:lar:corrections:etaLog}. One cell in the detector spans from $\eta=-0.005$ to $\eta=0.005$.}
\end{figure}

The result of this position calculation for 50\,GeV electrons is presented in Fig.~\ref{fig:layout:lar:corrections:sshape}. It can be seen that there are systematic differences between the calculated and generated pseudorapidity of the particles as a function of pseudorapidity. This 'S-shaped' difference can be minimised by taking advantage of the exponential transverse profile of the shower and using the logarithms of cell energies as weights, as shown in Eq.~\eqref{eq:layout:lar:corrections:etaLog}. $E_\mathrm{layer}$ is the energy deposited in the given layer, $w_0^\mathrm{layer}$ is a parameter that needs to be carefully chosen.

\begin{eqnarray}
 \eta_{\mathrm{rec}}^\mathrm{layer} &=& \frac{\sum_{i}w_i\eta_i}{\sum_{i}w_i}~,~\text{where}\nonumber\\
  w_i &=& max(0, w_0^\mathrm{layer}+\log{\frac{E_i}{E_\mathrm{layer}}})~.
\label{eq:layout:lar:corrections:etaLog}
\end{eqnarray}

\begin{figure}[ht]
  \centering
  \includegraphics[width=0.55\textwidth]{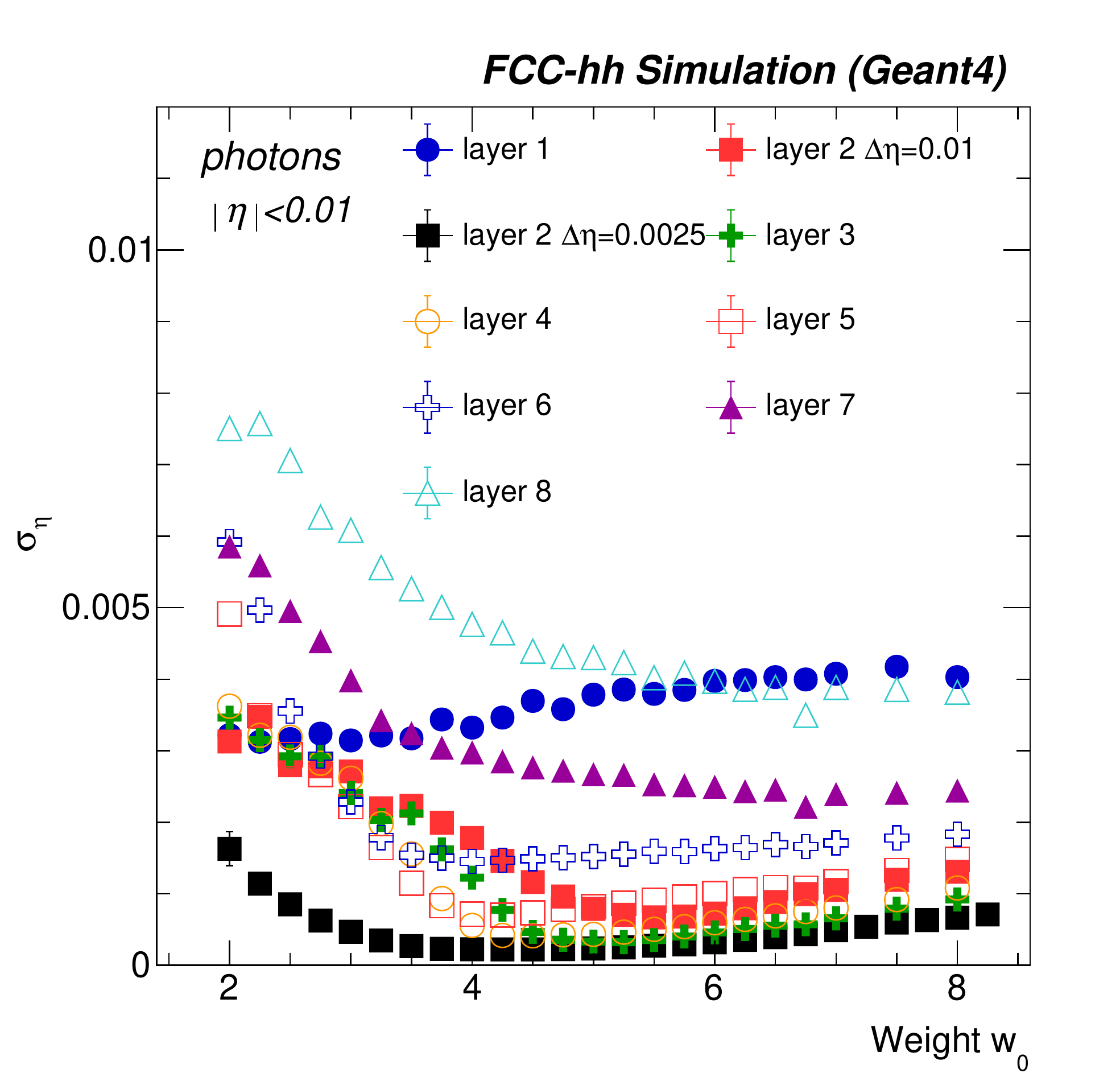}
  \caption{Dependence of the pseudorapidity resolution on the weight parameter $w_0$ for 50\,GeV photons. A resolution is shown for each of the detector layers. The minimum of the pseudorapidity resolution derived the choice of the weight parameter $w_0$. A value of $w_0=5$ is very close to the optimum for all layers and was chosen in this study. The segmentation of all layers used here is $\Delta\eta=0.01$, with the exception of the second layer where both  $\Delta\eta=0.01$ (red full squares) and $\Delta\eta=0.0025$ (black squares) are presented.}
  \label{fig:layout:lar:corrections:sshapeParams}
\end{figure}

The parameter $w_0$ is optimised for each layer in order to minimise the position resolution. Effectively, $w_0$ defines a threshold on the fraction of deposited energy per layer which a cell must exceed to be included in the calculation. It hence adjusts the relative importance of the tails of the shower transverse profile: for $w_0\rightarrow\infty$ all cells are weighted equally, and for too small $w_0$ only few cells dominate in the calculation, making it again position-sensitive. The position resolution for 50\,GeV photons for different values of $w_0$ is presented in Fig.~\ref{fig:layout:lar:corrections:sshapeParams}. It can be seen, that $w_0=5$ is very close to the optimum in all layers apart from the first layer which has little energy deposit. A large improvement due to the fine segmentation of the second layer is also visible (resolution for both $\Delta\eta=0.01$ and $\Delta\eta=0.0025$ in the second layer is presented).

Finally, particle position $\eta_\mathrm{rec}$ is calculated as the weighted mean of the layer positions. Weight could be for instance energy deposited in the given layer (analogously to Eq.~\ref{eq:layout:lar:corrections:etaGravity}).  The calculation of the pseudorapidity using those parameters significantly mitigates the 'S-shaped' systematic differences as can be seen in Fig.~\ref{fig:layout:lar:corrections:sshapeFixed}. In order to improve further the pseudorapidity resolution as a function of energy, instead of energy, resolution in a given layer could be used. This approach, however, requires a prior knowledge about the particle resolution. It has been presented in the results summarised in Sec.~\ref{sec:performance:egamma:position}.

\clearpage

\subsubsection{Energy Resolution and Linearity}

The energy resolution of a calorimeter can be parameterised using Eq.~\eqref{equation:intro:eneResolution} with $a$ as the sampling term, $c$ as the constant term and $b$ as the noise term.

\begin{figure}[ht]
  \centering
  \includegraphics[width=0.5\textwidth]{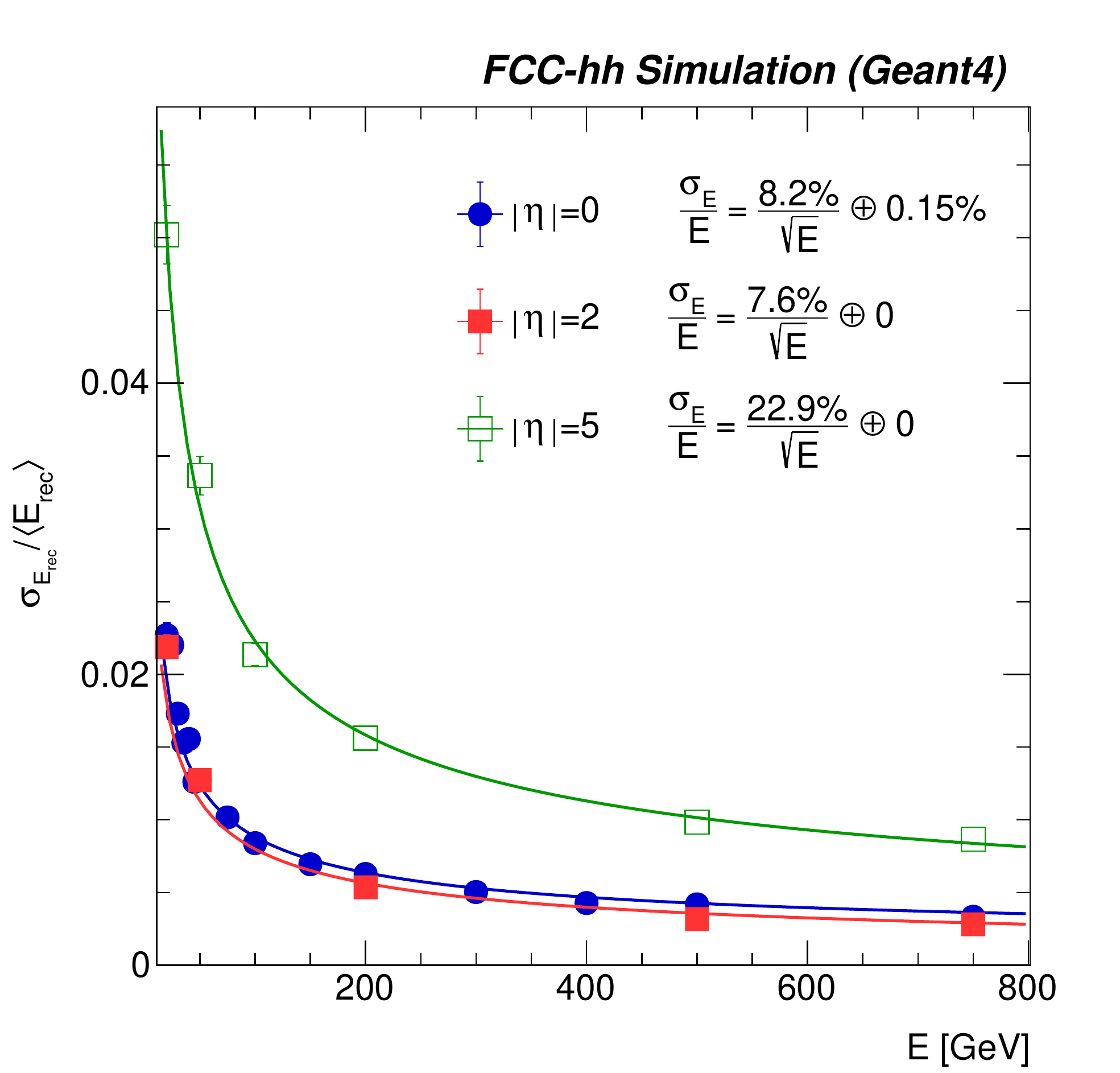}
  \caption{Energy resolution of response for electrons in barrel ($\eta=0$), endcaps ($\abseta=2$) and forward detector ($\abseta=5$).}
  \label{fig:performance:egamma::energyRes:compareEta}
\end{figure}

Energy resolution for single electrons for all calorimeter sub-systems is presented in Fig.~\ref{fig:performance:egamma::energyRes:compareEta}. Those results do not take into account neither electronic nor pile-up noise ($b=0$ for the fit with Eq.~\eqref{equation:intro:eneResolution}). A similar performance is achieved for barrel ($\eta=0$) and for endcaps ($\abseta=2$), with the sampling term equal to $a=8.2\%$ for barrel and $a=7.6\%$ for endcaps. The constant term is $c=0.15\%$ for barrel due to the increasing thickness of liquid argon gap, and is equal to 0 for endcaps, where the ratio of liquid argon to absorber is constant. The obtained energy resolution matches the design goal resolution of \eres{10}{0.7} Eq.~\eqref{eq:em-resolution}. In the forward region for $\abseta>4$ sampling term of the energy resolution increases to $a=23\%$ due to the decrease of the ratio of liquid argon to absorber.

\begin{figure}[ht]
  \centering
  \begin{subfigure}[b]{0.49\textwidth}
  \includegraphics[width=\textwidth]{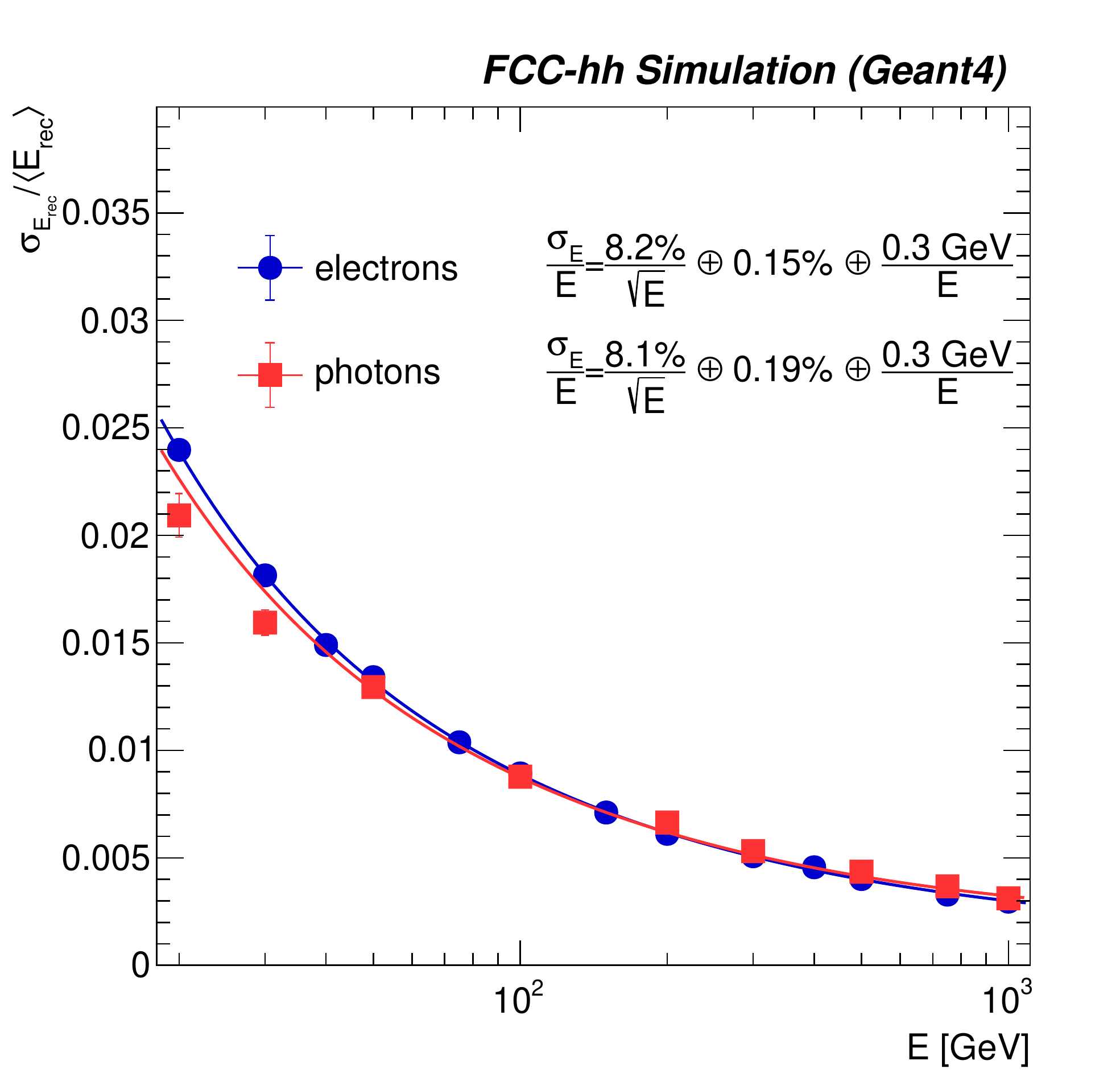}\caption{}
  \end{subfigure}
  \begin{subfigure}[b]{0.49\textwidth}
  \includegraphics[width=\textwidth]{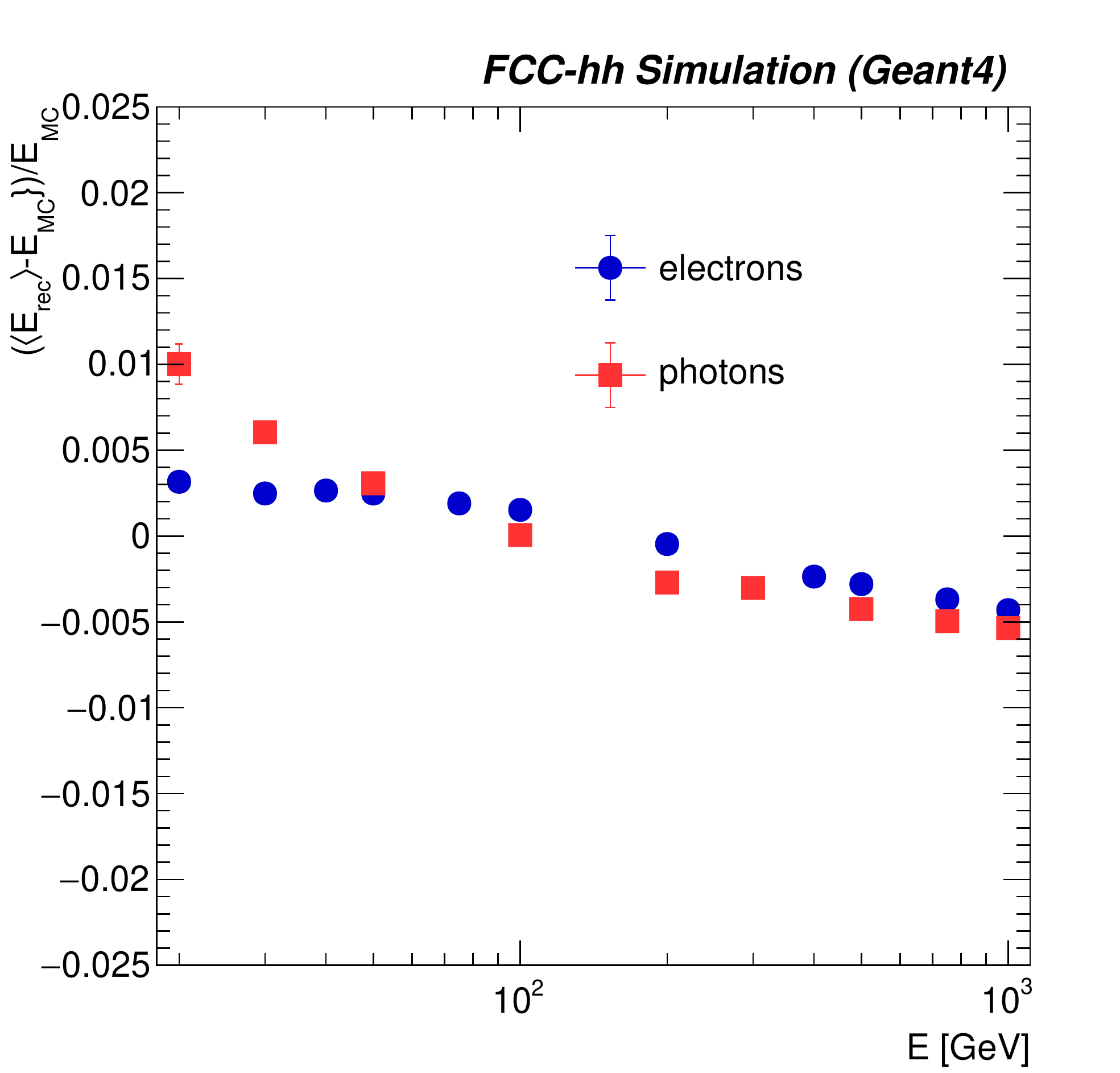}\caption{}
  \end{subfigure}
  \caption{(a) Energy resolution  and (b) linearity of response for single electrons and photons in barrel ($\eta=0$) for no pile--up environment. Same correction factors are applied in reconstruction of both particles.}
  \label{fig:performance:egamma::energyRes:noPU}
\end{figure}

Figure~\ref{fig:performance:egamma::energyRes:noPU} presents energy resolution and linearity of single photons and electrons in barrel detector for no pile-up environment, with the electronic noise included. The obtained energy resolution is similar for photons and electrons. The sampling and constant terms of the energy resolution are fixed to the values obtained without presence of electronic noise,
and noise term $b=0.3$\,GeV matches the estimation of the noise level per cluster that is described in Sec.~\ref{sec:software:noise:electronics:lar}. Same correction factors are applied in the reconstruction of both particles: the upstream material correction and the response scaling. The upstream material correction parameters are extracted from the simulation of electrons and hence the overestimation of the energy deposited by low-energetic photons as can be seen in the linearity plot. Regarding the energy scaling factor, the cluster energy is scaled with an energy-independent factor of $1/0.96$ to compensate for energy deposited outside of the reconstructed cluster. This factor has been extracted from the response of 100\,GeV photons.

\begin{figure}[ht]
  \centering
  \begin{subfigure}[b]{0.49\textwidth}
    \includegraphics[width=\textwidth]{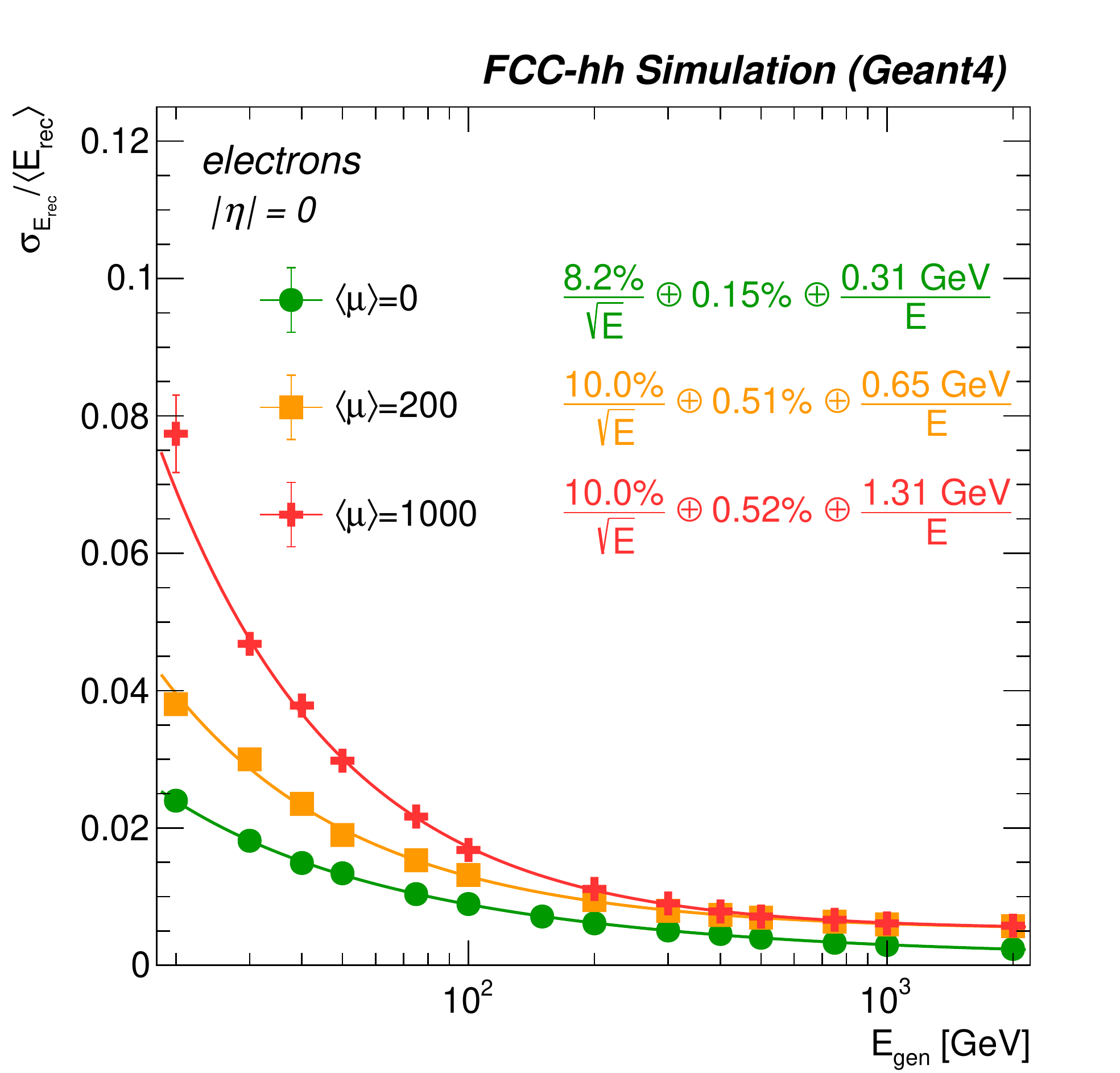}
    \caption{}\label{fig:performance:egamma::energyRes:PU}
  \end{subfigure}
  \begin{subfigure}[b]{0.49\textwidth}
    \includegraphics[width=\textwidth]{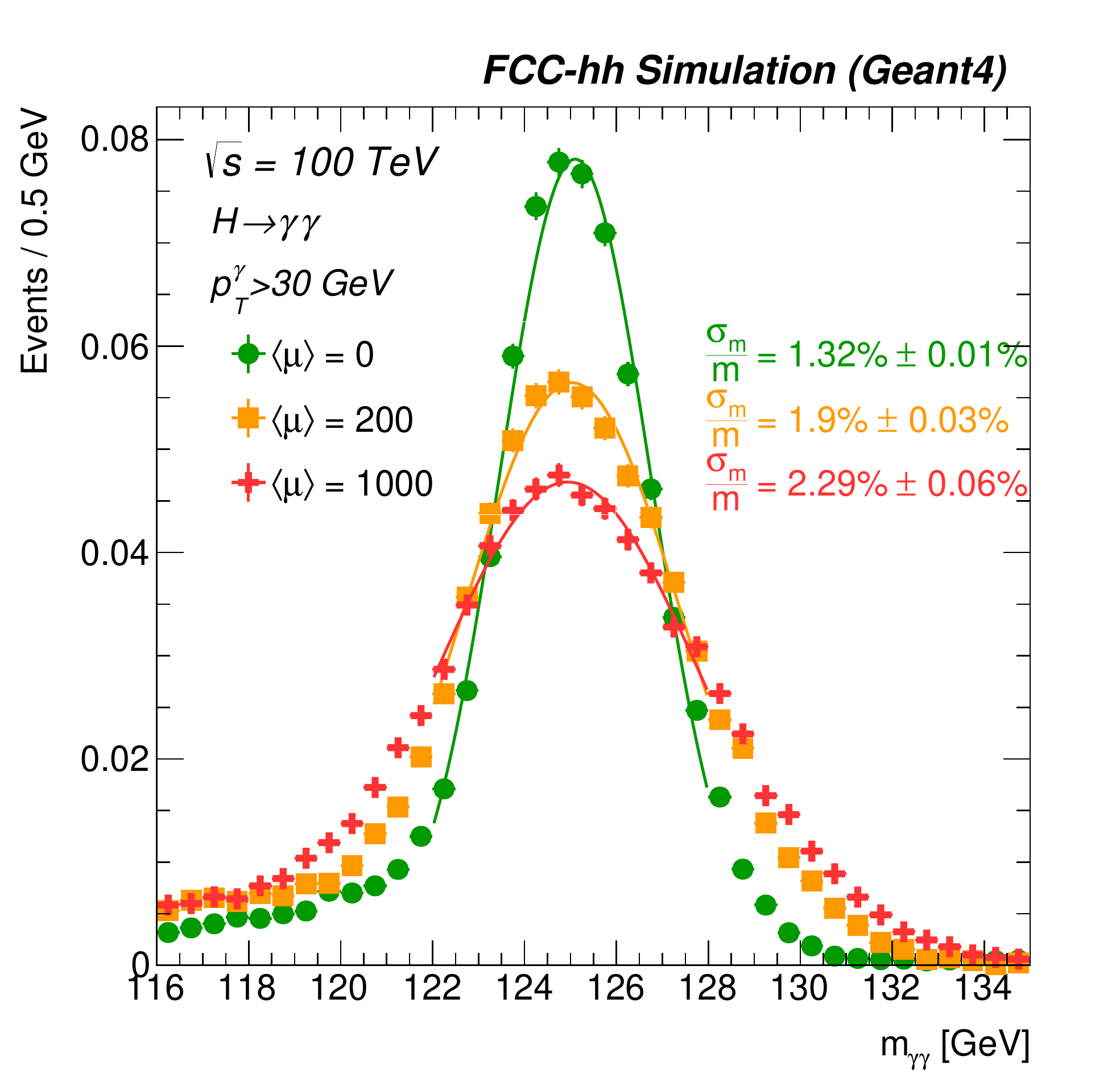}
    \caption{}\label{fig:performance:egamma::mass:PU}
  \end{subfigure}
  \caption{\textbf{(a)} Energy resolution of single electrons for different levels of pile-up at $\eta=0$. The no pile-up configuration uses a cluster size of $\Delta \eta { \times } \Delta \phi { = } 0.07 {\times } 0.17$ while in presence of pile-up the optimised cluster size is $\Delta \eta { \times } \Delta \phi { = } 0.03 {\times } 0.08$. \textbf{(b)} Effect of pile-up on the Higgs invariant mass distribution by selecting two electromagnetic clusters with $p_T^\gamma > 30$\,GeV.}
\end{figure}

\begin{figure}[ht]
  \centering
  \begin{subfigure}[b]{0.49\textwidth}
    \includegraphics[width=\textwidth]{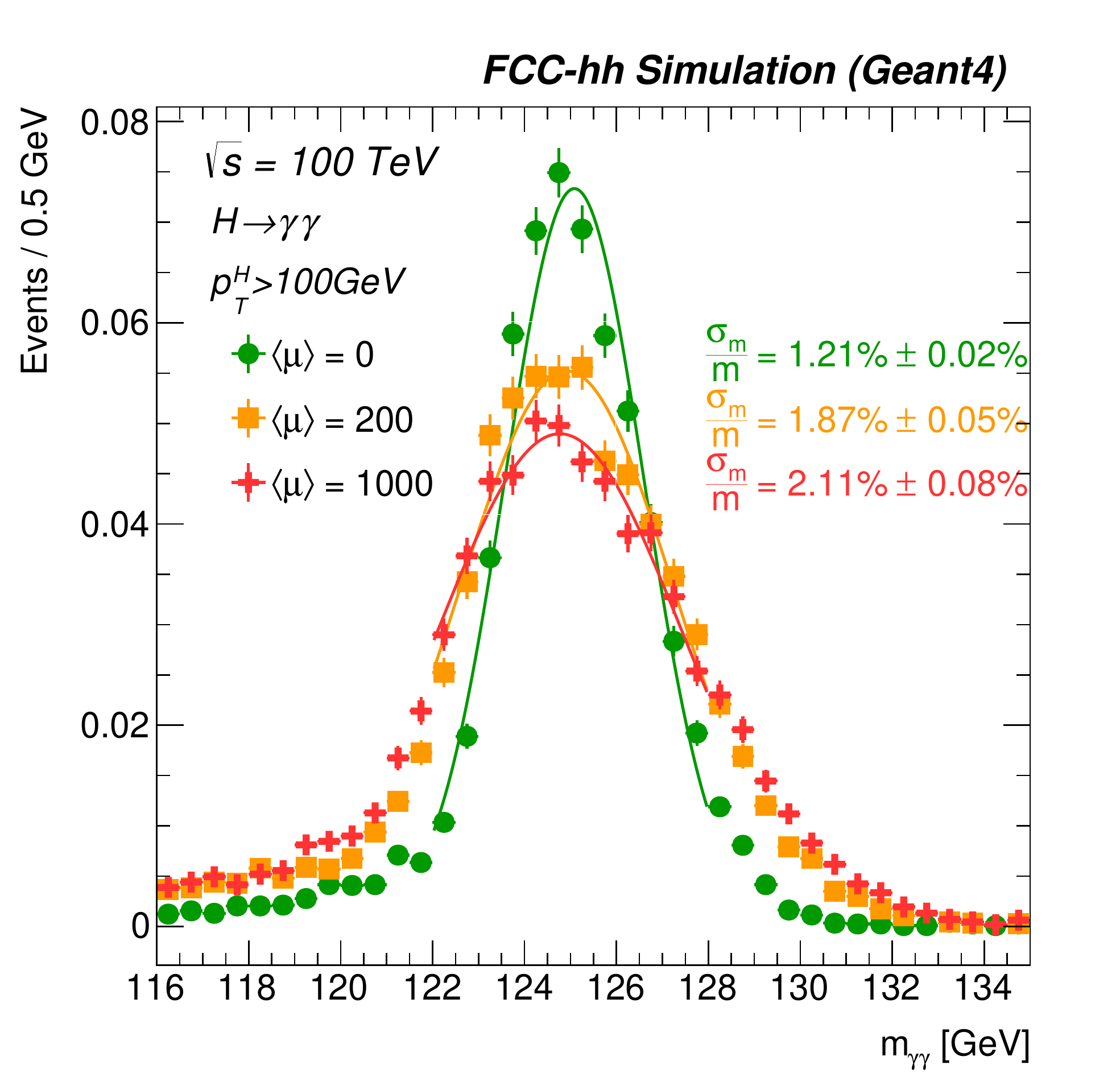}
    \caption{}
  \end{subfigure}
  \begin{subfigure}[b]{0.49\textwidth}
    \includegraphics[width=\textwidth]{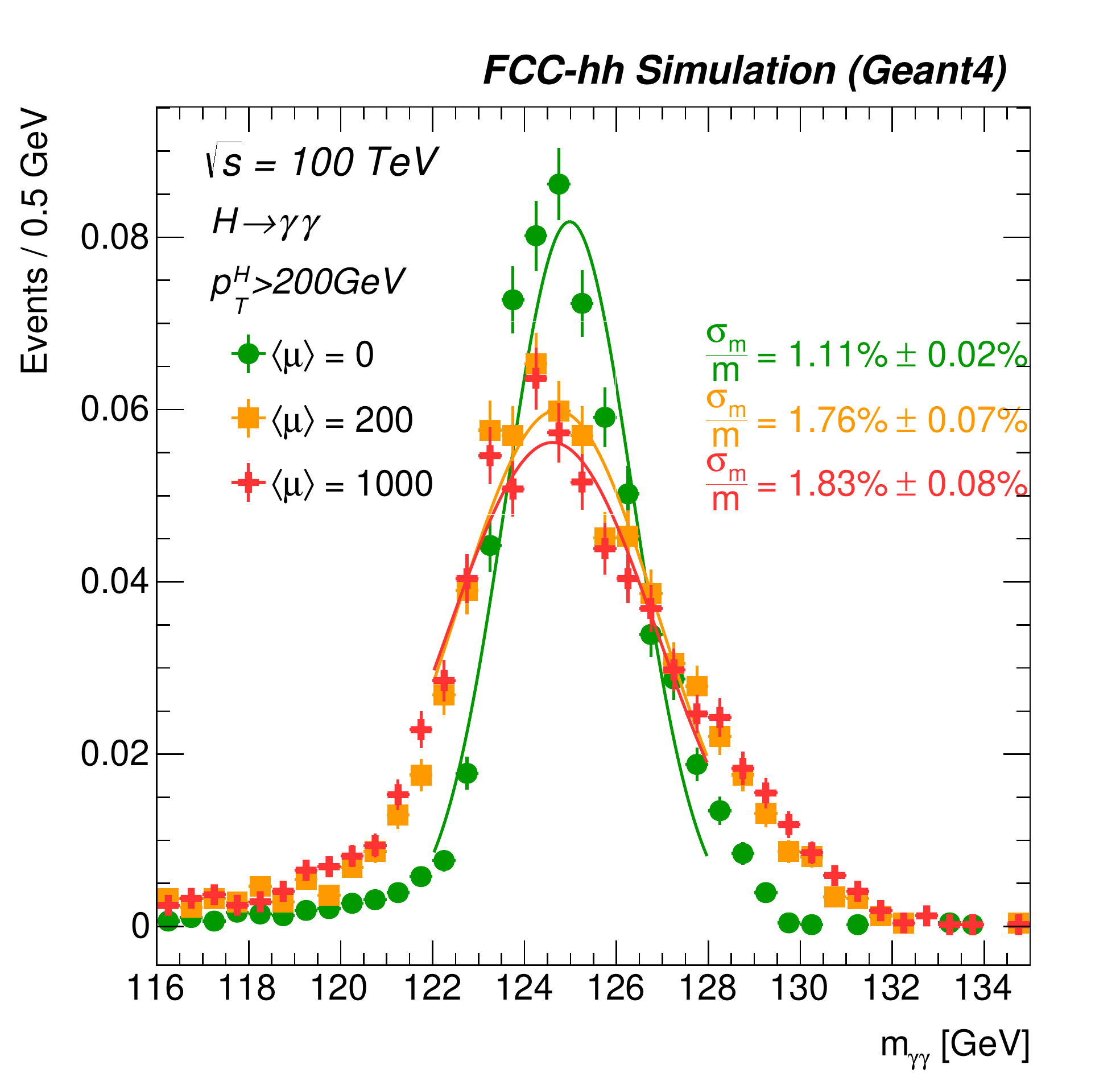}
    \caption{}
  \end{subfigure}
  \caption{Higgs invariant mass distribution by selecting two electromagnetic clusters with $p_T^\gamma > 30$\,GeV in no pile-up environment, for $\left<\mu\right>=200$ and $\left<\mu\right>=1000$. Additional cut on the transverse momentum of the reconstructed Higgs is applied, improving the mass resolution: \textbf{(a)} {$p_T^{H} > 100$\,GeV}, and \textbf{(b)} {$p_T^{H} > 200$\,GeV.} }\label{fig:performance:egamma::mass:PU:HiggsPt}
\end{figure}

In the presence of pile-up, the energy resolution deteriorates. For the sliding window reconstruction with the elliptic window of size $\Delta\eta\times\Delta\varphi=0.03\times 0.08$ the noise term increases to $b=0.65$~GeV for $\left<\mu\right>=200$ and $b=1.31$~GeV for $\left<\mu\right>=1000$, as can be seen in Fig.~\ref{fig:performance:egamma::energyRes:PU}. It has a direct impact on the width of the invariant mass peak for Higgs decaying to two photons generated with Pythia8. Using the current reconstruction with sliding window and calculating invariant mass for all pairs of electromagnetic clusters with energy above $E_\gamma>30$~GeV (no particle identification and no isolation cuts), the width of the invariant mass peak is $1.3\%$ in a no pile-up environment, as can be seen in Fig.~\ref{fig:performance:egamma::mass:PU}. The width increases to $1.9\%$ in presence of pile-up  $\left<\mu\right>=200$ and to $2.3\%$ for $\left<\mu\right>=1000$. This results show the importance of the pile-up mitigation. First of all, information from calorimeters should be complemented with information from tracking detectors. This should allow to estimate the contribution from the charged particles from the pile-up events thus reducing the contribution of the energy deposited in calorimeters originating from the pile-up events. It is important to notice that for physics analysis where Higgs bosons with high transverse momenta are considered, the mass resolution improves, as can be seen in Fig.~\ref{fig:performance:egamma::mass:PU:HiggsPt}. For pile-up  $\left<\mu\right>=1000$ mass resolution improves from $2.3\%$ for an inclusive sample to $2.1\%$ for the transverse cut $p_T^{H} > 100$\,GeV and further to $1.8\%$ for $p_T^{H} > 200$\,GeV.

\subsubsection{Position Resolution}
\label{sec:performance:egamma:position}

\begin{figure}[ht]
  \centering
  \begin{subfigure}[b]{0.49\textwidth}
    \includegraphics[width=\textwidth]{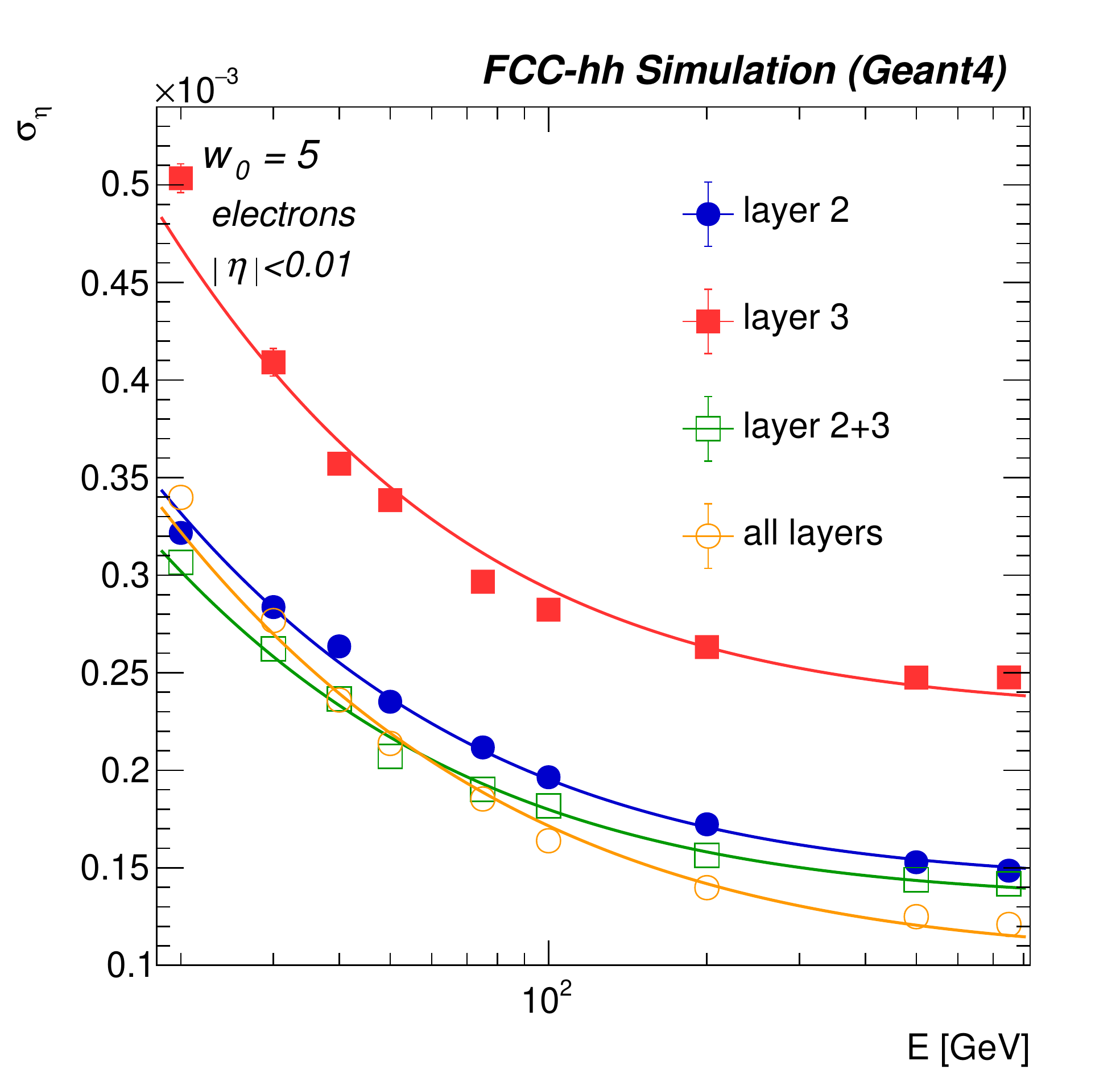}
    \caption{}
  \label{fig:performance:egamma:etaRes}
  \end{subfigure}
  \begin{subfigure}[b]{0.49\textwidth}
    \includegraphics[width=\textwidth]{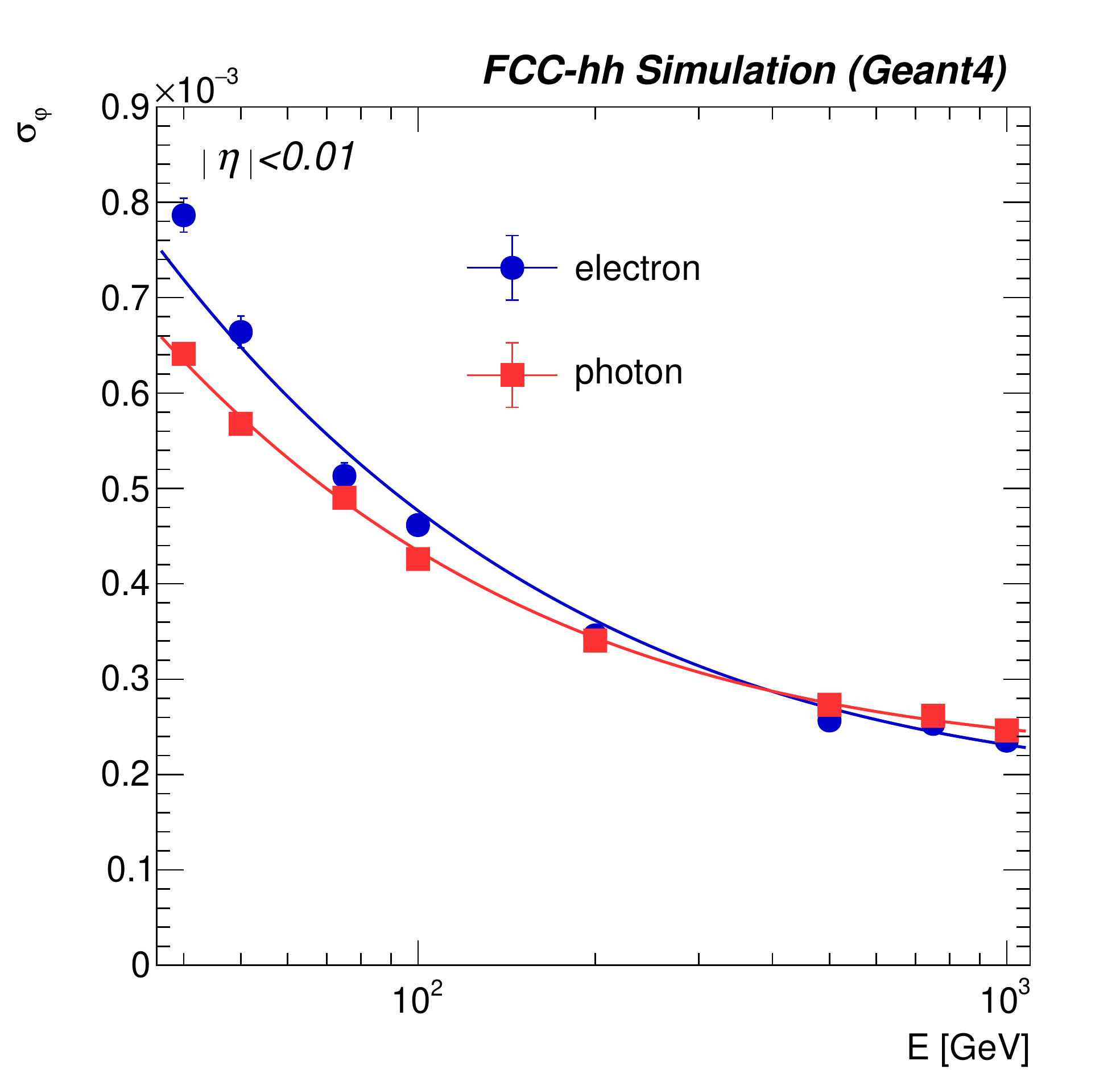}
    \caption{}
  \label{fig:performance:egamma:phiRes}
  \end{subfigure}
  \caption{(a) Pseudorapidity resolution for two best calorimeter layers: second (red full circles) and third (blue full squares), as well as combined measurements of those two layers (green hollow squares) and from all EMB layers (yellow hollow circles). (b) Azimuthal angle resolution for electrons (blue circles) and photons (red squares).}
\end{figure}

Using logarithmic weights and the optimisation described in~\ref{sec:layout:lar:corrections:eta} leads to a pseudorapidity resolution as presented in Fig.~\ref{fig:performance:egamma:etaRes}. The position resolution can be described with Eq.~\eqref{eq:performance:egamma:etaRes}, where energy $E$ is expressed in GeV and parameters $a$ and $c$ for simulation of photons summarised in Tab.~\ref{tab:performance:egamma:etaRes}. The layer combined measurement is calculated according to Eq.~\eqref{eq:performance:egamma:etaCombined}, where $i$ indicates the layers of the detector used in the combined calculation. Two layers that yield the best pseudorapidity resolution are the second layer (finely segmented) and the third one (usually containing the shower maximum). Combined $\eta$ measurement obtained with those layers ($i=2,3$) results in an improvement of $\eta$ resolution. Further improvement is obtained for photons with energies above 50\,GeV once all layers ($i=1, ..., 8$) are used. For low energetic particles, which deposit most of the energy in the first three (four) layers, the resolution degrades and it is more beneficial to use only those layers in the $\eta$ position calculation.

  \begin{equation}
  \sigma = \frac{a}{\sqrt{E}} \oplus c
    \label{eq:performance:egamma:etaRes}
  \end{equation}

  \begin{equation}
  \eta = \frac{\sum_{i}\eta_i\cdot\sigma_i^{-1}}{\sum_{i}\cdot\sigma_i^{-1}}
    \label{eq:performance:egamma:etaCombined}
  \end{equation}

  \begin{table}
    \centering
    \begin{tabular}{|c| c c|}
      \hline
      layer & a $\left(\cdot 10^{-3}\right)$ & c $\left(\cdot 10^{-3}\right)$  \\
      \hline
      2 & 1.34 & 0.14\\
      3 & 1.82 & 0.23\\
      2+3 & 1.21 & 0.13\\
      all & 1.36 & 0.10\\
      \hline
    \end{tabular}
    \caption{Summary of the pseudorapidity resolution presented in Fig.~\ref{fig:performance:egamma:etaRes}. It includes two layers with best resolution: second (finely segmented) and third (usually containing the shower maximum), as well as combined measurements of those two layers and combined measurement of all EMB layers.}
  \label{tab:performance:egamma:etaRes}
  \end{table}

Resolution of the azimuthal angle $\varphi$ for photons and electrons is presented in Fig.~\ref{fig:performance:egamma:phiRes}. Obtained result for photons is $\sigma_{\varphi} = \left(\frac{3.76}{\sqrt{E}} \oplus 0.22 \right) \cdot 10^{-3}$, and for electrons $\sigma_{\varphi} = \left(\frac{4.39}{\sqrt{E}} \oplus 0.18 \right) \cdot 10^{-3}$. For low energetic electrons there is a clear degradation observed due to presence of the magnetic field.

\subsubsection{Timing Resolution}
\label{sec:performance:egamma:timing}

As described in Sec.~\ref{sec:intro:requirements}, the exact measurement of the time of arrival of particles at the calorimeter will be very be necessary to help mitigate pile-up. Since the expected number of proton collisions every bunch crossing (up to 1000) will not happen simulataneously, but, depending on the exact beam parameters, will take place in a time window of 50 to 500\,ps, a time measurement with $\cal{O}$(30\,ps) resolution could help to reduce pile-up substantially by rejecting all particles which arrival time is not compatible with the time of the primary vertex. Such a timing measurement in front of the calorimeters is planned to be introduced for HL-LHC for both ATLAS~\cite{Collaboration:2623663} and CMS~\cite{Collaboration:2296612}. Since the measurement is performed before showers develop, single charged particles will be measured and each track from the inner tracker will get its time tag. 

A timing measurement inside the calorimeter could in addition supply timing information for neutral particles, that would help to identify the primary vertex for e.g. $\mathrm{H}\rightarrow\gamma\gamma$ events as planned for the HL-LHC upgrade of CMS~\cite{Barria:2273277}.
The high granularity will help to obtain separate clusters for each incoming particle and keep merging of clusters at a minimal level. A time tag for each calorimeter cluster would then be a strong handle to reject energy deposits coming from pile-up vertices. Furthermore, within merged clusters the timing measurement of single cells could be used to disentangle parts of the cluster containing energy deposits of different particles.  On top of that, the timing information could also be used to obtain higher connection efficiency between tracks and calorimeter clusters. Again, a timing resolution of  $\cal{O}$(30\,ps) per cluster would be a good target for a FCC-hh calorimeter. 

The timing resolution of a LAr calorimeter will depend on the signal rise time of the ionisation signal after preamplification and shaping and the electronic noise. The signal rise time will be determined by the time constants of the preamplifier (defined by the product of the preamplifier's input impedance and the cell capacitance, see Sec.~\ref{layout:barrel}) and the shaper, as well as the signal amplitude. Whereas the electronic noise depends mainly on the signal attenuation along the signal traces and read-out cables, and the cell capacitance.  
The ATLAS LAr calorimeter was not optimised for a timing measurement, it nevertheless achieves a timing resolution of $\cal{O}$(65\,ps) for high energetic clusters. It is expected that a careful optimisation of all parameters will allow a more precise time measurement of the proposed FCC-hh LAr calorimeter. 

%\subsubsection{Outlook}

%% - Ideas on combined reconstruction with inner tracker \\
%% - First ideas on particle ID, $\pi^0$ rejection (need pre-shower?), jet rejection \\
%% - impact of material of photon conversion (combined tracker study)

%% file: tex/performance/hadronic.tex
\subsection{Hadronic Showers}
\label{sec:performance:hadronic}
The 100\,TeV proton-proton collisions inside the FCC-hh detector will produce jets with transverse momenta of up to 50\,TeV. In such jets 10\,\% of the hadrons are expected to have an energy of at least 1\,TeV~\cite{Carli:2016iuf}. Thus a precise hadron energy measurement up to very high energies is essential for several key physics channels. The single hadron reconstruction is a crucial first step towards the full jet performance.
%%High energetic hadron showers in the full simulation of the FCC-hh detector has been performed with different configurations. 

\subsubsection{Reconstruction at the Cell Level}
\label{sec:performance:hadronic:cellLevel}

\begin{figure}[ht]
  \centering
    	\includegraphics[width=0.65\textwidth]{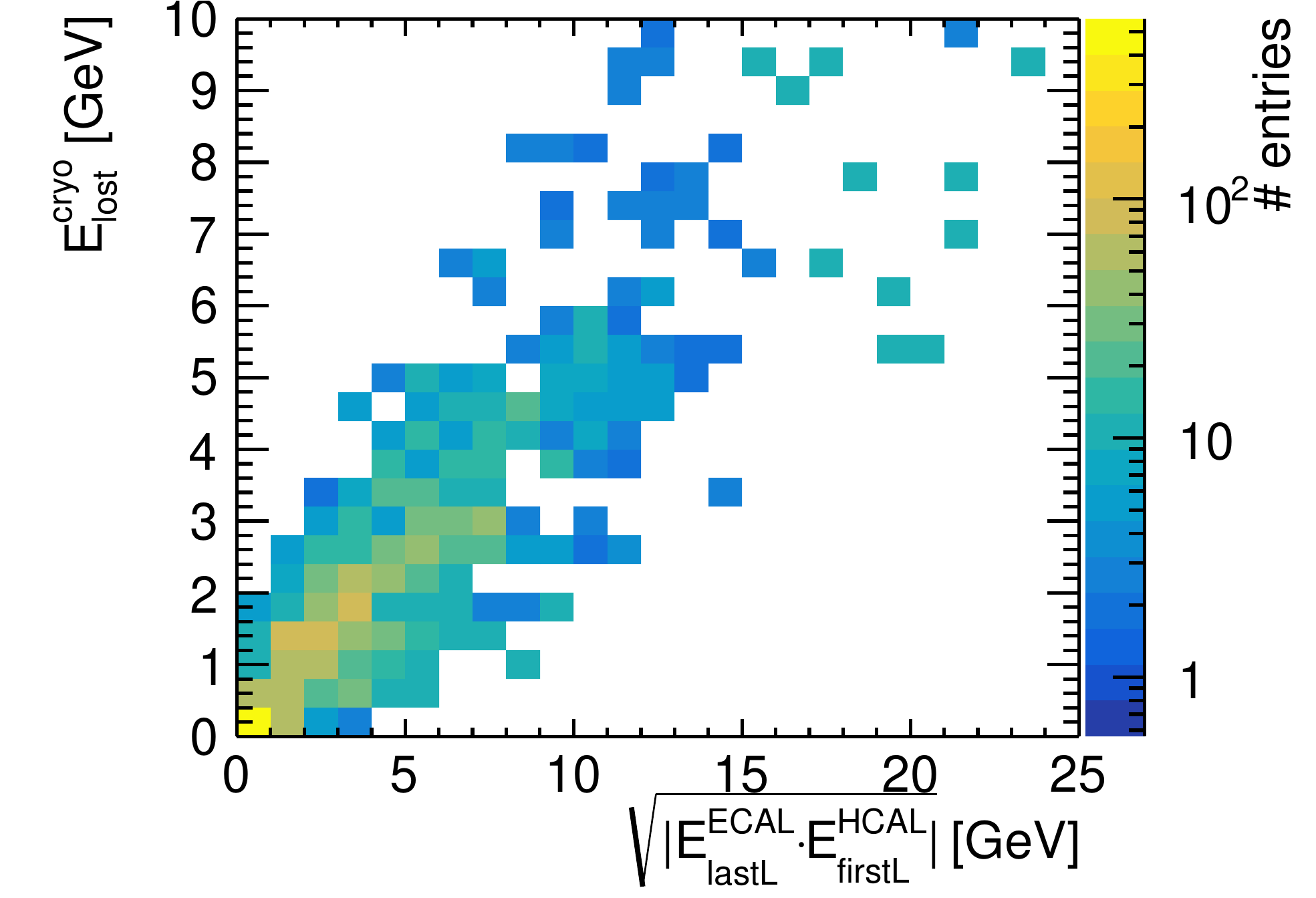}	
        \caption{Correlation of the energy loss of a pion in the LAr cryostat between the EMB and the HB, and the geometric mean of the energy deposited in the last EMB and first HB layer.}\label{fig:performance:hadronic:corrEloss}
\end{figure}

The reconstruction of hadrons in the central barrel region has to combine the energy deposits in the EMB and HB. This can be done simply by summing up all deposits on EM scale. However, this leads to sub-optimal results due to the different hadronic energy scales of the two calorimeters (different $e/h$ ratios), and due to the significant energy loss between the calorimeters in the passive material of the LAr cryostat. To recover the energy losses, the correlation between the energy in the cryostat and the energy deposits in the last EMB and first HB layer can be used as depicted in Fig.~\ref{fig:performance:hadronic:corrEloss}. The energy reconstruction performed in this subsection is done without considering yet the electronics noise.

\subsubsubsection{Benchmark Method}
\label{sec:performance:hadronic:cellLevel:benchmark}
The so-called benchmark method has been developed for ATLAS test-beam measurements. It applies a correction for the lost energy between EMB and HB and calibrates the energy deposits to the hadronic scale. 
The total energy is reconstructed as following
\begin{equation}
	E^{\text{bench}}_{\text{rec}}  = E^{\text{EM}}_{\text{EMB}}\cdot p_{0} + E^{\text{had}}_{\text{HB}} + p_{1}\cdot \sqrt{\left|E^{\text{EMB}}_{\text{last~layer}}\cdot p_{0}\cdot E^{\text{HB}}_{\text{first~layer}} \right|} + p_{2}\cdot \left(E^{\text{EM}}_{\text{EMB}}\cdot p_{0}\right)^{2}
\end{equation} 
The first term, $E^\mathrm{EM}_\mathrm{EMB}$, is the energy sum in the electromagnetic barrel calorimeter. This energy is by default calibrated to EM scale and thus needs to be corrected by the first parameter $p_{0}$. The second term, $E^\mathrm{had}_\mathrm{HB}$, is the energy sum in the hadronic calorimeter which is calibrated to the hadronic scale using instead of the sampling fraction a constant scaling of $f_{\mathrm{hadronic}}=2.4\,\%$. This value has been determined for the HB using a linear fit, similarly as to determine the EM fraction, but comparing the mean pion instead of electron response to the true particle energy. 
The third term determines and corrects the amount of energy lost between EMB and HB as a function of $\sqrt{\left|E^{\text{EMB}}_{\text{last~layer}}\cdot p_{0}\cdot E^{\text{HB}}_{\text{first~layer}}\right|}$, and the fourth term is a correction for the non-compensation of the EMB. 

The parameters $p_{0}$, $p_{1}$, and $p_{2}$ are determined minimising $\xi^{2} = \sum_{i}\frac{\left(E_{\text{true}}^i-E^{\text{bench,i}}_{\text{rec}}\right)^2}{E_{\text{true}}^i}$ with a sample simulation set of 2,000 pion events with energies of 10, 100, 1000, and 10000\,GeV each. Due to the energy dependence of hadronic shower shapes, and the constant parameters $p_{0}$, $p_{1}$, and $p_{2}$, a non-linearity of up to 10\,\% remains. In a last step this non-linearity is corrected, using a simple power law fit to the response. Thus, the final energy is measured following:
\begin{equation}
E^{\text{final}}_{\text{rec}} =
%\begin{cases}
    \left(\frac{E^{\text{bench}}_{\text{rec}}-p_{3}}{p_{4}}\right)^{1/p_{5}} 
%& \quad \text{if E}_{\text{bench}} \geq \left(-\frac{d}{e}\right)^{1/f} \approx 1\,\text{GeV} \\
%     \text{E}_{\text{bench}} & \quad \text{if E}_{\text{bench}} < 1\,\text{GeV}
%  \end{cases}
\end{equation}
which is valid for particle energies between 10\,GeV and 1\,TeV.

\begin{figure}[ht]
  \centering
    	\includegraphics[width=0.5\textwidth]{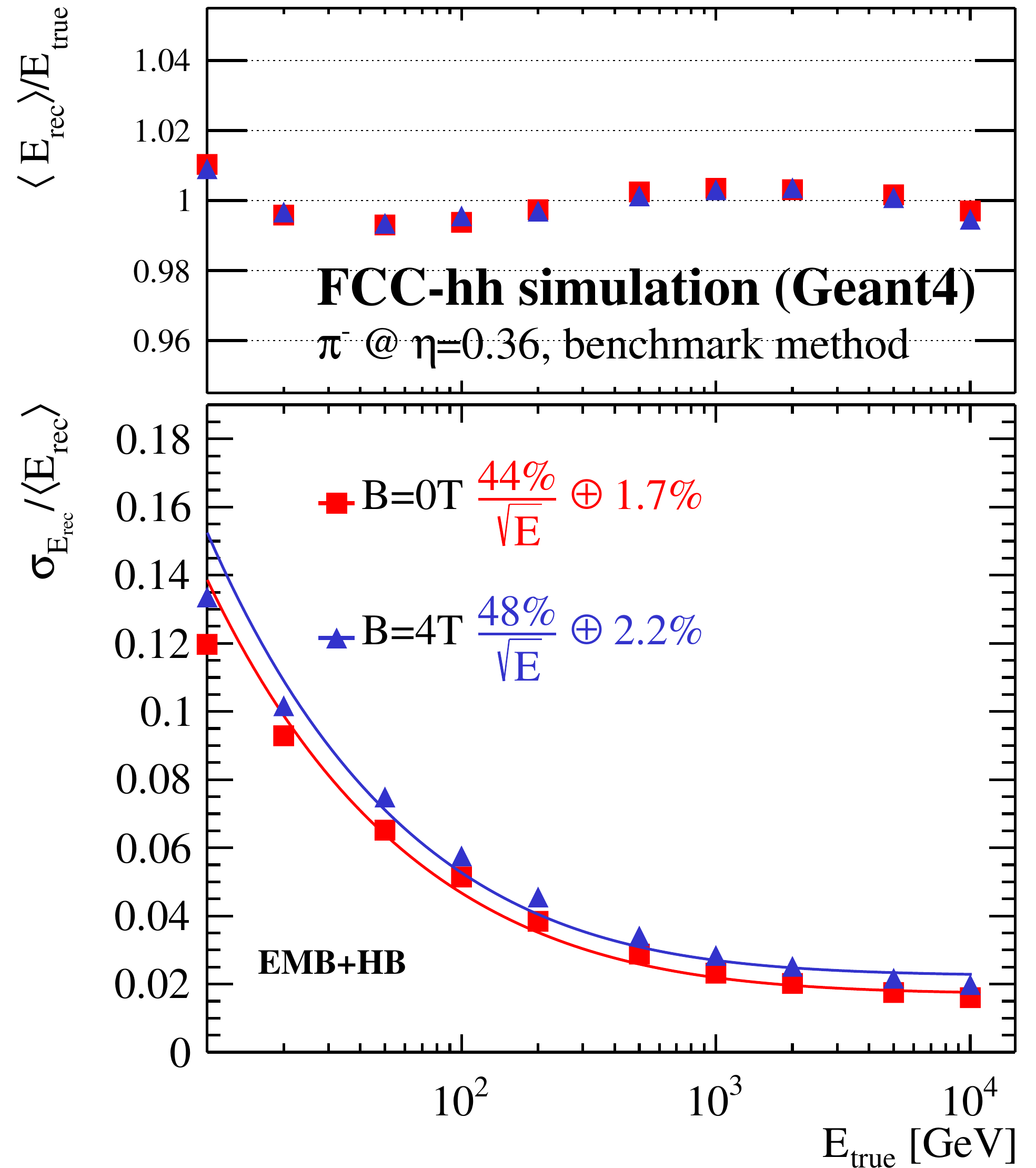}
	\caption{Energy resolution (bottom) and linearity (top) for single $\pi^{-}$ at $\eta=0.36$ using the benchmark reconstruction on all barrel calorimeter cells. The electronic noise is not considered here. \label{fig:performance:hadronic:benchmark}}
\end{figure}

The reconstruction parameters are summarised in Table~\ref{tab:performance:hadronic:benchmark}. The resulting energy resolution and linearity obtained from the sum over all cells without considering electronics noise, with and without magnetic field are shown in Fig.~\ref{fig:performance:hadronic:benchmark}. After the response correction, a linearity at the level of one percent is obtained.

\begin{table}[htp]
\begin{center}
  \begin{tabular}{|c|c|c|c|c|c|c|}
    \hline
B [T] & $p_{0}$ & $p_{1}$ & $p_{2}$\,[1/$\text{GeV}$] & $p_{3}$\,[GeV] & $p_{4}$ & $1/p_{5}$ \\
\hline 
% B=0T: 1.06191,1,0.659853,-6.31231e-06 , d = -0.837742, e = 0.983387, f = 1.00269
% B=4T: 0.996434,1,0.564584,-7.41524e-06, 
0			& $1.062\pm0.001$	& $0.659\pm0.005$	& $-(6.31\pm0.16)\times10^{-6}$ & $-0.83$ & $0.9834$ & $0.9973$  \\
4			& $0.996\pm0.002$	& $0.565\pm0.006$ 	& $-(7.42\pm0.22)\times10^{-6}$ & $-0.92$ & $0.9697$ & $0.9956$\\
\hline
\end{tabular}
\end{center}
\caption{Benchmark parameters for single pions, determined at $\eta=0.36$.}
\label{tab:performance:hadronic:benchmark}
\end{table}%

The single pion energy resolution of the FCC-hh combined calorimeter system results in a stochastic term of 44\,\% without magnetic field and 48\,\%  with magnetic field respectively and a constant term of approximately 2\,\%, well inside the requirements described in Sec.\ref{sec:intro:requirements}. %The electronic noise is not considered as already mentioned before.

\subsubsubsection{Comparison to Energy Reconstruction using Deep Neural Networks}
\label{sec:performance:hadronic:compDNN}

\begin{figure}[ht]
  \centering
  \begin{subfigure}[b]{0.49\textwidth}
    \includegraphics[width=\textwidth]{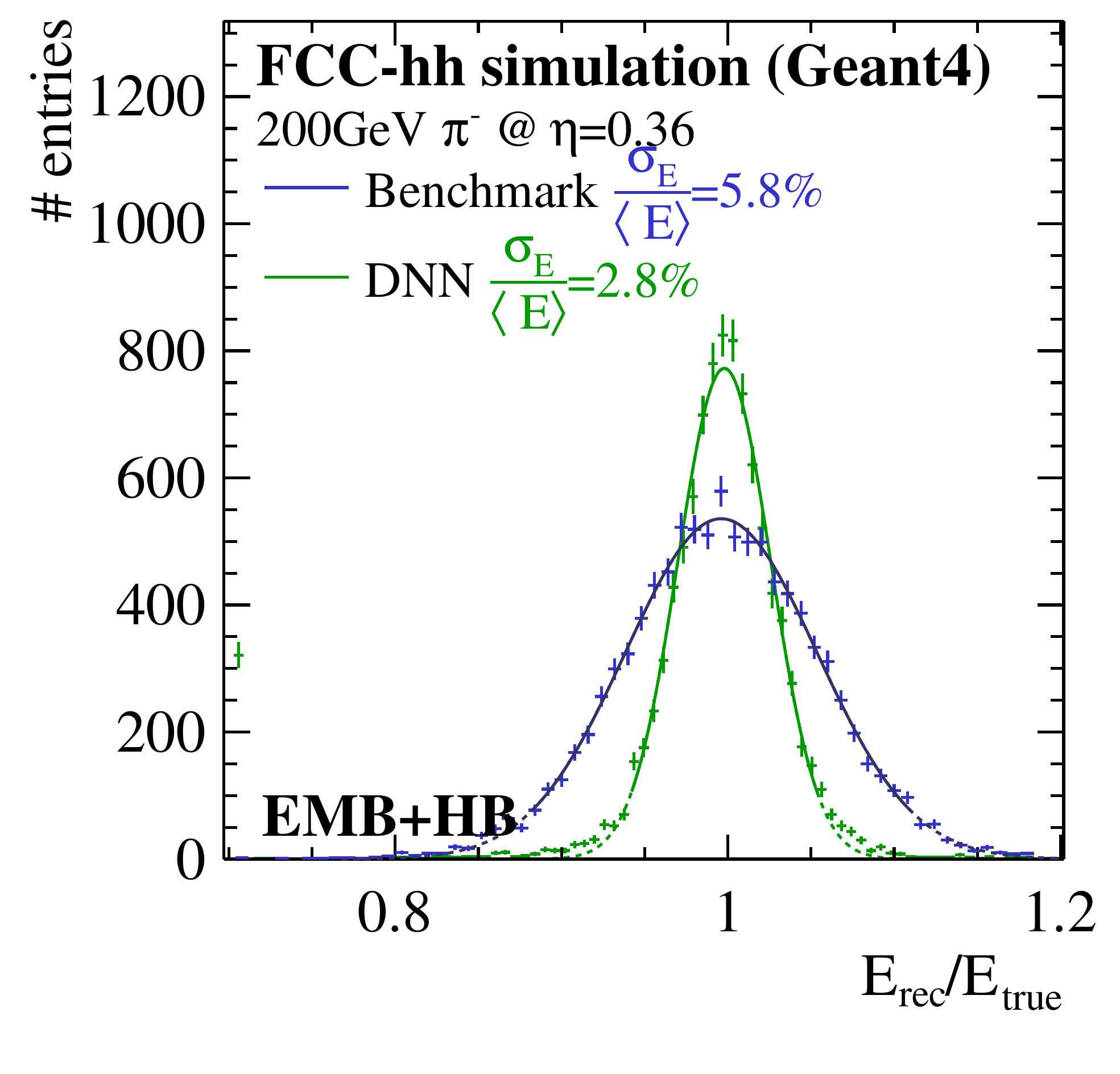}\caption{} \label{fig:performance:hadronic:dnn:200GeV}
  \end{subfigure}
  \begin{subfigure}[b]{0.49\textwidth}
    \includegraphics[width=\textwidth]{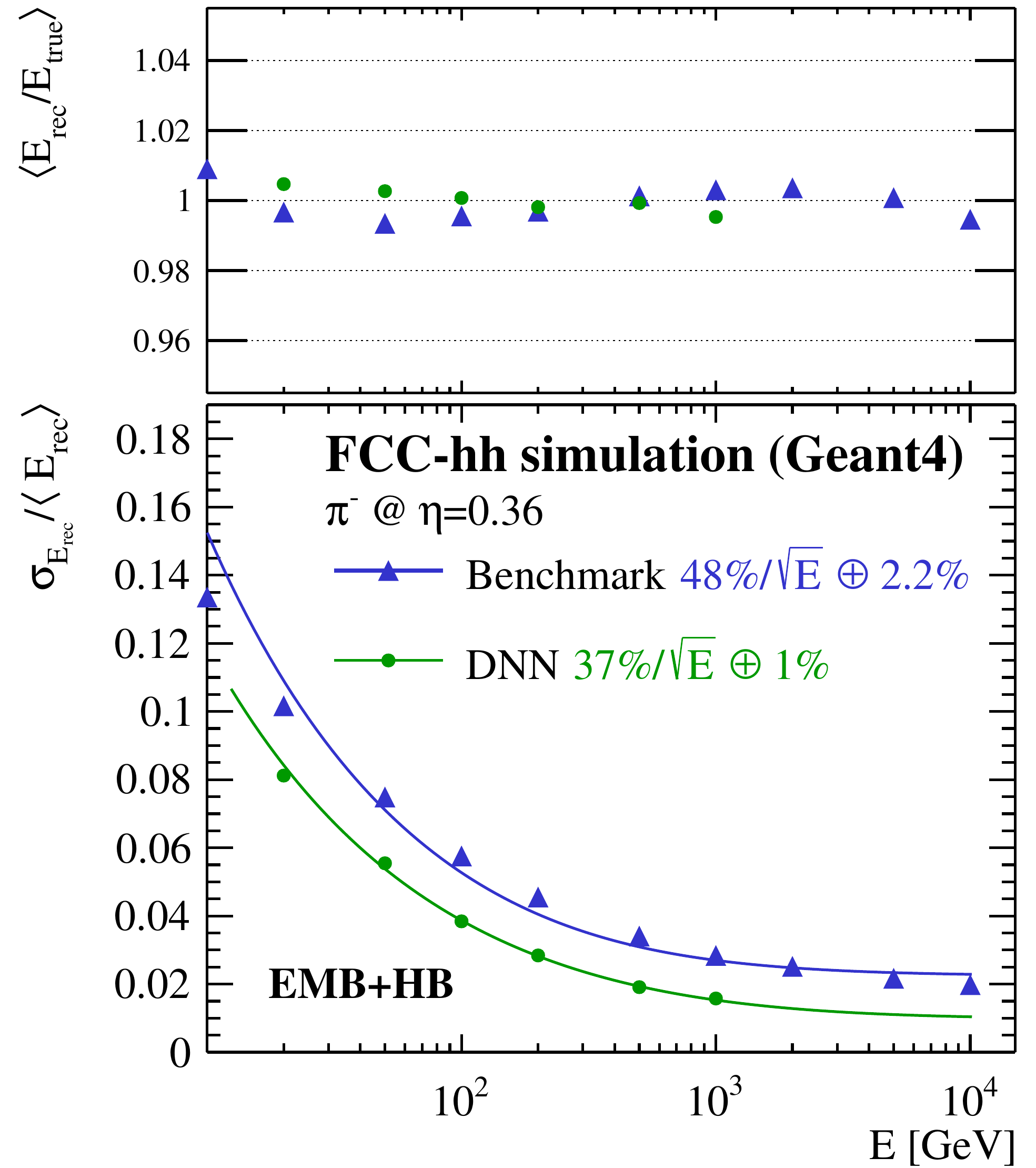}\caption{} \label{fig:performance:hadronic:dnn:compare}
  \end{subfigure}
  \caption{(a) Energy distributions, (b) resolution and linearity for charged pions using the DNN (green) and benchmark (blue) energy reconstruction, with the magnetic field of 4 Tesla applied. \label{fig:performance:hadronic:dnn}	}
\end{figure}

Deep Neural Networks can bring benefits for the energy reconstruction particularly for hadronic showers. As described in Sec.~\ref{sec:software:dnn}, a part of the energy reconstruction network is designed similar to an object identification network in computer vision. Therefore, parts of the shower can be identified on an event-by-event basis, such as e.g. the EM fraction, provided sufficiently granular information is available. As a result, the energy resolution can potentially be much better than for simpler approaches such as the benchmark method described in the previous section. An example of the energy distributions after applying the DNN (green circles) and benchmark method (blue triangles) to 200\,GeV pion showers in the presence of the magnetic field, is shown in Fig.~\ref{fig:performance:hadronic:dnn:200GeV}. The better performance of the DNN is evident. As shown in the top plot of Fig.~\ref{fig:performance:hadronic:dnn:compare}, the response is very linear between 20 and 1,000\,GeV, and the resolution (bottom plot) is strongly improved compared to the results obtained with the benchmark method. The DNN reconstruction achieves a stochastic term of 37\% and the constant term of 1\%. It should be noted that these results are obtained with calorimetry information only (no tracker information), without electronics noise and without pile-up. It is expected that this approach can also be applied for more realistic assumptions of noise and pile-up and will profit from additional information from the tracker. 

\begin{figure}[ht]
  \centering
  \begin{subfigure}[b]{0.49\textwidth}
    	\includegraphics[width=\textwidth]{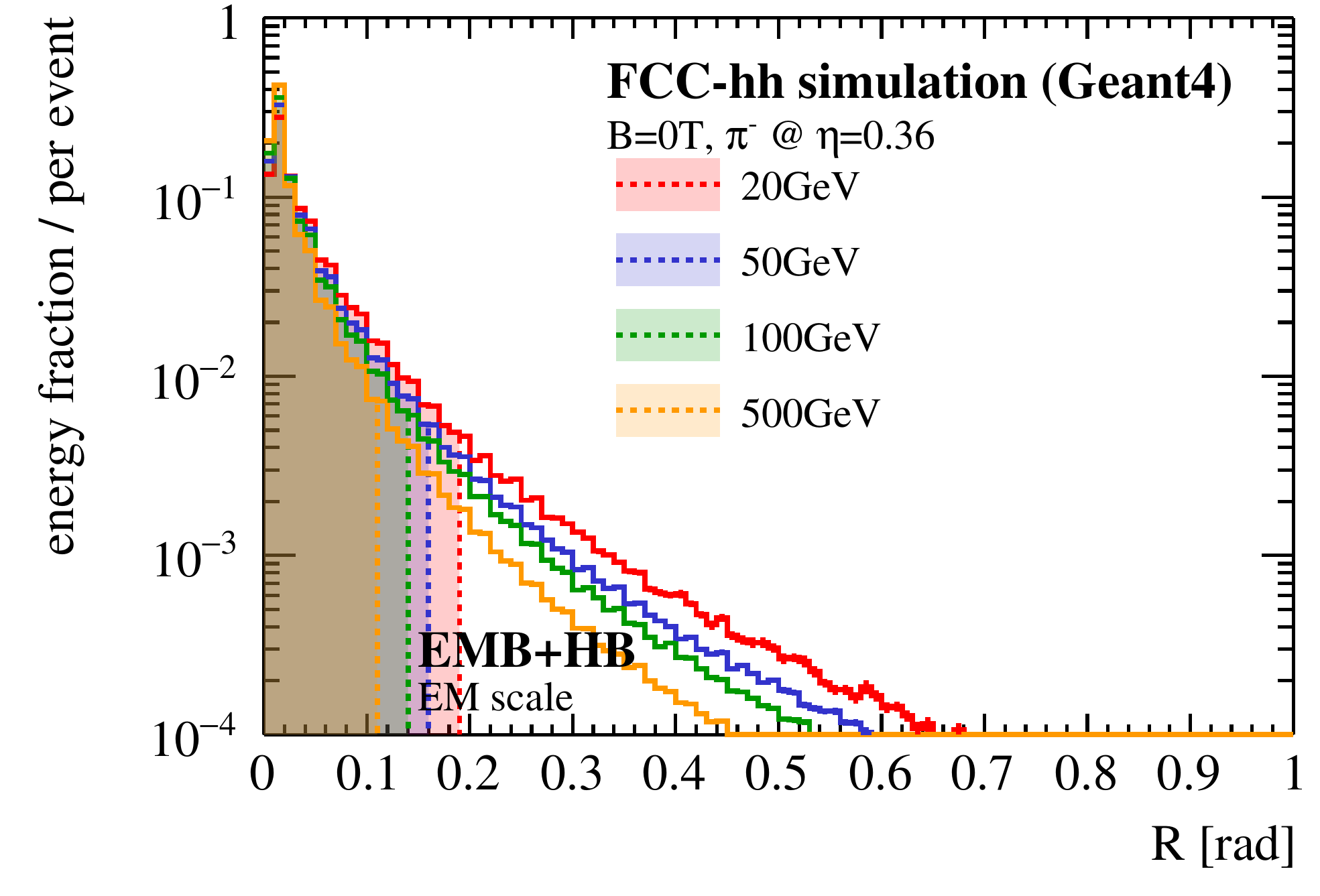}\caption{}
  \end{subfigure}
  \begin{subfigure}[b]{0.49\textwidth}
   	\includegraphics[width=\textwidth]{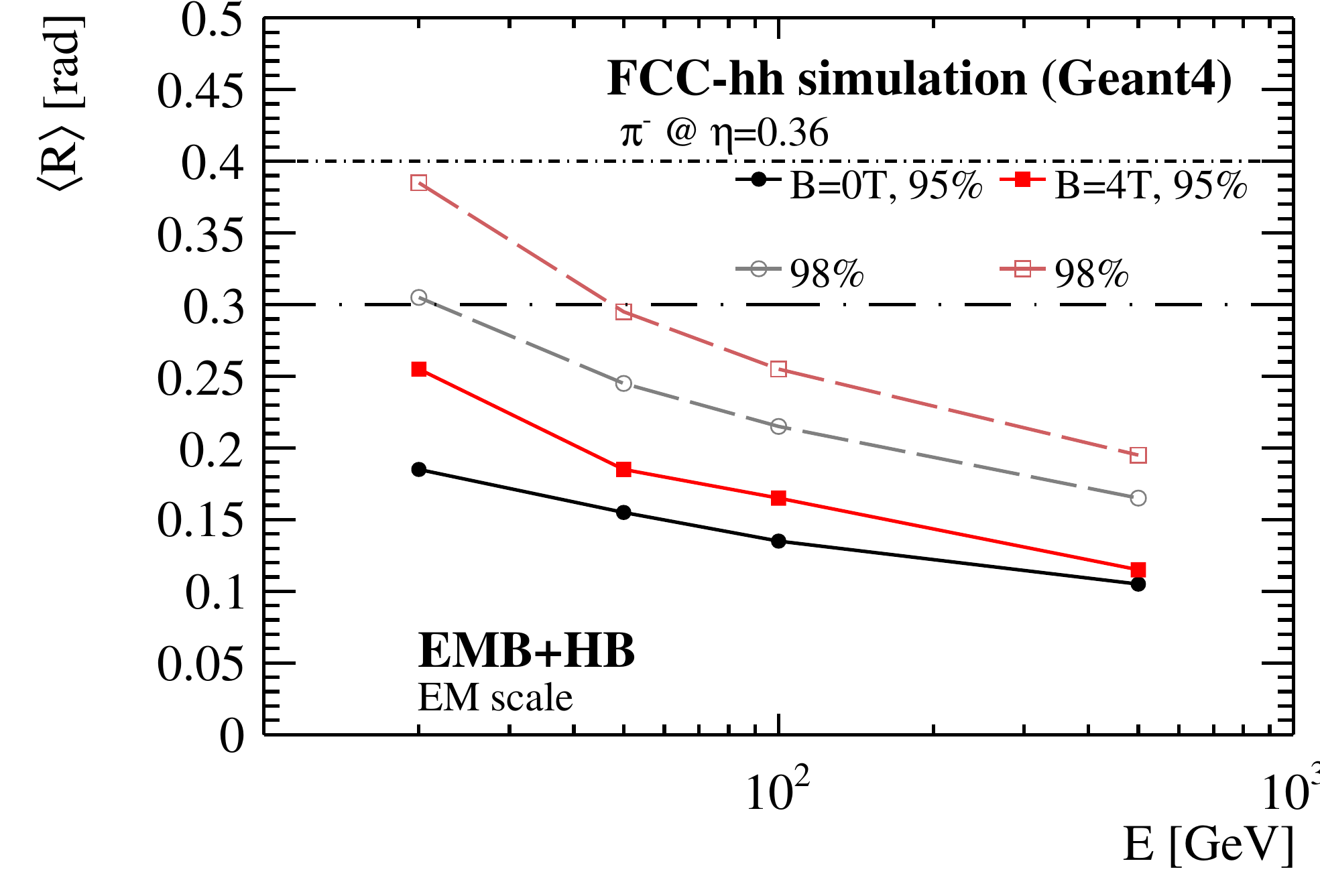}\caption{}
  \end{subfigure}
  \caption{a) Energy fractions of 20-500\,GeV pions within R in the EMB+HB on the EM scale. Filled area corresponds to 95\,\% containment. Lateral containment of 95, and 98\,\% of pion showers in the EMB+HB as a function of the particle energy. }
 	\label{fig:performance:hadronic:showerContainment}
\end{figure}

\subsubsubsection{Including Electronic Noise at the Cell Level Reconstruction}
Realistic estimates of the expected electronic noise have been added to the cells of the EMB and HB calorimeters. To test the impact on the single pion energy resolution, the included cells have been limited to a range within a cone in $\eta$ and $\phi$ of radius $R=\sqrt{\left(\Delta\eta\right)^2+\left(\Delta\phi\right)^2}$ with $\Delta\eta=\eta_{cell}-\eta_{true}$, $\Delta\phi=\phi_{cell}-\phi_{true}$. The expected noise contribution $\sigma_{noise}$ to the single particle resolution is estimated by adding in quadrature the noise contributions per cell.
The number of cells within the cone, the noise in EMB and HB, and the total noise for different cone radii are summarised in Table~\ref{tab:performance:hadronic:cellNoise}. 
In the following, the cone radius has been chosen to contain 98\,\% of the single pion energy over the full energy range, see Fig.~\ref{fig:performance:hadronic:showerContainment}, and found to be 0.3/0.4 for B=0/4\,T, respectively.

\begin{table}[htp]
\begin{center}
 \begin{tabular}{|c|p{2cm}|p{2cm}|p{1.9cm}|p{1.9cm}|p{1.5cm}|p{1.5cm}|}
 % \begin{tabular}{|c|c|c|c|c|c|c|}
  
    \hline
    cone/window & EMB $\left<noise\right>$/cell & HB $\left<noise\right>$/cell & EMB cells & HB cells & EMB noise & HB noise \\
    %\hline
    %units 			&	MeV					& MeV 					& \# 			& \#			& GeV	&	GeV	& GeV \\
    	&	[MeV]					& [MeV] 					&  		[\#]	& 	[\#]		& [GeV]	&	[GeV] \\
\hline
\multirow{2}{*}{$R<0.40$}			& \multirow{2}{*}{3.5-40}		& \multirow{2}{*}{10}		& 45,080	& 8,150 & 2.05 	& 0.90 \\
							&			&		& \multicolumn{2}{c|}{\textbf{53,230}} & \multicolumn{2}{c|}{\textbf{2.24}} \\
\hline
\multirow{2}{*}{$R<0.30$}			& \multirow{2}{*}{3.5-40}		& \multirow{2}{*}{10}		& 25,344	& 4,610 & 1.52 	& 0.68 \\
							&			&		& \multicolumn{2}{c|}{\textbf{29,954}} & \multicolumn{2}{c|}{\textbf{1.66}}  \\
%\hline
%$R<0.24$	(r$_{HCalFront}$~$\widehat{=}~0.7$~\,m)	&	true		& 10		& 16,442	&  2,830	& 1,21 & 0.53 & 1.32 \\
\hline
\multirow{2}{*}{$R<0.17$}			& \multirow{2}{*}{3.5-40}		& \multirow{2}{*}{10}		& 8,088	& 1,480 	& 0.86 & 0.38 \\
							&			&		& \multicolumn{2}{c|}{\textbf{9,568}} & \multicolumn{2}{c|}{\textbf{0.94}} \\
\hline
$0.34\times0.34$ 				& \multirow{2}{*}{3.5-40}		& \multirow{2}{*}{10}		& 41,344	& 7,290 & 1.98 & 0.84 \\
(DNN)							& 			&		& \multicolumn{2}{c|}{\textbf{48,634}} & \multicolumn{2}{c|}{\textbf{2.15}} \\
\hline
\end{tabular}
\end{center}
\caption{Summary of the noise expected for different cone radii, thus different number of cells around $\eta=0.36$.
  \label{tab:performance:hadronic:cellNoise}}
\end{table}%

\begin{figure}[ht]
 % \centering
 \begin{minipage}{0.49\textwidth}
  \begin{subfigure}[b]{1\textwidth}
    	\includegraphics[width=1\textwidth]{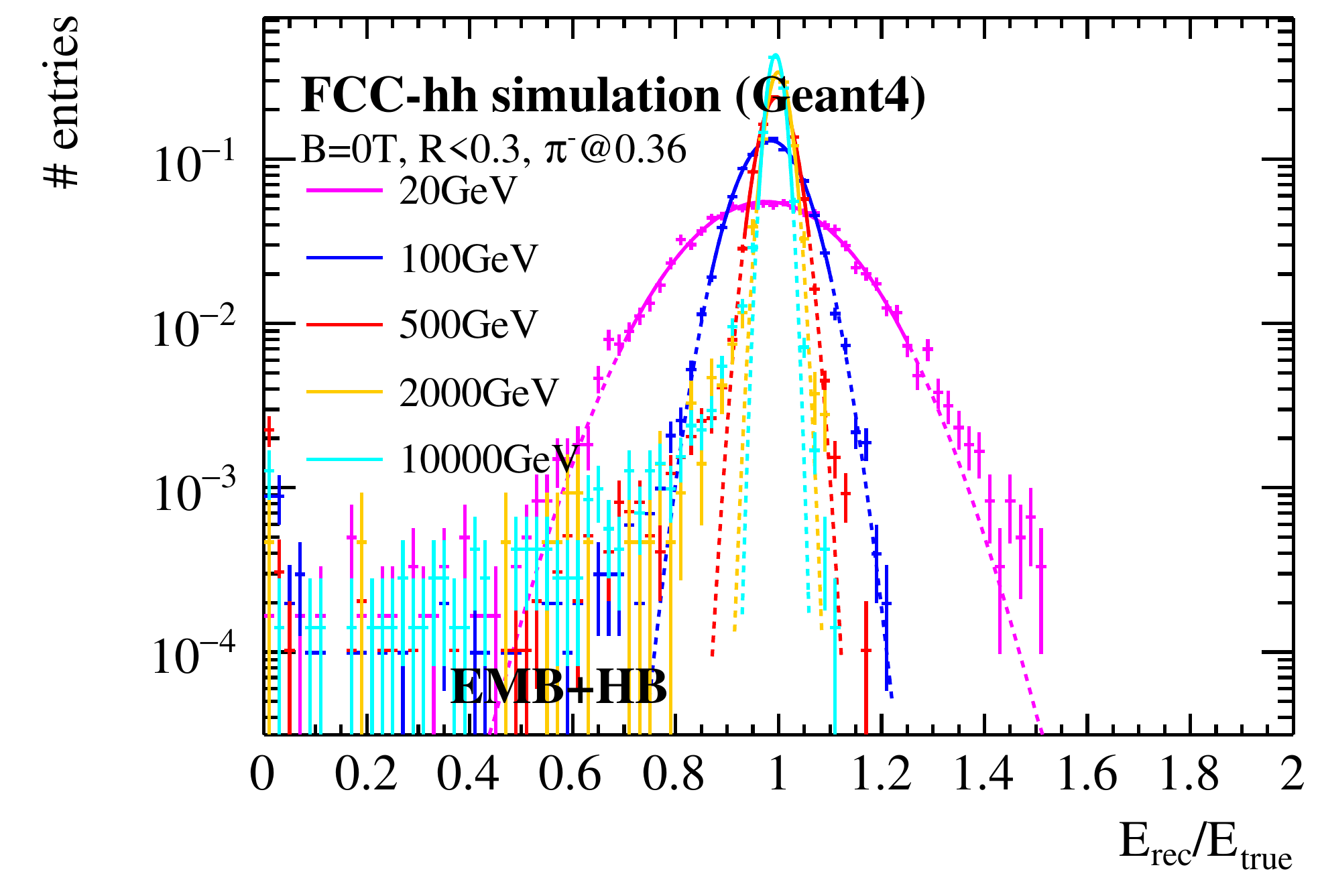}\caption{}
  \end{subfigure}
%  \begin{subfigure}[b]{1\textwidth}
%    	\includegraphics[width=1\textwidth]{hadrons/pionResponse_benchmarkStep1_ecalCone017_hcalCone017_bfieldOff_log.pdf}\caption{}
%  \end{subfigure}
\end{minipage}
\hfill
 \begin{minipage}{0.49\textwidth}
  \begin{subfigure}[b]{1\textwidth}
   	\includegraphics[width=1\textwidth]{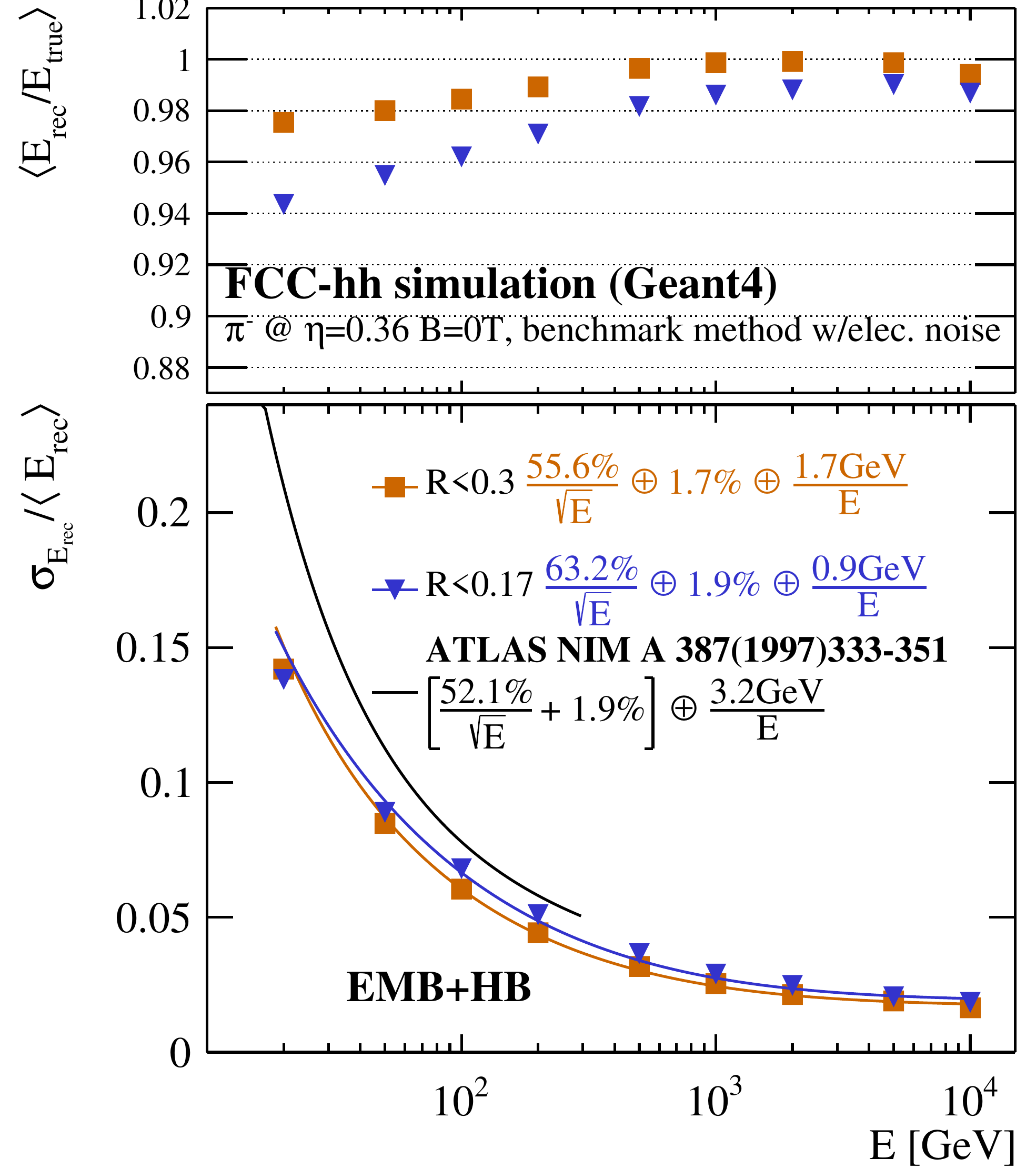}\caption{}
  \end{subfigure}
\end{minipage}
 \caption{Left plot (a): Energy distributions for a cone cut of $\mathrm{R}<0.3$. Right plot (b): The energy resolution of pions at $\eta=0.36$, with $B=0$\,T is shown for 98\,\%, and 95\,\% lateral shower containment, corresponding to $\mathrm{R}<0.3$ and 0.17 respectively. The combined performance of EMB+HB is compared to ATLAS testbeam results~\cite{Ajaltouni:1996fw}.
 	\label{fig:performance:hadronic:benchmarkWNoise}	 
		}	
\end{figure}

The impact of the cone selection on the single pion energy resolution has been studied and is summarised in Fig.~\ref{fig:performance:hadronic:benchmarkWNoise}. The obtained FCC-hh resolutions are given for B=0\,T and compared to ATLAS testbeam results~\cite{Ajaltouni:1996fw}. The benchmark method was also used by ATLAS in the combined testbeam for the energy combination of an electromagnetic liquid argon accordion calorimeter and an hadronic scintillating-tile calorimeter, however featuring a different ratio of active to passive absorber material with a Sci:Fe ratio of 1:4.7. The cone size of $R$ = 0.17 corresponds approximately to the selection used in the analysis of the ATLAS testbeam. The parameters of the energy resolution are listed in Table~\ref{tab:performance:hadronic:resoConeSize} for different cone sizes, that correspond to different lateral shower containments $C_{20\,\text{GeV}}$ (values for 20\,GeV pions).
Figure~\ref{fig:performance:hadronic:benchmarkWNoise} shows that the FCC setup achieves a better resolution than the ATLAS testbeam over the full energy range. An improvement with respect to ATLAS is expected due to the optimised absorber of the HB. However, part of the difference has to be attributed to the fact that testbeam measurements are compared to simulations assuming perfect calibration and uniformity of the calorimeter response.

\begin{table}[htp]
\begin{center}
  \begin{tabular}{|c|c|c|c|c|c|}
    \hline
    $R$ 				& B 	& C$_{20\,\text{GeV}}$	& $a$ 				& $b$ 		& $c$ 	\\
    {[rad]} 				& [T] 	& [\%]		& [$\%\sqrt{\text{GeV}}$] 	& [GeV]	& [\%]	\\
\hline
	0.3				& 0	& 98			& 55.6			& 1.66	& 1.7 \\
	0.17				& 0	& 95			& 64.5			& 0.94	& 1.9	 \\
\hline
	0.4		 		& 4	& 98			& 59.9			& 2.24	& 2.3 \\
	0.3				& 4	& 95			& 61.8			& 1.66	& 2.3 \\
	0.17				& 4	& 92			& 74.1			& 0.94	& 2.6 \\
\hline
%   $R$ 		& r$_{\text{HBFront}}$	& B 	& C$_{20\,\text{GeV}}$	& $a$ 				& $b$ 		& $c$ 	\\
%    {[rad]} & [m]		& [T] 	& [\%]		& [$\%\sqrt{\text{GeV}}$] 	& [GeV]	& [\%]	\\
%\hline
%	0.3		& $\widehat{=}~0.9$ 		& 0	& 98			& 55.6			& 1.66	& 1.7 \\
%	0.17	& $\widehat{=}~0.5$		& 0	& 95			& 64.5			& 0.94	& 1.9	 \\
%\hline
%	0.4		& $\widehat{=}~1.2$ 		& 4	& 98			& 59.9			& 2.24	& 2.3 \\
%	0.3		& $\widehat{=}~0.9$		& 4	& 95			& 61.8			& 1.66	& 2.3 \\
%	0.17	& $\widehat{=}~0.5$		& 4	& 92			& 74.1			& 0.94	& 2.6 \\
%\hline
\end{tabular}
\end{center}
\caption{Impact of selection cut in $R$ on the stochastic $a$, noise $b$, and constant term $c$ for benchmark reconstruction on cell level. The lateral shower containment, corresponding to the different cone sizes, is given for 20\,GeV pions as $C_{20\,\text{GeV}}$. \label{tab:performance:hadronic:resoConeSize}}
\end{table}

\clearpage

\subsubsection{Performance of Topological Cell Clustering Algorithm}
\begin{figure}[ht]
  \centering
    	\includegraphics[width=1\textwidth]{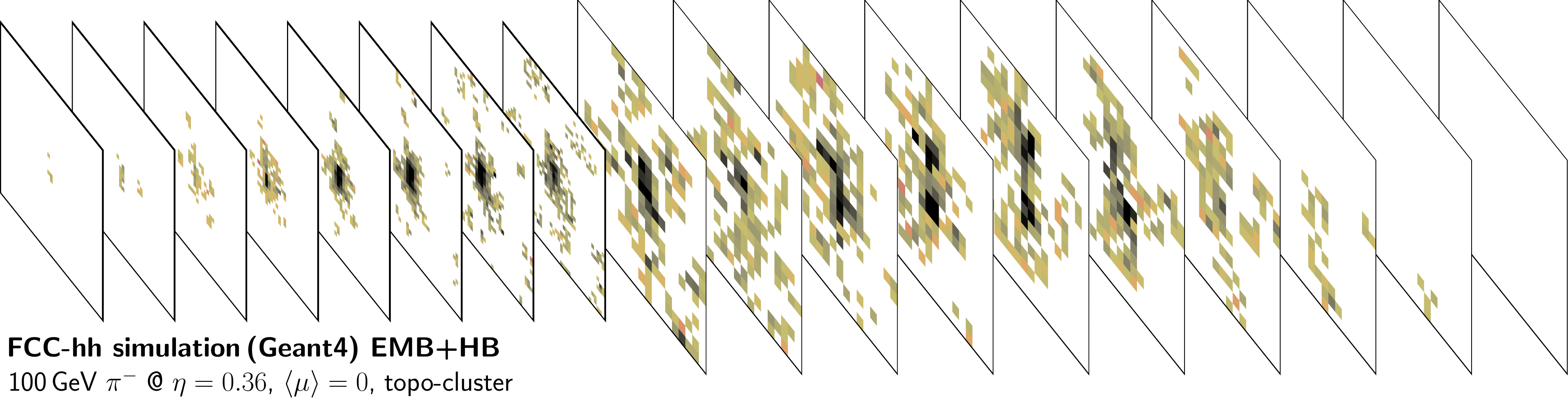}
  \caption{The energy deposited per calorimeter layer of single 100\,GeV pion shower in the EMB+HB. The colour code visualises the amount of energy per calorimeter cell in log scale. }
 	\label{fig:performance:hadronic:shower3D}	
\end{figure}

An example of one pion shower, after topo-clustering is presented in Fig.~\ref{fig:performance:hadronic:shower3D} and visualises the use of the calorimeters granularity as well as the effective noise suppression by the reconstruction algorithm. The electronic noise is always included in the case of the reconstruction using the topo-clustering algorithm. 

In the following, clusters within a cone around the generated particles with radius 
\begin{equation}
  \text{R}=\sqrt{\left(\eta^{cluster}_{rec}-\eta_{true}\right)+\left(\phi^{cluster}_{rec}-\phi_{true}\right)}
\end{equation}
are summed up. This cut on the radius is different to the cell selection, previously introduced in Sec.~\ref{sec:performance:hadronic:cellLevel}, since the cluster positions are based on the centre of gravity of the contained cells, thus the contained cells can extend the selection cone. This selection criterion is used to substitute the missing information from the tracker, which would effectively preselect a certain calorimeter area to match the track with calorimeter cluster. This will be crucial especially in the high pile-up environment.

\subsubsubsection{Optimisation of the Topo-Clustering Algorithm}
\label{sec:performance:hadronic:topoClusterEMscale}

The thresholds of the topological clustering algorithm, introduced in Sec.~\ref{sec:software:reco:topo}, are optimised for single pions on the EM scale.

\begin{figure}[ht]
  \centering
  \begin{subfigure}[b]{0.49\textwidth}
    	\includegraphics[width=\textwidth]{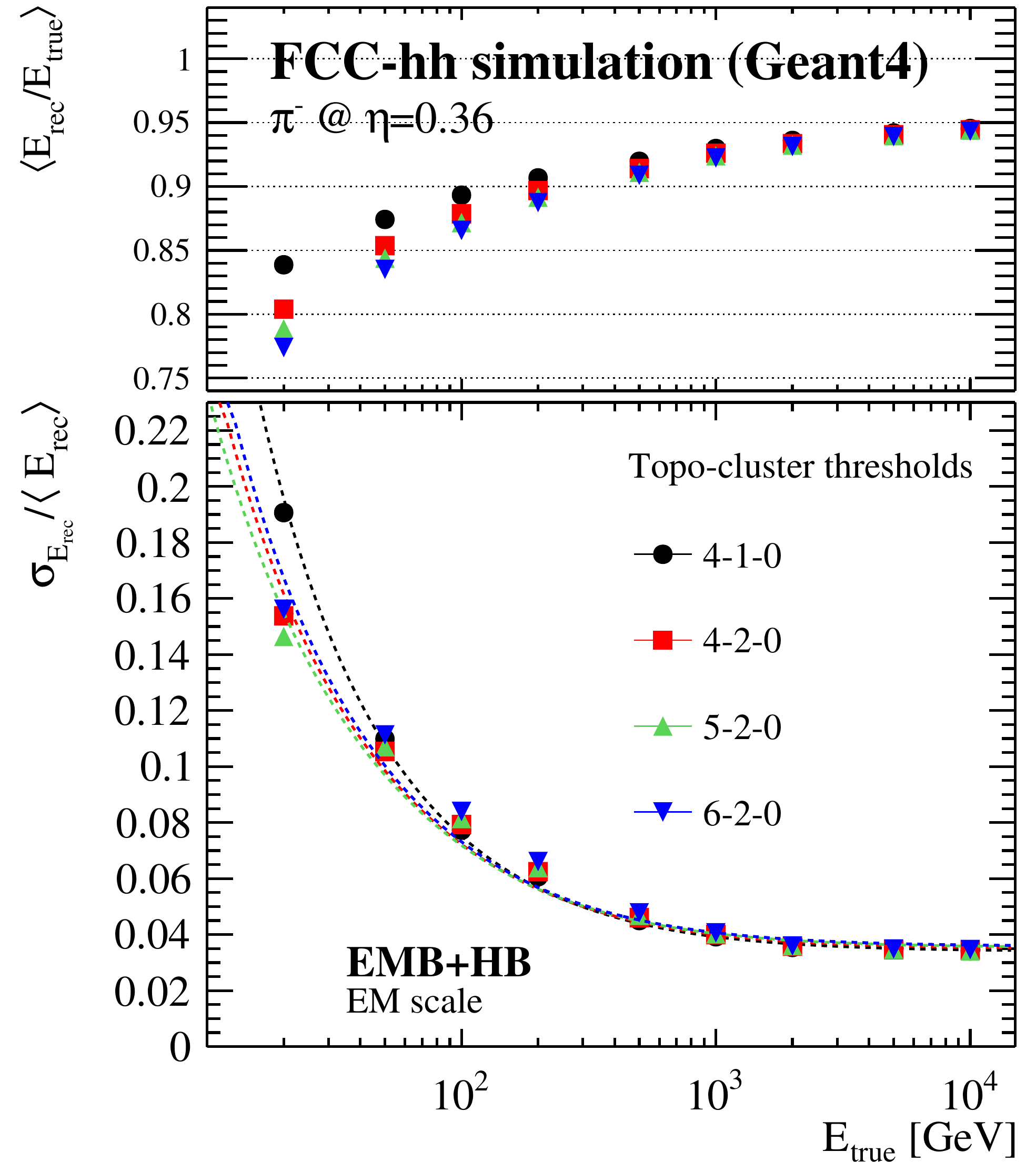}\caption{}\label{fig:performance:hadronic:resoClustersElecNoiseThresholdsNocone}
  \end{subfigure}
  \begin{subfigure}[b]{0.49\textwidth}
   	\includegraphics[width=\textwidth]{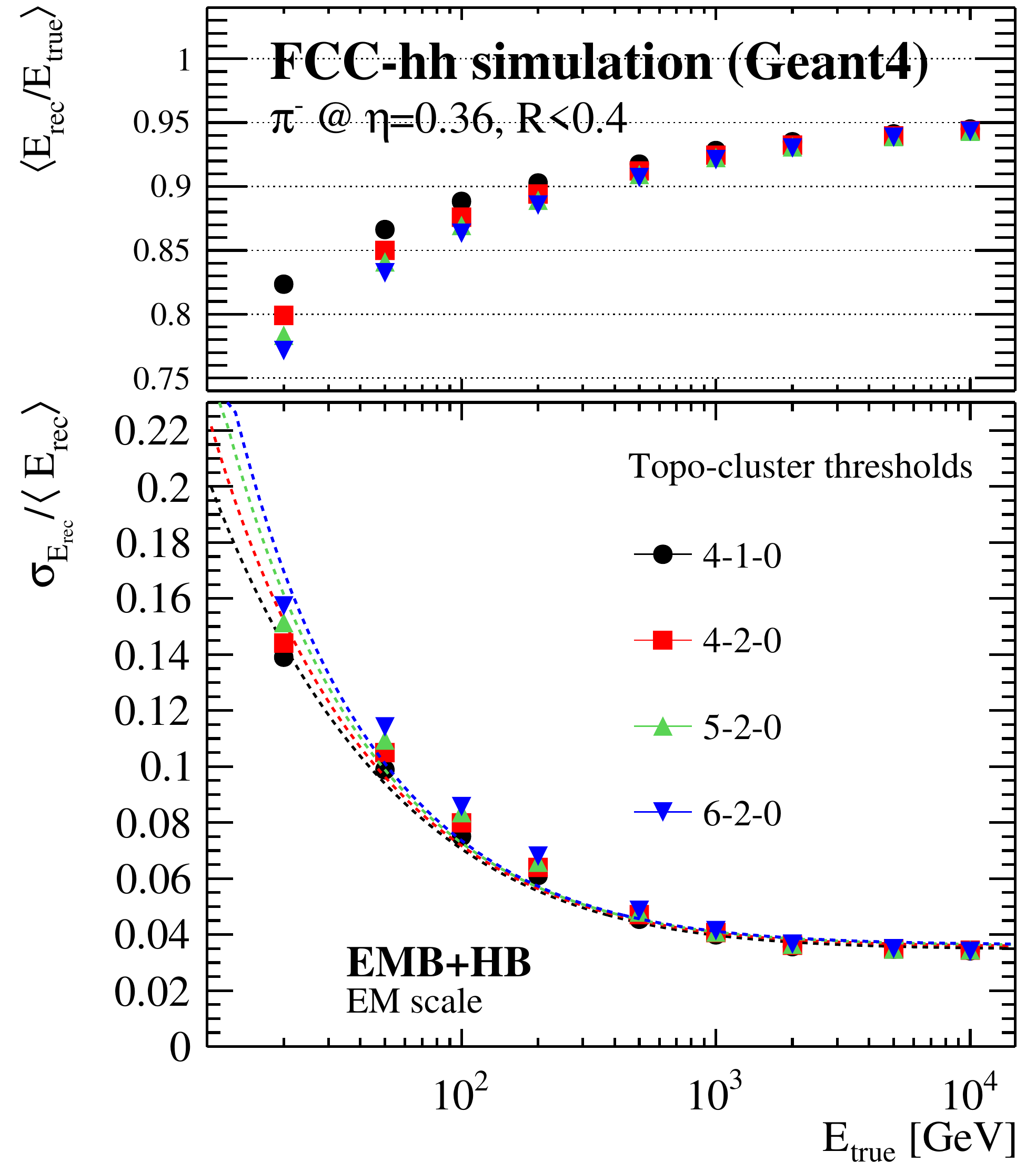}\caption{}\label{fig:performance:hadronic:resoClustersElecNoiseThresholdsCone04}
  \end{subfigure}
 \caption{Energy resolution and linearity of pion showers at the EM scale for different thresholds of the topo-clustering algorithm with (a) no cut on R applied and (b) with cut R<0.4 on the barycentre of the topo-cluster.}
 	\label{fig:performance:hadronic:resoClustersElecNoiseThresholds}	
\end{figure}

The linearity of the response and the energy resolution curves for different parameters of the topo-clustering algorithm were tested. The results without any cut on $\text{R}$ and those obtained when applying a cut of $\mathrm{R}<0.4$ around the barycentre of the topo-cluster were compared, see Fig.~\ref{fig:performance:hadronic:resoClustersElecNoiseThresholdsNocone} and~\ref{fig:performance:hadronic:resoClustersElecNoiseThresholdsCone04}, respectively. The difference between various thresholds is obvious at low energies only, the different values are in good agreement in the high energy regime, as expected. The thresholds $S = 4$ (seed threshold), $N = 2$ (threshold for neighbours) and $P = 0$ (final step), noted in units of $\sigma_\mathrm{noise}$, are found to be optimal. These parameters result in a good linearity as well as energy resolution for both cases - with and without the cut on the radius, and they are consistent with the choices made by the ATLAS collaboration~\cite{Aad:2016upy}.

\subsubsubsection{Calibration of Topo-Cluster}
\label{sec:performance:hadronic:topoClusterCalibration}
The clusters are built of cells which have been calibrated to the EM scale, nevertheless, the performance can be improved if a correction for the lost energy within the LAr cryostat is included. To correct for the energy lost between EMB and HB, as described in Sec.~\ref{sec:performance:hadronic:cellLevel:benchmark}, a calibration is applied using the deposited energy in the last EMB and first HB layer as a measure of the energy deposited in the cryostat. 
The benchmark correction on topo-cluster is based on the clustered cells, whereas the total energy per event is given by the sum over all clusters: 
\begin{equation}
	E^{\text{topo-cluster}}_\text{rec} = \sum_{i}^\text{cluster}E_i,
\end{equation}
with cluster energies $E_\text{i}$
\begin{equation}
	E_\text{i}= \sum_{j}^\text{cells} E^\text{EM}_{\text{EMB,}\,j}\cdot p_{0} +  \sum_{j}^\text{cells} E^\text{had}_{\text{HB,}\,j} + p_{1}\cdot\sqrt{\left|E^\text{EM}_\text{last~layer}\cdot p_{0}\cdot E^\text{HB}_\text{first~layer}\right|} + p_{2}\cdot\left(  \sum_{j}^\text{cells} E^\text{EM}_{\text{EMB,}\,j}\cdot p_{0} \right)^2.
\end{equation}
The parameters $p_{0}$, $p_{1}$ and $p_{2}$ are determined on the cell level, and the performance has been validated. Further tests have been performed to allow for an energy dependence of parameters $p_{0}$, $p_{1}$ and $p_{2}$. However, only little further improvement was achieved. Therefore the simplest approach without energy dependence was used in the following.

\begin{figure}[ht]
  \centering
 \begin{minipage}{0.49\textwidth}
  \begin{subfigure}[b]{1\textwidth}
    	\includegraphics[width=1\textwidth]{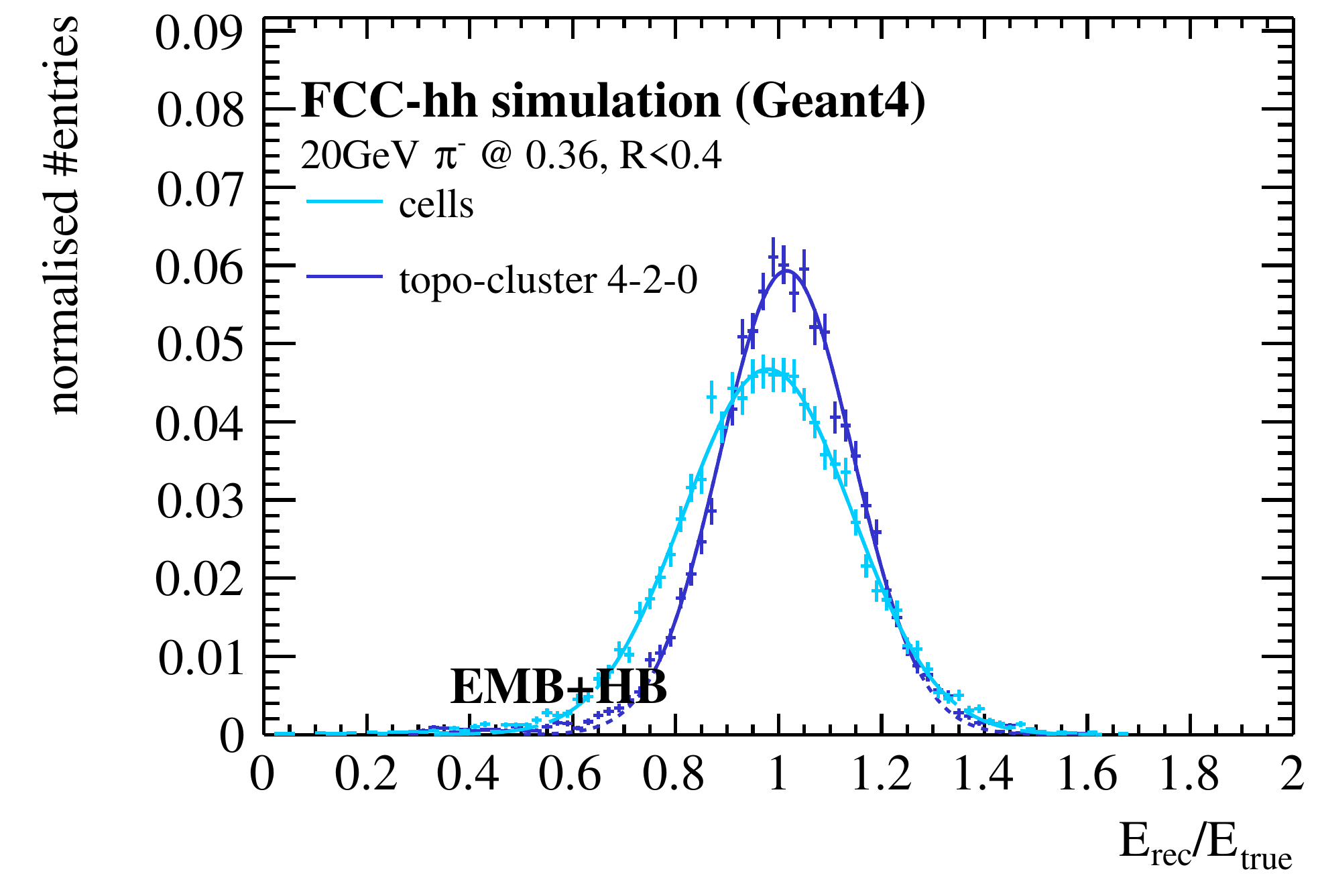}\caption{}
  \end{subfigure}
  \end{minipage}
 \begin{minipage}{0.49\textwidth}
  	\begin{subfigure}[b]{1\textwidth}
    \includegraphics[width=\textwidth]{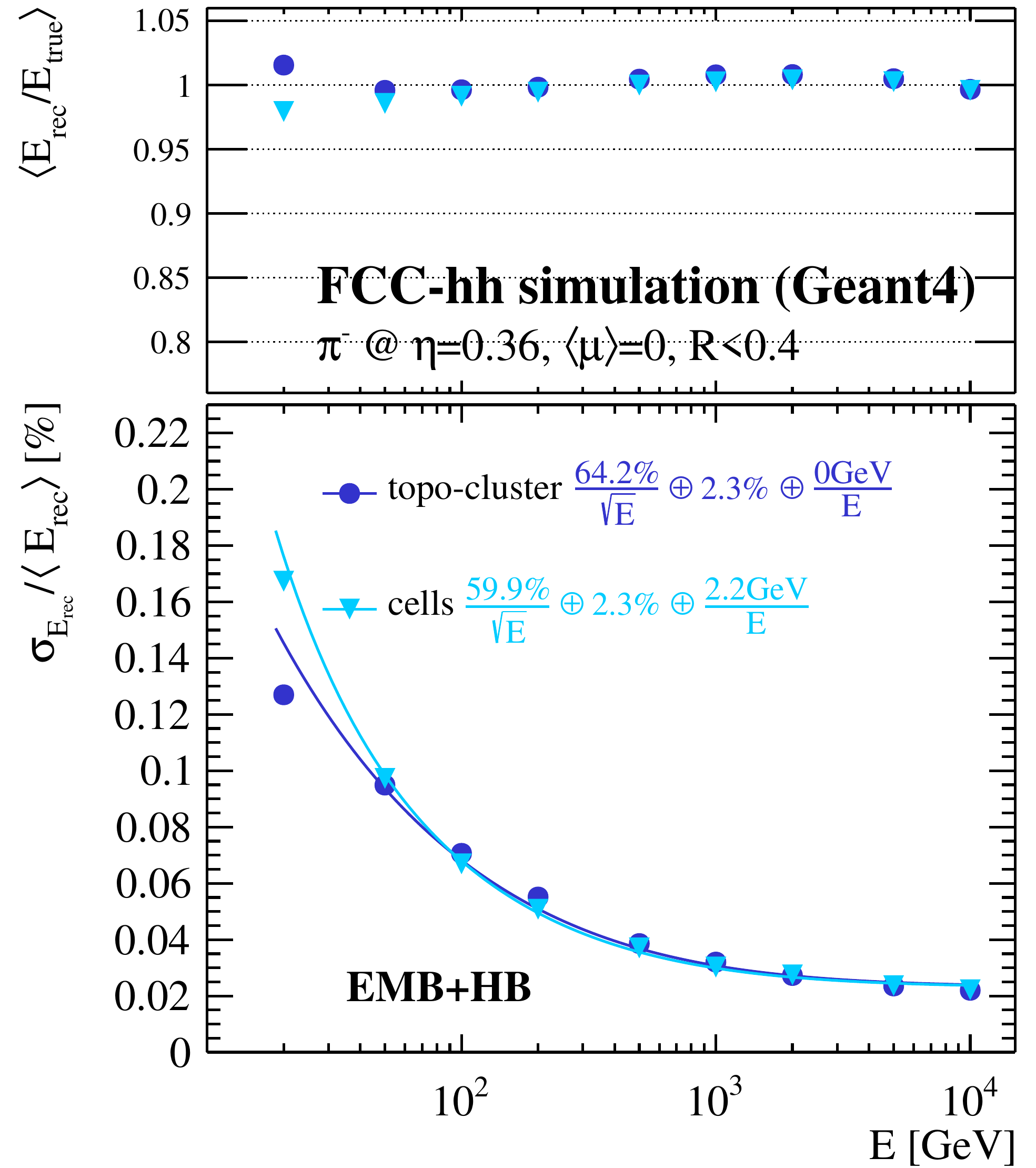}\caption{}
  \end{subfigure}
\end{minipage}
  \caption{Energy distributions of single pions (a) when adding up all cells with $\mathrm{R}<0.4$ and after topo-clustering, both including the benchmark corrections. (b) The energy resolution and linearity of pion showers in full EMB+HB simulations at $\eta=0.36$. The resolution is obtained from topo-cluster with 4-2-0 thresholds within $\mathrm{R}<0.4$. Again, the values with and without topo-cluster are shown.        
 \label{fig:performance:hadronic:resoTopoClustersCalib}}
\end{figure}

The impact of the cluster calibration on the reconstructed energy can be seen in Fig.~\ref{fig:performance:hadronic:resoTopoClustersCalib}. The linearity as well as the resolution is improved over the full energy range. 
In comparison to the performance obtained when adding up all cells, the improvement due to the decreased number of cells is most pronounced at 20\,GeV. 

\subsubsubsection{Angular Resolutions and Impact of HB Granularity}
\label{sec:performance:hadronic:angular}
The highest HB granularity in $\eta$ of $\Delta\eta<0.006$ could be obtained if each scintillating tile is read out separately. However, to reduce the number of readout channels and thus cost, several scintillating tiles will be summed together resulting in an effective granularity of $\Delta\eta=0.025$. In this way the number of effective readout channels is reduced from 1,305,600 to 226,307, still providing a 5 times higher granularity than the current ATLAS Tile calorimeter~\cite{CERN-LHCC-96-042}. The effect on the angular resolution for single pions is discussed in the following, while the angular resolutions for jets is shown in Sec.~\ref{sec:performance:jets}.

\begin{figure}[htp]
  \centering
  \begin{subfigure}[b]{0.49\textwidth}
	  \includegraphics[width=\textwidth]{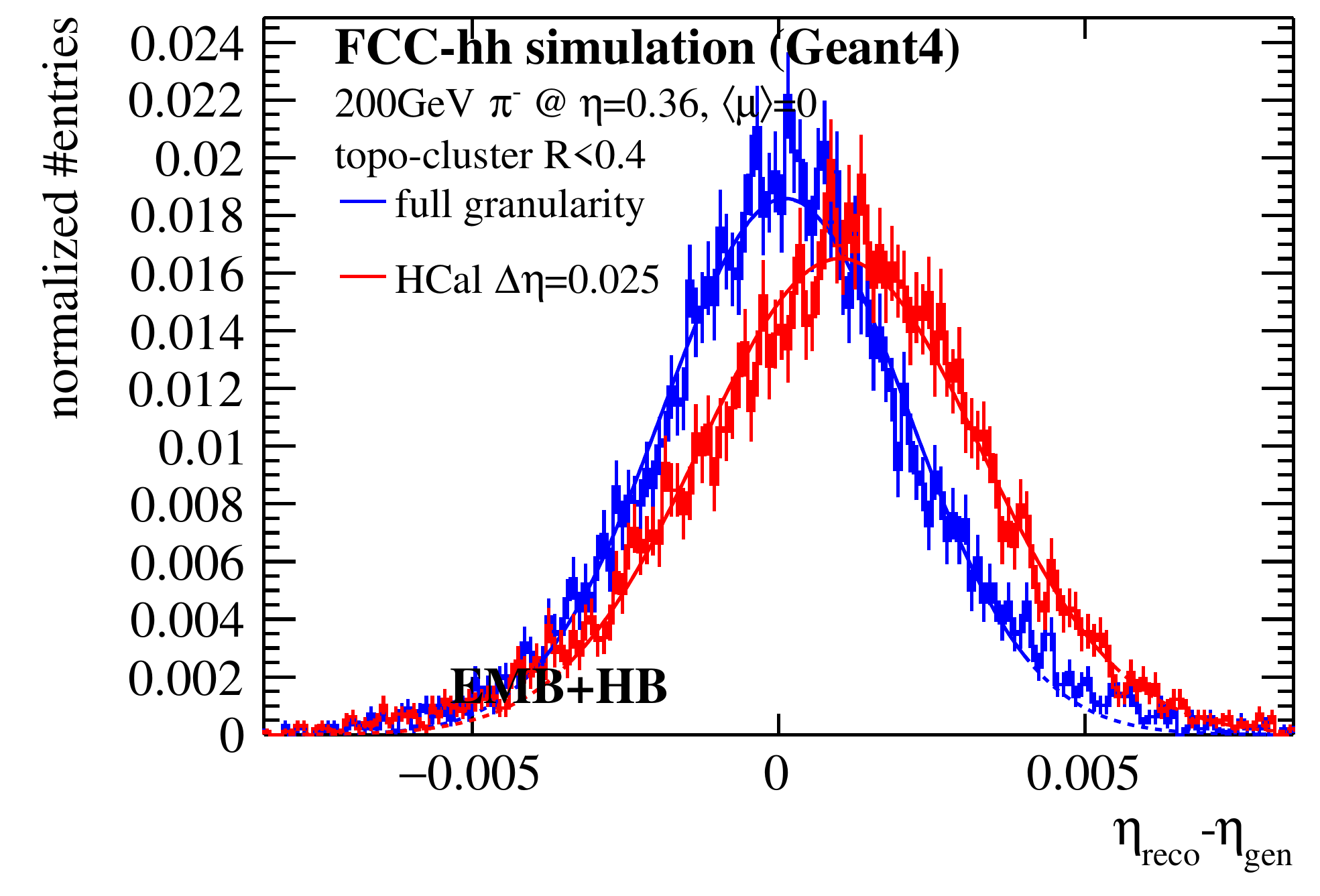}\caption{}\label{fig:performance:hadronic:eta:200pions}
  \end{subfigure}
  \begin{subfigure}[b]{0.49\textwidth}
	  \includegraphics[width=\textwidth]{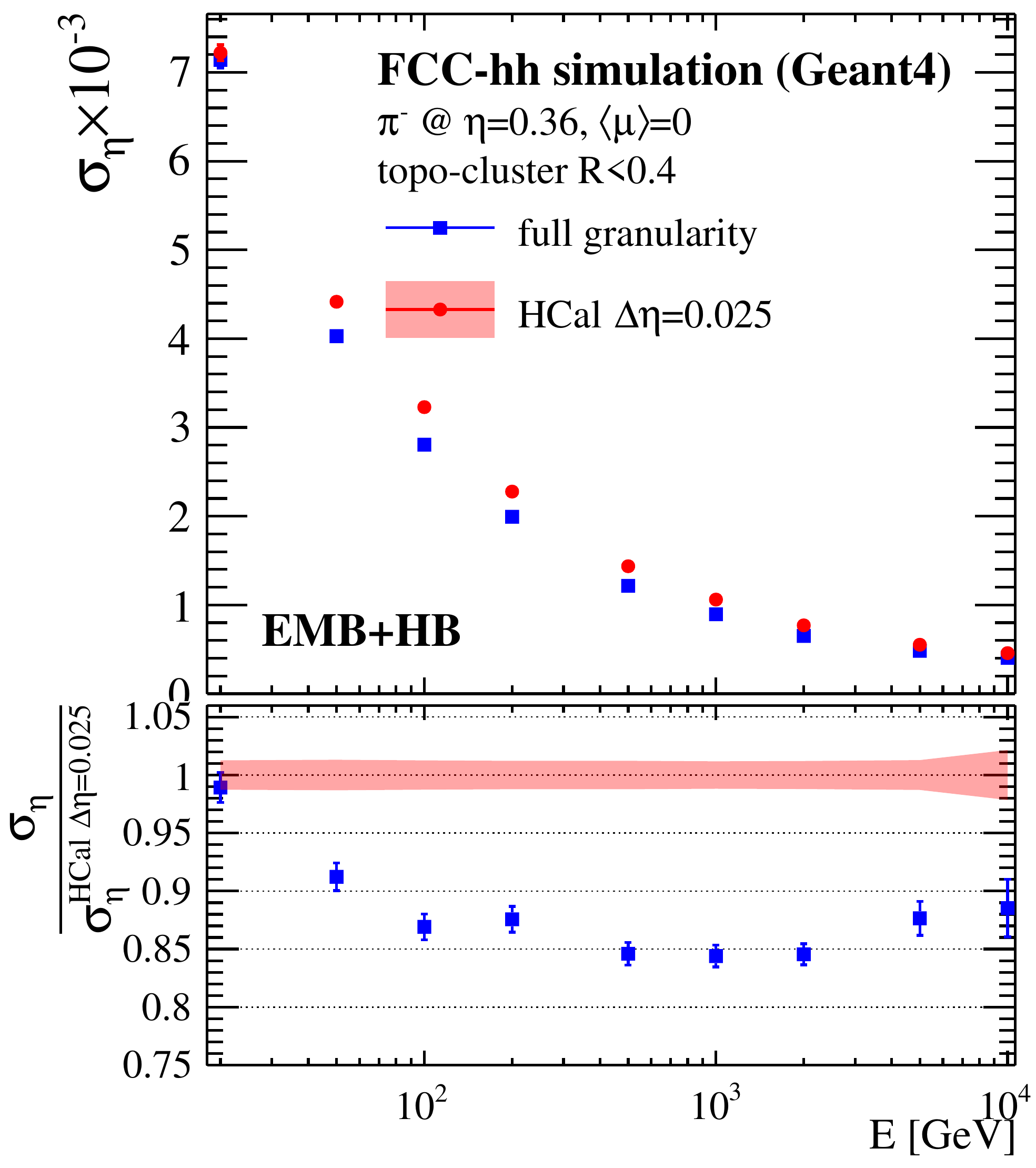}\caption{}\label{fig:performance:hadronic:eta:pions}
  \end{subfigure}
  \caption{(a) Difference between reconstructed and generated $\eta$ for topo-cluster within $\mathrm{R}<0.4$ for 200\,GeV pions generated at $\eta=0.36$, shown for the finest possible HB granularity of $\Delta\eta<0.006$ (full granularity, blue) compared to the baseline granularity of the HB ($\Delta\eta=0.025$, red). (b) Impact of the HB granularity on the $\eta$ resolution for single pions.}
 	\label{fig:performance:hadronic:eta}	
\end{figure}

\begin{figure}[ht]
  \centering
  \begin{subfigure}[b]{0.49\textwidth}
	  \includegraphics[width=\textwidth]{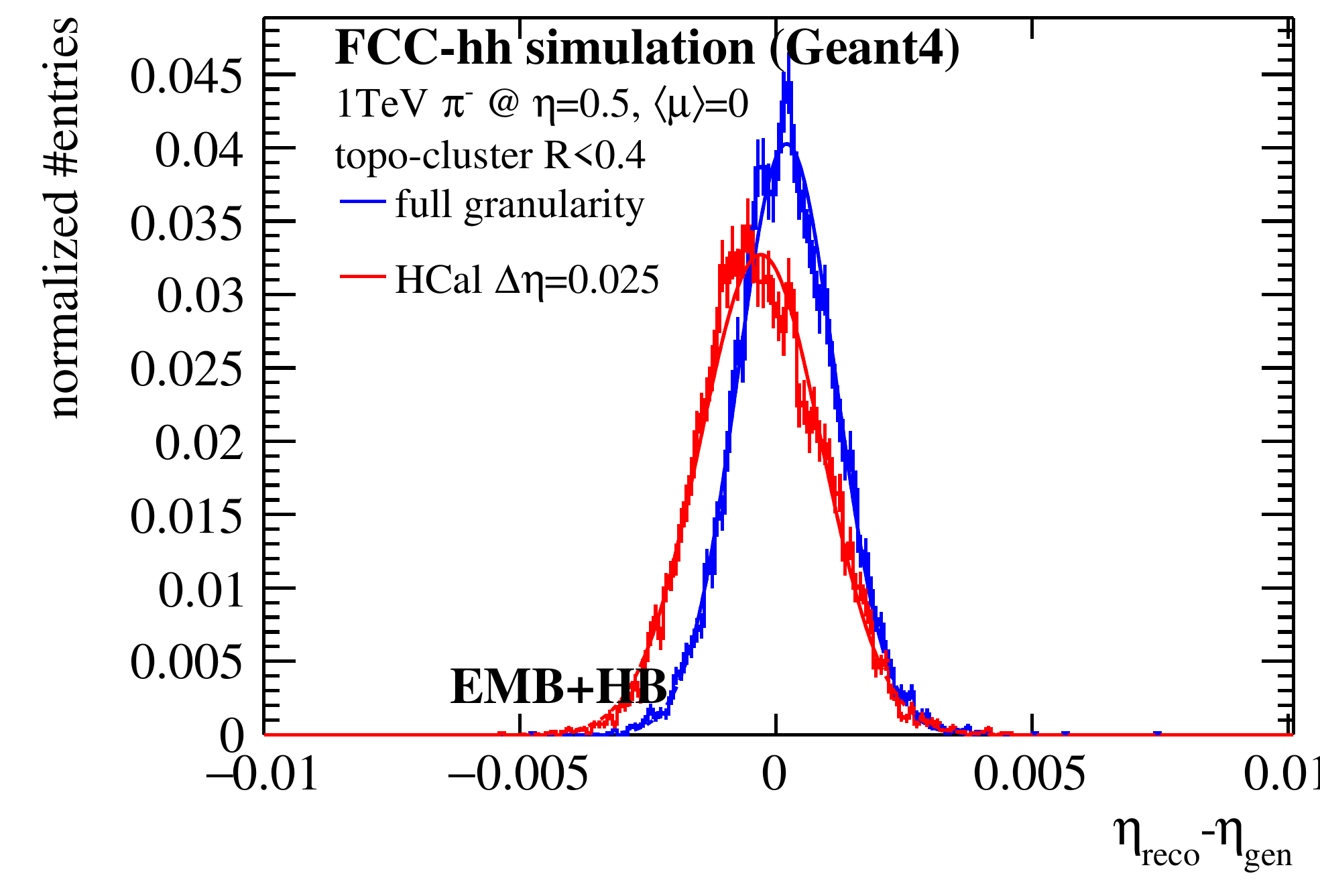}\caption{}\label{fig:performance:hadronic:diffEta:pions}
  \end{subfigure}
  \begin{subfigure}[b]{0.49\textwidth}
	  \includegraphics[width=\textwidth]{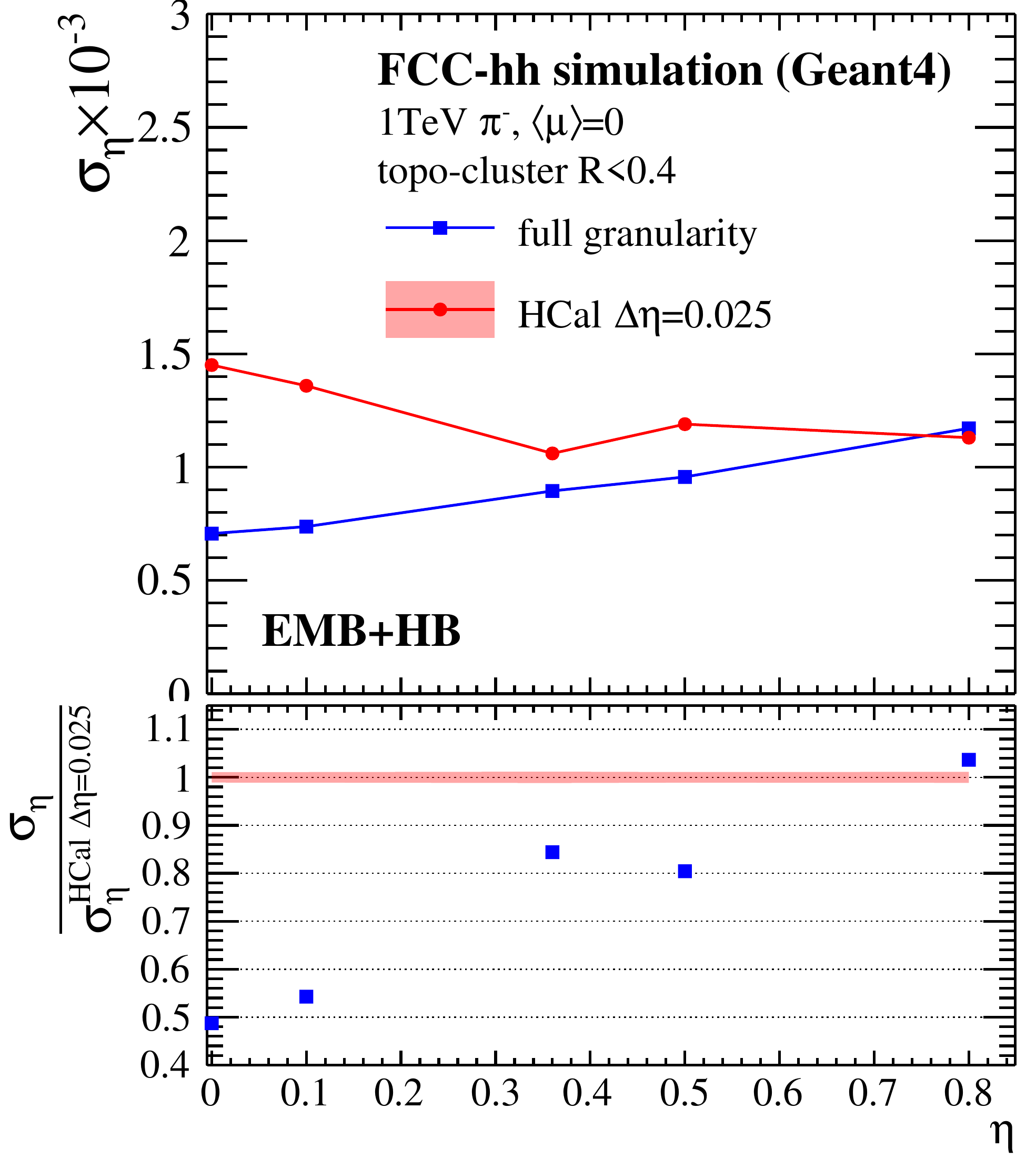}\caption{}\label{fig:performance:hadronic:diffEta:1TeV}
  \end{subfigure}
 \caption{(a) Difference between reconstructed and generated $\eta$ for topo-cluster within $\mathrm{R}<0.4$ for 1\,TeV pions generated at $\eta=0.5$, shown for the finest possible HB granularity of $\Delta\eta<0.006$ (full granularity, blue) compared to the baseline granularity of the HB ($\Delta\eta=0.025$, red). (b) Angular resolution as a function of $\eta$ for 1\,TeV pions, shown for the full HB granularity (blue) compared to the baseline granularity of the HB (red).}
 	\label{fig:performance:hadronic:diffEta}	
\end{figure}

As shown in Fig.~\ref{fig:performance:hadronic:eta:pions}, the $\eta$-resolution for single pions falls steeply with energy and improves up to 15\,\% if the full granularity of the HB is exploited. 
%However, even with the baseline HB granularity of 0.025 the achieved $\eta$ resolution of $<0.007$ is well below the intrinsic calorimeter resolutions of 0.01 in the EMB and $<0.006$, and $0.025$ of the HB, respectively. 
Figure~\ref{fig:performance:hadronic:diffEta:1TeV} shows the angular resolution for the two HB granularities as a function of the generated particle rapidity. The position resolution decreases for the default HB granularity, and increases for the full HB granularity with increasing $\eta$. These tendencies can be explained by the increasing sampling fraction in the EMB. The HB granularity of $\Delta\eta<0.006$ in the full granularity configuration is finer than the granularity of the EMB with $\Delta\eta=0.01$, thus the larger the shower fraction in the EMB, the worse the resolution. And vice versa, in case of a broader HB segmentation of $0.025$, the larger energy fraction in the EMB improves the resolution.

To determine whether the full granularity at scintillator level should be exploited, further studies including particle flow algorithms and particle ID algorithms will be needed. Additional studies are as well needed to evaluate the impact of the very high pile-up environment and the power of the high granularity for pile-up suppression.

\begin{figure}[ht]
  \centering
  \begin{subfigure}[b]{0.49\textwidth}
    	\includegraphics[width=1\textwidth]{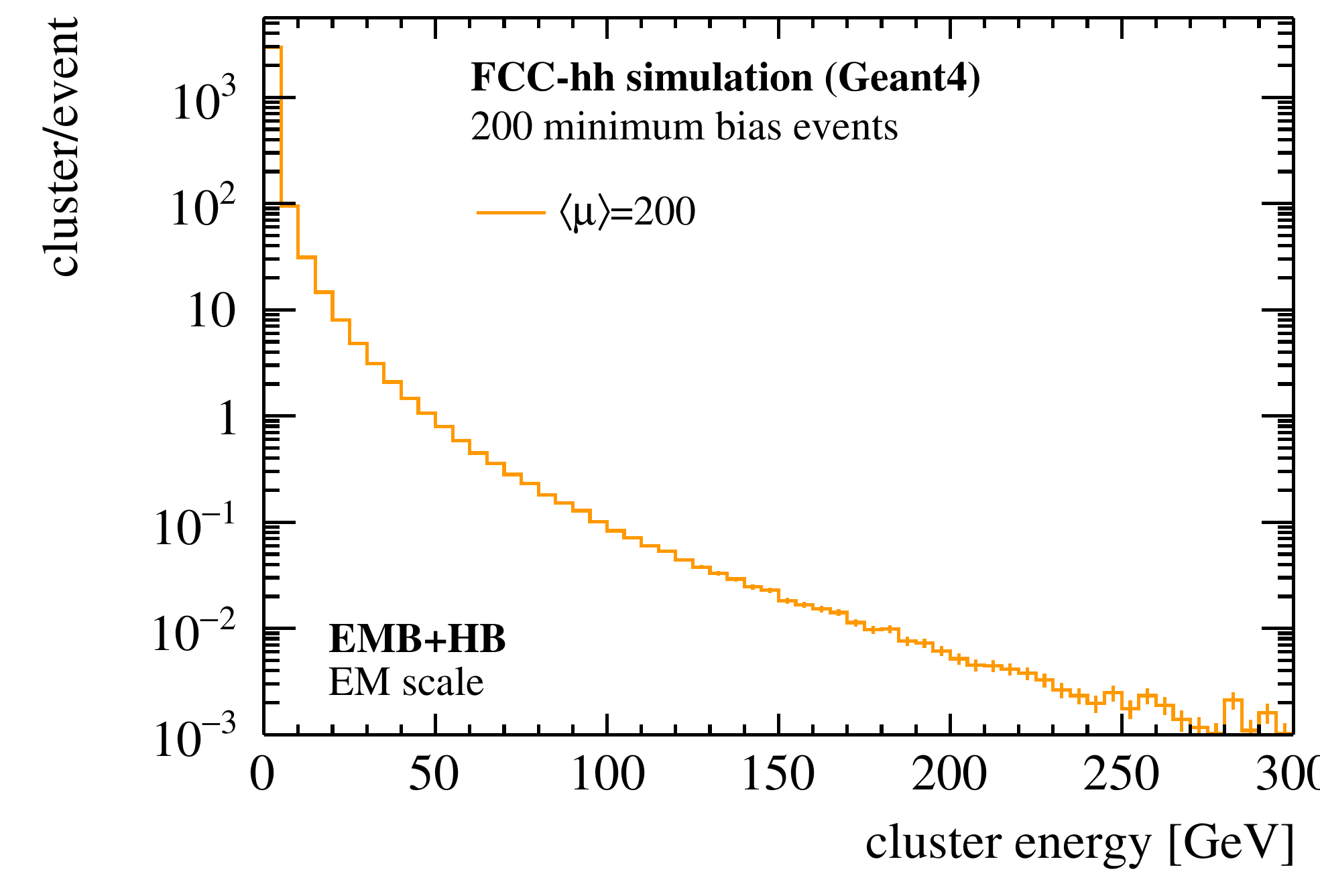}\caption{}
 	\label{fig:performance:hadronic:clusterEnergyPU200}	
\end{subfigure}
  \begin{subfigure}[b]{0.49\textwidth}
	\includegraphics[width=1\textwidth]{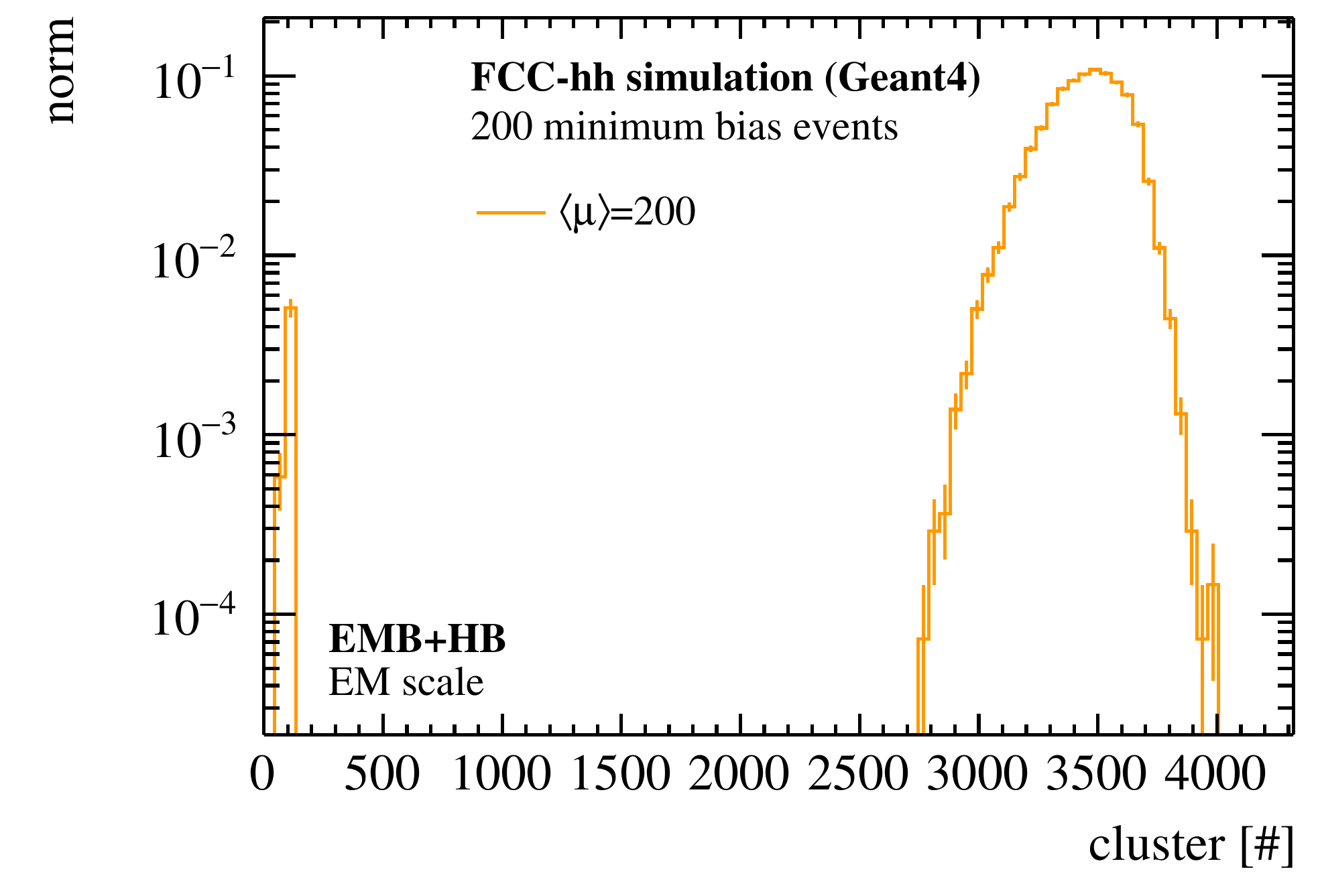}\caption{}
 	\label{fig:performance:hadronic:numClusterPU200}	
\end{subfigure}
        \caption{(a) Energy distribution and (b) the number of topo-cluster for events with 200 collisions per bunch crossing ($\left<\mu\right>=200$), on EM scale, and 4-2-0 topo-cluster threshold configuration.}
\end{figure}

\clearpage
\subsubsubsection{Energy Reconstruction of Single Hadrons in a High Pile-Up Environment}
\label{sec:performance:hadronic:pileup}
To determine the performance of the topo-cluster algorithm, 200 simulated minimum bias events are overlaid to mimic a realistic pile-up environment. It is expected that - due to the bipolar shaping - the out-of-time pile-up will in average compensate the positive energy deposit due to in-time pile-up. Therefore, due to the fact that out-of time pile-up is neglected, the mean energy deposit per calorimeter cell is shifted to positive values. In order to emulate the effect of out-of time pile-up, the mean value is therefore corrected\footnote{It should be noted that this approach neglects the impact of out-of-time pile-up on the event-to-event fluctuations due to energy deposits in previous bunch-crossings. An event-to-event correction of the out-of-time pile-up through the application of advanced filtering algorithms using the full event history is being studied for the HL-LHC upgrades.} and shifted back to 0, see also Section~\ref{sec:software:noise:pileup}. However, due to the very steep energy spectrum and small cell sizes this correction is relatively small with maximum shifts by 25 and 9\,MeV for $\left<\mu\right>=200$ of the EMB and HB cell energies, respectively. The expected noise per cell, needed to determine the cells significance, are estimated by the quadratic sum of the pile-up and electronics noise contribution. The pile-up noise contributions are estimated as the RMS of the energy distribution of 200 and 1,000 minimum bias events per cell, above the respective electronics noise level, as also described in Sec.~\ref{sec:software:noise:pileup}. The energy reconstruction within the high pile-up environment is especially challenging for low energetic particles, due to the number of pile-up cluster with energies of up to multiple hundreds of \,GeV (topo-cluster in 4-2-0 mode, on EM scale) for $\left<\mu\right>=200$. Fig.~\ref{fig:performance:hadronic:clusterEnergyPU200} shows the energy distribution for topo-cluster (EM scale) for events with 200 collisions per bunch crossing ($\left<\mu\right>=200$) and Fig.~\ref{fig:performance:hadronic:numClusterPU200} shows the corresponding number of cluster that are build from 200 minimum bias events in the full EM and hadronic barrel calorimeters.

\begin{figure}[ht]
  \centering
    	\includegraphics[width=0.5\textwidth]{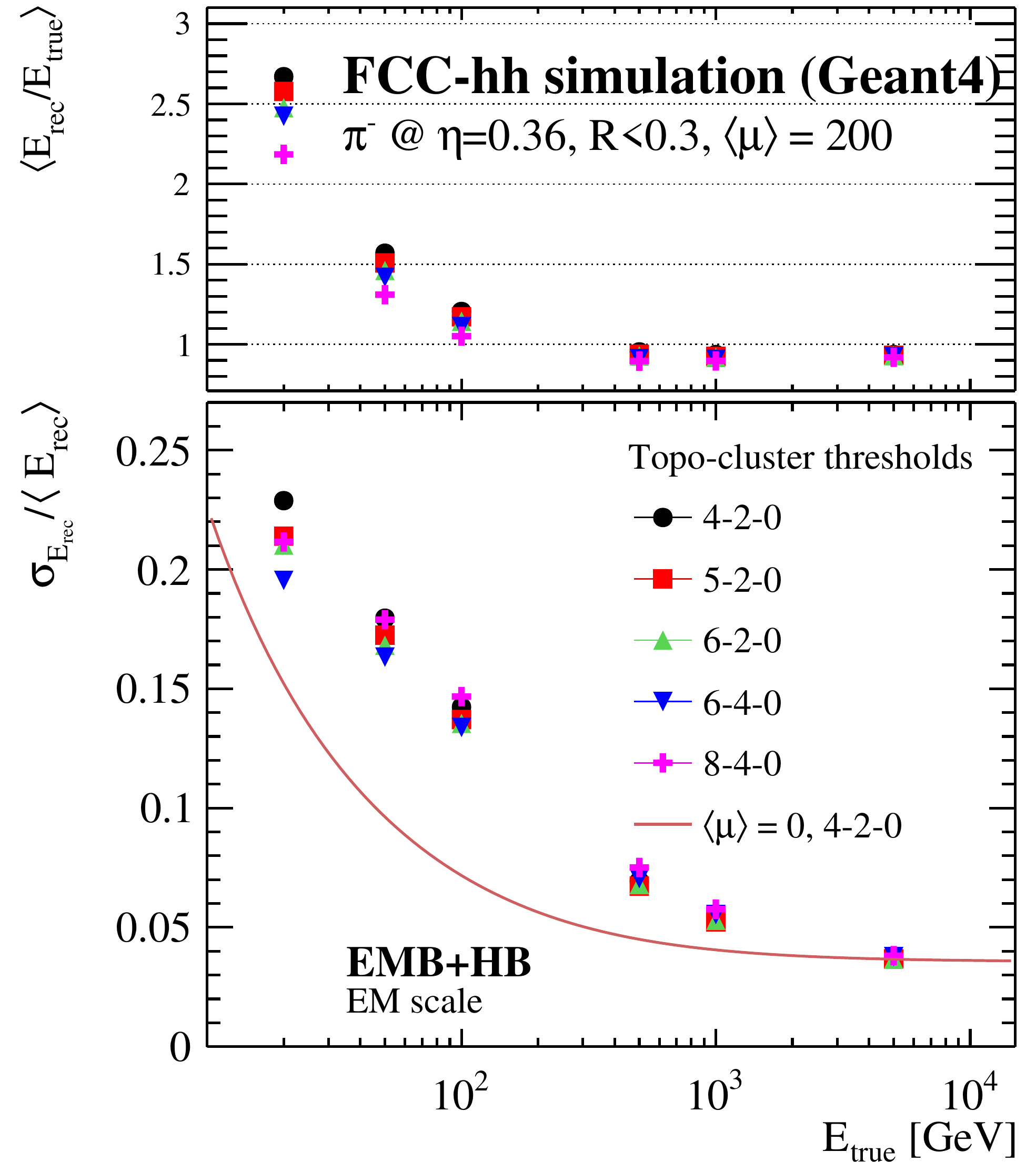}
        \caption{Energy resolution and linearity of pion showers at the EM scale for different thresholds of the topo-clustering algorithm at $\left<\mu\right>=200$ with a cut of $\mathrm{R}<0.3$ around the barycentre of the topo-cluster. The solid line corresponds to the resolution curve for the default threshold (4-2-0) without pile-up. Configurations with $P$ = 1 were also tried, but showed no difference from cases with $P$ = 0, therefore not shown in the plot. }
 	\label{fig:performance:hadronic:resoClustersElecNoise_PU200}	
\end{figure}

The linearity and energy resolution for different thresholds at the EM scale with an in-time pile-up of $\left<\mu\right>=200$ are shown in Fig.~\ref{fig:performance:hadronic:resoClustersElecNoise_PU200}. To reduce the effect of pile-up, a cylindrical region with $R<0.3$ around the barycentre of the topo-cluster is used to calculate the energy. The linearity of the energy response is heavily degraded, the ratio of the reconstructed and true energy exceeds 2.5 for 20\,GeV pions. This result demonstrates that an energy reconstruction at $\left<\mu\right>=200$ needs to incorporate pile-up suppression techniques.  The comparison to the energy resolution without pile-up (solid line) shows the strong impact on the performance for $\left<\mu\right>=200$, up to high energies of 5\,TeV. It is obvious that such high pile-up cannot be handled with a simple clustering algorithm alone, but more sophisticated rejection techniques and the information from the tracker will be needed for an accurate energy measurement of the products of the hard scatter. It was not possible to develop such techniques in the scope of this study. 
Figure~\ref{fig:performance:hadronic:resoClustersElecNoise_PU200} shows that a small improvement can be achieved by optimising the thresholds of the topo-cluster algorithm.
The thresholds $S$ = 6, $N$ = 2 and $P$ = 0 are chosen for the following plots.

\begin{figure}[ht]
  \centering
  \begin{subfigure}[b]{0.49\textwidth}
    	\includegraphics[width=1.\textwidth]{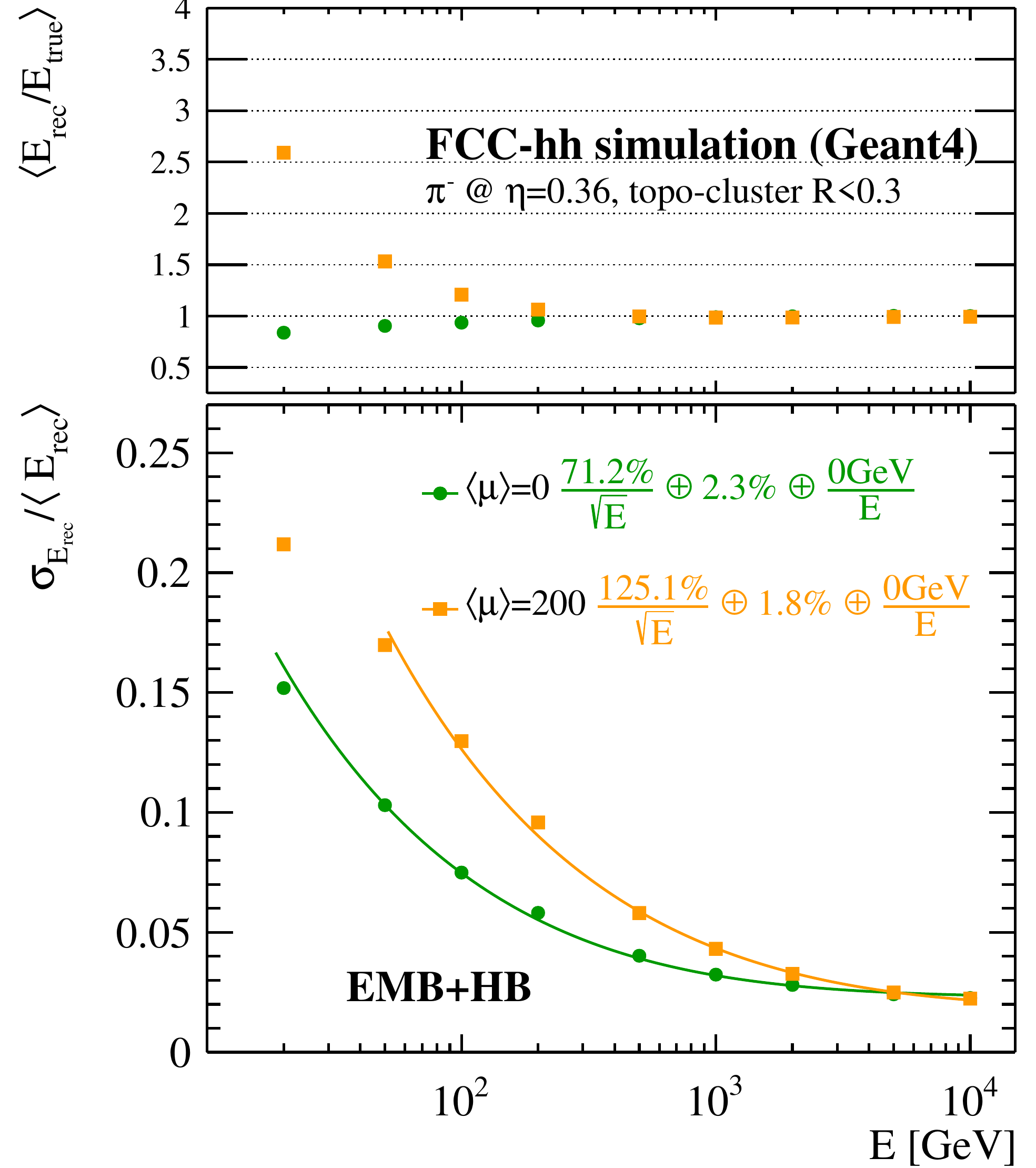}\caption{}	\label{fig:performance:hadronic:resoPU200:PU}	
  \end{subfigure}
  \hfill
  \begin{subfigure}[b]{0.49\textwidth}
    	\includegraphics[width=1.\textwidth]{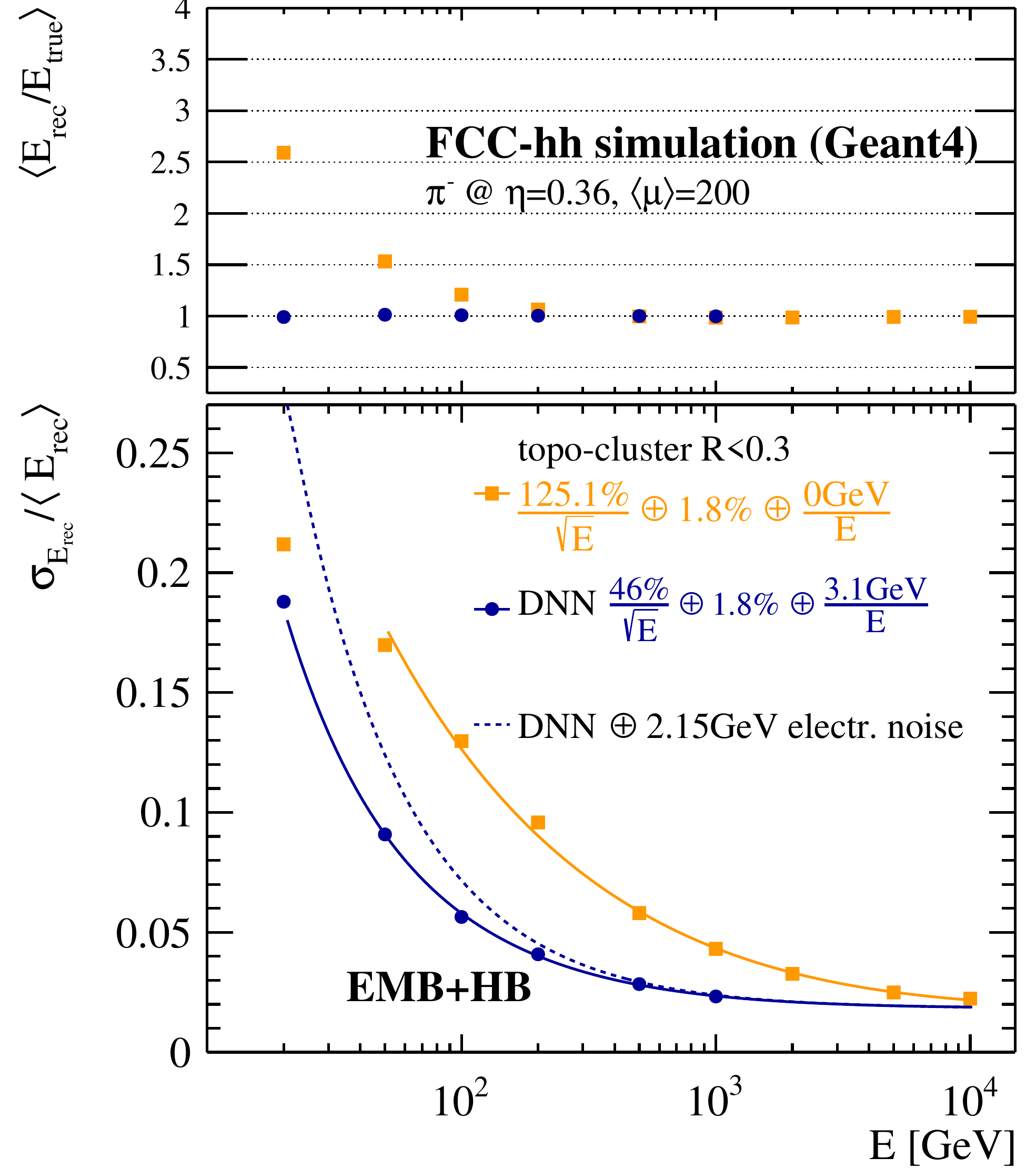}\caption{}	\label{fig:performance:hadronic:resoPU200:DNN}	
  \end{subfigure}
  \caption{(a) Energy resolution and linearity of pion showers at $\eta=0.36$ for pile-up of $\left<\mu\right>=0/200$. The resolution is obtained after topo-clustering in 4-2-0/6-2-0 mode after calibration, with $\text{R}<0.3$, and (b) compared for $\left<\mu\right>=200$ to the performance of the Deep Neural Network (DNN).}
 	\label{fig:performance:hadronic:resoPU200}	
\end{figure}

\begin{figure}[ht]
  \centering
  \begin{subfigure}[b]{0.49\textwidth}
    	\includegraphics[width=1.\textwidth]{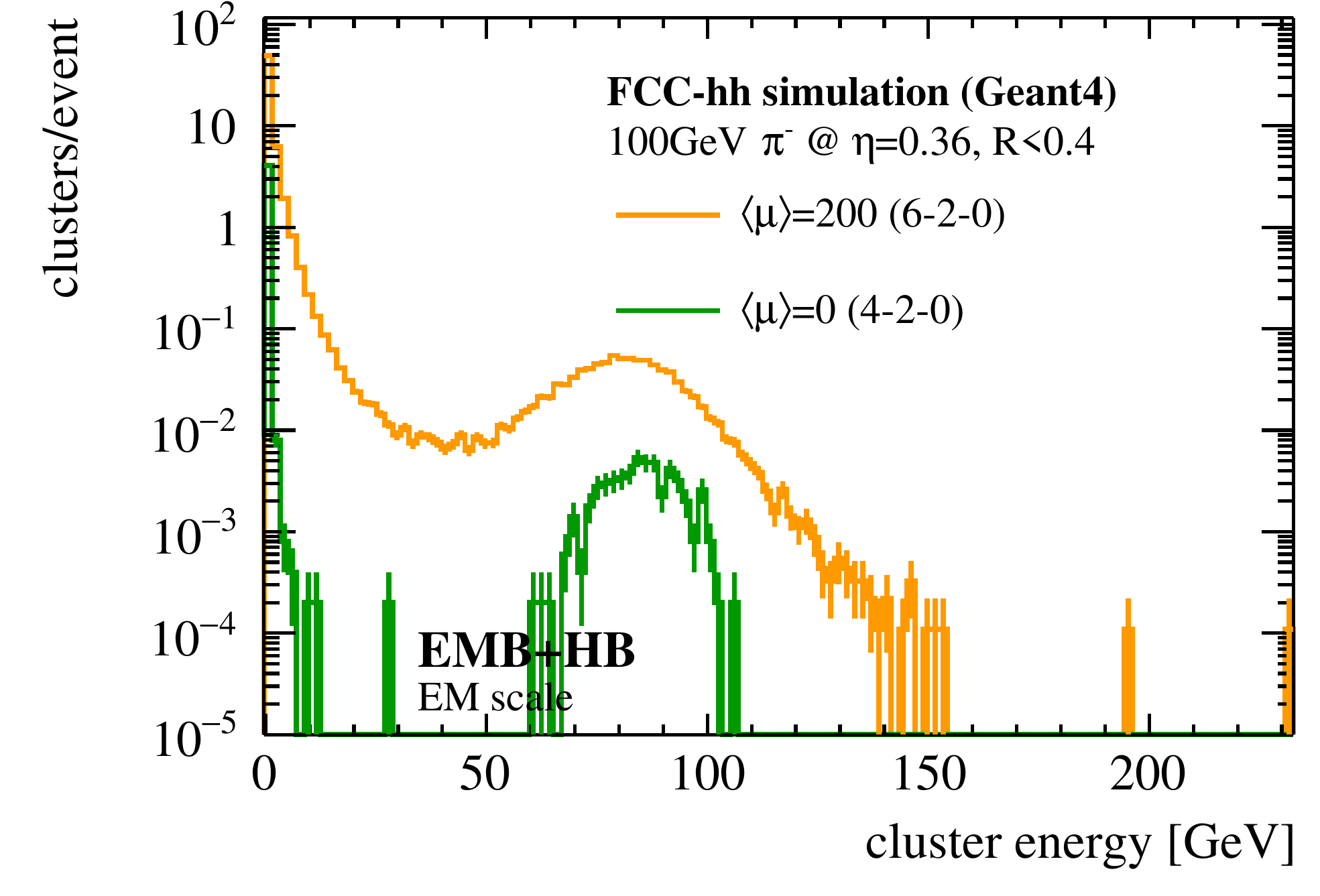}\caption{}
    	\includegraphics[width=1.\textwidth]{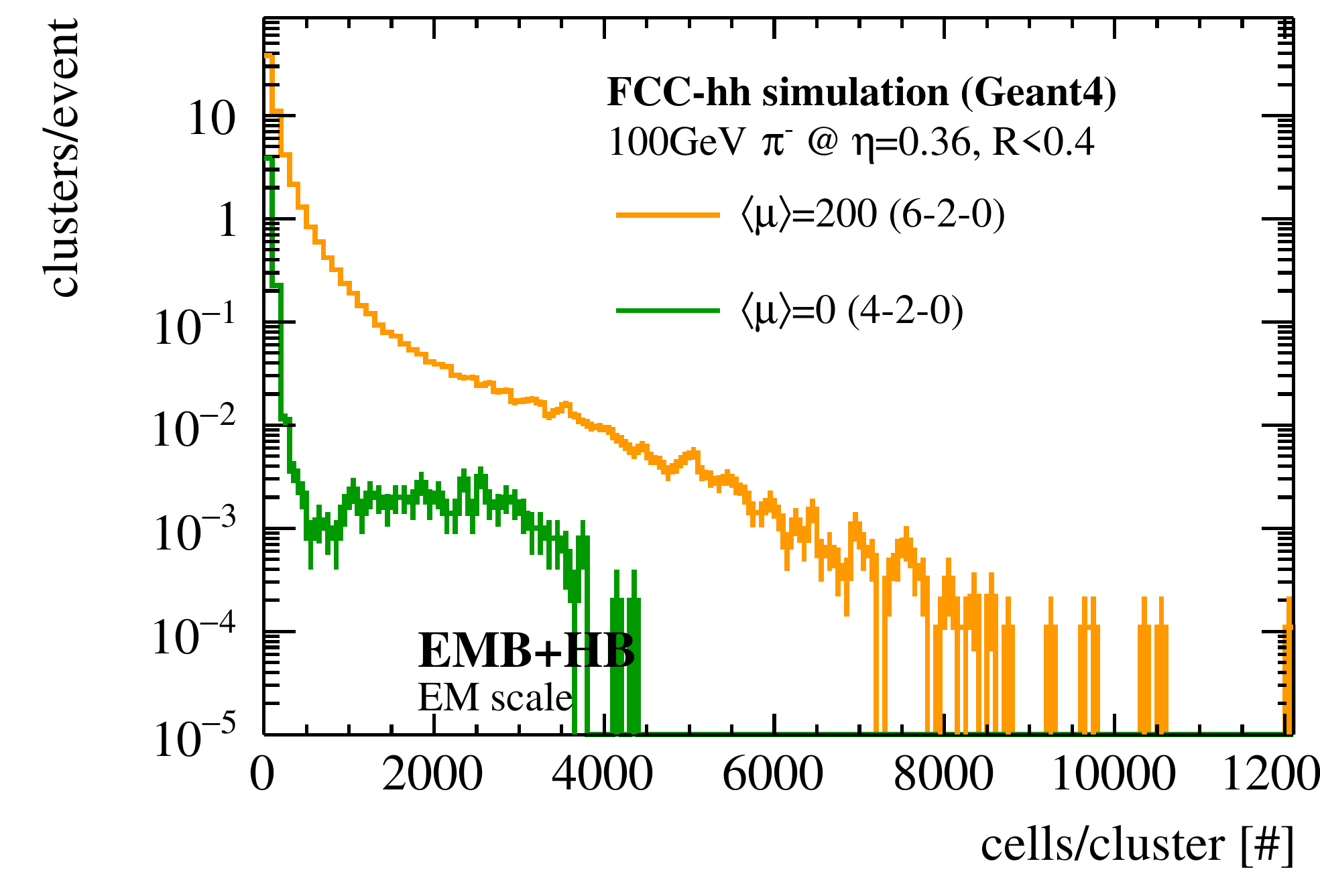}\caption{}    	  
	\end{subfigure}
  \hfill
  \begin{subfigure}[b]{0.49\textwidth}
	\includegraphics[width=1.\textwidth]{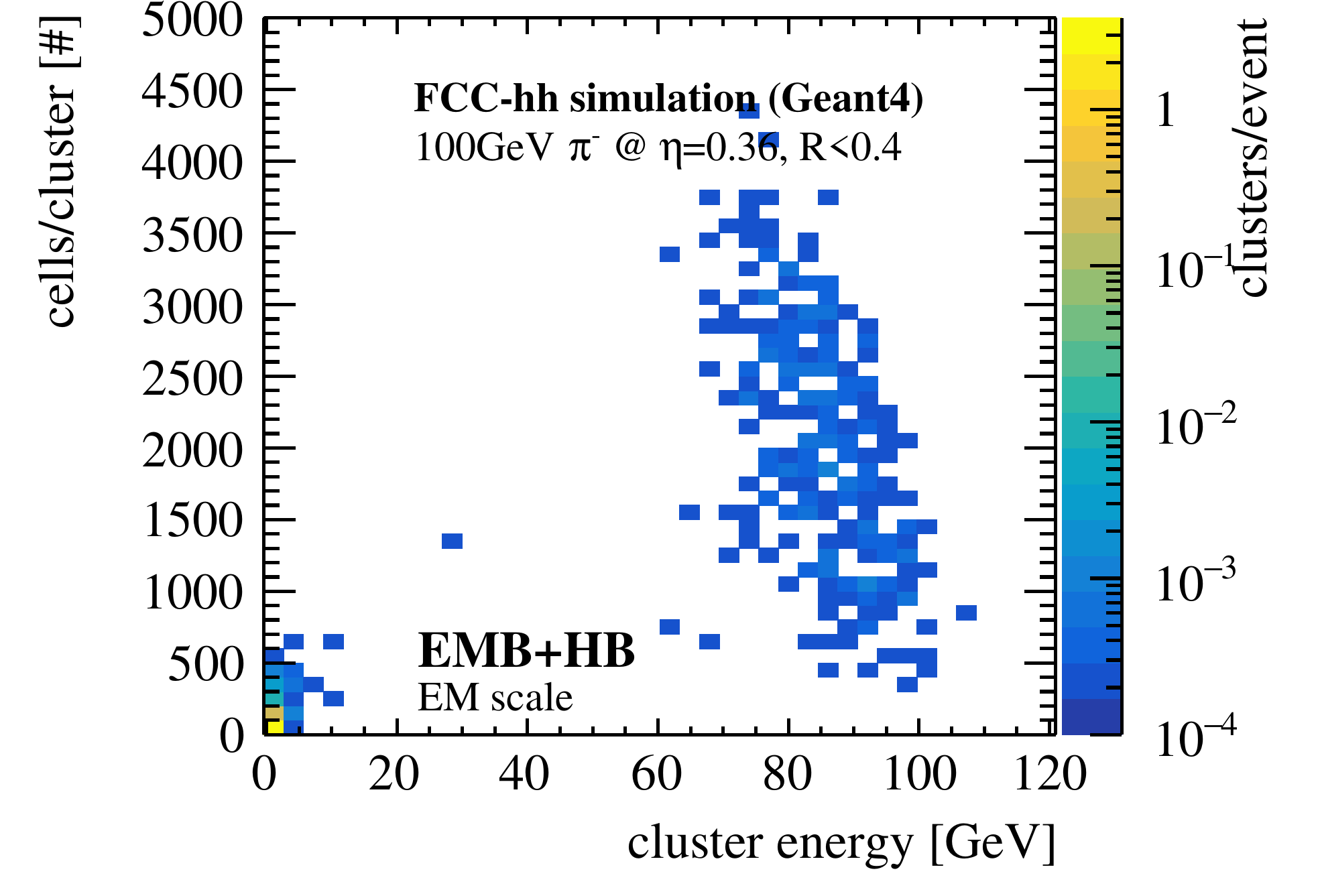}\caption{$\left<\mu\right>=0$}
    	\includegraphics[width=1.\textwidth]{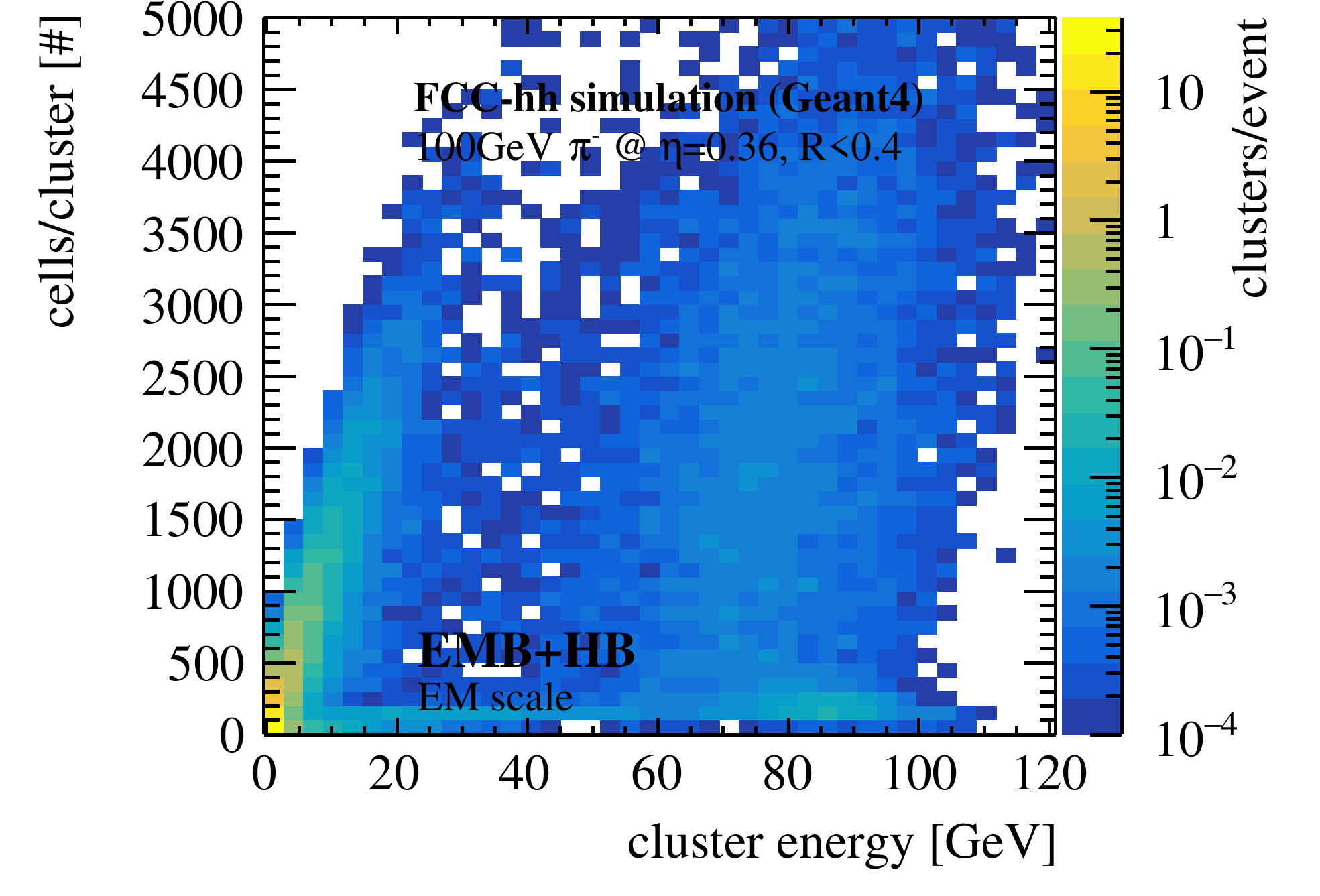}\caption{$\left<\mu\right>=200$}	
  \end{subfigure}
  \caption{(a) Cluster energies and (b) number of cells per cluster for 100\,GeV pion showers at $\eta=0.36$, in EMB+HB for $\left<\mu\right>=0$ and $200$. (c) and (d) show the strong increase in the total number of topo-cluster, and their energy distribution.}
 	\label{fig:performance:hadronic:PU200cluster}	
\end{figure}

The energy resolution for single pions and 200 pile-up events after topo-cluster reconstruction, including a simple calibration as described in~\ref{sec:performance:hadronic:topoClusterCalibration}, is shown in Fig.~\ref{fig:performance:hadronic:resoPU200:PU}. A strong degradation of the resolution is observed especially at energies below 500\,GeV, which increases the stochastic term from 70\, to 125\,\%. Additionally, the linearity cannot be ensured in this scenario due to the addition of pure pile-up clusters, see top plot in Fig.~\ref{fig:performance:hadronic:resoPU200:PU}. Due to the bad linearity below 50\,GeV, the resolution fit does not include the low energy points. The topo-cluster algorithm alone is not able to reject pile-up, which illustrates the need for either smart cluster energy correction like the jet-area-based offset correction~\cite{TheATLAScollaboration:2013pia}, or particle flow algorithms that match the particle tracks in the tracker to the calorimeter clusters~\cite{Bertolini:2014bba}. The effect of pile-up on the number of clusters and the number of clustered cells is shown in Fig.~\ref{fig:performance:hadronic:PU200cluster}. It is observed that the total number of clusters is increased by one order of magnitude moving from  $\left<\mu\right>=0$ to 200. The energy spectra of these topo-cluster in $\left<\mu\right>=200$ range from 0 to 260\,GeV and include up to 12,000 cells. 

The comparison of the resolution with the results of the DNN for $\left<\mu\right>=200$ is shown in Fig.~\ref{fig:performance:hadronic:resoPU200:DNN}. As expected, the DNN is able to reconstruct the particle energy with a much better linearity, and an energy resolution of 46\,\% in the stochastic and 1.8\,\% constant term is achieved. However, it needs to be mentioned, that electronic noise is in this case not included. The effect can be estimated by adding in quadrature another 2.15\,GeV to the noise term to the original fit, see the dashed blue line in Fig.~\ref{fig:performance:hadronic:resoPU200:DNN}. The noise value has been determined from cells within a window of $\Delta\eta\times\Delta\phi=0.34\times0.34$, see bottom row in Table~\ref{tab:performance:hadronic:cellNoise}.

The impact of $\left<\mu\right>=1,000$ in-time pile-up on the hadron performance has been tested as well. However, as the results for $\left<\mu\right>=200$ already show, are more sophisticated algorithms needed to reject and suppress pile-up events and contributions, respectively. A first step towards improving the topo-cluster algorithm has been recently developed~\cite{FCChhBoostedPaper}.

%%\subsubsection{Conclusions}
%%\label{sec:performance:hadronic:summary}

%% file: tex/performance/id.tex
%%%%%%%%%%%%%%%%%%%%%%%%%%%%%%%%%%%
\subsection{Pion and photon identification using Multivariate Analysis (MVA) Techniques}
\label{sec:performance:idMVA}
In the following, the ability to distinguish single photons from $\pi^0$ mesons in the EMB of the FCC-hh detector is tested. This is an important property to be considered for optimising the design of the detector as it is key for reducing the background for the important $H \to \gamma\gamma$ decay. Due to the small mass of the $\pi^0$, the two photons coming from a $\pi^0\rightarrow\gamma\gamma$ decay are very close to each other and therefore can be misidentified as a single photon. With sufficient fine granular calorimetry, however, it is possible to detect the separation between the two photons and therefore distinguish the $\pi^0$ signal from a single $\gamma$ signal. In this study we analyse the $\pi^0$ rejection in the EMB for transverse momenta $p_T\in{[10 - 80]}$\,GeV at $|\eta| = {0}$.

\subsubsection{Methodology}
\begin{figure}[htp]
\begin{center}
\includegraphics[width=0.48\textwidth]{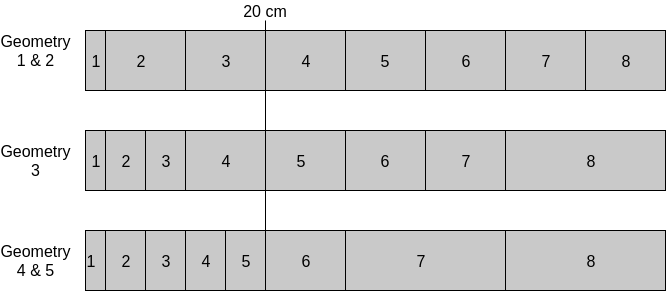}
\caption{Schematic of the distribution of EMB layers in different geometries that were analysed. }
\label{fig:id:Layers}
\end{center}
\end{figure}

\subsubsubsection{Monte Carlo Simulations}
\label{sec:id:samples}
The single particle simulations of $\pi^0$ and $\gamma$ in the EMB were produced without considering pile-up or electronic noise. The number of events analysed for every data point was at least $(1 \pm 0.05) \times 10^5$ for each particle in order to minimise random fluctuations. The following geometries were explored:
\begin{enumerate}
\item The simplest geometry with cell size $\Delta\eta = 0.01$ and $\Delta\phi = 0.009$ in all layers.
\item The EMB layout with cell size $\Delta\eta = 0.0025$ and $\Delta\phi = 0.009$ in the $2^{nd}$ Layer and $\Delta\eta = 0.01$ in all other layers.
\item Geometry 1 with the $2^{nd}$ layer split in half (in longitudinal direction) while combining the $7^{th}$ and $8^{th}$ layers to maintain a constant number of readout channels.
\item Geometry 2 with the now $4^{th}$ layer (layer 3 in geometry 1) also halved while combining the new $6^{th}$ and $7^{th}$ layers (layers 5,6,7 and 8 in geometry 1).
\item Geometry 4 with also finer segmentation in $\phi$, $\Delta\phi = 0.0045$ in all layers, which effectively doubles the number of layers in the $\phi$ direction.
\end{enumerate}
The individual characteristics of all analysed data sets is summarised in Table~\ref{tab:id:Samples} and the layer structure for all geometries is illustrated on Fig.~\ref{fig:id:Layers}. 

\begin{table}[htp]
\begin{center}
\begin{tabular}{|c|c|c|c|c|c|}
\hline
& $\pt$ [GeV]
& $\#$ used layer
& $\eta$
& $\Delta\eta^{*}$
& $\Delta\phi$ \\		
\hline
Sample 1 & 10 to 50 & 3 & 0 & 0.01 & 0.009\\
Sample 2 & 10 to 50 & 3 & 0 & 0.0025 & 0.009\\
Sample 3 & 10 to 50 & 4 & 0 & 0.0025 & 0.009\\
Sample 4 & 10 to 80 & 5 & 0 & 0.0025 & 0.009\\
Sample 5 & 10 to 80 & 5 & 0 & 0.0025 & 0.0045\\
\hline
\end{tabular}
\end{center}
\caption{Properties of the data sets used in the analysis. The segmentation in pseudo-rapidity is set to $\Delta\eta = 0.0025$ only in the second layer for samples 2 to 5. The "used layer" refers to the number of longitudinal layers used }
\label{tab:id:Samples}
\end{table}

\subsubsubsection{Discriminating Variables}
The energy and cell data from the single particle simulations was then used to calculate a set of variables that are expected to distinguish between $\pi^0$ and $\gamma$ energy deposits. These are inspired by the previous study done on $\gamma/\pi^0$ separation in the $1^{st}$ compartment of the ATLAS EMB and adjusted to accommodated more layers provided by the FCC detector. The variables are defined as follows:
\begin{enumerate}
\item $E_{max}$ - Maximal cell energy deposit for all the cells of the second layer.

\item $E_{2^{nd}max}$ - A second energy maximum separate from $E_{max}$ by at least one cell.

\item $E_{ocore}$ - Fraction of energy deposited outside the shower's centre where $E(\pm n)$ is the energy deposited in $\pm n$ cells around the cell with the maximal energy deposit:
\begin{equation}
E_{ocore} = \frac{E(\pm n) - E(\pm 1)}{E(\pm 1)}\label{eq:eocore} 
\end{equation}
with $n=3$

\item $E_{dmax}$ - Difference between the second energy maximum and the minimal energy deposit in the valley between the maximal energy deposit and the second energy maximum:
\begin{equation}
E_{dmax} = E_{2ndmax} - E_{min} \label{eq:edmax} 
\end{equation}

\item $W_{nst}$ - The shower width summed over $n$ central cells along $\eta$ that are within $\pm 1$ of the $\phi$ coordinate of the maximal energy deposit. Here $i$ denotes the cell number and $i_{max}$ the cell with the maximal energy deposit. It is always computed in the same layer.
\begin{equation}
W_{nst} = \frac{\sum E_i \cdot (i-i_{max})^2}{\sum E_i} \label{eq:width} 
\end{equation}

\item $E_{nT}$ - Energy deposited in the $n^{th}$ layer divided by the total energy deposited in the EMB:
\begin{equation}
E_{nT} = \frac{E_n}{E_T} 
\end{equation}

\item $E_{n1}$ - Energy deposited in the $n^{th}$ layer divided by the total energy deposited in the $1^{st}$ layer of the EMB.
\begin{equation}
E_{n1} = \frac{E_n}{E_1} 
\end{equation}
\end{enumerate}

\begin{figure}[hpt]
\begin{center}
\includegraphics[width=0.8\textwidth]{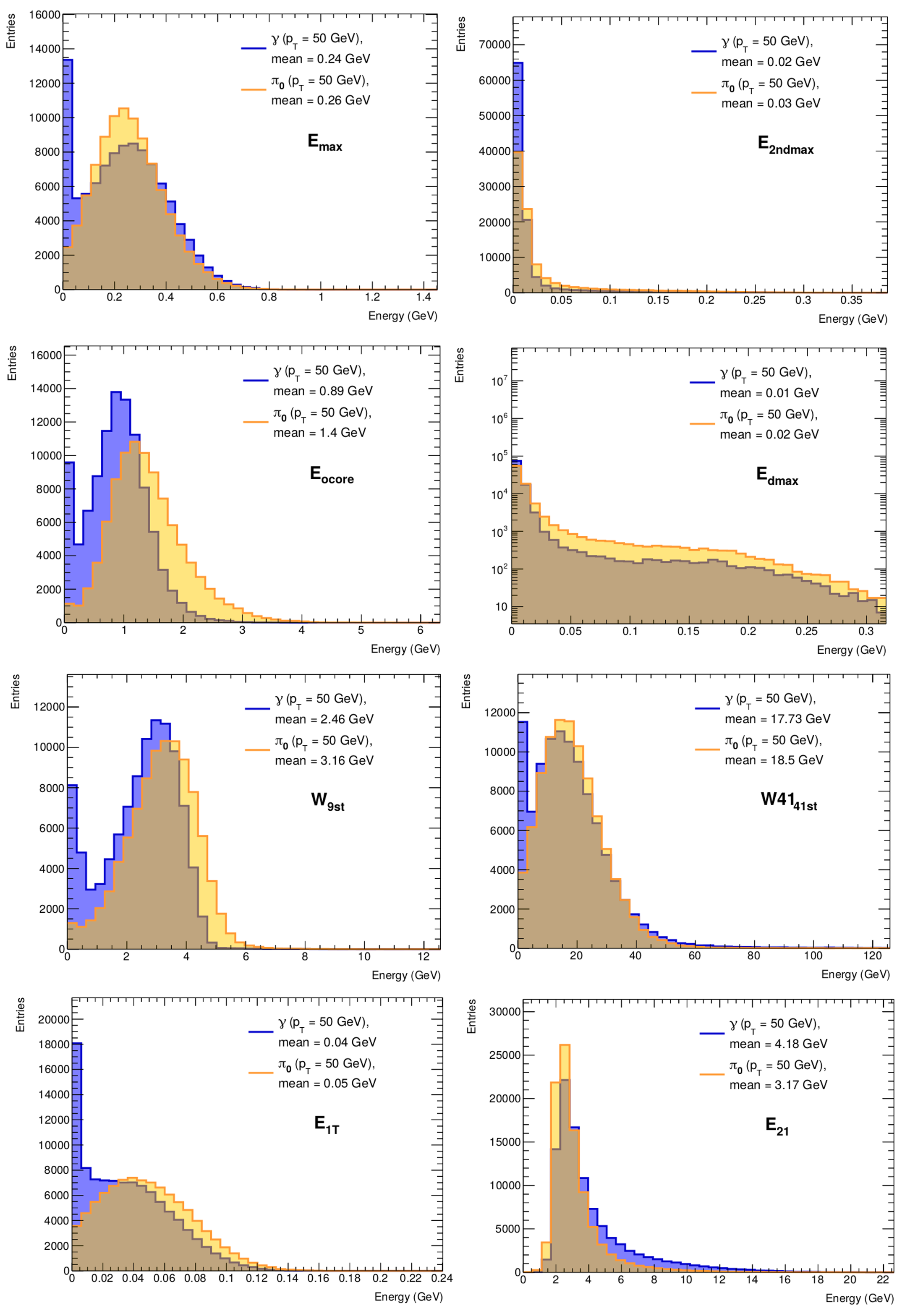}
\caption{Distributions of discriminating variables for both $\gamma$ and $\pi^0$ with $p_T = 50$ GeV and $\eta = 0$. No pile-up or electronic noise is included.}
\label{fig:id:variables}
\end{center}
\end{figure}

The shower width was calculated twice in each layer, using a low and high number of cells to sum over. $E_{nT}$ and $E_{n1}$ were calculated for all layers but other variables were capped at 20\,cm depth of the EMB since the energy considered here is low and the particles do not penetrate too far. The corresponding final layer can be seen in Fig.~\ref{fig:id:Layers}. Also, because of the varying properties of the calorimeter layers such as finer granulation in $\eta$ in the $2^{nd}$ layer, the variables were adjusted for each layer individually to obtain the best discrimination. For example, the shower width in the $1^{st}$ layer was summed over 9 and 41 cells whereas it was summed over 3 and 21 cells in other layers. Figure~\ref{fig:id:variables} shows the distributions of the discriminating variables calculated for the $1^{st}$ layer of sample 5.

\subsubsubsection{Multivariate Data Analysis}
The multivariate classification using the discrimination variables was done using the boosted decision tree (BDT) algorithm of TMVA. This takes repeated decisions on every single variable individually until a criterion is fulfilled and this creates multiple regions in the phase space which are classified as either signal ($\gamma$) or background ($\pi^0$). 
Half of the simulated events are used for training the analysis methods and obtaining the importance of each variable in distinguishing between the events i.e. how often they were used to split decision tree nodes. The other half of events is then analysed with the trained BDT. From this the $\pi^0$ rejection factor, $R_{\pi}$, is calculated using Equation~\ref{eq:id:rejection}, where $B$ is the fraction of $\pi^0$s rejected at a given $\gamma$ signal efficiency.
\begin{equation}
\label{eq:id:rejection}
R_{\pi} = \frac{1}{1 - B} 
\end{equation}

\begin{figure}[htp]
\begin{center}
\includegraphics[width=0.48\textwidth]{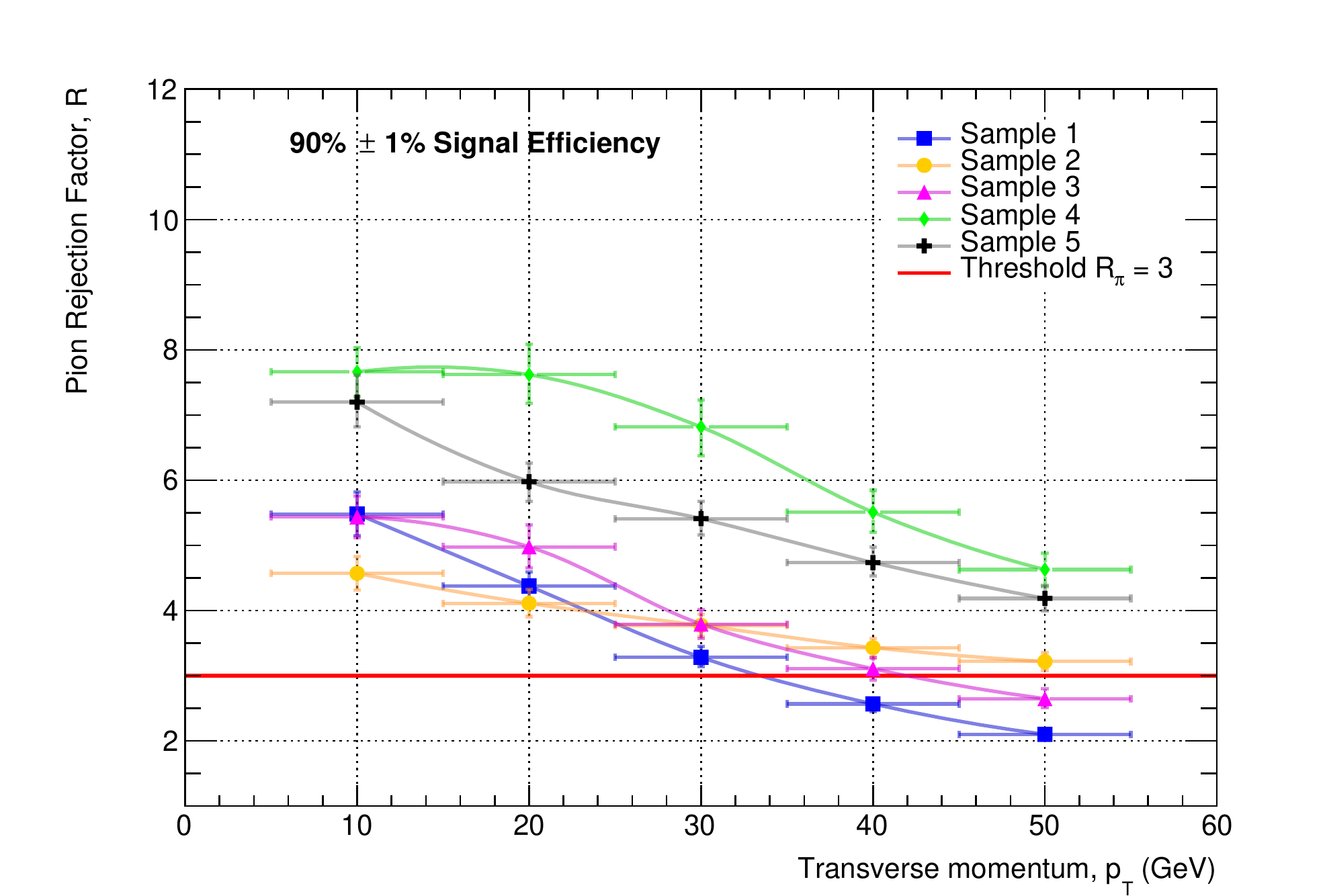}
\caption{$\pi_0$ rejection of all samples at $p_T$ up to 50\,GeV for a signal efficiency of 90\,\%. The threshold value for the $\pi_0$ rejection factor is 3.}
\label{fig:id:geometries}
\end{center}
\end{figure}

\subsubsubsection{Optimal Geometry}
The different geometries of the EMB that were tested are laid out in Sec.~\ref{sec:id:samples}. All of these were investigated to find highest $\pi^0$ rejection while considering the same calorimeter depth of 20\,cm. For sample 3 the $2^{nd}$ layer was halved and for samples 4 and 5 also the $3^{rd}$ layer was halved therefore these contain more data and variables and are expected to perform better. Figure~\ref{fig:id:geometries} shows the $\pi^0$ rejection obtained with each geometry at various $p_T$ for a 90\% single photon reconstruction efficiency. 

There is significant improvement in $\pi^0$ rejection at $p_{T}>$ 40\,GeV when fine segmentation in $\eta$ is introduced in the $2^{nd}$ layer. This effect alone increases the rejection above threshold value as is shown by the yellow line on Fig.~\ref{fig:id:geometries} which describes the baseline geometry of the FCC EMB. The best choice of geometry is $\#4$ or $\#5$  with also finer segmentation in $\phi$. These both performed similarly and produced $R_{\pi} > 4$ at 50\,GeV while other geometries were close to $R_{\pi} = 3$ at this energy but geometry $\#5$ was chosen for the rest of the study because it contains more information and has potential to be better at higher energies. The variables were calculated identically for samples 4 and 5 which could be the reason in the low performance of geometry $\#5$ and could be improved by adjusting the number of $\phi$ bins to consider when calculating each variable. For example, the shower width is calculated by summing over 3 bins in $\phi$ in both cases but for geometry $\#5$ this will contain less hits. Therefore, more optimization needs to be done to get the full $\pi^0$ rejection potential out of finer segmentation in $\phi$ but this preliminary study shows that there does not appear to be much to gain in this $p_{T}$ range.

\begin{figure}[htp]
\begin{center}
\includegraphics[width=0.55\textwidth]{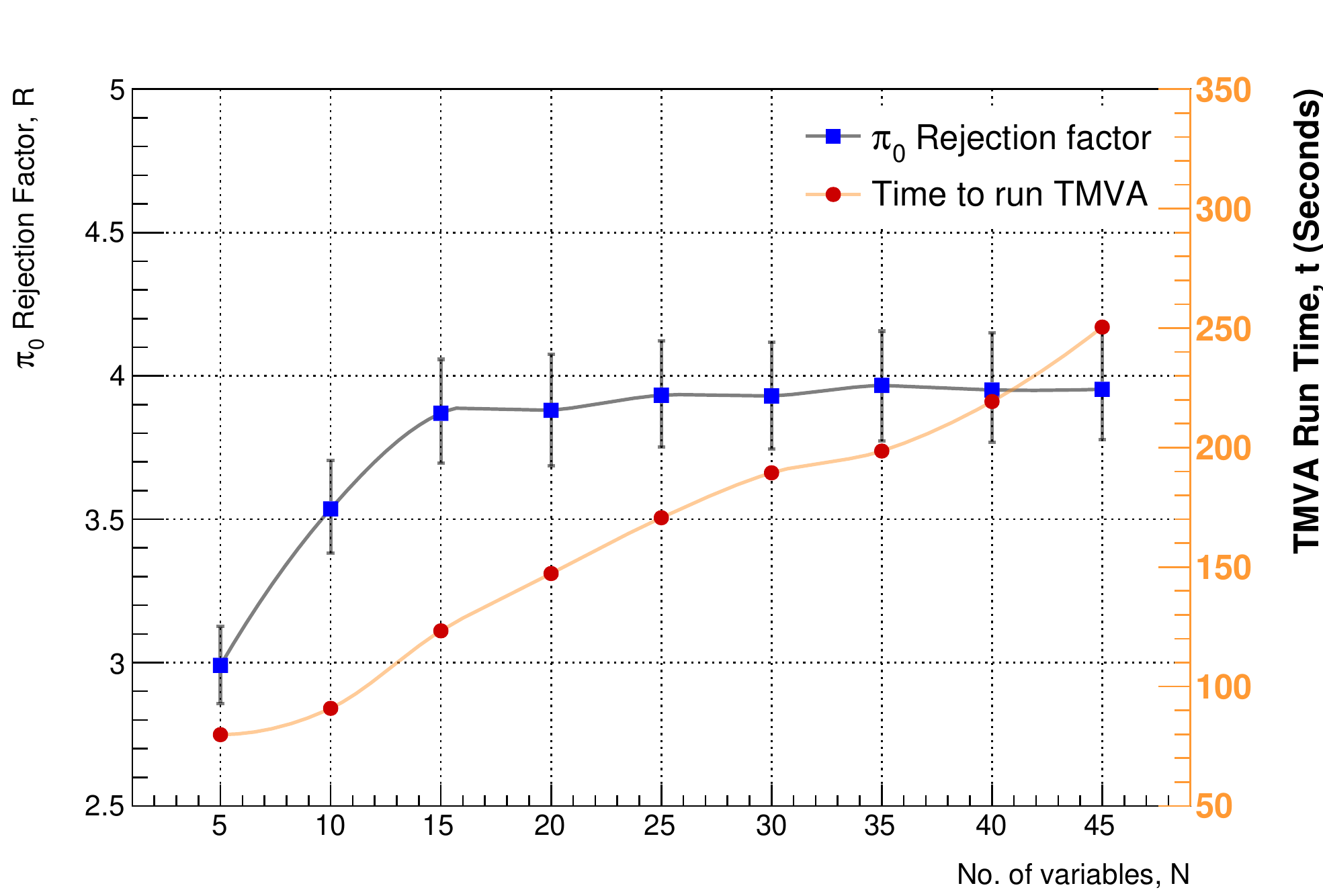}
\caption{$\pi^0$ rejection calculated at 50\,GeV and 90$\,\%$ signal efficiency with different number of discriminating variables chosen based on their separating power and correlation. The right axis shows the time needed for training the BDT.}
\label{fig:id:numbers}
\end{center}
\end{figure}

Sample 5 uses the highest granularity and up to 45 variables may be considered for the analysis. However, to improve calculation time some variables were removed from subsequent analysis while maintaining the largest possible $\pi^0$ rejection at higher energies. These were chosen based on their separating power and correlation with other variables at $p_T \geq $ 50\,GeV. Figure~\ref{fig:id:numbers} shows how the $\pi^0$ rejection factor changes when trained with different number of variables. The separating power of each variable changes with the energy as the $\pi^0$ decay kinematics become different. Table~\ref{tab:id:weights} shows the ranking of variables based on their separating power at 10\,GeV and 80\,GeV.

\begin{table}[htp]
\begin{center}
\begin{tabular}{|c|c|c|c|}
\hline
rank      
& variable
& variable
& variable\\
		
& $\pt = 10$\,GeV
& $\pt = 50$\,GeV
& $\pt = 80$\,GeV\\	
\hline
1 & $W_{3st}$ (layer 3)    & $W_{3st}$ (layer 3) & $W_{3st}$ (layer 3)\\
2 & $W_{3st}$ (layer 2)    & $W_{3st}$ (layer 2) & $E_{ocore}$ (layer 1)\\
3 & $E_{max}$ (layer 2)    & $E_{max}$ (layer 1) & $E_{max}$ (layer 1)\\
4 & $E_{max}$ (layer 3)    & $E_{ocore}$ (layer 1) &$W_{3st}$ (layer 2)\\
5 & $E_{3T}$               & $E_{max}$ (layer 2) & $W_{3st}$ (layer 4)\\
6 & $W_{21st}$ (layer 3)   & $W_{3st}$ (layer 4) & $E_{max}$ (layer 2)\\
7 & $W_{3st}$ (layer 4)    & $E_{max}$ (layer 3) & $E_{max}$ (layer 3)\\
8 & $E_{max}$ (layer 1)    & $E_{4T}$ & $E_{ocore}$ (layer 2)\\
9 & $E_{ocore}$ (layer 1)  & $E_{ocore}$ & $E_{1T}$\\
10 & $W_{21st}$ (layer 2)  & $W_{9st}$ (layer 1) & $W_{9st}$ (layer 1)\\
11 & $E_{1T}$ & $E_{8T}$   & $E_{3T}$\\
12 & $E_{8T}$ & $E_{21}$   & $E_{5T}$\\
13 & $E_{dmax}$ (layer 1)  & $e_{01}$ & $E_{8T}$\\
14& $E_{2T}$               & $E_{max}$\_$l04$ & $E_{21}$\\
15 & $E_{ocore}$ (layer 3) & $E_{max}$\_$l00$ & $E_{max}$ (layer 4)\\
\hline
%\multicolumn{4}{l}{}
\end{tabular}
\caption{The top 15 discriminating variables ranked by their method specific separating power at different $p_T$ for sample 5.}
\label{tab:id:weights}
\end{center}
\end{table}

In subsequent analysis the number of discriminating variables is lowered to 15 as this produces a similar result to 45 variables but in a shorter time. The difference is most prominent at lower energies (10 - 40\,GeV) and negligible at higher energies (50 - 80\,GeV). Since the $\pi^0$ rejection factor is very high at the lower energy region and although $\pi^0$ rejection is crucial at lower energies as well, this effect can be ignored because the rejection is large regardless. It is possible to optimize the training at every simulated $p_T$ point to obtain a better overall rejection. From Table~\ref{tab:id:weights} it is apparent that the variables have different separating power at different $p_T$ when comparing the most important variables at $p_T = 10$\,GeV and $p_T = 80$\,GeV. Adjusting the variables for training of each data set will improve the results obtained in this study since only the top 15 variables at $p_T = 50$\,GeV are considered for training at every $p_T$.

\subsubsection{${\pi^0}$ Rejection}

\begin{figure}[ht]
\begin{center}
\includegraphics[width=0.48\textwidth]{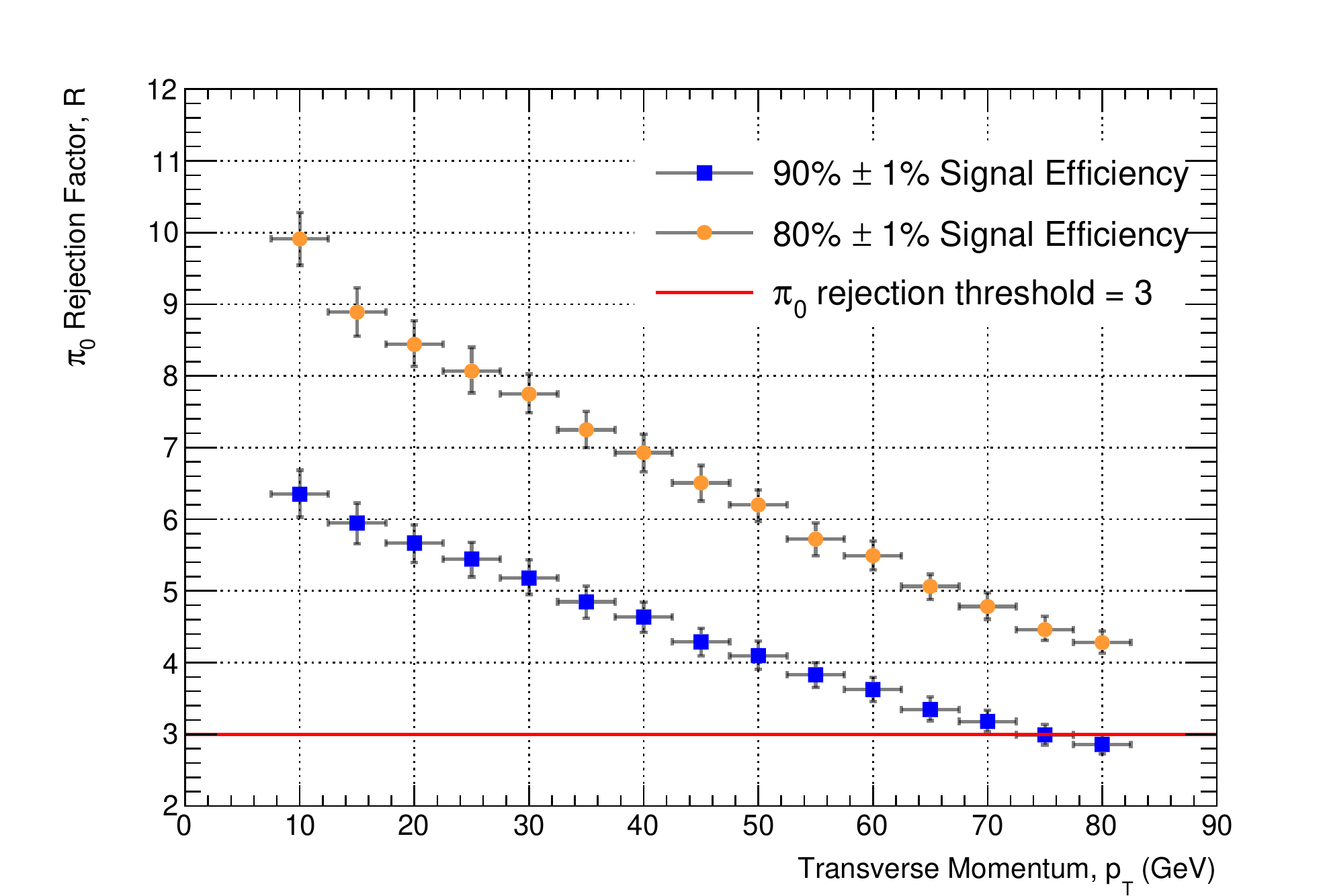}
\caption{$\pi^0$ rejection factor of sample 5 calculated for $p_T$ from 10 to 80\,GeV in steps of 5\,GeV with 80$\%$ and 90$\%$ signal efficiencies.}
\label{fig:id:Rejection}
\end{center}
\end{figure}

The $\pi^0$ rejection for transverse momenta $10 \leq p_T \leq 80$\,GeV at pseudorapidity $\eta = 0$ was investigated using sample 5 for $\pi^0 \to \gamma\gamma$ events. It was found that a $\pi^0$ rejection factor above 3 can be obtained for up to $p_T = 75$\,GeV in this regime, see Figure~\ref{fig:id:Rejection}. This result along with the one shown on Fig.~\ref{fig:id:geometries} suggest that there is reason to investigate finer segmentation in ${\phi}$ further. In this analysis, $\Delta\phi = 0.0045$ for all layers was assumed in geometry $\#5$, but it is likely sufficient to have $\Delta\phi=0.0045$ in one or two layers only while maintaining a similar performance in terms of $\pi^0$ rejection. The best segmentation would have to be optimized with simulations. Compared to the previous study done on the ATLAS EMB, the number of discriminating variables is increased because of multiple layers and higher granularity in the detector. Owing to this the $\pi^0$ rejection factor obtained here is significantly improved. The mean value over all $p_T$ was found to be $R_{\pi} = 3.58 \pm 0.16$ without pile-up which is considerably higher than $R_{\pi} = 2.82 \pm 0.19$ found in ATLAS for $p_T \in [20 - 75]$. It is important to note the difference in the number of events analyzed in each study, the statistics here is much higher ($\mathcal{O}({10^5})$ vs $\mathcal{O}({10^3})$ events) which gives a more accurate but possibly lower rejection value.

%\subsubsubsection{Conclusion}
%
%The ability to distinguish between single photons and neutral pions in the EM calorimeter of the FCC was investigated considering different geometries of the detector. It was found that increasing the segmentation in $\eta$ from $\Delta\eta = 0.01$ to $\Delta\eta = 0.0025$ considerably increases $\pi^0$ rejection at $p_{T} > 30$ GeV and dividing the $2^{nd}$ $\&$ $3^{rd}$ layer of the baseline EM calorimeter layout in half improves this even further. A $\pi^0$ rejection factor $R_{\pi} = 3.58 \pm 0.16$ was obtained for $p_T \in [10 - 80]$ GeV and $|\eta| = 0$ with this setup considering the first 5 layers and also finer segmentation in $\phi$: $\Delta\phi = 0.0045$ in all layers. The rejection factor was found to be above the threshold value of $3$ for $p_{T}$ up to $75$ GeV for $|\eta| = 0$. It may be possible to increase the rejection factor by changing the calculation of discriminating variables for each layer to find the optimal properties. Due to time constraints this was not done in this project. More importantly, a study in realistic conditions that considers also the noise and pileup in the detector needs to be done to fully evaluate the $\pi^0$ rejection ability. 

%%%%%%%%%%%%%%%%%%%%%%%%%%%%%%%%%%%%%%%%%
\subsection{DNN based Particle Identification at $\left<\mu\right>=1000$}
\label{sec:performance:dnn}

Pattern recognition is a strong domain of DNNs, that are trained to discriminate between individual patterns exploiting symmetries of the problem such as translation invariance. Similar techniques are used to separate individual particles from pile-up and identify them using calorimeter information by interpreting the showers as 3 dimensional images, as described in Section~\ref{sec:software:dnn}. Individual DNNs are trained for 0 and 1000 pile-up, each discriminating simultaneously between electrons, photons, muons, charged and neutral pions.

\begin{figure}[ht]
  \centering
        \includegraphics[width=0.48\textwidth]{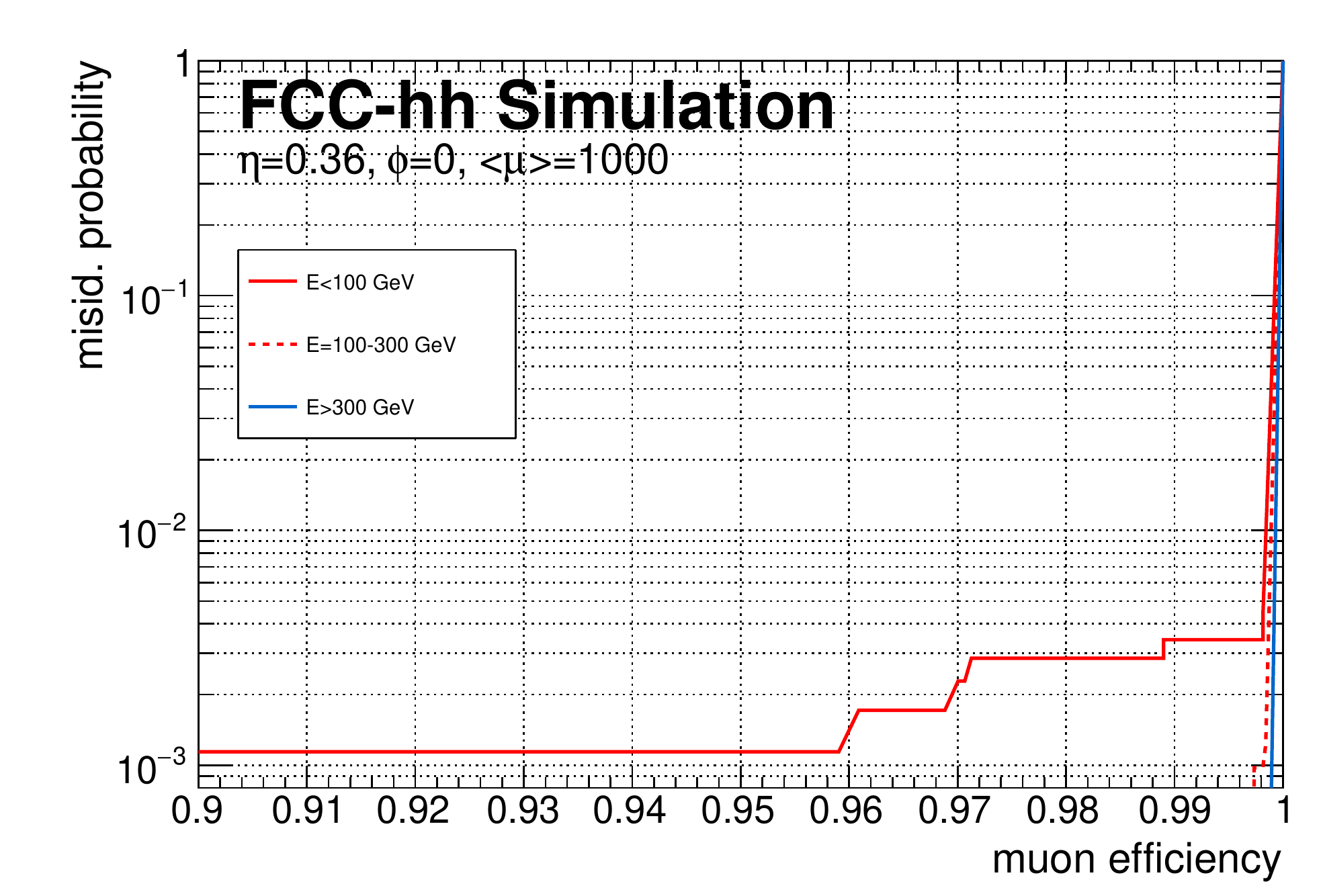}
        \includegraphics[width=0.48\textwidth]{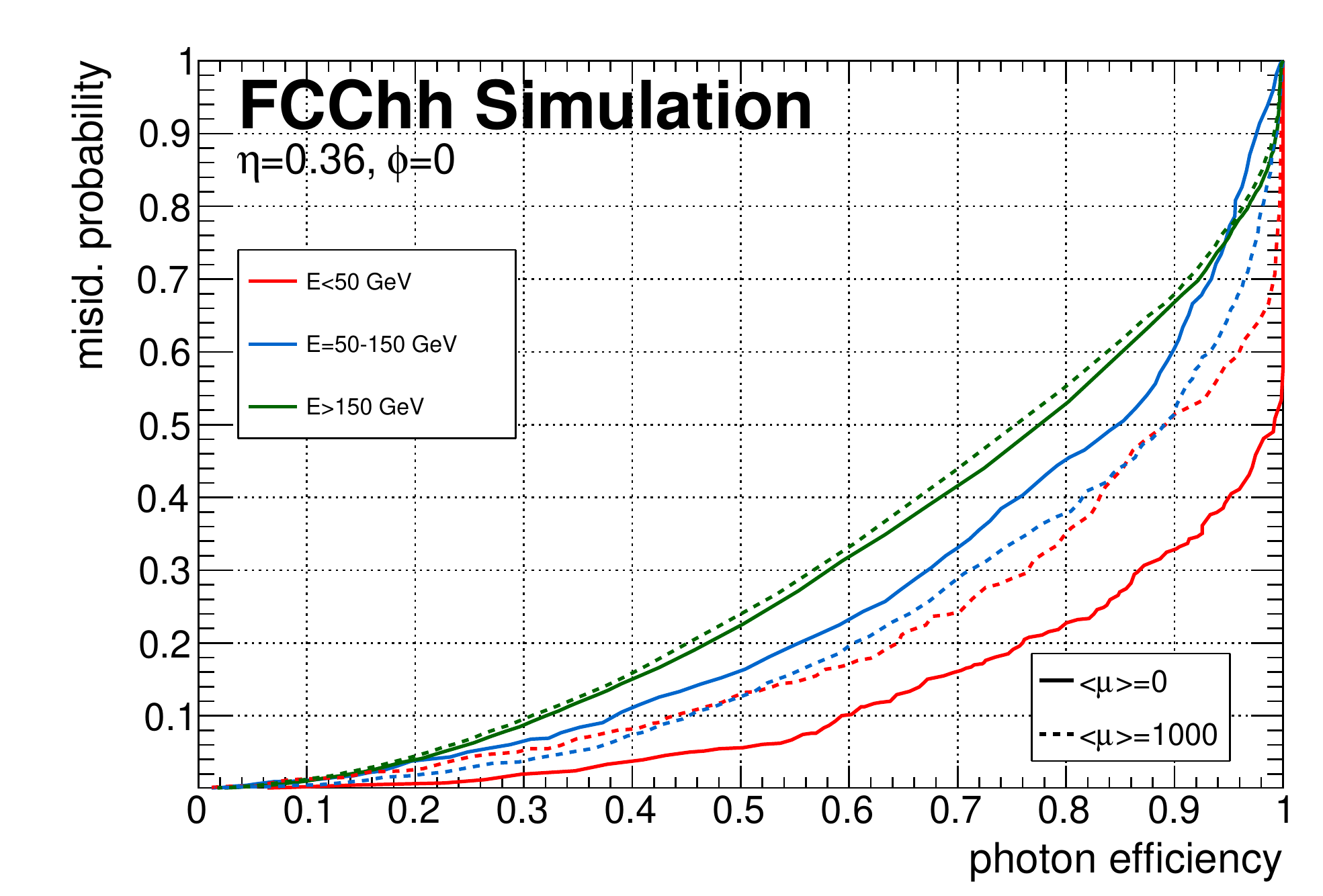} 
  \caption{Discrimination of muons from charged pions (left) and photons (right) from neutral pions for different particle energies and pile-up conditions. 
  \label{fig:performance:id:rocs}      }
\end{figure}

As shown in Fig.~\ref{fig:performance:id:rocs}, the discrimination between muons and charged pions is excellent using calorimeter information only, even with 1000 pile-up. Without pile-up or when discriminating against electromagnetic showers, the performance shown here is further exceeded. The mild energy dependence is less pronounced without pile-up. 
 
An important ingredient for the identification of photons is their separation from neutral pions that decay into two non-prompt photons, as discussed in the previous section, leaving two electromagnetic showers in the calorimeter. With increasing energy, these two showers merge, making the distinction between prompt photons and neutral pions even more challenging. In addition, energy deposits from pile-up make it harder to separate the showers. As shown in Fig.~\ref{fig:performance:id:rocs}, this also applies to the DNN based identification. Photons with energies below 50\,GeV can 
be well disentangled from neutral pions in particular without pile-up. For  pile-up and higher energies, the discrimination power decreases, which could only be mitigated by a higher EMB granularity and requires further investigations. It has to be mentioned that the used simulation samples featured the basic EMB geometry, which has been named as sample 1 in Sec.~\ref{sec:id:samples}. Thus the TMVA study displays better performance without pile-up and the optimised EMB geometry of sample 4 and 5. 

%% file: tex/performance/jets.tex
\newcommand*{\tauto}{\ensuremath{\tau_{2,1}}}

\subsection{Jets}
\label{sec:performance:jets}
\begin{figure}[ht]
  \centering
  \begin{subfigure}[b]{0.49\textwidth}
    \includegraphics[width=\textwidth]{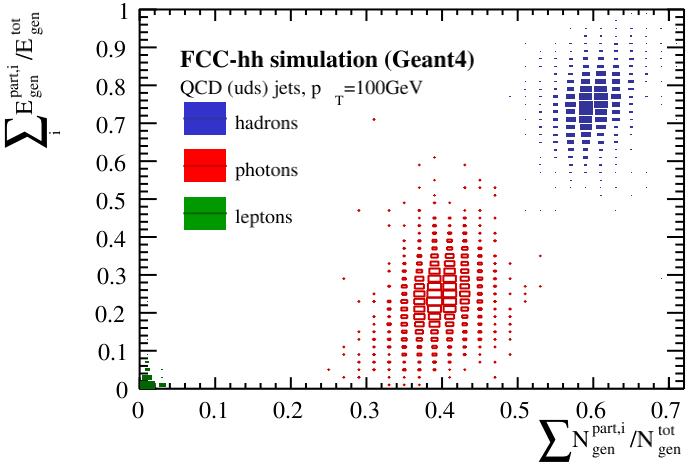}\caption{}
   \label{fig:performance:jets:contents2D}
  \end{subfigure}
  \begin{subfigure}[b]{0.49\textwidth}
	  \includegraphics[width=\textwidth]{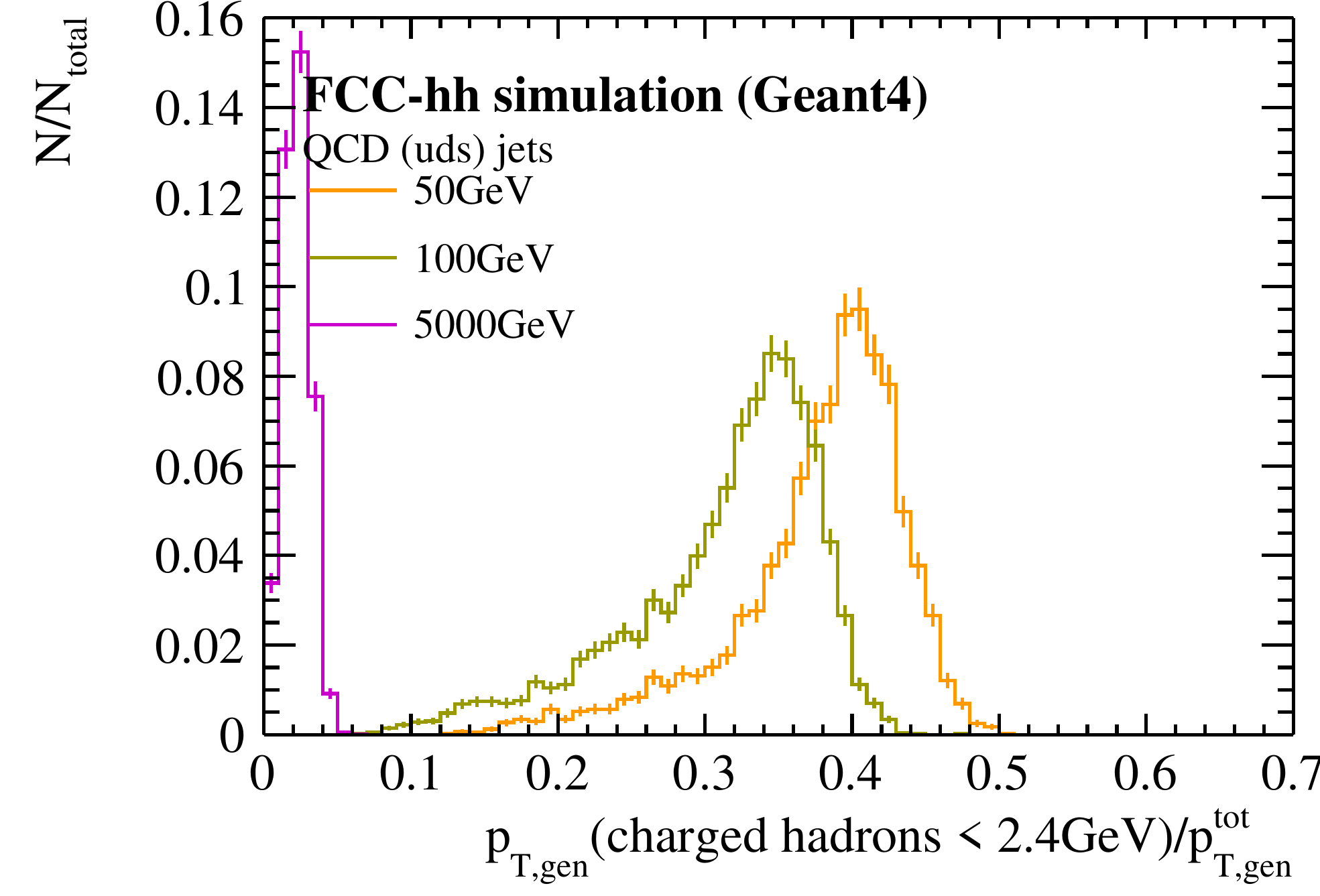}\caption{}
	   \label{fig:performance:jets:contents}
  \end{subfigure}
  \caption{(a) Particle and energy fraction of hadrons, leptons, and photons within QCD jets of p$_{\text{T}}=100$\,GeV. (b) Fraction of charged hadrons with $\pt<2.4$\,GeV in jets of transverse momenta p$_{\text{T, gen}}= 50$, $100$, and $5,000$\,GeV.}
\end{figure}

Hadronizing quarks form particle jets and produce hadronic and electromagentic cascades in the calorimeters. The main components of these jets are photons,  hadrons and marginally leptons that share the primary quarks' momentum. The jet content in the number of particles and particle energy is shown in Fig.~\ref{fig:performance:jets:contents2D}. While only $\thicksim60$\,\% of the particles within a jet are hadrons, they carry around 75\,\% of the total energy. The other 25\,\% of the energy is carried by photons, which are measured in the EM calorimeter~\footnote{It should be noted that the fractional energies dependent on the energy of the jet}. \\
The fraction of the total transverse momentum carried by charged single hadrons within these jets is shown in Fig.~\ref{fig:performance:jets:contents}. These fractions corresponds to the particles in the FCC-hh reference detector that do not reach the calorimeters but instead curl up within the tracking system due to the 4\,T magnetic field. The minimum p$^{min}_{\text{T}}$ necessary is estimated to
\begin{equation}
	\text{p}^{\text{min}}_{\text{T}}=0.3 \cdot 4\,\text{T} \cdot r_{0},
\end{equation}
with $r_{0}$ corresponding to the radial distance of the second barrel ECal layer from the interaction point of 1.97\,m at $\eta=0$.\\ 
This corresponds to a minimum of p$_{T}>2.4$\,GeV to reach the calorimeter in the presence of a 4\,T magnetic field. 
Thus the jet performance in the presence of a 4\,T magnetic field is expected to be seriously impacted by this effect if the calorimeter system alone is used, without any tracker information. Therefore, the performance of the calorimeter system is shown in the following without the magnetic field. Future studies are planned to combine the tracking information with the calorimeter clusters using particle flow algorithms.  

The jet reconstruction of the FCC-hh calorimeter system is based on the standard anti-k$_{\text{T}}$ algorithm described in Sec.~\ref{sec:software:reco:jets}, which uses topological clusters as input. 

\subsubsection{Jet energy resolution and energy scale}
\begin{figure}[ht]
  \begin{center}
  \begin{subfigure}[b]{0.49\textwidth}
    \includegraphics[width=\textwidth]{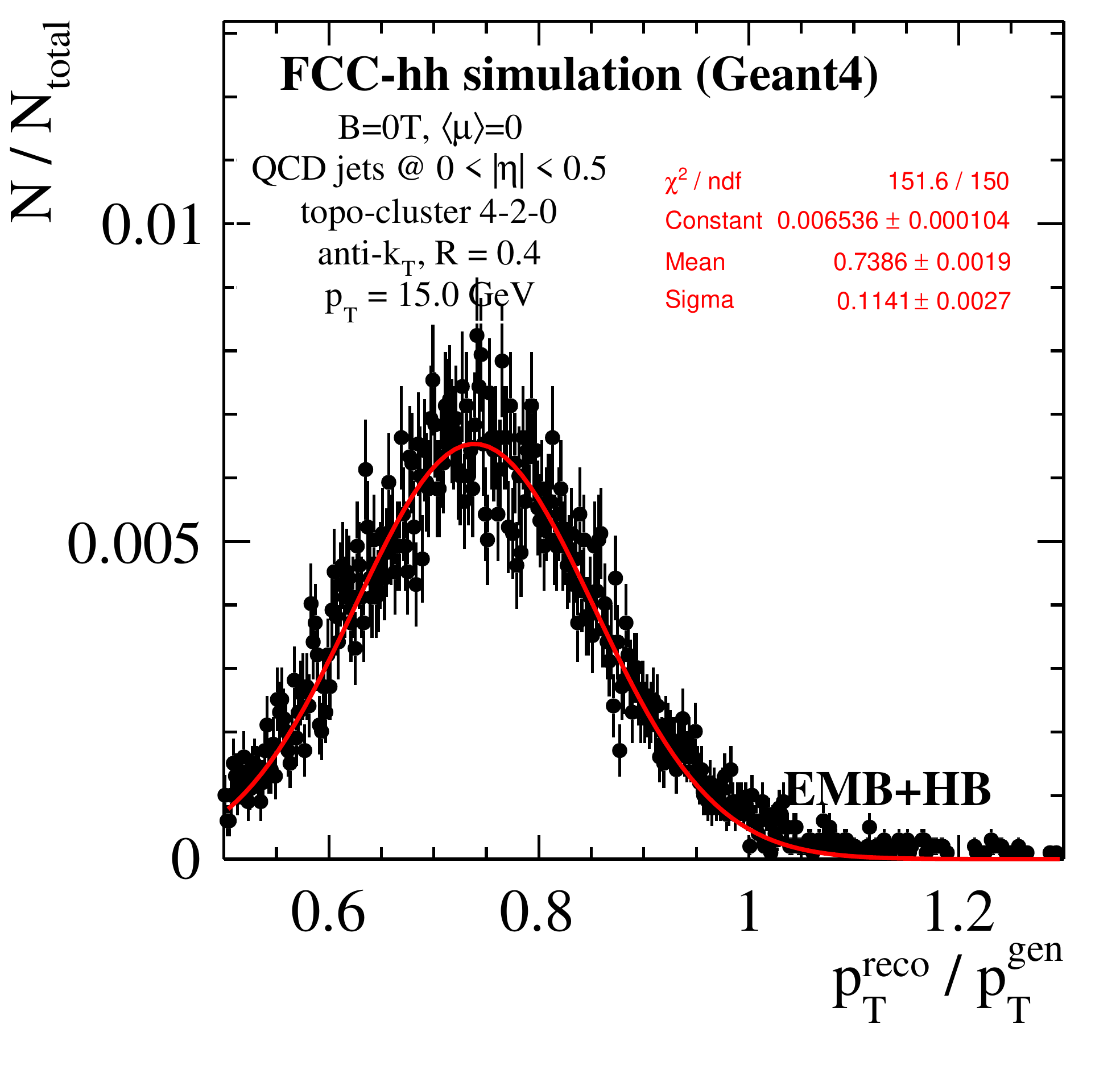}\caption{}
  	\label{fig:performance:jets:ptDist_15}
  \end{subfigure}
    \begin{subfigure}[b]{0.49\textwidth}
    \includegraphics[width=\textwidth]{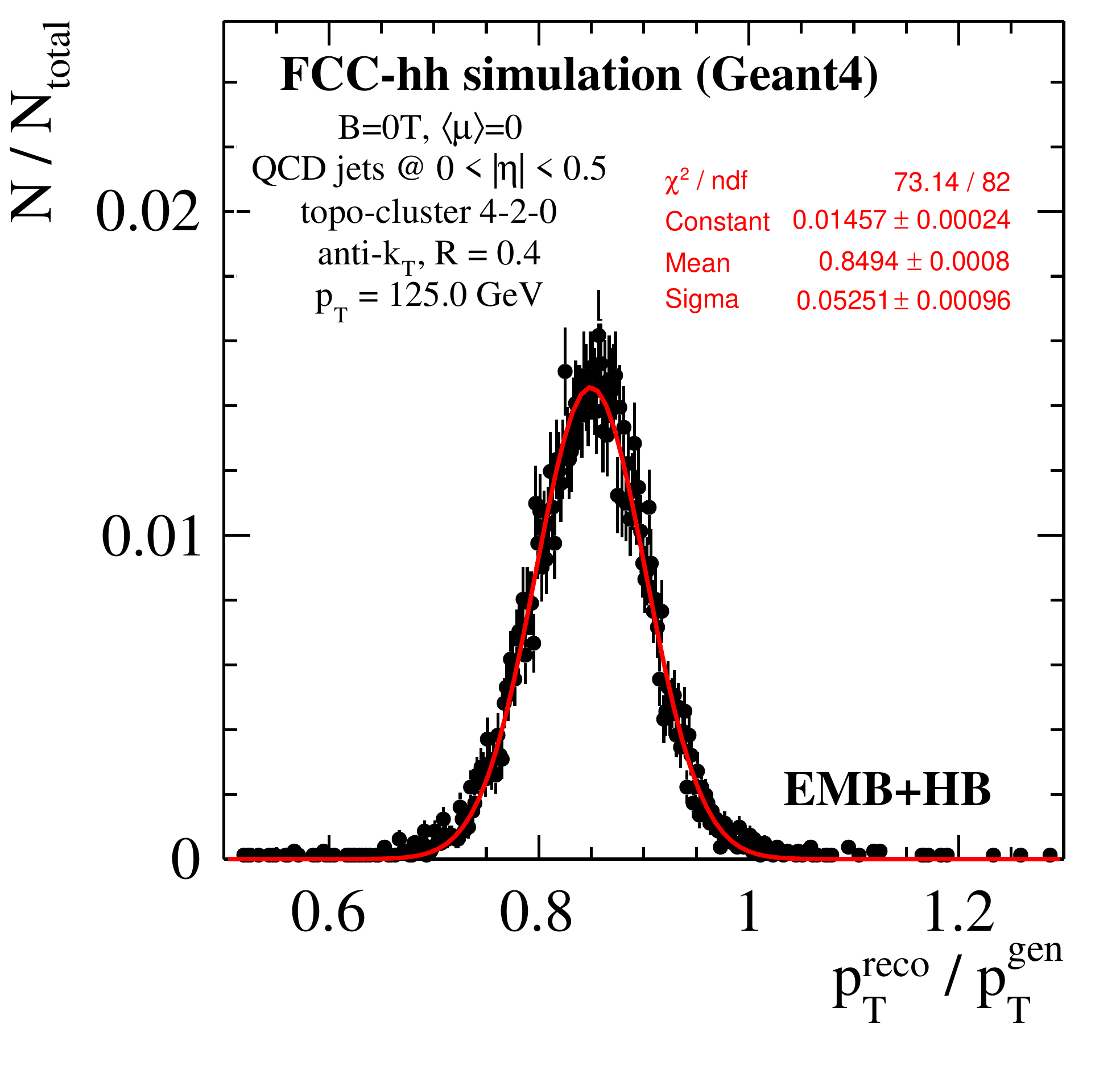}\caption{}
  	\label{fig:performance:jets:ptDist_125}
  \end{subfigure}
\end{center}
  \caption{Jet $\pt$ distributions for (a) 15 and (b) 125\,GeV jets. Reconstructed from topo-cluster in 4-2-0 mode after calibration, using the jet clustering algorithm for $\left<\mu\right>=0$. }
\end{figure}

In the following, the jet $\text{p}_{\text{T}}$ resolution, in the absence of magnetic field, is shown for di-jet events of up, down and strange quarks with transverse momenta of 20\,GeV to 10\,TeV. 
The jets are measured in the FCC-hh barrel calorimeters using so-called "calibrated" topo-clusters that are calibrated to the hadronic scale if they contain cells in the HB, or cells in both the EMB and HB. Additionally, the calibration corrects for the lost energies within the LAr cryostat between the EM calorimeter and the HCAL, see more details in Sec~\ref{sec:performance:hadronic:topoClusterCalibration}. To determine the performance, additionally to the \textit{rec}-jets built from clusters, the jet reconstruction is also run on stable, final-state, generated particles which represent the so-called \textit{truth/generated}-jets. For the determination of the resolution, the reconstructed and truth jets are matched within a distance of $R<0.3$. In case of the reconstructed jets, only the two leading jets with the highest transverse momentum are selected and considered for the $\text{p}^{\text{rec}}_{\text{T}}/\text{p}^{\text{gen}}_{\text{T}}$ distributions. The momentum resolution is determined in 16 $\text{p}_{\text{T}}$ bins, and an example of one distribution and the corresponding Gaussian fit within $\pm2\,\sigma$ is shown in Figure~\ref{fig:performance:jets:ptDist_15} and ~\ref{fig:performance:jets:ptDist_125}. 

\begin{figure}[ht]
  \begin{center}
  \begin{subfigure}[b]{0.49\textwidth}
    \includegraphics[width=\textwidth]{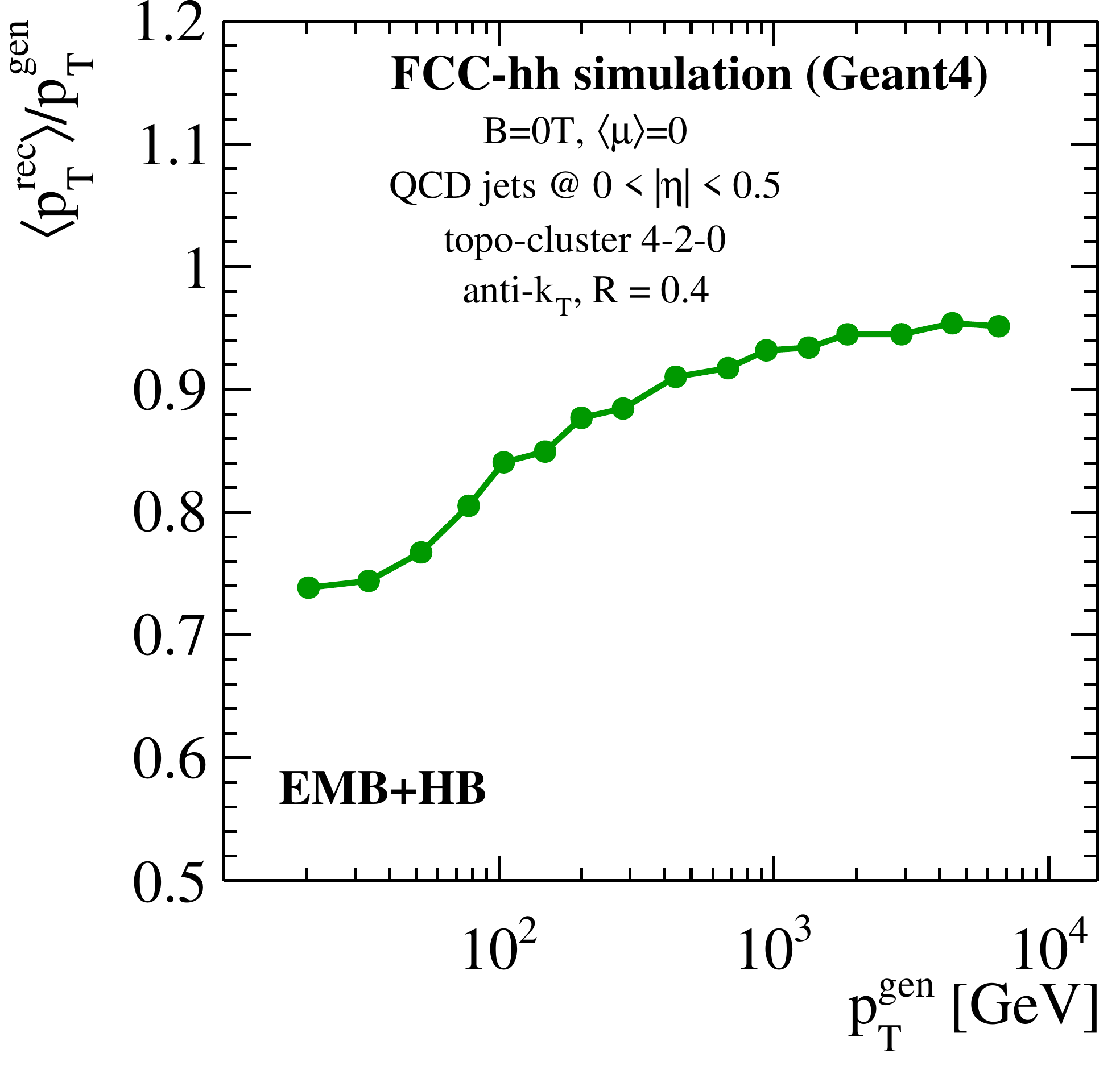}\caption{}
  	\label{fig:performance:jets:resp}
  \end{subfigure}
  \begin{subfigure}[b]{0.49\textwidth}
	\includegraphics[width=\textwidth]{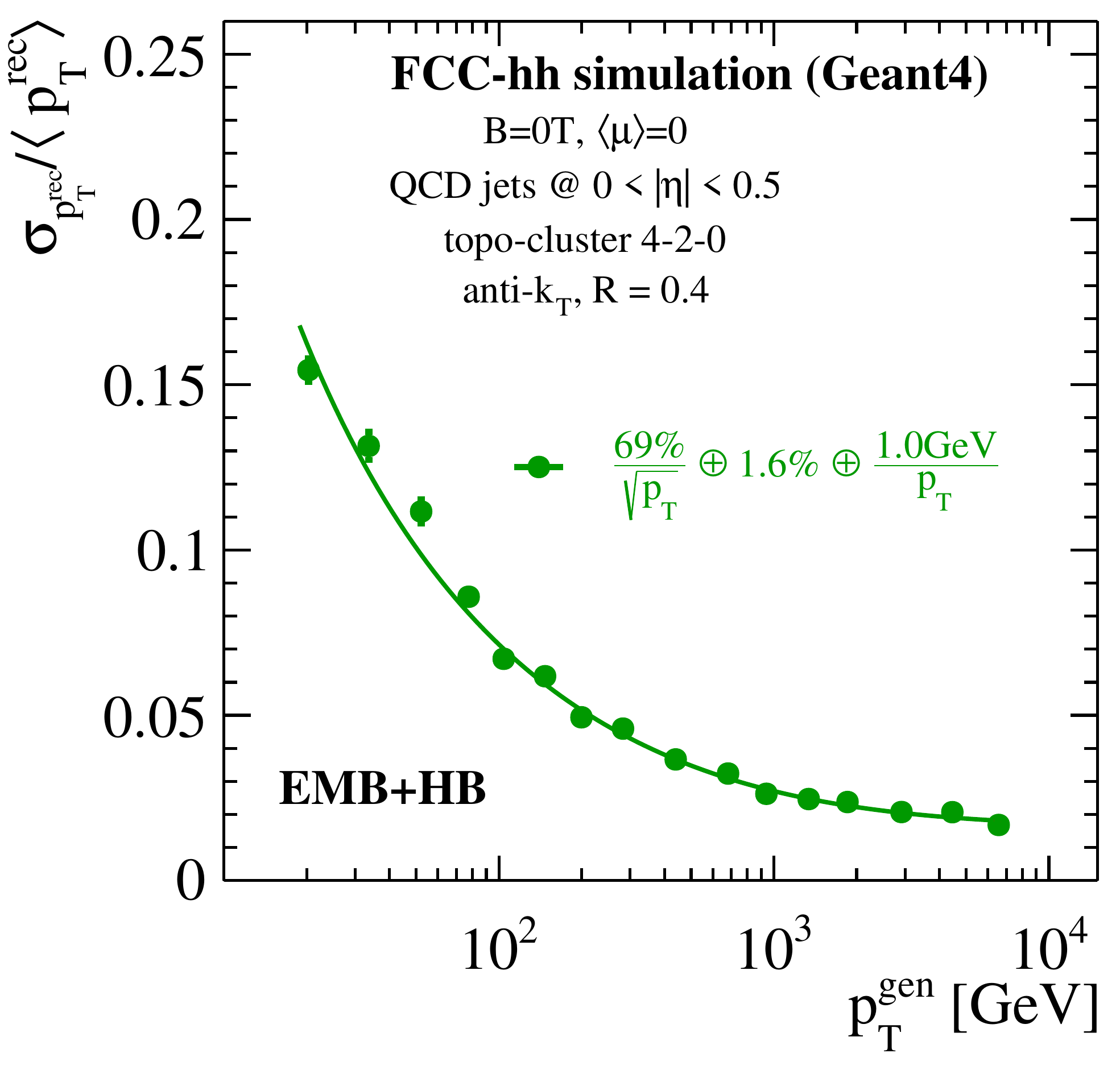}\caption{}
  	\label{fig:performance:jets:reso}
  \end{subfigure}
  \end{center}
  \caption{Jet $\text{p}_{\text{T}}$ (a) response and (b) resolution after topo-cluster reconstruction in 4-2-0 mode after calibration, using the jet clustering algorithm for $\left<\mu\right>=0$. }
\end{figure}

Without B-field, the FCC-hh barrel calorimeters alone achieve a jet energy resolution with a constant term $<2\,\%$, see Fig.~\ref{fig:performance:jets:reso}. Further development of reconstruction techniques like particle-flow algorithms, are expected to improve the jet energy measurement in the medium and low $p_{\text{T}}$ range by using tracking information for jet constituents. Due to the large spread of particles for the case of $B=4$\,T, a combined reconstruction of jets with the tracker is compulsory. Even without B-field a non-linearity of the mean transverse momentum of up to 25\,\% at low $p_{\text{T}}$ remains after the simplistic topo-cluster energy reconstruction as shown in Fig.~\ref{fig:performance:jets:resp}.

\begin{figure}[ht]
  \centering
  \begin{subfigure}[b]{0.49\textwidth}
    \includegraphics[width=\textwidth]{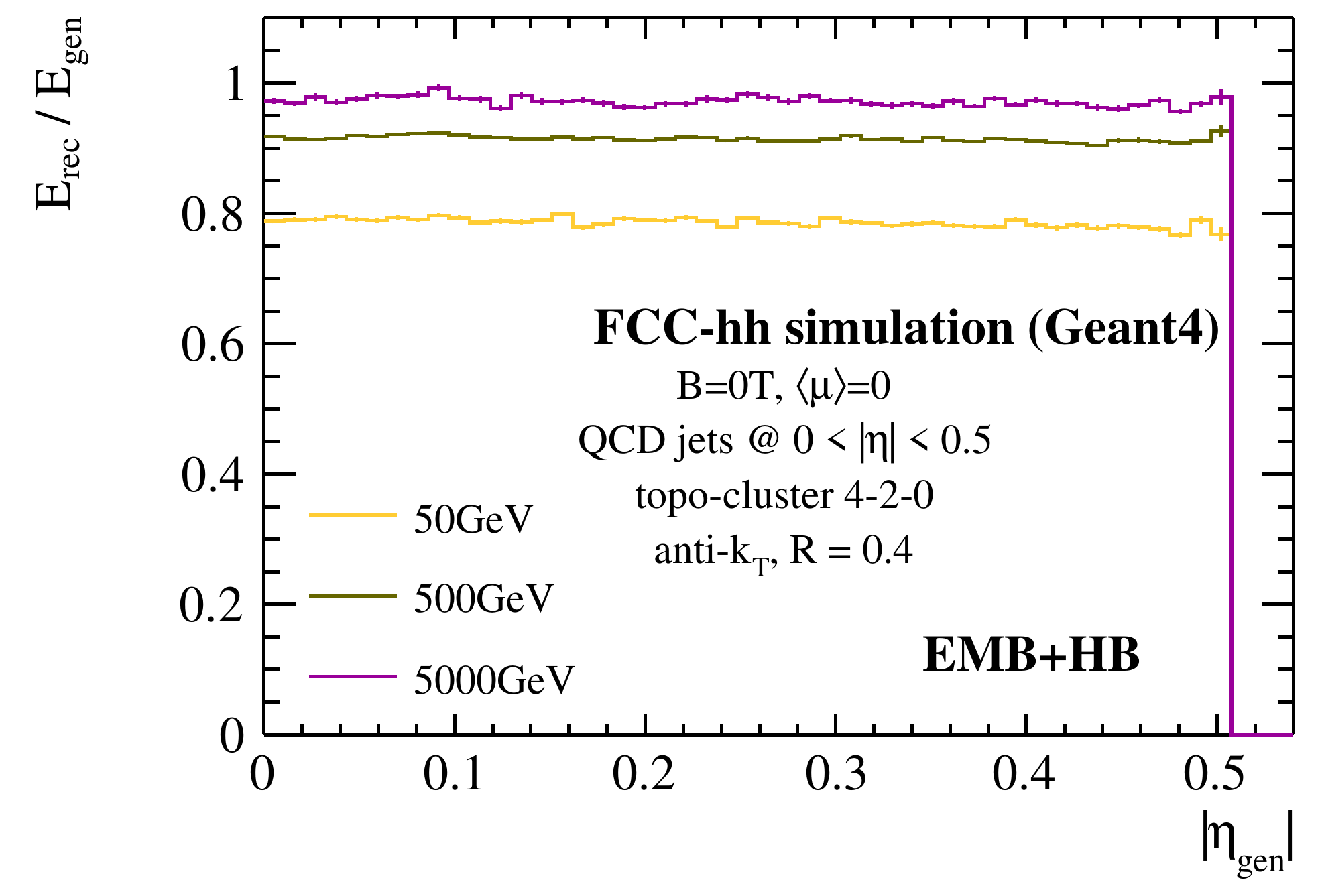}\caption{}\label{fig:performance:jets:resp_eta}
   \end{subfigure}
  \begin{subfigure}[b]{0.49\textwidth}
   \includegraphics[width=\textwidth]{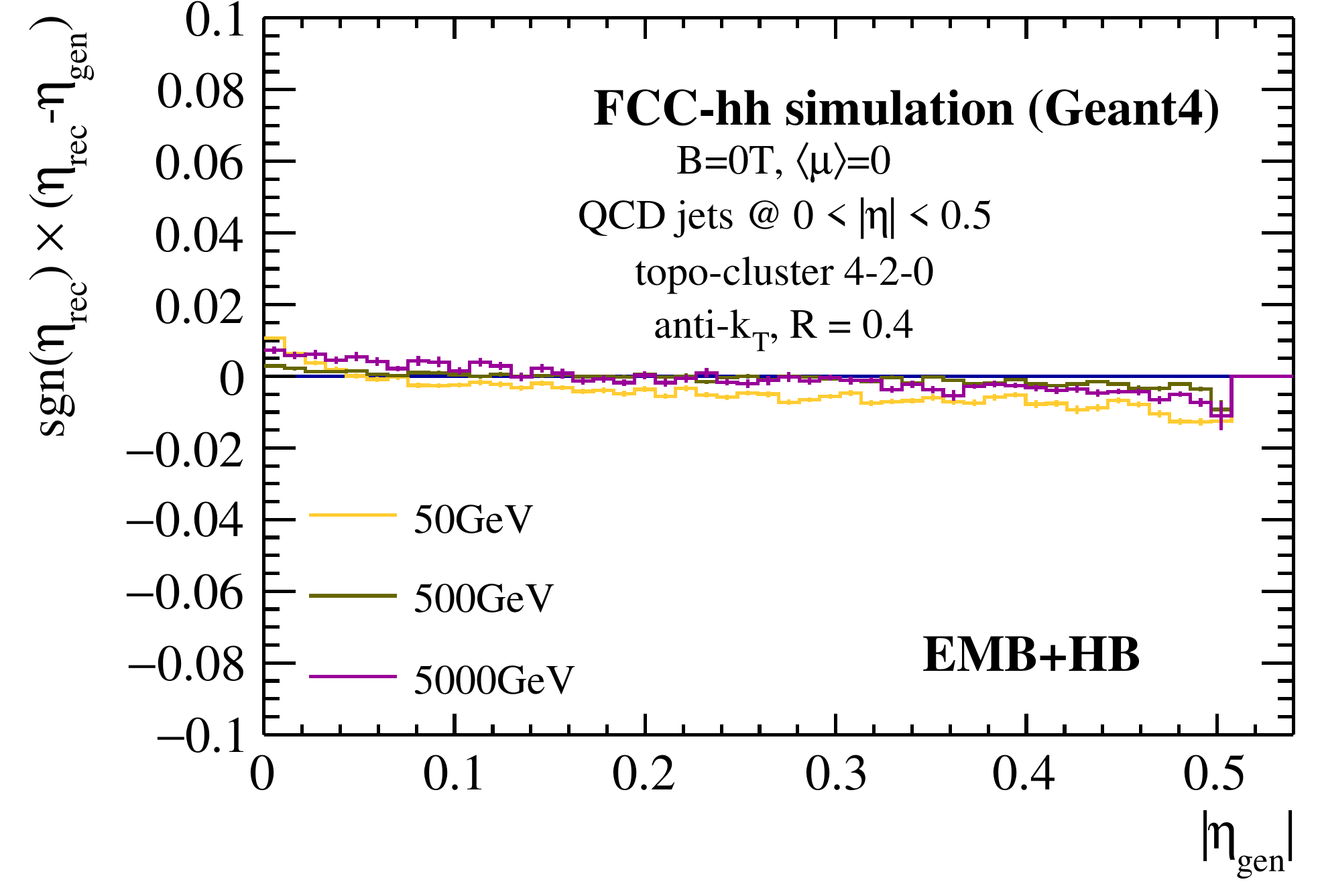}\caption{}\label{fig:performance:jets:gravity_eta}
  \end{subfigure}
  \caption{(a) Average energy response as a function of the pseudo-rapidity of the jets with p$_{\text{T}}$ 50, 500, and 5000\,GeV. (b) The signed difference of $\eta_{gen}$ and $\eta_{rec}$. }
\end{figure}

The average energy response to jets is shown in Fig.~\ref{fig:performance:jets:resp_eta}, and shows a constant response in pseudo-rapidity for the central barrel of the FCC-hh calorimeters. In a next step, a method to correct the jet energy scale, as done for the ATLAS experiment~\cite{Aaboud:2017jcu}, could be applied using a numerical inversion procedure similarly to the second step of the benchmark method. 
The center of gravity of the reconstructed jets has been tested along $\eta_{gen}$, see Fig.~\ref{fig:performance:jets:gravity_eta}, and do show a slight bias towards smaller $\eta_{rec}$ for increasing pseudo-rapidity. However, this effect is on the sub-percent level. 

\subsubsection{Angular resolutions}
First tests of the chosen angular segmentation of the EM and hadronic calorimeters have started. One adjustment made for optimising the separation of photons and $\pi^{0}$ by a highly segmented 2nd ECal layer (see Sec.~\ref{sec:performance:idMVA}). The impact of the full $\eta$ granularity of the HB has been tested with single pions in Sec.~\ref{sec:performance:hadronic:angular}, and an improvement in the angular resolution of up to 15\,\% was determined. However, this comes with the cost of a 6 times higher number of readout channels.

\begin{figure}[ht]
  \centering
  \begin{subfigure}[b]{0.49\textwidth}
    	\includegraphics[width=\textwidth]{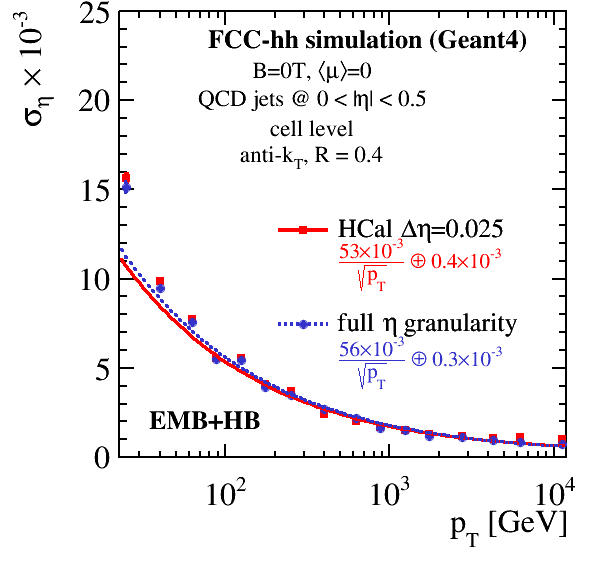}\caption{}
   \label{fig:performance:jets:resoEta}
  \end{subfigure}
  \begin{subfigure}[b]{0.49\textwidth}
    	\includegraphics[width=\textwidth]{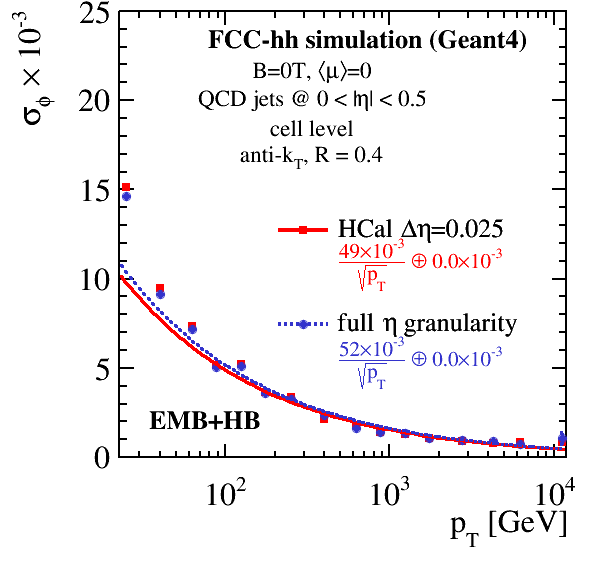}\caption{}
	 \label{fig:performance:jets:resoPhi}
  \end{subfigure}
  \caption{Angular resolution in pseudo-rapidity $\eta$ and azimuthal angle $\phi$ for jets up to $|\eta|<0.5$. The red curve show the results on cell level with the HB tiles merged in $\Delta\eta=0.025$, and the blue points/curves show the resolution obtained with ultimate HB granularity of $\Delta\eta<0.006$. }
\end{figure}

The $\eta$ granularity of the HB has been tested on di-jet events and the precision on which the jet angle can be measured. The results are presented in Fig.~\ref{fig:performance:jets:resoEta} and \ref{fig:performance:jets:resoPhi} for the pseudo-rapidity and the azimuthal angle, respectively. The determined precision of $<0.01$ in both $\eta$ and $\phi$, lies well below the HCAL granularity of $0.025$ for jet p$_{\text{T}}$ larger than 40\,GeV. This indicates that the intrinsic calorimeter segmentations is still being exceeded by the combination of cells. However, a further increase of the $\eta$ granularity of the HB is not improving the jet angular resolution significantly which supports the choice for the reference detector design. Studies of the cell granularity of the hadronic calorimeter, optimised for jet sub-structure variable, have shown similar results~\cite{Yeh:2019xbj} and support the chosen granularity of $\Delta\eta\times\Delta\phi=0.025\times0.025$ for the hadronic calorimeter. Including more sophisticated jet reconstruction techniques like particle flow algorithms or DNNs could be however more sensitive to the granularity and thus point into another direction. 

\subsubsection{Outlook}
\begin{figure}[ht]
  \centering
  \begin{subfigure}[b]{0.295\textwidth}
    \includegraphics[width=\textwidth]{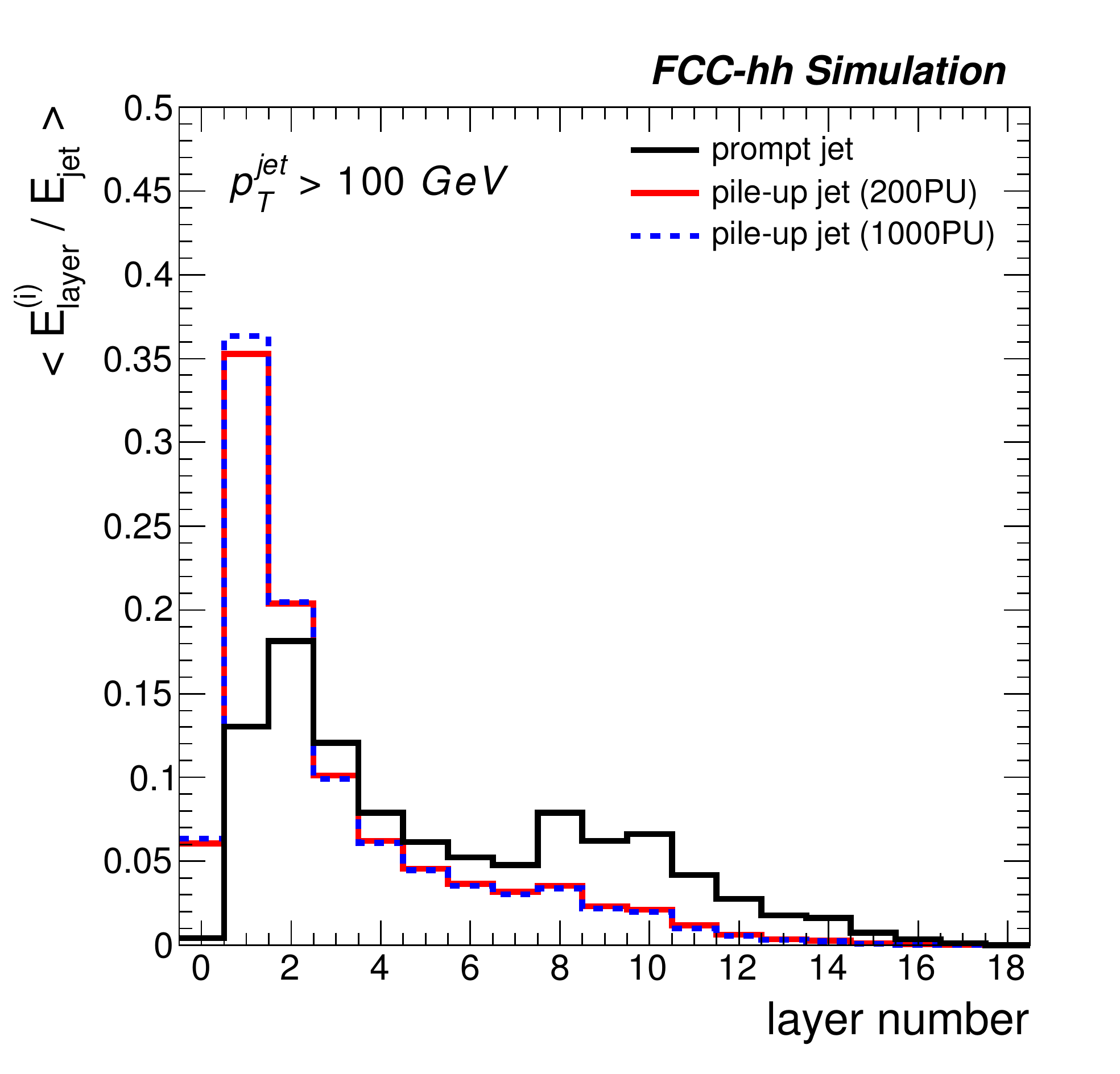}
  \end{subfigure}
  \begin{subfigure}[b]{0.295\textwidth}
    \includegraphics[width=\textwidth]{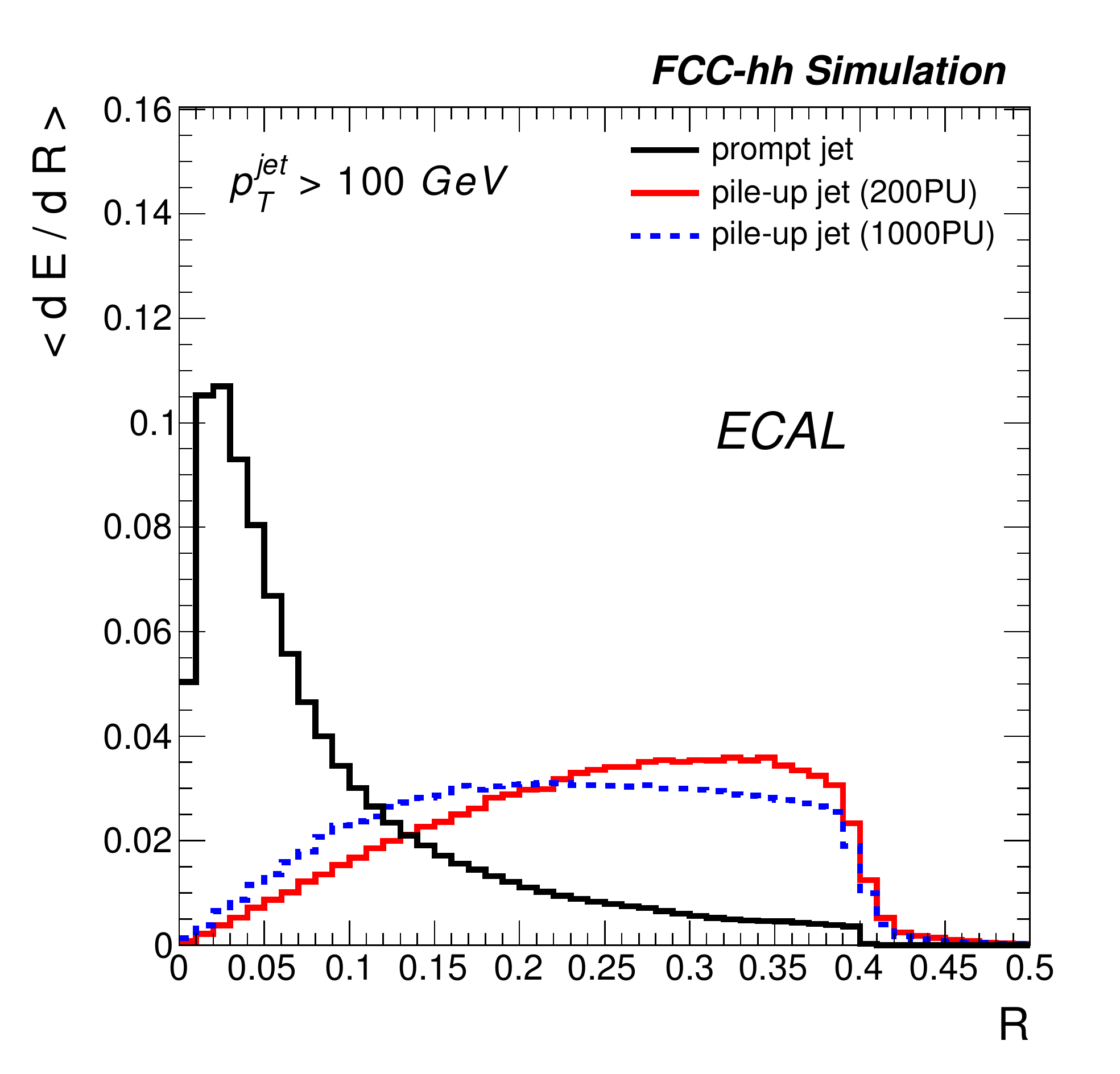}
  \end{subfigure}
  \begin{subfigure}[b]{0.295\textwidth}
    \includegraphics[width=\textwidth]{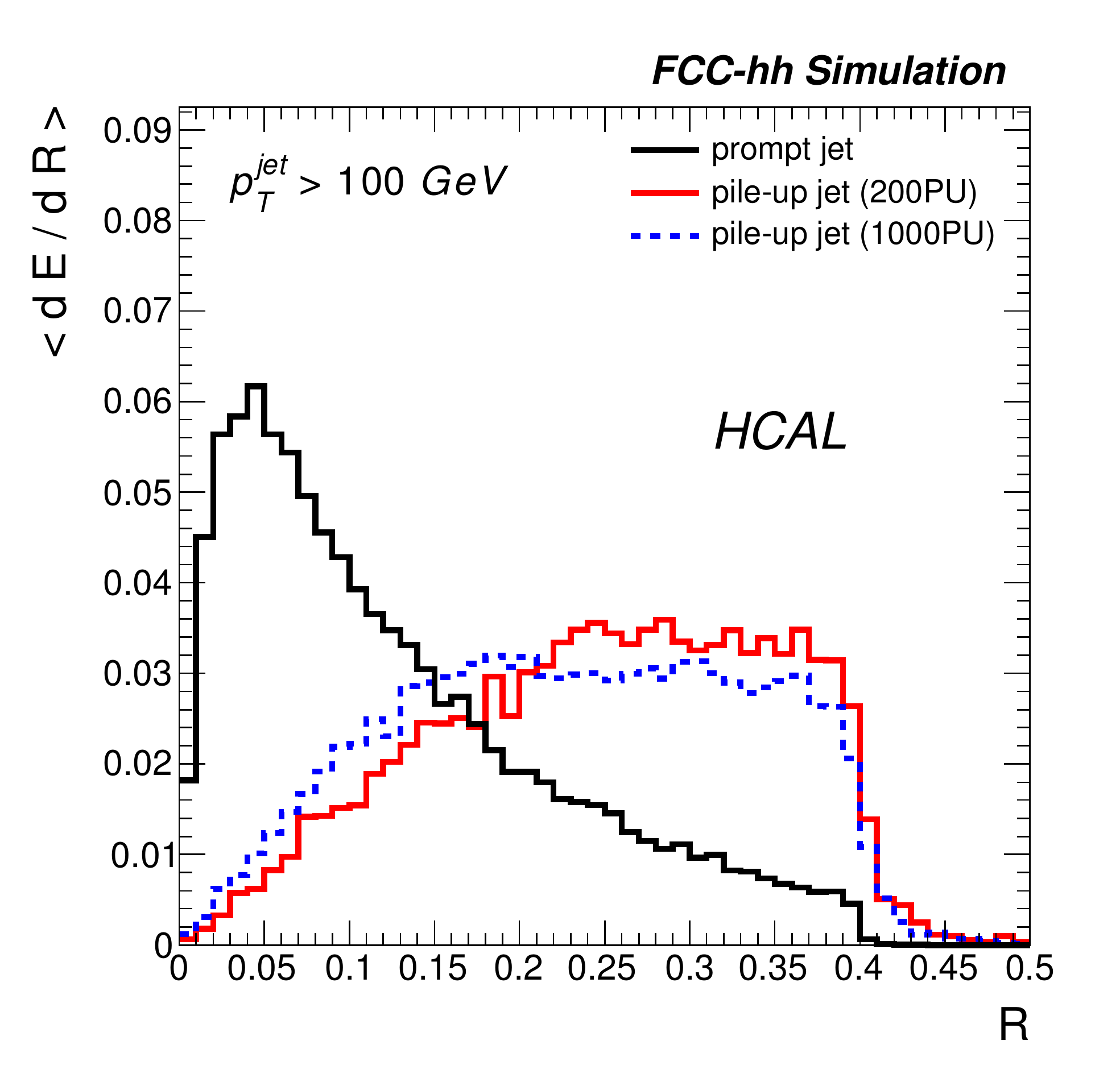}
  \end{subfigure}
  \caption{Left: Transverse energy deposited as a function of calorimeter layer by QCD jets and by jets reconstructed from pile-up. Centre/Right: Radial profile of QCD jets and jets made from pile-up in the EMB (centre) and in the HB (right). %\protect\todo[inline]{CN: Add Geant4 label}
\label{fig:performance:jets:puid} }
\end{figure}

\subsubsubsection{Pile-up jet identification}
Pile-up interactions can affect the global event reconstruction in many ways. In extreme pile-up regimes (PU$>200$), random associations of low energy showers can fake prompt jets, especially in the forward region of the detector where large amounts of energy are deposited. This can have large effects on measurements of processes that feature the presence of forward jets such as vector boson fusion Higgs production. Pile-up jets can be disentangled from prompt jets by exploiting the difference in the longitudinal and transverse energy profile. In Fig.~\ref{fig:performance:jets:puid} (left) the energy of the jet per layer normalized to the total jet energy is shown as a function of the layer number. It can be seen clearly that a large fraction of the energy is deposited in the first layers for pile-up jets. The explanation is that pile-up jets feature a uniform soft energy distribution among its consituents that penetrate few layers of the calorimeter, as opposed to a prompt QCD jet that is typically made up of fewer and harder long lived hadrons. Similarly the transverse energy profile, integrated over all layers of the ECAL and HCAL subdetectors can be seen respectively in Fig.~\ref{fig:performance:jets:puid} (center and right). Prompt jets concentrate their deposited energy on a well-defined center whereas pile-up jets feature a uniform diffuse transverse energy profile. Having at disposal such handles, provided by a high longitudinal and tranverse segmentation will clearly improve the identification of pile-up jets. Finally we note that an optimal pile-up rejection can be performed with the so-called particle flow approach~\cite{Sirunyan:2017ulk} that aims at combining optimally calorimetric and tracking information into single particle candidates. Since particle-flow does rely on extrapolating and matching reconstructed tracks to calorimeter deposits, it is clear that in order to achieve an optimal performance with such an approach the highest possible transverse and longitudinal granularity should be aimed for.

\subsubsubsection{Boosted objects, substructure}

\begin{figure}[ht]
  \centering
  \begin{subfigure}[b]{0.45\textwidth}
    	\includegraphics[width=\textwidth]{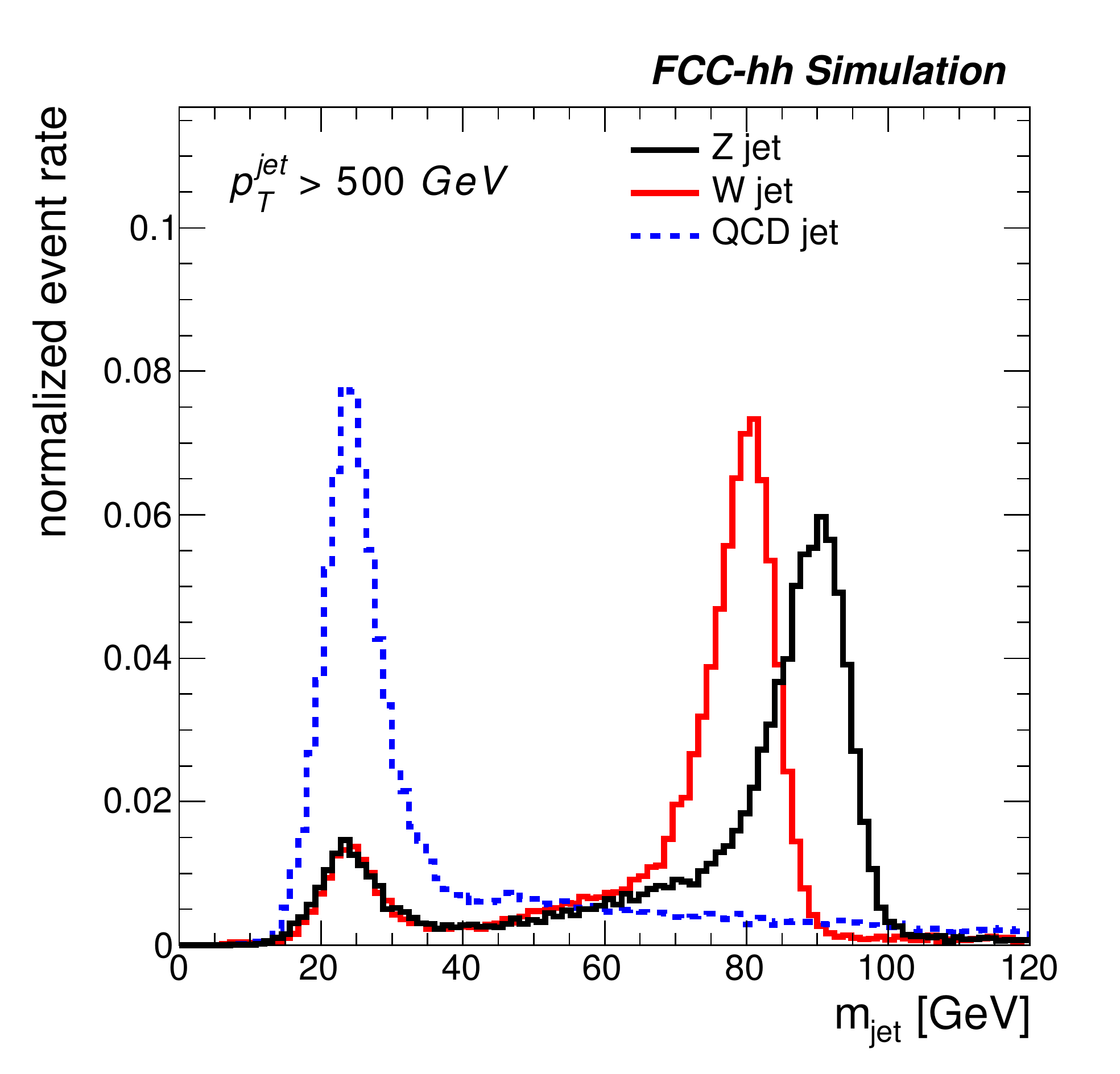}
  \end{subfigure}
  \begin{subfigure}[b]{0.45\textwidth}
    \includegraphics[width=\textwidth]{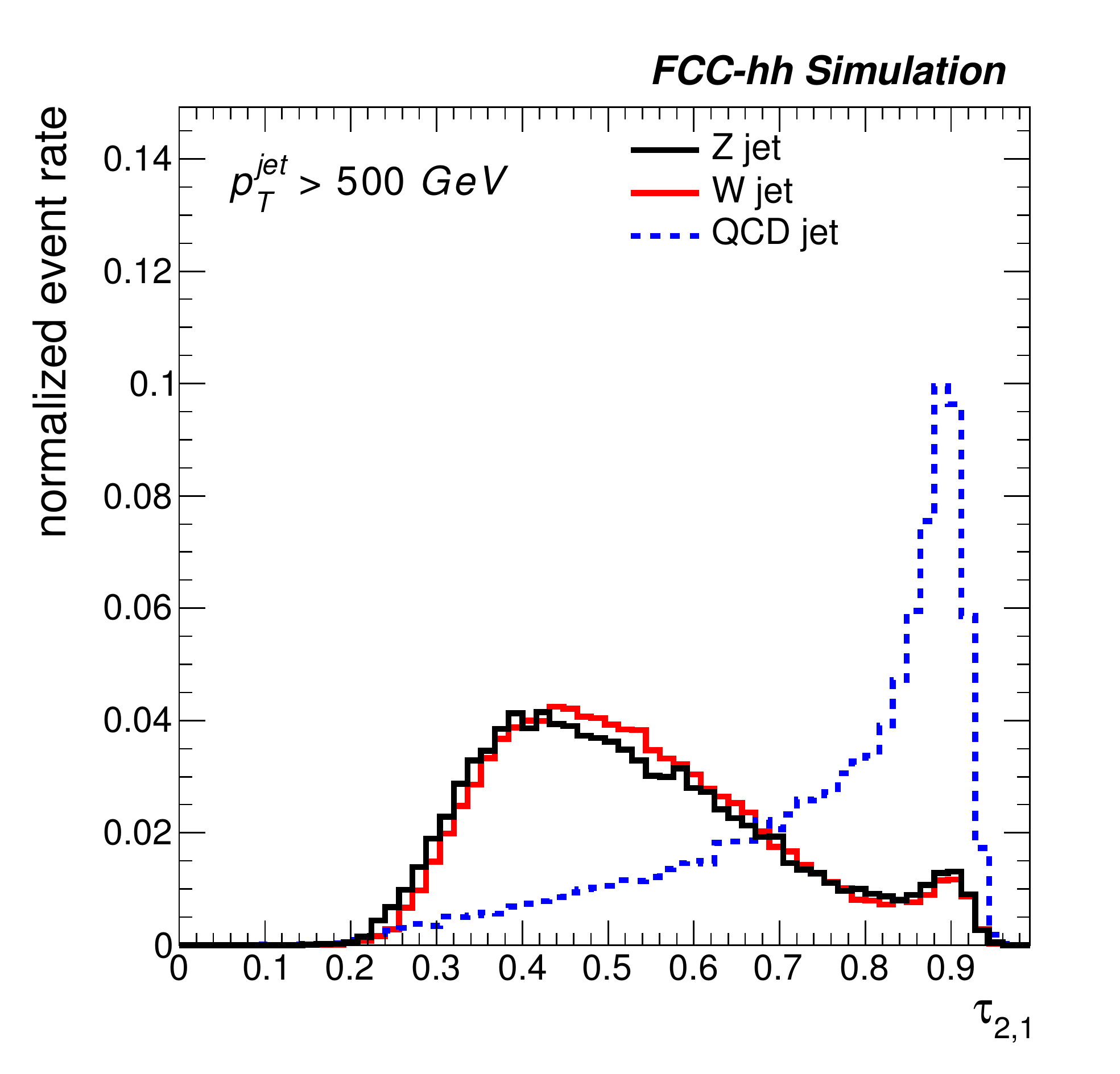}
  \end{subfigure}
  \caption{Distributions of the jet mass (left) and \tauto\ (right) for boosted W, Z and QCD jets with $\pt>500$~GeV.%\protect\todo[inline]{CN: Add Geant4 label.} 	
  \label{fig:performance:jets:boosted}}
\end{figure}

The impact of a high lateral segmentation can be seen on observables that are sensitive to the angular separation in jets. The mass of a highly boosted jet depends both on the energy and the angular separation of hadrons and can be used for such an investigation. Another useful variable is the N-subjettiness ratio $\tau_{2,1}$. A detailed description of this complex observable can be found here~\cite{Thaler:2010tr}. We simply point out that this variable is also built from the energy-momentum vector of the jets constituents. It is expected to peak at values close to 0 if the jet features a 2-prong structure (such as W, Z of Higgs jets) and close to 1 if the jet substructure is one prong-like. Jets are reconstructed with anti-kT algorithm~\cite{Cacciari:2008gp} with R=0.2 directly from calorimeter hits. No magnetic field was applied in the simulation implying that charged and neutral hadrons are treated equally and no pile-up was assumed. In Figure~\ref{fig:performance:jets:boosted} (left) we show the reconstructed jet mass for W, Z and QCD with $p_T=500$~GeV. A good separation between QCD and V=W,Z jets can be observed, as well as decent discrimination between W and Z bosons. In Figure~\ref{fig:performance:jets:boosted} (right) we show the $\tau_{2,1}$ variable. Although W and Z jets can hardly be discriminated with $\tau_{2,1}$ (both feature a two-prong structure), it is clear that this observable provides a handle versus background QCD jets.
It should be noted that this preliminary study does not make use of tracking, that is expected to provide additional angular separation power for jets, especially in combination with calorimetric information using the particle-flow approach. In such paradigm, high (transverse) granularity is indeed crucial in order to uniquely assign tracks to calorimeter deposits.
%Jet imaging: \\
%- Ideas of using timing information (4D clustering), Dependence on segmentation.

%% file: tex/alternative/layout.tex
\subsection{Silicon Tungsten Calorimeter}
\label{sec::layout::alt}

In addition to the LAr baseline, the feasibility of a sampling electromagnetic calorimeter using silicon as the sensitive layer and tungsten as the absorber (SiW) has been studied for the FCC-hh. Two distinct readout modes have been investigated: a conventional analogue readout Si calorimeter such as those proposed for the ILD electromagnetic calorimeter within CALICE~\cite{Anduze:2008hq}, and the CMS HGCAL~\cite{Collaboration:2293646}; and an ultra granular digital electromagnetic calorimeter that counts the number of particles in a shower, rather than the energy they deposit, first investigated by the SPiDeR collaboration~\cite{SPIDER} and later adopted as a potential technology for the ALICE FoCal~\cite{Inaba:2016mpb}. 

\begin{figure}[h]
  \centering
    \includegraphics[width=\textwidth]{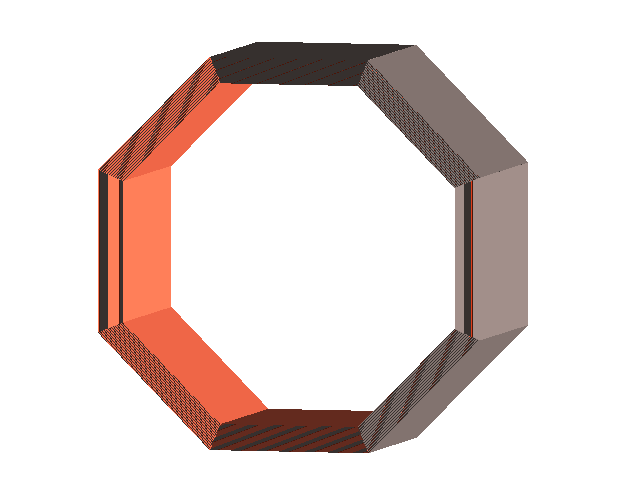}
  \caption{The cross section of the silicon electromagnetic barrel calorimeter. The centre of each of the eight staves was fixed to the inner radius of the baseline design to ensure the calorimeter remained within the required envelope. }
  \label{fig:layout:alt:barrel}
\end{figure}

%\subsubsection{Barrel}
%%~\cite{PERNEGGER201992}->~\cite{Munker:2672506}
Current developments in Depleted MAPS have demonstrated a radiation tolerance of at least 10$^{15}$\,n$_{eq}$/cm$^2$~\cite{Munker:2672506} and is assumed that the required radiation tolerance in the ECAL barrel (as stated in Tab.~\ref{tab:layout:dimensions}) will be achieved on the timescales of the project. However, the radiation tolerance required in the forward regions at the FCC-hh are an order of magnitude higher, and as such, we do not envisage the use of silicon in these regions of the calorimeter. The silicon based electromagnetic calorimeter barrel consists of five modules in $z$, each with eight staves arranged in an octagonal configuration as shown in Fig~\ref{fig:layout:alt:barrel}. Each stave is segmented longitudinally into 50 layers of alternating silicon and 0.6\,$X_0$ of absorbing material to achieve the necessary calorimeter depth of 30\,$X_0$. The impact on the number of layers, cell size, and choice of absorbing material are detailed here.%~\cite{}\todo{AZ:missing ref}. 
Tungsten, with a radiation length of 3.5\,mm, was chosen as the absorbing material for two main reasons: firstly, its Molie$\mathrm{\grave{r}}$e radius of 9.3\,mm leads to compact electromagnetic showers and allows better separation of nearby showers and pile-up events; and secondly, the calorimeter itself can be much more compact, reducing the size and cost of all the detector components outside of it. An air gap of 3\,mm for services was included between each alternating pair of silicon and tungsten. The silicon electromagnetic calorimeter operates at room temperature and as such, there is no need for a cryostat that would introduce additional passive material in front of the calorimeter systems.

\subsubsection{Analogue Readout}

For the analogue variant of the SiW, a layer of 300\,$\mathrm{\mu{m}}$ silicon is assumed. This value is consistent with the pad detectors used for CALICE and the CMS HGCAL and leads to a sampling fraction, $f_{\mathrm{sampl}}=5.3\times10^{-3}$. The silicon is divided in to 5$\times$5\,mm$^2$ pads, leading to $\sim$10$^8$ readout channels. As the main focus of these studies was the digital readout, the analogue calorimeter has a simplistic clustering algorithm to find the total energy of the shower as there has been no consideration of pile-up removal. However, as the granularity of the analogue and digital readouts (following reconfiguration from pixels to pads) are both 5$\mathrm{\times}$5\,mm${^2}$, and as detailed below, the digital case allows for excellent suppression of pile-up events, it can be assumed that the same will be possible for the analogue silicon ECAL.

\subsubsection{Digital Readout}

\begin{figure}[h]
  \centering
    \includegraphics[width=0.49\textwidth]{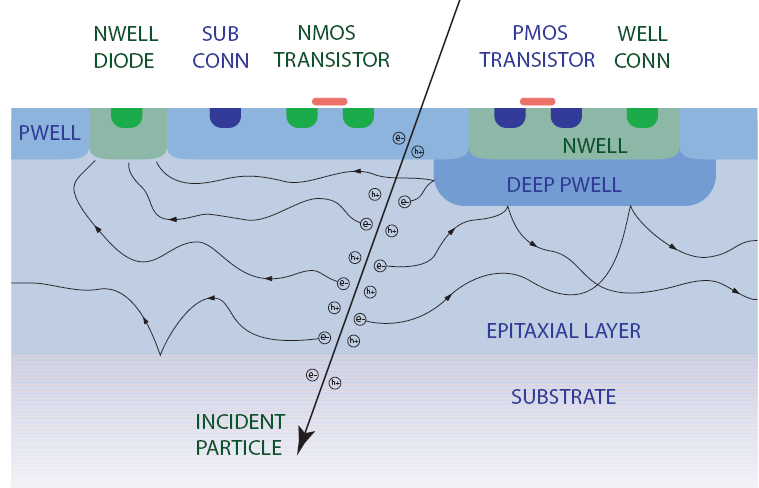}
  \caption{The cross section of a typical MAPS device, taken from~\cite{PRICE2012932}}
  \label{fig:layout:alt:maps}
\end{figure}

The basic premise of the digital electromagnetic calorimeter (DECal) is to count the number of particles in the shower rather than the energy they deposit. The currently envisaged DECal pixels only register one hit even if multiple particles pass through the given pixel. Therefore, the DECal begins to saturate should more than one particle traverse each pixel per readout cycle. In order to prevent this situation even in the very dense shower cores, the cells must be very small. A cell size of 50$\times$50\,$\mu\mathrm{m}^2$ was therefore used for previous studies for the International Large Detector (ILD) at the ILC~\cite{Price:2013rlt}. To achieve a calorimeter with the required granularity, CMOS Monolithic Active Pixel Sensors (MAPS) are proposed. Each pixel of a MAPS contains the required readout electronics, can be read out every 25\,ns bunch crossing, and, due to recent developments, can withstand the required radiation levels expected in the FCC-hh barrel region (see Tab.~\ref{tab:layout:dimensions}). A typical cross section of a MAPS device is shown in Fig.~\ref{fig:layout:alt:maps}. When a MIP traverses the sensor, the deposited energy liberates electron-hole pairs in the epitaxial layer which diffuse (or drift in the case of depleted MAPS) to the collection electrodes. 

The studies in this document assume an 18\,$\mu\mathrm{m}$ epitaxial thickness, on a substrate of 300\,$\mu\mathrm{m}$. The 50$\times$50\,$\mu\mathrm{m}^2$ pitch pixels are summed into 5$\times5\,\mathrm{mm}^{2}$ pads to reduce the data rate. A depleted MAPS device, capable of collecting all the charge within the pixel and summing the number of pixels in the pad within 25\,ns has been designed, fabricated and tested for the proposed FCC-hh DECal~\cite{IKopsalis}.

%The performance results shown in Chap.~\ref{sec:performance:egamma} show the potential of a digital ECAL for FCC-hh.

%% file: tex/alternative/software.tex
\subsection{Software Implementation}
\label{sec:alternative:software}

\subsubsection{Hit Generation and Digitisation}

\begin{figure}[ht]
  \centering
     \includegraphics[width = 0.8\textwidth]{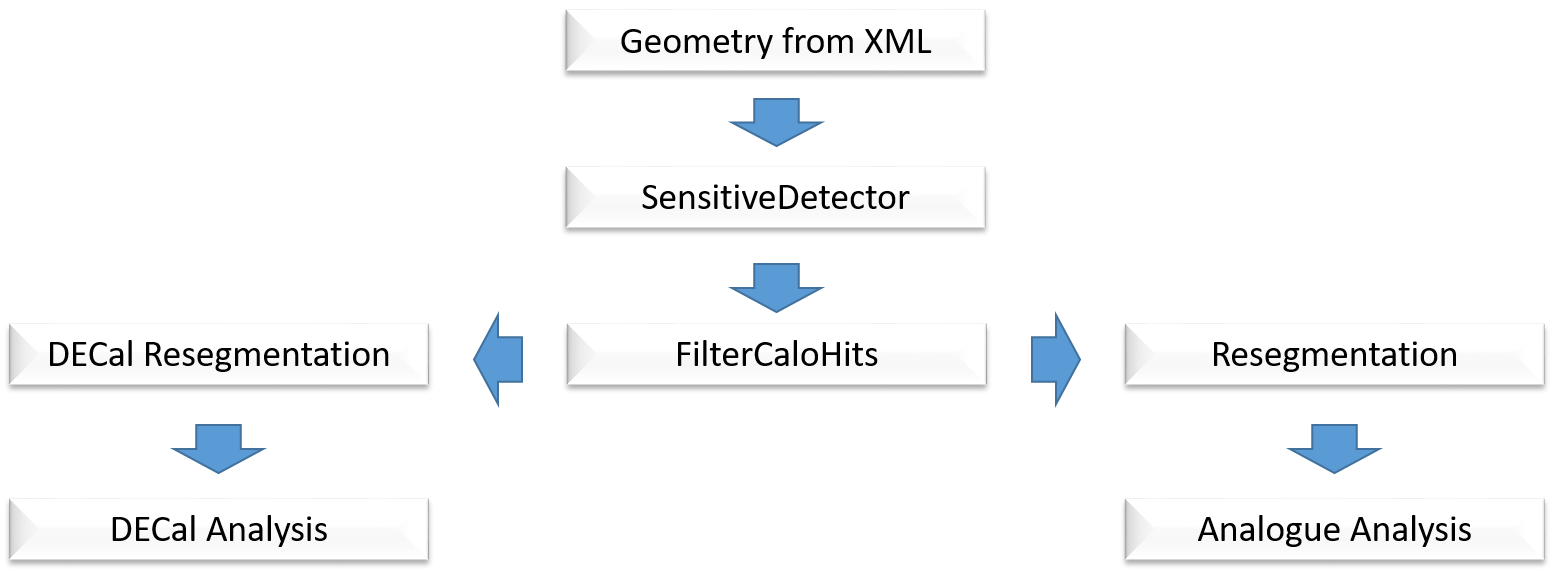}
     \caption{The workflow of the SiW calorimeter system as implemented in FCCSW}
     \label{fig:alt:software:flow}
\end{figure}

Both of the SiW geometries are implemented and scored in the simulation at the same time to allow for direct comparisons to be made between the two technologies. Each layer of the calorimeter consists of an 18\,$\mu\mathrm{m}$ silicon epitaxial sensitive volume, used for scoring the digital hits, and a 300\,$\mu\mathrm{m}$ silicon substrate for scoring the analogue deposits. A 2.1\,mm thick tungsten absorber is located directly after the two silicon layers, followed by a 3\,mm air gap. The standard 4\,T field is used for all studies and the inner radii of the first layer is consistent with the baseline design. In the first instance both the epitaxial and substrate layers are segmented in to 50$\times$50\,$\mu\mathrm{m}^2$ pixels to allow a single detector object to score all energy deposits. The energy deposits are then filtered depending on the analysis stream as outlined in Fig.~\ref{fig:alt:software:flow}. For the analogue SiW, all the energy deposits which occur within the substrate are selected, the pixels are grouped into 5$\times5\,\mathrm{mm}^{2}$ pads, the energy deposited in all pixels in a pad are summed together, and finally all pads which contain an energy deposit are summed to yield the total energy deposited in an event. For these studies, there is no threshold applied to the analogue case and all energy deposits are combined. This makes the analogue results optimistic due to the electronic noise anticipated with a 5$\times$5\,mm$^2$ pad~\cite{Collaboration:2293646}. The DECal analysis flow begins by selecting the hits which occur in the epitaxial layer and summing the energy deposits in each pixel and applying a threshold of 480 electrons to each pixel (corresponding to a 6$\sigma$ noise cut in the DECal reconfigurable MAPS sensor device level simulations). The pixels which remain after thresholding are then reconfigured and resegmented in to 5$\times5\,\mathrm{mm}^{2}$ pads and the number of pixels above threshold in each pad, and the mean $\mathrm{\eta}$ and $\mathrm{\phi}$ are found. The sum of all pixels in all pads is then found to yield the signal in the calorimeter.

\subsubsection{Noise and Pile-up}

In the DECal there are approximately 10$^{12}$ pixels covering the barrel region, each with an estimated noise level of 80 electrons prior to irradiation from device simulation and measurements of the new reconfigurable DECal sensor~\cite{IKopsalis}. A MIP transversing 18\,$\mu\mathrm{m}$ of silicon has a most probable energy deposit of 1400 electrons. Applying a threshold of 480 electrons maintains an excellent particle detection efficiency and yields a probability of a pixel firing due to noise of 10$^{-8}$ which translates to 10$^{4}$ pixels firing every bunch crossing. As these are caused by random noise fluctuations they will be evenly distributed, with a low hit density throughout the barrel region and are easily mitigated by searching for clusters. The electronic noise is generated in the simulations on an event by event basis by randomly generating 10$^{4}$ pixel addresses and merging these with the pixel addresses prior to the reconfiguration of the sensor into pads. Whilst this creates an excess of pixels in a shower, the fraction is negligible due to the low hit density and can thus be ignored in these studies.

\begin{figure}[ht]
  \centering
		\includegraphics[width = .48\textwidth]{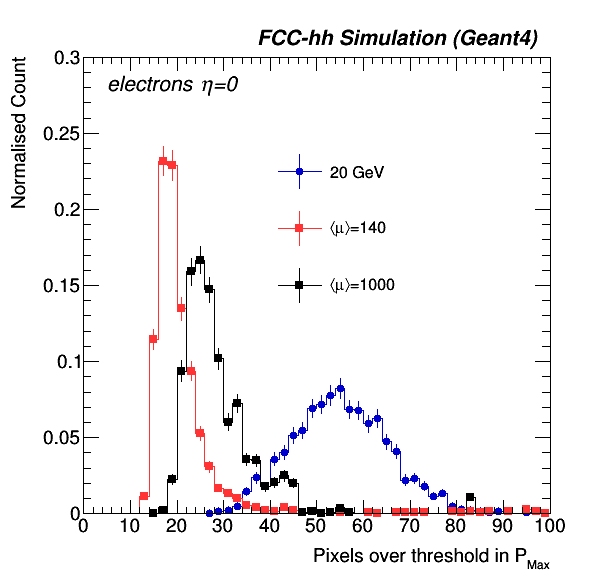}
		\caption{The maximum number of pixel hits per pad for different levels of pile-up and 20\,GeV electrons. \label{fig::software::reco::Dclust::padmax}}
\end{figure}

A larger and altogether more difficult source of noise hits to identify and remove arises from pile-up events. Whilst most of the additional hits will be caused by low energy particles which can be easily mitigated in the same way as the random noise hits, the presence of high energy particles which cause dense showers in the calorimeter will require additional information from other detector sub-systems to fully remove. The energy deposits for pile-up events were generated on a single pile-up event basis and then randomly selected and merged with the signal and the electronic noise pixels, before thresholds were applied. In these studies, we remove these dense secondary showers by requiring that the pad corresponding to the cluster seed (described in the following section) has at least 30 pixels that are above threshold. Distributions of the maximum pad occupancy for 20 GeV electrons, and for 140 pile-up and 1000 pile-up events can be seen in Fig.~\ref{fig::software::reco::Dclust::padmax}. Whilst it is clear that a higher cut value will remove a greater proportion of the pile-up clusters, the cut has been optimised for maximum efficiency for 20\,GeV electron showers.

\subsubsection{DECal clustering}
\label{sec:alt:soft:clustering}

The DECal clustering algorithm is a modified version of the sliding window algorithm as detailed in Section~\ref{sec:software:reco:slidingWindow}. The entire barrel region is scanned for the $5\times5$\,mm$^2$ pad containing the most pixels above threshold, Pad$_\mathrm{max}$, which is then used as the cluster seed.  The mean $\eta$ and mean $\phi$ of all pixels in Pad$_\mathrm{max}$ are calculated, and $\Delta\eta$, $\Delta\phi$ found for all hits in the barrel relative to these values. The total number of pixels above threshold in a cone, originating from the interaction point, is summed for each layer. The mean value of Pad$_\mathrm{max}$ with no pile-up, increases with incident particle energy due to a larger particle density in the shower core. This relationship has been parameterised with a second order polynomial which allows a first estimate of the incident energy to be made. Using the first estimate of energy, the cone width, in $\eta$, and $\phi$ is then extracted from a second parameterisation of the shower width versus incident particle energy. Typical values for a 100\,GeV shower are $\Delta\eta=0.020$, $\Delta\phi=0.015$, which highlights the compactness of the showers due to the use of tungsten as the absorber.

%\begin{figure}[ht]
%  \centering
%    \includegraphics[width = \textwidth]{decal/decal_cut_counts}\caption{}\label{fig::software::reco::Dclust::fraction}
  
%  \caption{DECal clustering plots. FIX CAPTION}
%\end{figure}

\begin{figure}[ht]
  \centering
  \begin{subfigure}{0.49\textwidth}
   \includegraphics[width =\textwidth]{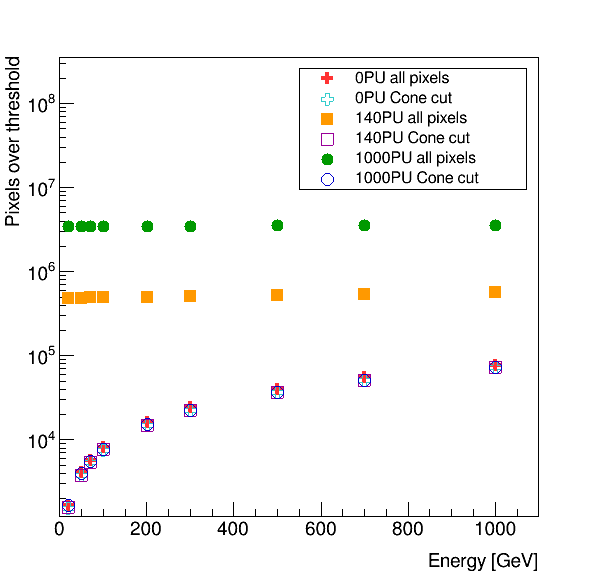}
  \end{subfigure}
  \begin{subfigure}{0.49\textwidth}
   \includegraphics[width =\textwidth]{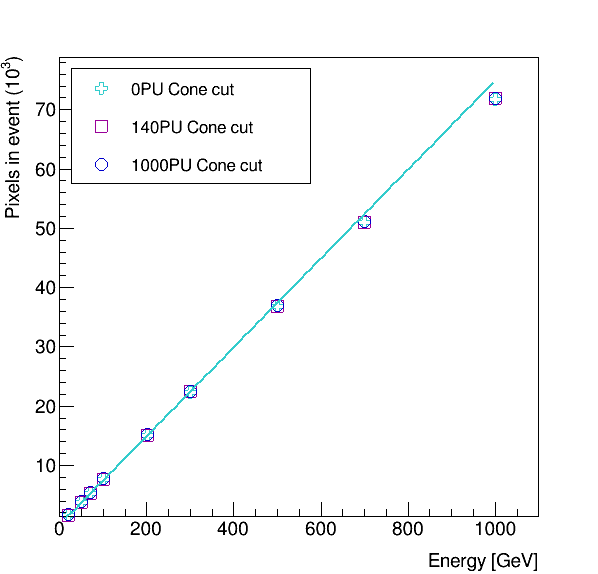}
  \end{subfigure}   
  \caption{The mean total number of pixels over threshold in an event for the different pile-up scenarios both before and after clustering (a) and only after clustering to highlight the performance (b).  \label{fig::software::reco::Dclust}}
\end{figure}

Fig.~\ref{fig::software::reco::Dclust} shows the number of pixels over threshold in the entire DECal for just the signal, signal with 140 minimum bias events, and signal with 1000 minimum bias events both before and after clustering has been applied. It is clear that a huge fraction of the pile-up is removed using the clustering. There is a finite number of events where the cluster seed is associated with a pile-up event, for completeness of this study these events are included in the final sample. However, these should be removed with more sophisticated analyses using other detector components and particle flow algorithms. The fraction of pixels in a cluster originating from pile-up events rises to 9\,\% for a 20\,GeV electron in the presence of 1000 minimum bias events. These pixels cannot be removed as it is not possible to determine whether a pixel is over threshold due to the signal, electronic noise, pile-up interactions, or a combination of the three. However, this fraction quickly falls to the percent level for showers of electrons greater than 100\,GeV.

\subsubsection{Non-Linearity Corrections}

In an ideal calorimeter the response as a function of incident energy should be linear. As highlighted in Fig.~\ref{fig::software::reco::Dclust} the response of the DECal becomes non linear for energies above 300\,GeV. This is due to the particle density in these showers being greater than 1 particle / pixel and yielding an under counting by the DECal. To correct for this a second order polynomial is used to convert the number of pixels in an event to the incident energy. Interestingly, the addition of the air gap for services between the absorbing and silicon of the next layer improves the linearity of the DECal. This is in part due to the showers having a greater distance over which they develop before reaching the next layer. Counter-intuitively the number of pixels recorded in an event with an increasing air gap decreases rather than increasing as one would expect from the increased inter particle separation in the shower. This is due to very low energy particles exiting the tungsten layer which in the presence of a magnetic field do not reach the silicon layer when they are seperated by the air gap. The behaviour and implication of these low energy particles need further study for the DECal concept.

A further non-linear correction which needs to be applied for the DECal is the number of pixels as a function of $\mathrm{\eta}$. In a conventional calorimeter, as the angle of the incident particle increases, so does the distance that the particle traverses in both the absorber and sampling material. In the DECal, there is no increase to the signal as a function of angle as we are counting pixels over threshold rather than the sum of all energy deposits. This effect has been studied previously for the ILC and is a small effect along the majority of the barrel; the current studies focus on $\mathrm{\eta=0}$ so the impact is negligible.

%% file: tex/alternative/performance.tex
\subsection{Performance}

\subsubsection{Comparison in the absence of pile-up}

\begin{figure}[ht]
  \centering
%  \begin{subfigure}[b]{0.49\textwidth}
%    \begin{tikzpicture}
%     \node[anchor=south west,inner sep=0] (image) at (0,0) {\includegraphics[width=\textwidth]{decal/enResolution_compareCalorimeters.png}};
%     \begin{scope}[x={(image.south east)},y={(image.north west)}]
%%        \draw[red,rounded corners] (0.63,0.55) rectangle (0.88,0.75);
%        \fill[white] (0.63,0.55) rectangle (0.88,0.75);
%      \end{scope}
%    \end{tikzpicture}
% instead of the line below
%    \includegraphics[width=\textwidth]{decal/enResolution_compareCalorimeters.png}\caption{}
%  \end{subfigure}
%  \begin{subfigure}[b]{0.49\textwidth}
  \includegraphics[width=0.5\textwidth]{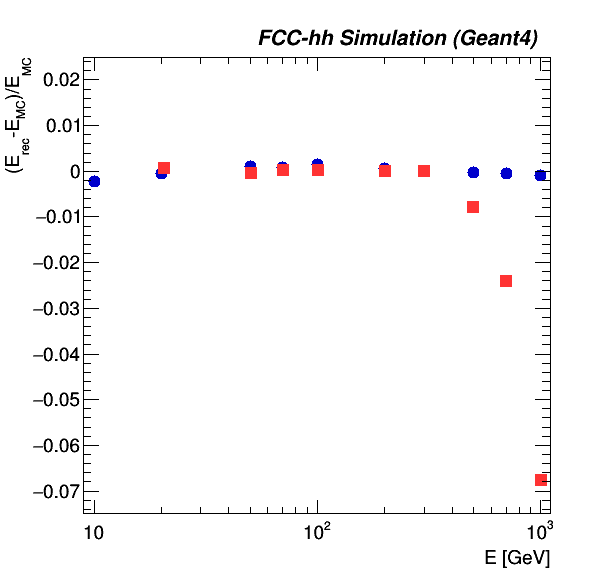} 
%\caption{}
%  \end{subfigure}
  \caption{The
% energy resolution (a)  and 
linearity of response for single electrons in the barrel ($\eta=0$) for the analogue SiW ECal (blue dots) and the DECal (red squares) in the absence of any pile-up.}
    \label{fig::alt:nopileup}
\end{figure}

Single electron showers were simulated in the absence of pile-up for the detector geometry detailed in the previous sections, for both the analogue and DECal read out methods. This allows a direct comparison to be made between the modalities and not the reconstruction methods. The linearity of response for the two modalities can be seen in Figure~\ref{fig::alt:nopileup}. The non-linearity of the DECal above 300\,GeV can be clearly seen, with a non-linearity of 2.5\,${\%}$ observed at 700\,GeV. Below this point, where the calorimeter is not saturating, the linearity is excellent as is the performance of the analogue case. 
%The energy resolution fits are restricted to the linear region and extrapolated to higher energies to allow comparisons between the detector concepts in this region. The stochastic term in the energy resolution for the analogue and DECal are 17.8\,$\%$ and 16.6\,$\%$ respectively with the DECal having a slightly higher constant term of 0.3\,$\%$ compared to 0.1\,$\%$ for the analogue case. Due to the non-linear trend of the DECal at higher energies we need to consider the resolution as a whole as the stochastic term no longer dominates. 
Table\,\ref{tab::alt::nopileup} highlights the resolution for 20\,GeV, 100\,GeV, 500\,GeV, and 1000\,GeV electrons for both the analogue and DECal. The DECal has a total energy resolution of 1.1\,$\%$ for a 1000\,GeV electron compared to 0.8\,$\%$ for the analogue case. The bottom of the table displays the results of fits of the resolution function Eq.~\ref{equation:intro:eneResolution}.

\begin{table}[htp]
\begin{center}
\begin{tabular}{|c|c|c|c|c|}
\hline
Energy
& Analogue
& DECal
& DECal
& DECal\\		
$[\text{GeV}]$ 
& $\langle\mu\rangle$=0
& $\langle\mu\rangle$=0
& $\langle\mu\rangle$=140
& $\langle\mu\rangle$=1,000\\
\hline
20 & 0.042 & 0.041& 0.043& 0.058\\
100 & 0.019 &  0.020 & 0.021& 0.023\\
500 & 0.009 &  0.012& 0.012& 0.012\\
1000 & 0.008 &  0.011& 0.011& 0.011 \\
\hline
$a$ $[\%\sqrt{\mathrm{GeV}}]$& 17.9& 17.0 & 17.9 & 17.9\\
$b$ $[\text{GeV}]$& 0.249 & 0.231 & 0.249 & 0.808\\
$c$ $[\%]$ & 0.47& 0.98 & 0.97 & 0.96\\
\hline
\multicolumn{5}{l}{}
\end{tabular}
\caption{The total energy resolution for the DECal and analogue readout for various energies of interest at the FCC-hh. In the bottom part of the table the fit parameters $a$, $b$, and $c$ determining the stochastic term, the noise term and the constant term of the energy resolution as defined in Eq.~\protect\ref{equation:intro:eneResolution}, are listed. }
\label{tab::alt::nopileup}
\end{center}
\end{table}

\subsubsection{Impact of pile-up on DECal}

The impact on performance of the DECal for two pile-up configurations, $\langle\mu\rangle$=140 and $\langle\mu\rangle$=1000, was investigated. It should be noted that the clustering algorithm is now applied. 
%hence the minor differences in the $\langle\mu\rangle$=0 resolutions between Figure\,\ref{fig::alt:nopileup} and Figure\,\ref{fig:alt:decal:res}. 
At higher energies, the fraction of pile-up hits remaining after clustering is small compared to the very dense showers and as such the energy resolution tends to the same value. However, at lower energies the remaining fraction becomes significant and the resolution degrades with pile-up. At 20\,GeV, 9\,$\%$ of all pixels over threshold originate from pile-up. As can be seen in Tab.~\ref{tab::alt::nopileup}, the addition of pile-up deteriorates the stochastic term and the noise term. The total energy resolution is again shown in Table\,\ref{tab::alt::nopileup} for each scenario. Interestingly, the linearity of the calorimeter at higher energies appears to improve for higher pile-up. It should be noted that this cannot be the case and is an artefact arising from an increased number of pile-up hits at lower energies causing a distortion to the regime in which we fit over.

\begin{figure}[ht]
 \centering
  \begin{subfigure}[b]{0.49\textwidth}

    \includegraphics[width=\textwidth]{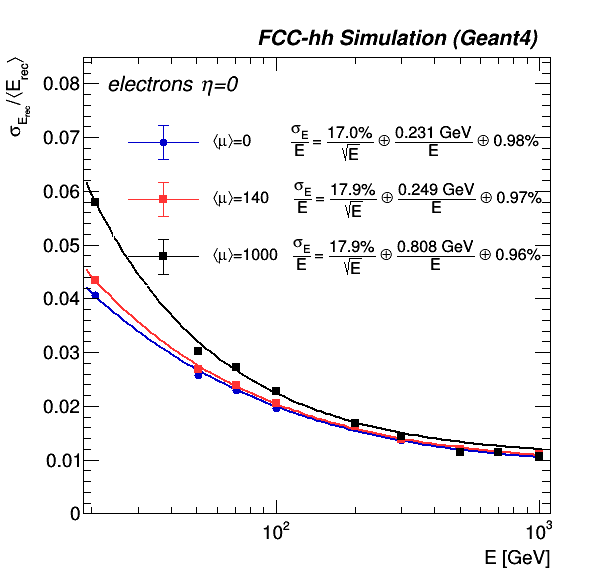}\caption{}
  \end{subfigure}
  \begin{subfigure}[b]{0.49\textwidth}
  \includegraphics[width=\textwidth]{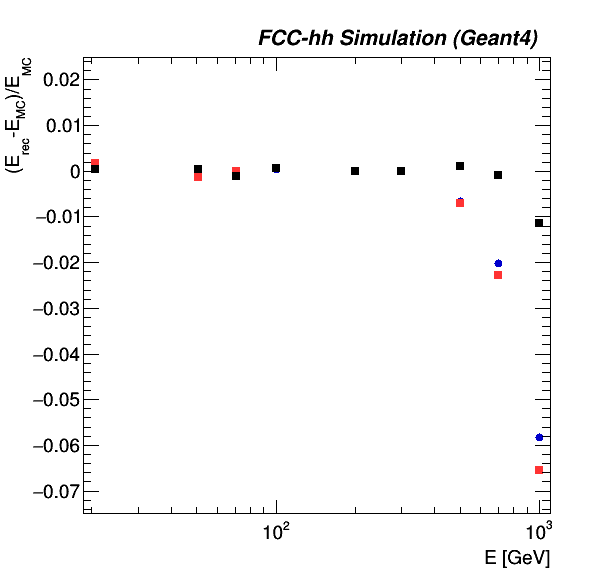}\caption{}
  \end{subfigure}
     \caption{Single electron DECal energy resolution (a) and linearity (b) for $\langle\mu\rangle$=0, $\langle\mu\rangle$=140, and $\langle\mu\rangle$=1000.}
          \label{fig:alt:decal:res}
\end{figure}

\subsection{Discussion}

The energy resolution presented here of $\sim 17\,\%\sqrt{\mathrm{GeV}}$/$\sqrt{E}$ for the FCC-hh SiW ECAL are consistent with previous studies in CALICE and for the CMS HGCAL. It has been shown that the DECal concept holds for showers up to 300\,GeV before the saturation of multiple particles traversing a single pixel becomes significant. Therefore, the impact on the physics from the use of the DECal is heavily dependant on the processes involved and beyond the scope of these studies.  The assumed performance of the DECal pixel for these studies matches that of the prototype reconfigurable CMOS MAPS for outer tracking and calorimetry described in~\cite{IKopsalis}. Reduced noise and additional functionality to cope with multiple hits in a pixel would be easily realised in future versions of the chip to extend the linearity regime.

The LAr baseline offers an improved energy resolution compared to either of the SiW options. However, the intrinsic standalone energy resolution of the ECal is not the only parameter for consideration in future detectors. Detectors that utilise Particle Flow Algorithms (PFA) will offer improved performance as the detector sub-system best suited for measuring each particle type is used. An essential aspect to PFA is the granularity and longitudinal segmentation of the detectors. The SiW options offer an improvement over the LAr for both of these and as such could offer improved PFA performance. In addition, the granularity of the SiW options are capable of measuring the internal structure of the denser showers induced in the tungsten and therefore are expected to offer increased $\mathrm{\pi^{0}}$ discrimination from single photons and better $\mathrm{\tau}$ identification. In these studies the most basic reconstruction has been used for the DECal where the total number of pixels corresponds directly to the energy of the particle. However, the ultra high granularity and large number of longitudinal layers of the DECal would also allow complex pattern recognition to be implemented to measure the energy. One such study using the number of pixels and ratio of these per layer was able to improve the linear response of the DECal~\cite{pricedecal:2018}.

The SiW options would operate at room temperature and as such would not require a cryostat, not only would this reduce the complexity of the design and the amount of material traversed before the calorimeters, but the inner radius of the calorimeter systems could also be reduced. The use of tungsten as the absorbing material reduces significantly the depth of the ECAL and therefore reduces the size of all the sub systems outside of the ECAL. It is anticipated that the cost of silicon detector systems will undergo significant reductions on the envisaged timescales before the FCC-hh construction begins, and whilst tungsten is expensive, the combination of silicon and smaller radii of external detector components promises significant cost saving benefits for the detector as a whole. 

 Finally, a hybrid approach, using a very high granularity pre-shower detector complemented by a more conventional energy measuring calorimeter, may offer an alternative optimisation but at the cost of a higher radial space required for the calorimeter and subsequent cost increases for outer sub-detectors and magnet system.

%% file: tex/summary/summary.tex
%Feasibility, cost, outlook R\&D programs, FCC-22

The goal of the international FCC study was to develop a conceptual design of a future circular collider including possible experiments to exploit its full physics potential, in time to serve as an input for the update of the European Strategy for Particle Physics that started in 2019. The Conceptual Design Report (CDR) that appeared in four volumes, with the third volume describing a possible hadron collider~\cite{Abada2019}, FCC-hh, summarises the results of this study. 

In order to demonstrate that the full physics potential of such a hadron collider could be exploited, a conceptual design of a possible FCC-hh experiment was developed. Many studies were conducted, that led to the short summary in~\cite{Abada2019}. This report presents the details of those studies conducted to develop the conceptual design of the calorimeter system and explains the reasoning behind the various design decisions that have been made. In general, the main focus was to demonstrate the feasibility of a calorimeter system that could fulfil the physics requirements which are briefly outlined at the beginning of this note. Beyond that, some promising alternative technologies are also described, but it goes without saying, that other designs to realise such a FCC-hh calorimeter system might exist. 

The general strategy has been followed: 
\begin{itemize}
\item determine a calorimeter concept from a pen and paper detector, and develop a possible design of a calorimeter system for FCC-hh (see Sec.~\ref{sec:layout});
\item implement this conceptual calorimeter system into a realistic simulation of a realistic FCC-hh experiment (see Sec.~\ref{sec:implementation}) and 
\item evaluate if the performance of such a calorimeter system would meet the expected performances (see Sec.~\ref{sec:performance}). 
\end{itemize}

Whereas final state particles in a future FCC-hh experiment will be identified and measured by combining the information of several detectors, the main focus of this document was to evaluate the standalone performance of the calorimeter system. 
As demonstrated in Sec.~\ref{sec:performance:egamma}, most required performance benchmarks for electromagnetic showers can already be achieved by a standalone measurement in the proposed electromagnetic calorimeter, based on LAr as active material with lead-steel absorbers, even at highest expected pile-up. Nevertheless, a combination with the measurement in the tracker will further improve this performance, mainly through pile-up suppression. For the hadronic calorimeter a sampling ``Tile'' calorimeter with scintillating tiles and passive stainless steel and lead absorbers is proposed in the barrel and extended barrel, and a LAr/Cu calorimeter in the forward region. The performance for hadrons and jets, discussed in detail in Sections~\ref{sec:performance:hadronic} and~\ref{sec:performance:jets} respectively, shows that the proposed calorimeter system achieves very good results at low pile-up and without magnetic field, but will have to be combined with the inner tracker via particle flow and pile-up suppression algorithms to achieve the required performance at a 4\,T magnetic field and highest pile-up regime.
In a nutshell, it has been shown that the proposed calorimeter concept performs as expected, but only a combination with the inner tracker will achieve the ultimate performance in the highest pile-up scenario. The high granularity of the proposed calorimeter system will facilitate this combination with the inner tracker through the use of particle flow and will provide 3D imaging information for machine learning algorithms that will be used for energy reconstruction, particle ID and pile-up suppression.  

A promising alternative technology is presented in Sec.~\ref{sec:alternative} based on Si sensors as active material (digital or analogue read-out) and tungsten absorber plates, which is inspired by the on-going R\&D for the upgrades of the current LHC detectors. This detector technology focuses even more on facilitating the combination with the inner tracker by even higher lateral granularity and more longitudinal layers, while accepting a slightly worse standalone performance for electromagnetic showers. 

For all the described calorimeters concepts, future studies will be needed to determine the best solution. Additional R\&D in two main directions will be crucial to prepare these concepts for more technical designs: 
\begin{itemize}
\item R\&D on the detector concepts, technical designs and prototypes demonstrating the assumed performance.
\item Further development of the FCC software towards a fully functional particle-flow algorithm to combine the tracker  and the calorimeter measurements to evaluate the necessary granularity of the calorimeter system. 
\end{itemize}

Several proposals for research projects on calorimetry and software for future colliders have been accepted by the CERN EP R\&D program which will start providing resources in January 2020. On top of that expressions of interest for the H2020 Innovation Pilot (AIDA++) have been submitted, awaiting a funding decision in early 2020. It is crucial that the effort that has started for the FCC CDR continues to develop the existing concepts into more solid designs of future experiments. In parallel investigations have started, whether the presented calorimeter system could be adapted for an electron-positron collider experiment. 